\documentclass[a4paper,fleqn]{article}

\usepackage{cancel}
\usepackage{titlesec}
\usepackage{graphicx}
\usepackage{dcolumn}
\usepackage{bm}
\usepackage{amsmath}
\usepackage{amssymb}
\usepackage{amsfonts}
\usepackage{esvect}
\usepackage[dvipsnames]{xcolor}
\usepackage{fix-cm}
\usepackage{cite}
\usepackage{soul}
\usepackage{mdframed}
\usepackage{lipsum}
\usepackage{multicol}
\usepackage{float}
\usepackage{longtable}
\usepackage{setspace}
\usepackage{tabularx}
\usepackage{ltablex}
\usepackage{newfloat}
\DeclareFloatingEnvironment[fileext=frm,placement={!ht},name=Box]{boxfloat}
\usepackage{capt-of}
\usepackage[htt]{hyphenat}
\usepackage{hyperref}
\usepackage{url}
\usepackage{afterpage}
\usepackage[utf8]{inputenc}
\usepackage{longtable}
\usepackage[capitalise, english]{cleveref}
\usepackage{makecell}
\usepackage{verbatim}
\usepackage{graphicx}
\usepackage{subcaption}
\usepackage{tikz}
\usetikzlibrary{decorations.pathreplacing}
\usepackage{scalerel}

\titleclass{\subsubsubsection}{straight}[\subsection]

\newcounter{subsubsubsection}[subsubsection]
\renewcommand\thesubsubsubsection{\thesubsubsection.\arabic{subsubsubsection}}

\titleformat{\subsubsubsection}
  {\normalfont\normalsize\bfseries}{\thesubsubsubsection}{1em}{}
\titlespacing*{\subsubsubsection}
{0pt}{3.25ex plus 1ex minus .2ex}{1.5ex plus .2ex}

\makeatletter
\renewcommand\paragraph{\@startsection{paragraph}{5}{\z@}
  {3.25ex \@plus1ex \@minus.2ex}
  {-1em}
  {\normalfont\normalsize\bfseries}}
\renewcommand\subparagraph{\@startsection{subparagraph}{6}{\parindent}
  {3.25ex \@plus1ex \@minus .2ex}
  {-1em}
  {\normalfont\normalsize\bfseries}}
\def\toclevel@subsubsubsection{4}
\def\toclevel@paragraph{5}
\def\toclevel@subparagraph{6}
\def\l@subsubsubsection{\@dottedtocline{4}{7em}{4em}}
\def\l@paragraph{\@dottedtocline{5}{10em}{5em}}
\def\l@subparagraph{\@dottedtocline{6}{14em}{6em}}
\makeatother

\usepackage[theorems,skins]{tcolorbox}
\newtcolorbox{mymathbox}[1][]{colback=white, sharp corners, #1}

\usepackage{graphicx}
\usepackage{dcolumn}
\usepackage{bm}
\usepackage{amsmath}
\usepackage{amssymb}
\usepackage{amsfonts}
\usepackage{esvect}
\usepackage[dvipsnames]{xcolor}
\usepackage{fix-cm} 
\usepackage{cite}
\usepackage{soul}
\usepackage{mdframed}
\usepackage{lipsum}
\usepackage{multicol}
\usepackage{float}
\usepackage{longtable}
\usepackage{setspace}
\usepackage{tabularx}
\usepackage{ltablex}
\usepackage{newfloat}
\usepackage{capt-of}
\usepackage[htt]{hyphenat}
\usepackage{hyperref}
\usepackage{url}
\usepackage{afterpage}
\usepackage[utf8]{inputenc}
\usepackage{longtable}
\usepackage[capitalise, english]{cleveref}
\usepackage[normalem]{ulem}

\usepackage{tikz}
\usetikzlibrary{decorations.pathreplacing}

\newcommand{\conf}[1]{\tilde{#1}}
\newcommand{\prgv}[1]{{\bar #1}}

\newcommand{\Eq}[1]{Eq.~\eqref{#1}}
\newcommand{\EEq}[1]{Equation~\eqref{#1}}

\newcommand{\Eqq}[1]{Eqs.~\eqref{#1}}
\newcommand{\Eqs}[2]{Eqs.~\eqref{#1} and \eqref{#2}}
\newcommand{\Eqss}[2]{Eqs.~\eqref{#1}--\eqref{#2}}
\newcommand{\EEqss}[2]{Equations~\eqref{#1}--\eqref{#2}}

\newcommand{\kk}{\vec{k}}

\newcommand{\ArtII}{{\tt The Art\,-\,II}}

\usepackage{listings}
\usepackage{minted}
\newminted{shell-session}{
    frame=lines, breaklines=true}
\definecolor{mintbackground}{RGB}{246, 246, 246}

\newcommand{\fl}{{\rm Fl}}
\newcommand{\f}{f}

\newcommand{\Sec}[1]{section~\ref{#1}}
\newcommand{\Secs}[2]{sections~\ref{#1} and \ref{#2}}

\newcommand{\SSecs}[2]{Sections~\ref{#1} and \ref{#2}}

\newcommand{\App}[1]{appendix~\ref{#1}}

\newcommand{\EQb}{\begin{mymathbox}[ams equation]}
\newcommand{\EQgather}{\begin{mymathbox}[ams gather]}
\newcommand{\EQalign}{\begin{mymathbox}[ams align]}
\newcommand{\ENb}{\end{mymathbox}}

\newcommand{\bra}[1]{\langle #1\rangle}

\newminted{C++}{framesep=2mm,
baselinestretch=1.2,
bgcolor=mintbackground,
fontsize=\small,breaklines=true}

\usepackage{empheq}

\usepackage[symbol]{footmisc}

\newcommand{\CL}{$\mathcal{C}${\tt osmo}$\mathcal{L}${\tt attice}~}
\newcommand{\CLns}{${\mathcal C}${\tt osmo}${\mathcal L}${\tt attice}}

\newcommand{\dx}{\ensuremath{\delta x}}

\newcommand{\bn}{{\bf n}}
\newcommand{\dd}{\text{d}}
\newcommand{\mpl}{m_p}

\newcommand{\be}{\begin{equation}}
\newcommand{\ee}{\end{equation}}
\newcommand{\bea}{\begin{eqnarray}}
\newcommand{\eea}{\end{eqnarray}}

\newcommand{\mn}{{\mu\nu}}
\newcommand{\ab}{{\alpha\beta}}

\newcommand{\HH}{{\cal H}}
\newcommand{\pf}{{\rm pf}}
\newcommand{\ipf}{{\rm ipf}}
\newcommand{\cs}{c_{\rm s}}

\newcommand{\vA}{v_{\rm A}}
\newcommand{\rhs}{{\em rhs}\ }
\newcommand{\lhs}{{\em lhs}\ }
\newcommand{\LL}{{\rm L}}
\newcommand{\dev}{\delta T}
\newcommand{\cdev}{\delta \tilde{T}}
\newcommand{\pdev}{\delta \bar{T}}

\usepackage{textgreek}
\DeclareSymbolFont{ttgreek}{LGR}{cmtt}{m}{n}
\DeclareMathSymbol{\ttrho}{\mathord}{ttgreek}{`r}

\def\blue{\textcolor{blue}}

\newcommand{\affil}[2]{%
  \parbox[c]{0.95\linewidth}{\centering
    \textit{$^{#1}$~}\ignorespaces#2%
  }%
}

\newcommand{\addressIFIC}{\it Instituto de F\'isica Corpuscular (IFIC), Universitat de Val\`encia (UV) and\\
\it Consejo Superior de Investigaciones Cient\'ificas (CSIC), 46980 Paterna, Valencia, Spain}

\newcommand{\addressUNIGE}{\it D\'epartement de Physique Th\'eorique, Universit\'e de Gen\`eve, CH-1211 Gen\`eve,
Switzerland}

\begin{document}

\begin{figure}
\vspace*{-0.5cm}
\hspace{0.0cm}
\includegraphics[width = 8.6cm]{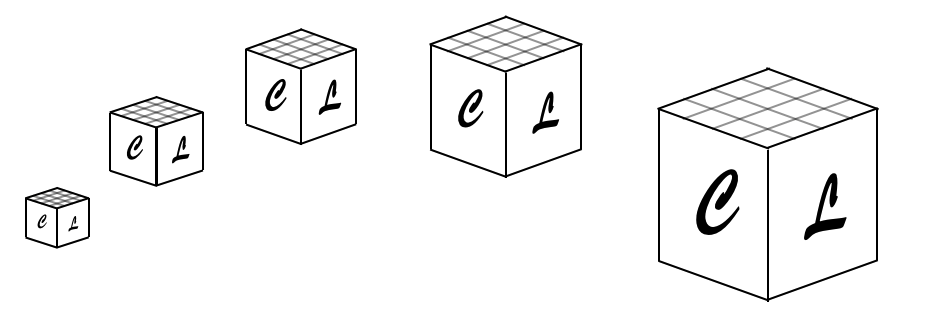}~
\includegraphics[width = 8.6cm]{CL_iconSequence.png}
\end{figure}

\vspace*{0.5cm}

\begin{center}
\vspace*{-0.5cm}
{\fontsize{25}{0} \bf\textsf{The art of simulating the early Universe}}
\\[0.15cm]
$\hspace*{2mm}\left(~
\makecell{
\text{\LARGE \it A dissertation series about lattice techniques for simulating} \vspace{1mm}\\\text{\LARGE \it the dynamics of interacting fields in an expanding Universe}
} ~~\right)$
\vspace*{2.0cm}\\

{\fontsize{19.2}{0}\bf  Part III.\,Scalar-Gauge-Fluid Dynamics}
\\[2.25cm]

{\Large  Daniel~G.~Figueroa}\\
\addressIFIC
\\[0.75cm]
{\Large \rm Kenneth Marschall}\\
\addressIFIC
\\[0.75cm]
{\Large \rm Antonino S. Midiri}\\
\addressUNIGE
\\[0.75cm]
{\Large \rm Alberto Roper Pol}\\
\addressUNIGE

\footnotetext{Corresponding authors: kenneth.marschall@ific.uv.es, antonino.midiri@unige.ch}

\vspace{.3cm}
\end{center}

\thispagestyle{empty}
\addtocounter{page}{-1}
\newpage
\,
\thispagestyle{empty}
\addtocounter{page}{-1}
\newpage

\setcounter{page}{1}

\title{\bf The art of simulating the early Universe. Part III}

\author{Daniel G. Figueroa\,$^{1}$,  Kenneth Marschall\,$^{1}$, Antonino S. Midiri\,$^{2}$, Alberto Roper Pol\,$^{2}$\vspace*{0.35cm}\\
\affil{1}{\small\addressIFIC}\\
\affil{2}{\small\addressUNIGE}
}

\date{}
\maketitle

\begin{abstract}
We discuss lattice methods for the simulation of fluid dynamics in the early Universe. This review represents a third entry in the monographic series on lattice cosmology techniques, Refs.\cite{Figueroa:2020rrl,Baeza-Ballesteros:2025tme}, which previously covered canonical
and non-canonical 
field theory dynamics.
Here, we first review the continuum theory of fluid dynamics in flat spacetime, and then in an FLRW background.
We consider conservation and non-conservation forms of the equations
of motion for
fluids in isolation or coupled to scalar and/or gauge fields, and either
fully relativistic
or 
subrelativistic
regimes of fluid bulk motion. 
After reviewing 
basic lattice concepts, we introduce detailed discretization schemes for fluid dynamics in expanding backgrounds for: {\it i)} isolated perfect fluids,
{\it ii)} isolated imperfect (viscous) fluids,
{\it iii)} fluids coupled to gauge fields, 
and {\it iv)} fluids coupled to scalar fields. Our evolution algorithms
accommodate self-consistent expansion 
sourced by all scalar, gauge, and fluid sectors, preserving gauge invariance to machine precision in some cases.
We also review lattice methods to set up the initial conditions for fluids, and the implementation of gravitational wave dynamics sourced by all scalar, gauge, and fluid degrees of freedom. This document represents the theoretical basis for the scalar-gauge-fluid module that
will be publicly released as part of ${\mathcal C}$osmo${\mathcal L}$attice~{\tt v3.0} after publication of this monograph, check \href{http://www.cosmolattice.com}{{\color{blue}{http://www.cosmolattice.com}}} for updates. 
\end{abstract}

\tableofcontents

\newpage

\section*{Conventions and notation}
\label{sec:Conventions}
\addcontentsline{toc}{section}{Conventions and notation}

\vspace*{0.5cm}
Unless otherwise specified, we use the following conventions throughout the document:
\vspace*{0.3cm}

\begin{itemize}

\item We use natural units $c=\hbar=1$ and metric signature $(-1,+1,+1,+1)$.\vspace*{0.2cm}

\item We use interchangeably the Newton constant $G$, the full Planck mass $M_p \simeq 1.22\cdot 10^{19}$ GeV, and the reduced Planck mass $m_p \simeq 2.44\cdot 10^{18}$ GeV, all related through $M_p^2 = 8\pi m_p^2 = 1/G$.\vspace*{0.2cm}

\item Latin indices $i, j, k, ... = 1,2,3$ are reserved for spatial dimensions, and Greek indices $\alpha, \beta, \mu, \nu,... = 0,1,2,3$ for spacetime dimensions. We use the {\it Einstein convention} of summing over repeated indices {\it only in the continuum}.
Einstein summation is also assumed for both lower
indices when dealing with 3D spatial vectors, which are always
identified with lower indices.
{\bf In the lattice, unless stated otherwise, repeated indices do not represent summation}.\vspace*{0.2cm}

\item We consider a flat FLRW metric with squared line
element $ds^2 = -a^{2\alpha}(\eta)d\eta^2 + a^2(\eta) \, \delta_{ij} \, dx^i dx^j$ with $\alpha \in 
{\mathbb R}$ a constant chosen conveniently in each scenario. For $\alpha = 0$, $\eta$ denotes the {\it cosmic time} $t$, whereas for $\alpha = 1$, $\eta$ denotes the {\it conformal time} $\tau = \int {dt'\over a(t')}$. For arbitrary $\alpha$, we will refer to the time variable as the {\it $\alpha$-time}.\vspace*{0.2cm}

\item We reserve the notation $()^{\cdot}$ 
for derivatives with respect to cosmic time with $\alpha = 0$, and $()'$ for derivatives with respect to $\alpha$-time with arbitrary $\alpha$.\vspace*{0.2cm}

\item The $\alpha$-time Hubble rate is given by $\mathcal{H} = a'/a$, whereas the physical Hubble rate is denoted by $H = \mathcal{H}|_{\alpha = 0}$.\vspace*{0.2cm}

\item Our Fourier transform convention in the continuum is given by
\begin{eqnarray}\label{eq:FTcont}
f({\bf x}) = \frac{1}{(2 \pi)^3} \int d^3 {\bf k} \, f({\bf k}) \, e^{+i {\bf k} {\bf x}}\, ~~~ \Longleftrightarrow ~~~  f({\bf k}) = \int d^3 {\bf x} \, f ( {\bf x}) \, e^{-i {\bf k} {\bf x}}\,.\nonumber
\end{eqnarray}\vspace*{-0.1cm}

\item Given $N$ lattice points per spatial direction,
with ${\bf n}$ and ${\bf \tilde{n}}$ respectively indicating the configuration and reciprocal space lattice sites, our discrete Fourier transform (DFT) is defined by
\begin{eqnarray}\label{eq:FTdiscreteAux}
f({\bf n}) \equiv {1\over N^3}\sum_{\tilde n} e^{+i{2\pi\over N} {\bf \tilde n \cdot n}} f({\bf \tilde n}) ~~~~ \Leftrightarrow ~~~~  f({\bf \tilde n}) \equiv \sum_{n} e^{-i{2\pi\over N} {\bf n \cdot \tilde n} }f({\bf n})\,.\nonumber
\end{eqnarray}
\vspace*{-0.1cm}

\item A scalar field living in a generic lattice site $n = (n_o,\bn) = (n_o,n_1,n_2,n_3)$, i.e., $\phi_n = \phi(n)$, will be simply denoted as $\phi$. If the point is displaced in the $\mu-$direction by one unit lattice spacing/time step, $n + \hat\mu$, we will then use the notation $n+\mu$ or simply by $+\mu$ to indicate it, so that the field amplitude in the new point is expressed as $\phi_{+\mu} \equiv \phi(n+\hat\mu)$.
\vspace*{0.2cm}

\item When representing gauge fields on the lattice, it is usually understood that they live in between lattice points, half step away from each lattice site, i.e., 
$A_{\mu} \equiv A_{\mu}(n+{1\over2}\hat\mu)$.
It follows then that, e.g.,
$A_{\mu,+\nu} \equiv A_{\mu}\big(n + {1\over2}\hat\mu +  \hat\nu\big)$. 
{In the case of gauge fields coupled to a fluid, however, it is possible to consider an alternative formulation  where the gauge fields are \textit{collocated}, which means that they live at integer lattice sites $A_\mu \equiv A_\mu(n)$.}
\vspace*{0.2cm}

\item Even though the {\it lattice spacing} $\dx$ and the {\it time step} $\delta t$ do not need to be equal, sometimes we may speak loosely of corrections of order $\mathcal{O}(\dx)$, independently of whether we are referring to the lattice spacing or to the time step. 

\end{itemize}

\newpage
\,
\thispagestyle{empty}
\addtocounter{page}{-1}
\newpage

\section{Introduction and purpose of this monograph}

~~~~~An early phase of accelerated expansion in the early Universe, {\it inflation}, represents the leading paradigm for resolving the shortcomings of the {\it hot Big Bang} framework~\cite{Guth:1980zm,Linde:1981mu,Albrecht:1982wi,Brout:1977ix,Starobinsky:1980te,Kazanas:1980tx,Sato:1980yn}, while simultaneously providing a compelling mechanism for the generation of the cosmological perturbations~\cite{Mukhanov:1981xt,Guth:1982ec,Starobinsky:1982ee,Hawking:1982cz,Bardeen:1983qw} observed in the cosmic microwave background (CMB)~\cite{Planck:2018nkj,Aghanim:2018eyx,Akrami:2018odb}. In standard realizations, inflation is driven by a scalar singlet field, the \textit{inflaton}, endowed with a suitable potential capable of sustaining a sufficiently prolonged period of accelerated expansion. Subsequent to inflation, the Universe must undergo a phase of \textit{reheating}, during which the energy stored in the inflaton field is transferred into other particle species, which are expected to rapidly approach thermal equilibrium. These {\it daughter} particles (or their decay products) ultimately come to dominate the energy density of the Universe, forming a relativistic plasma. From this stage onward, the standard {\it hot Big Bang} description—characterized by a thermal bath of relativistic particles governing the cosmic expansion—becomes applicable.

The phenomenology of the early Universe is vast, often involving non-linear field dynamics. These include {\it preheating} effects and other mechanisms of particle production~\cite{Traschen:1990sw, Kofman:1994rk, Shtanov:1994ce, Kaiser:1995fb, Kofman:1997yn, Greene:1997fu, Kaiser:1997mp, Kaiser:1997hg, Greene:1998nh, Greene:2000ew, Peloso:2000hy, Berges:2010zv,Enqvist:2012tc,Figueroa:2015rqa}, the amplification of scalar metric perturbations~\cite{Bassett:1998wg, Bassett:1999mt, Bassett:1999ta, Finelli:2000ya, Chambers:2007se, Bond:2009xx,Linde:2012bt,Imrith:2019njf, Musoke:2019ima, Martin:2020fgl, Adshead:2023mvt},
which might collapse to form primordial black holes~\cite{Cotner:2019ykd, Martin:2019nuw, GarciaBellido:1996qt, Green:2000he, Cotner:2018vug}, magnetohydrodynamic
turbulent motion
in the primordial plasma \cite{Brandenburg:1996fc,Jedamzik:1996wp,Subramanian:1997gi,Christensson:2000sp,Kosowsky:2001xp,Caprini:2006jb,Gogoberidze:2007an,Caprini:2009yp,Kahniashvili:2010gp,Brandenburg:2014mwa,Brandenburg:2017neh,Brandenburg:2017rnt,RoperPol:2025lgc}, and other phenomena like the formation of soliton-like objects like oscillons~\cite{Gleiser:1993pt,Copeland:1995fq,Amin:2010dc,Amin:2011hj,Gleiser:2011xj,Antusch:2015ziz,Lozanov:2017hjm,Hasegawa:2017iay,Amin:2018xfe,Kitajima:2018zco,Antusch:2019qrr,Ibe:2019lzv,Sang:2019ndv,Kou:2019bbc,Nazari:2020fmk,Sang:2020kpd,Aurrekoetxea:2023jwd,Mahbub:2023faw,Piani:2023aof,Shafi:2024jig,Drees:2025iue,Piani:2025dpy}, the nucleation of bubbles in first-order phase transitions~\cite{Guth:1982pn,Witten:1984rs,Kosowsky:1991ua,Kosowsky:1992rz,Kamionkowski:1993fg,Rajantie:2000fd,Hindmarsh:2001vp,Copeland:2002ku,GarciaBellido:2002aj,Caprini:2007xq}, or the creation of cosmic defects~\cite{Hindmarsh:1994re, Felder:2000hj, Hindmarsh:2000kd, Rajantie:2001ps, Rajantie:2002dw, Donaire:2004gp, Copeland:2009ga, Hiramatsu:2012sc, Kawasaki:2014sqa, Fleury:2016xrz, Moore:2017ond}. These non-linear phenomena can actually yield relevant cosmological consequences, like the generation of dark matter relics~\cite{Garcia:2018wtq, Garcia:2021iag, Garcia:2022vwm, Lebedev:2022vwf, Zhang:2023xcd}, the realization of cosmological magnetogenesis~\cite{Turner:1987bw, Quashnock:1988vs, Vachaspati:1991nm, Ratra:1991bn, Garretson:1992vt, Dolgov:1993vg, Cheng:1994yr, Gasperini:1995dh, Cornwall:1997ms, Sigl:1996dm, Joyce:1997uy, Forbes:2000gr, Grasso:2000wj,Vachaspati:2001nb,Widrow:2002ud,DiazGil:2005qp, Campanelli:2005ye,DiazGil:2007qx, DiazGil:2007dy, DiazGil:2008tf, Durrer:2013pga, Subramanian:2015lua, Fujita:2016qab, Adshead:2016iae, Miniati:2017kah, Vilchinskii:2017qul,Zhang:2019vsb,Vachaspati:2020blt}, or the baryon asymmetry of the Universe~\cite{Kolb:1996jt, Kolb:1998he, GarciaBellido:1999sv, Allahverdi:2000zd, Rajantie:2000nj, Cornwall:2001hq, Copeland:2001qw, Smit:2002yg, GarciaBellido:2003wd, Tranberg:2003gi, Tranberg:2009de, Kamada:2010yz, Lozanov:2014zfa}. They can also affect the post-inflationary equation of state, with implications for the CMB~\cite{Podolsky:2005bw,Dufaux:2006ee, Lozanov:2016hid, Figueroa:2016wxr, Krajewski:2018moi, Maity:2018qhi, Antusch:2020iyq, Saha:2020bis, Antusch:2021aiw, Mansfield:2023sqp,Garcia:2023eol,Garcia:2023dyf,Antusch:2025ewc}, and generate potentially detectable gravitational wave backgrounds~\cite{Caprini:2018mtu,Grishchuk:1974ny,Hogan:1986dsh,Deryagin:1986qq,Krauss:1991qu,Khlebnikov:1997di,Dolgov:2002ra,Grojean:2006bp,Easther:2006gt,Easther:2006vd,Garcia-Bellido:2007nns,GarciaBellido:2007af,Dufaux:2007pt,Dufaux:2008dn,Dufaux:2010cf,Figueroa:2012kw,Hiramatsu:2013qaa,Hindmarsh:2013xza,Zhou:2013tsa, Bethke:2013aba,Bethke:2013vca,Hindmarsh:2015qta,Hindmarsh:2016lnk,Figueroa:2016ojl,Antusch:2016con, Hindmarsh:2017gnf,Antusch:2017flz,Antusch:2017vga,Figueroa:2017vfa,Cutting:2018tjt, Liu:2018rrt,Lozanov:2019ylm,Hindmarsh:2019phv ,Caprini:2019egz,  Adshead:2019lbr,Niksa:2018ofa,Adshead:2019igv, Cutting:2019zws,RoperPol:2018sap,RoperPol:2019wvy, Kahniashvili:2020jgm,Figueroa:2020lvo,Cutting:2020nla,RoperPol:2021xnd,RoperPol:2022iel,Athron:2023xlk,RoperPol:2023dzg,RoperPol:2023bqa,Figueroa:2022iho,Cosme:2022htl, Klose:2022knn,Cui:2023fbg, Baeza-Ballesteros:2023say,Baeza-Ballesteros:2024otj,Caprini:2024hue,Servant:2023tua,Caprini:2026nnk}. 

The non-linear dynamics of this high-energy physics landscape is often too complex to be captured accurately by analytical methods.
Developing numerical techniques to simulate nonlinear field dynamics becomes therefore essential to gain a comprehensive understanding of the non-linearities arising in these scenarios. As validating numerical results against analytical methods is often impossible (in the interesting non-linear regime), sanity checks of simulation results must come from adherence to physical principles, such as, e.g., energy conservation or gauge invariance, and from cross-checking the numerical outcome from different methods applied to the same given problem. 

The use of numerical techniques to study the non-linear field dynamics has increased considerably, as reflected by the number of specialized packages created over the last years~\cite{Felder:2000hq,Felder:2007nz,Frolov:2008hy,Sainio:2009hm,Easther:2010qz,Huang:2011gf,Sainio:2012mw,Daverio:2015ryl,Lozanov:2019jff,RoperPol:2018sap,Giblin:2019nuv,PencilCode:2020eyn,Figueroa:2021yhd,Andrade:2021rbd,Figueroa:2023xmq,Buschmann:2024bfj,Caravano:2025klk,Baeza-Ballesteros:2026uao,Florio:2026vde}. Only by developing increasingly efficient and reliable codes and techniques, we will achieve a solid foundation for the predictions of the early Universe. A new field---{\bf Lattice Cosmology}---has emerged on its own merits, thanks to its unique ability to capture the precise details of non-linear field dynamics in the early Universe, and its potential to provide accurate predictions for the observables of such dynamics. This latter point is of paramount importance in the current era of precision observational cosmology. {\bf Lattice Cosmology Techniques} (LCT) have established themselves as a clear pathway towards understanding the early Universe, and
we expect them to become increasingly influential in determining observational strategies to probe the early Universe. 
It is precisely in this spirit that the package \CL\cite{Figueroa:2021yhd} was originally created, see \href{http://www.cosmolattice.com}{\color{blue} http://www.cosmolattice.com}.
Contrary to standard codes focused on a given set of equations and observables, \CL is rather a {\it platform} for the implementation of any field theory system characterized by partial differential equations suitable for
discretization on a lattice~\cite{Figueroa:2023xmq}. Written in C++, \CL uses a modular structure to separate technical details from the physics, establishing a unique symbolic language, wherein field variables and their associated operations are defined in close resemblance to the continuum. 

The release of \CL {\tt v1.0} in 2021~\cite{Figueroa:2021yhd} was preceded by the publication of a monographic review on LCT: {\it The Art of Simulating the early Universe, Part I. Integration techniques and canonical cases}~\cite{Figueroa:2020rrl}. This monograph, which we will refer to, colloquially, as {\tt The Art\,-\,I}, constituted the theoretical basis for the physics capabilities initially released in {\tt v1.0} of the code: namely, {\it canonical} scalar-singlet and $SU(2)$$\times$$U(1)$ scalar-gauge theories in a spatially-flat expanding background. In particular, {\tt The Art\,-\,I} presented symplectic explicit-in-time evolution algorithms for theories with canonically normalized kinetic terms, with accuracies ranging from $\mathcal{O}(dt^{2})$ to $\mathcal{O}(dt^{10})$, and with self-consistent expansion of the Universe sourced by all the dynamical fields involved. The algorithms were designed to simulate singlet scalar and/or Abelian $U(1)$ and non-Abelian $SU(2)$ scalar-gauge interactions, preserving the Gauss constraint to machine precision, for either $U(1)$ or $SU(2)$. 

The recent release of \CL {\tt v2.0}~\cite{Baeza-Ballesteros:2026uao}
was also preceded by another monographic review on LCT: {\it The Art of Simulating the early Universe, Part II. Non-canonical cases \& gravitational waves}~\cite{Baeza-Ballesteros:2025tme} or, for short, {\tt The Art\,-\,II}. This second monograph served for a twofold purpose, on the one hand, constituting the theoretical basis for the new modules added in \CL{\tt v2.0}, and on the other hand,  extending the discussion on LCT to new physics cases beyond those considered in {\tt The Art\,-\,I}. In particular, {\tt The Art\,-\,II} introduced lattice implementations of non-canonical interactions, including scalars with a non-minimal coupling to gravity, $\phi^2R$, non-minimal scalar kinetic theories, $\mathcal{G}_{ab}(\lbrace\phi_c\rbrace)\partial_\mu\phi^a\partial^\mu\phi^b$, and axion-like particle (ALP) interactions with Abelian gauge fields, $\phi F_{\mu\nu}\tilde F^{\mu\nu}$. It also discussed methods to set up special field configurations, like the creation of cosmic defect networks  towards scaling (e.g., cosmic strings and domain walls), 
field configurations based on arbitrary power spectra or spatial profiles, 
and probabilistic methods as required, e.g., for thermal configurations. 
Furthermore, it extended the notion of non-canonical theories to the dimensionality of space, discussing the discretization of scalar field 
dynamics in $d + 1$ dimensions, with 
$d \neq 3$. Finally, unrelated to non-canonical aspects, {\tt The Art\,-\,II} also  discussed previous and novel implementation(s) of gravitational wave (GW) dynamics on the lattice.

{\tt The Art\,-\,I} and {\tt The Art\,-\,II} are the first and second monographic reviews in a series of dissertations on LTC that were originally planned at the conception of \CLns. 
The present document, {\it The Art of Simulating the early Universe - Part III. Scalar-Gauge-Fluid Dynamics}, or for short, {\tt The Art\,-\,III}, constitutes therefore a third entry in the monographic series. It serves again for a twofold purpose: on the one hand, it extends previous discussions on LCT, in this occasion to
fluid dynamics;
on the other hand, it constitutes the theoretical basis for new modules--- related to fluid dynamics--- added in \CL{\tt v3.0}, which will be publicly released after the publication of this monograph. Eventually, we expect to complete the dissertation series with at least one more entry, {\tt The Art-IV}, where we will discuss lattice formulations of general relativity as sourced by all degrees of freedom, scalars, gauge fields, and fluids.

Here, in {\tt The Art\,-\,III}, we present a discussion on lattice techniques for the simulation of fluid dynamics in a (spatially-flat) expanding Universe.
In \Sec{sec:cont} we review in full detail the theory of fluid dynamics in the continuum. We first discuss the dynamics on flat spacetime,
and then on a Friedmann-Lema\^itre-Robertson-Walker (FLRW) background.
We consider both conservation and non-conservation
forms of the equations of motion (EOM) governing the dynamics of the fluid in either the
fully relativistic or 
subrelativistic regimes of the fluid bulk velocity.
We describe the dynamics of an isolated fluid,
considering
perfect fluids in local thermal equilibrium (LTE),
and then imperfect (viscous) fluids, 
which present small
deviations with respect to LTE.
We extend our discussions
to fluids coupled to bosonic sectors, via Lorentz forces with gauge fields,
or through derivative couplings to scalar field sectors.
In \Sec{sec:generalLattice} we review briefly some lattice concepts, including basic and higher-order lattice derivatives, lattice momenta, program variables, and evolution algorithms to solve partial differential equations
in a computer. Next, we dwell into detailed discretization procedures for fluid dynamics on a FLRW background. In \Sec{sec:fluidLattice}, we present discretization and evolution schemes
for perfect and imperfect fluids, both for conservation and non-conservation forms of the EOM, while 
not yet considering 
interactions with any external field. In Section~\ref{sec:fluid_bosonic}, we couple the fluids to infrared (IR)--- i.e., long wavelength---bosonic sectors.
We present both conservation and non-conservation forms of the lattice EOM. Remarkably, all our evolution algorithms can be run
for self-consistent expansion of the Universe, as sourced by all scalar, gauge, and fluid degrees of freedom involved in the dynamics. 
In the case of gauge theories,
we consider three different discretization schemes: {\it collocated}, {\it semi-collocated}, or {\it staggered}, depending on whether gauge fields and (some) fluid variables live at the integer ($\bf n$), 
or semi-integer $({\bf n}+\hat{\imath}/2$) lattice sites. Some of these formulations preserve gauge invariance on the lattice (to machine precision) throughout evolution, whereas others allow for the implementation of arbitrary higher-order derivatives.
In \Sec{sec:init_cond} we discuss methods to set up the initial conditions of fluids and of bosonic fields on a lattice, and
in \Sec{sec:gravitational_waves} we discuss the lattice implementation of gravitational wave (GW) dynamics sourced by fluids (together with scalar and/or gauge fields).

The applications of the techniques presented in this review are very relevant for a broad range of early Universe situations, especially when a thermal distribution of particles is present. This is highlighted by the number of numerical studies in recent years that focus on non-linear fluid
dynamics of the
early Universe primordial plasma.
An important example is the development of
acoustic and vortical turbulence that
can lead to the production of GWs, e.g., due to
the interaction of first-order phase transition bubbles
with the primordial plasma
\cite{Hindmarsh:2013xza,Giblin:2014qia,Hindmarsh:2015qjv,Kahniashvili:2020jgm,RoperPol:2018sap,RoperPol:2019wvy,Cutting:2019zws,Dahl:2021wyk,Dahl:2024eup,Auclair:2022jod,Jinno:2022mie,Jinno:2020eqg,Caprini:2024gyk,Stomberg:2025kxf,Sharma:2023mao,Correia:2025qif}, or due to the nonlinear coupling of the plasma
with a primordial magnetic field, described by magnetohydrodynamic (MHD) turbulence \cite{RoperPol:2018sap,RoperPol:2019wvy,RoperPol:2021xnd,Brandenburg:2021bvg,RoperPol:2022iel,Auclair:2022jod,Sharma:2022ysf}.
Some of these numerical studies consider the dynamics
of the
fluid-scalar system in flat spacetime, e.g., 
\cite{Hindmarsh:2013xza,Hindmarsh:2015qjv,Hindmarsh:2017gnf,Cutting:2019zws,Correia:2025qif}, or evolve only the fluid
system in the so-called Higgsless approach \cite{Jinno:2020eqg,Jinno:2022mie,Caprini:2024gyk,Stomberg:2025kxf}, where the value of the scalar field is prescribed
at every location in the simulation.
On the other hand, some numerical studies have focused
on the production of GWs only by the fluid perturbations,
assumed to be either purely compressional \cite{RoperPol:2019wvy,Dahl:2021wyk,Dahl:2024eup,Sharma:2023mao} or purely vortical \cite{RoperPol:2019wvy,Neronov:2020qrl,RoperPol:2022iel,Auclair:2022jod}.
The formulation of scalar-gauge-fluid dynamics in the present work gives the possibility to extend these numerical
studies to fully relativistic bulk fluid motion
in an expanding FLRW background,
taking into account the dynamics of the fluid and its
interactions with both scalar and gauge sectors.\vspace{2mm}

\begin{mdframed}
{\bf Note -.} All the physics and technical aspects that we present here in {\tt The Art\,-\,III} are implemented in \CL{\tt v3.0}, which will be publicly released after this monograph is published, check the $\mathcal{C}${\tt osmo}$\mathcal{L}${\tt attice} website (\href{http://www.cosmolattice.com}{\color{blue} http://www.cosmolattice.com}) for updates. In any case, the presentation of the numerical techniques in this monograph constitutes
by itself a theoretical review on LCT for fluid dynamics on expanding backgrounds, complementing our previous reviews on LCT on canonical and non-canonical field theories, {\tt The Art\,-\,I} and {\tt The Art\,-\,II}. The techniques presented here are therefore general LCT for fluid dynamics, which can be 
implemented in any code.
\end{mdframed}


\newpage

\section{Continuum formulation}\label{sec:cont}

As a starting point, we consider a system with a large density of particles, corresponding to quanta of interacting scalar, gauge, and fermion fields. We assume that the particle 
{\it de Broglie} wavelengths are much smaller than the typical mean free path length scales of the problem, $\lambda_{dB} \ll l_{\rm mfp}$, so that the wavefunctions of the different particles are widely separated, and hence quantum interference of their wave packages can be neglected. The particles can then be considered as moving like classical particles, even if the collisions they sustain are set by interactions of quantum nature. Due to their interactions, the particle ensemble is expected to reach thermal equilibrium, so that within a certain volume element $\Delta V \sim l_{\rm cont}^3$,
the system
can be described in terms of averaged quantities, like the bulk
velocity $\vec u = (u_1,u_2,u_3)$, pressure $p$, energy density $\rho$, temperature $T$, or other macroscopic quantities that characterize the particle ensemble.
The consistency of this continuum description requires the element $\Delta V$ to be large enough to contain a high number of particles (which is always the case if one requires ~$l_{\rm cont} \gg l_{\rm mfp}$), 
but small enough to guarantee 
that the macroscopic quantities $\lbrace q_j \rbrace \equiv \lbrace u_1, u_2, u_3, p, \rho, T, ... \rbrace$ 
do not depend on $\Delta V$.
This ensemble, from now on, will be referred to as the {\it fluid}.
If the ensemble contains
scalars and/or gauge fields, a macroscopic length scale $L \gg l_{\rm cont}$, might also be present in the problem
(more on this below).
The fluid quantities $\lbrace q_j \rbrace \equiv \lbrace u_1, u_2, u_3, p, \rho, T, ... \rbrace$
are then allowed to vary over scales comparable to $L$.
The hierarchy of scales in the above description is summarized by $\lambda_{dB} \ll l_{\rm mfp} \ll l_{\rm cont} \ll L$. Our intention is then to describe the system at scales $\lambda$ 
much larger than $l_{\rm cont}$, but which could be either smaller, of the same order, and/or larger than $L$. In other words, as the scales of interest must fall within a finite range, say $\lambda \in (\lambda_{\rm min},\lambda_{\rm max})$, we will assume 
$\lambda_{\rm min} \gg l_{\rm cont}$, and
$\lambda_{\rm min} < L < \lambda_{\rm max}$.

We will naturally distinguish {\it infrared} (IR) and {\it ultraviolet} (UV) degrees of freedom ($dof$), as those corresponding to modes $k \ll 1/l_{\rm cont}$ and $k \gg 1/l_{\rm cont}$, respectively.
The macroscopic scale $L$ is typically introduced in the system due to some 
mechanism in the bosonic sector, like for example the nucleation of scalar field bubbles of radius $L \equiv R \gg l_{\rm cont}$ in a first-order phase transition, or the cosmological
correlation length $L \equiv l_{\rm corr} \gg l_{\rm cont}$
of a magnetic field created during inflation. 
Our aim is to describe a theory for the IR $dof$, where the UV $dof$ are integrated out. The IR $dof$ include
the macroscopic quantities $\lbrace q_j \rbrace$ that characterize the fluid,
as well as bosonic fields
with long wavelengths $\lambda_{\rm Bos} \gg l_{\rm cont}$.
We therefore consider a system
with a fluid sector ($\mathcal{L}_\fl$) coupled to a bosonic sector ($\mathcal{L}_{\rm Bos, IR}$).
In the latter,
we allow in full generality 
for the presence of interactive singlet, Abelian $U(1)-$charged, and non-Abelian $[SU(N)\times U(1)]-$charged scalar fields, denoted respectively as $\phi$, $\varphi$, and $\Phi$, and correspondingly for the $U(1)$ Abelian gauge fields $A_{\mu}$, and $SU(N)$ non-Abelian gauge fields $C_{\mu}$. Such a system is described by a Lagrangian like 
\EQalign
\label{full_lagrangian}
\hspace*{-1cm}\mathcal{L} = \mathcal{L}_\fl \,   \underbrace{-\frac{1}{4} F_{\mu \nu} F^{\mu \nu} \, - 
\frac{1}{4} G_{\mu \nu}^a G_a^{\mu\nu}
\, - \frac{1}{2}\partial_{\mu} \phi \, \partial ^{\mu}\phi \, - (D_\mu^A \varphi)^* (D^\mu_A \varphi)
\, -(D_\mu \Phi)^{\dagger} (D^\mu \Phi) \, - V_0(\phi, |\varphi|, |\Phi|)}_{\mathcal{L}_{\rm Bos, IR}}\,,
\end{mymathbox}
\noindent
where $V_0(\phi,|\varphi|,|\Phi|)$ is the tree-level potential, $F_{\mu \nu}\equiv \partial_\mu A_\nu - \partial_\nu A_\mu$ the Abelian field strength, and $G_{\mu \nu}^a = \partial_\mu C_\nu^a - \partial_\nu C_\mu^a + f_{abc} C_\mu^b C_\nu^c$ the non-Abelian field strength, with $f_{abc}$ the totally antisymmetric structure constants of $SU(N)$, and $a=1, ..., N^2-1$ the $SU(N)$ ``color" index. In \Eq{full_lagrangian}, $D_\mu^A  = \partial_\mu - i  \, g \, Q \, A_\mu$
is the Abelian gauge covariant derivative, with $g$ the $U(1)$
coupling constant, and $Q$ the $U(1)$ charge of the
scalar fields, $Q_\varphi$ or $Q_\Phi$, respectively;
$D_\mu \Phi = \mathcal{I} D_\mu^A \Phi - i \,
g' \, Q'_{\Phi} \, C_\mu^a \, T_a \Phi$ is the non-Abelian gauge covariant derivative,
with $g'$ the
$SU(N)$ coupling constant and $Q'_{\Phi}$ the
$SU(N)$ charge of $\Phi$. Here $\mathcal{I}$ is the $N \times N$ identity matrix, and $T_a$ are the $N^2-1$ group generators of $SU(N)$, which follow the Lie algebra $[T_a, T_b]=if_{abc} T_c$. 

In the following, we assume the presence of
a fluid at finite temperature characterized by $\mathcal{L}_\fl$, which is typically divided in three terms: a {\it perfect fluid} contribution $\mathcal{L}_{\rm pf}$ that accounts for the {\em equilibrium} part of the particle distribution, an {\it imperfect fluid} contribution $\mathcal{L}_{\rm ipf}$ that accounts for small departures from thermal equilibrium,
and a {\it fluid interaction} term $\mathcal{L}_{\rm int}$ that accounts for the coupling of the fluid to the long wavelength bosonic $dof$, characterizing the interactions of the fluid with the IR scalar and/or gauge fields. Proper definitions of the perfect and imperfect fluid descriptions
will be presented in \Sec{subsec:fld_dynamics_cont}, whereas the fluid-gauge and scalar-fluid interactions will be covered in \Sec{subsec:bosonic_inters}. 
At this point, we simply anticipate
that we will not provide a Lagrangian formulation for the aforementioned fluid terms, as it turns out to be more practical to describe instead the fluid dynamics in terms of its stress-energy tensor.

\subsection{Dynamics in flat spacetime}
\label{subsec:FlatSpace_Dynamics_Cont}

Assuming the fluid to be characterized by a finite number of macroscopic quantities $\{ q_j \, | \, j=1, \dots,N^{\rm dof}_\fl \}$, which we do not need to specify yet, its interaction with the gauge and scalar IR sectors is expected to be expressed throughout gauge-invariant terms depending on such fluid variables $\{q_j\}$, as well as on the bosonic field amplitudes $\{ f_i^{\rm IR} \} $ $\equiv \{ A_\nu, C_\nu^a, \phi, \varphi, \Phi \}$ and their derivatives $\{ \partial f^{\rm IR}_i \} $ $\equiv \{ \partial_\mu A_\nu, \partial_\mu C_\nu^a, \partial_\mu\phi, \partial_\mu \varphi, \partial_\mu \Phi \}$.
The classical equations of motion (EOM) can then be obtained from the total variation of the action $S \equiv \int \dd^4 x \, \mathcal{L}$,
with the Lagrangian
of the system given by \Eq{full_lagrangian}, and
$\mathcal{L}_\fl = \mathcal{L}_\fl(\{q_j\},\{f_{i}^{\rm IR}\},\{\partial f^{\rm IR}_i\})$.
Under these circumstances, the variation of the action yields
\begin{eqnarray}\label{eq:deltaS1}
&& 
\hspace*{-1.25cm}
\delta S = \int \dd^4 x
\Biggl\{ \sum_{i=1}^{N^{\rm dof}_\fl} \frac{\partial \mathcal{L}_\fl}{\partial q_i} \delta q_i + \biggl( \frac{\partial \mathcal{L}_\fl}{\partial \phi} - \frac{\partial V_0}{\partial \phi} \biggr) \delta \phi + \biggl( \frac{\partial \mathcal{L}_\fl}{\partial (\partial_\mu \phi)}  - \partial^\mu \phi \biggr) \, \delta (\partial_\mu \phi) \nonumber \\
&& 
\hspace*{-0.75cm}
+\, \biggl[ \biggl( \frac{\partial \mathcal{L}_\fl}{\partial \varphi^*} - \frac{\partial V_0}
{\partial \varphi^*} - i g Q_{\varphi} A_\mu D^\mu_A \varphi \biggr) \delta \varphi^* + \biggl( \frac{\partial \mathcal{L}_\fl}{\partial (\partial_\mu \varphi^*)} - D^\mu_A \varphi \biggr) \delta (\partial_\mu \varphi^*) + c. c. \biggr]
\nonumber \\
&& 
\hspace*{-0.75cm}
+\, \biggl[ \biggl( \frac{\partial \mathcal{L}_\fl}{\partial \Phi^{\dagger}} - \frac{\partial V_0}
    {\partial \Phi^{\dagger}} - (i g Q_{\Phi} A_\mu \mathcal{I} - i g' Q'_{\Phi} C_{\mu}^a T_a) D^\mu \Phi \biggr) \delta \Phi^{\dagger} +
    \biggl( \frac{\partial \mathcal{L}_\fl}{\partial (\partial_\mu \Phi^{\dagger})} - D^\mu \Phi \biggr) \delta (\partial_\mu \Phi^{\dagger}) + c. c.
    \biggr] \nonumber  \\
&& 
\hspace*{-0.75cm}
+\, \biggl( \frac{\partial \mathcal{L}_\fl}{\partial A_\nu} + J^\nu_{\varphi} + J^{\nu}_{\Phi} \biggr) \delta A_\nu + \biggl( \frac{\partial \mathcal{L}_\fl}{\partial (\partial_\mu A_\nu)} - F^{\mu \nu} \biggr) \delta (\partial_\mu A_\nu) \nonumber \\
&& 
\hspace*{-0.75cm}
+\, \biggl( \frac{\partial \mathcal{L}_\fl}{\partial C_\nu^a} + {\cal J}^\nu_{\Phi, a} - f_{abc} C_{\mu}^c G^{\mu \nu}_b \biggr) \delta C_\nu^a + \biggl( \frac{\partial \mathcal{L}_\fl}{\partial (\partial_\mu C_\nu^a)} - G^{\mu \nu}_a \biggr) \delta (\partial_\mu C_\nu^a) \Biggr\} \nonumber\\
&& 
\hspace*{-1.1cm}
=\, \int \dd^4 x \Biggl\{  \sum_{i=1}^{N^{\rm dof}_\fl}
    \frac{\partial \mathcal{L}_\fl}{\partial q_i} \delta q_i + \Bigl( \partial_\mu \partial^\mu \phi - \frac{\partial V_0}{\partial \phi}
    + \Omega_\phi \Bigr) \delta \phi +
    \biggl[ \Bigl( D_\mu^A D^\mu_A \varphi -  \frac{\partial V_0}{\partial \varphi*}
    + \Omega_\varphi
    \Bigr) \delta \varphi^* + c.c. \biggl]
    \nonumber\\
    && 
    \hspace*{-0.75cm}
    + \biggl[ \Bigl( D_\mu D^\mu \Phi - \frac{\partial V_0}{\partial \Phi^\dag}
    + \Omega_\Phi
    \Bigr) \delta \Phi^{\dagger} + c. c. \biggr]
    + \bigl( \partial_\mu F^{\mu \nu} + J^\nu_{\varphi} +
    J^\nu_{\Phi} +
    J^\nu_\fl
    \bigr) \delta A_\nu 
    + \bigl[({\cal D}_\nu)_{ab} G^{\mu \nu}_b + \mathcal{J}^\nu_{\Phi, a}+\mathcal{J}^{\nu}_{\fl,a} \bigr] \delta C^a_{\nu} 
    \Biggr\}\,,
    \label{eq:deltaS1} 
\end{eqnarray}
where, in the last step, we performed an integration by parts, set to zero the boundary terms, and collected all the terms proportional to $\delta q_i$,
$\delta A_\nu$, $\delta C_\nu^a$, $\delta \phi$, $\delta \varphi^*$, and $\delta \Phi^{\dagger}$, respectively. In the process, we introduced $({\cal D}_\nu)_{ab} = \delta_{ab} \partial_\nu - f_{abc} C^c_\nu$, an Abelian current ($J^\mu_{\varphi}$) associated to $\varphi$, and Abelian ($J^\mu_{\Phi}$) and non-Abelian (${\cal J}^\mu_{\Phi, a}$) currents associated to $\Phi$, as 
\begin{eqnarray}\label{currents}
J^\mu_{\varphi} \equiv 2 \, g \, Q_{\varphi} \, {\rm Im}[\varphi^* (D^\mu_A \varphi)]\,, \qquad
    J^\mu_\Phi\equiv2 \, g \, Q_{\Phi} \,  {\rm Im}[\Phi^{\dagger} (D^\mu \Phi)]\,,~~~ {\rm and} ~~~ {\cal J}^\mu_{\Phi, a} \equiv 2 \, 
    g' \, Q'_{\Phi} \,  {\rm Im}[\Phi^{\dagger} T_a (D^{\mu} \Phi)]\,.
\end{eqnarray}
The variational process, led us  naturally to define {\it fluid current densities} associated to the Abelian ($J_\fl^\mu$) and 
non-Abelian (${\cal J}_{\fl, a}^\mu$) gauge fields, as
\begin{equation}
    J^\mu_\fl \equiv \frac{\partial \mathcal{L}_\fl}{\partial A_\mu} - \partial_\nu \frac{\partial \mathcal{L}_\fl}{\partial (\partial_\nu A_\mu)}\,,
    \qquad
    {\cal J}^\mu_{\fl, a} \equiv \frac{\partial \mathcal{L}_\fl}{\partial C^a_\mu} - \partial_\nu \frac{\partial \mathcal{L}_\fl}{\partial (\partial_\nu C^a_\mu)}
    \,,
    \label{gauge_fluid_couplings}
\end{equation}
as well as {\it source terms}, associated to the scalar field sectors,
as
\begin{equation}
    \Omega_{\phi} \equiv \frac{\partial \mathcal{L}_\fl}{\partial \phi} - \partial_\mu \frac{\partial \mathcal{L}_\fl}{\partial (\partial_\mu \phi)}\,,
    \qquad
    \Omega_{\varphi} \equiv \frac{\partial \mathcal{L}_\fl}{\partial \varphi^*} - \partial_\mu \frac{\partial \mathcal{L}_\fl}{\partial (\partial_\mu \varphi^*)}\,,
    \qquad
    \Omega_{\Phi} \equiv \frac{\partial \mathcal{L}_\fl}{\partial \Phi^{\dagger}} - \partial_\mu \frac{\partial \mathcal{L}_\fl}{\partial (\partial_\mu \Phi^{\dagger})}
     \,.
    \label{friction_Lagrangian}
\end{equation}
From \Eq{eq:deltaS1}, the EOM for the IR bosonic sectors are obtained
from setting $\delta S = 0$,
\begin{subequations}
\label{EOM_flat}
\EQalign
    -\partial_\mu F^{\mu\nu} &  = J^\nu\,, \qquad \text{with} \quad J^\nu = J^\nu_{\varphi} 
     + J^\nu_{\Phi} + J^\nu_\fl \,,
    \label{scalar_fluid_gauge_minkowski1}
    \\
    -({\cal D}_\mu)_{ab} G^{\mu\nu}_b & = \mathcal{J}^\nu_{a}\,, \qquad \! \text{with} \quad
    \mathcal{J}^\nu_a = \mathcal{J}^\nu_{\Phi, a} + 
    \mathcal{J}^\nu_{\fl, a}\,, \quad
    ({\cal D}_\nu)_{ab} =
    \delta_{ab} \partial_\nu - f_{abc} C^c_\nu\,,
    \label{scalar_fluid_gauge_minkowski2}
    \\
    -\partial_\mu \partial^\mu \phi
    +\frac{\partial V_0}{\partial \phi}
    &=
     \Omega_\phi\,,
    \label{scalar_fluid_gauge_minkowski3}\\
    -D_\mu^A D^\mu_A \varphi
    + \frac{1}{2} \frac{\partial V_0}{\partial |\varphi|} \frac{\varphi}{|\varphi|} &= 
    \Omega_{\varphi}\,,
    \qquad  \text{with} \quad D_\mu^A \varphi = (\partial_\mu - i\,  g \, Q_\varphi A_\mu) \varphi \,, 
    \label{scalar_fluid_gauge_minkowski4}\\
    -D_\mu D^\mu \Phi
    + \frac{1}{2} \frac{\partial V_0}{\partial |\Phi|} \frac{\Phi}{|\Phi|} &= 
     \Omega_{\Phi}\,,
    \qquad \text{with} \quad D_\mu \Phi = \bigl[ \mathcal{I} (\partial_\mu - i \, g \, Q_\Phi A_\mu) - i \,
g' \, Q'_{\Phi} \, C_\mu^a \, T_a\bigr] \Phi \,,
\label{scalar_fluid_gauge_minkowski5}
\end{mymathbox}
\end{subequations}
\noindent
where we have defined a {\it total Abelian
current} as $J^\mu = J^\mu_\varphi + J^\mu_{\Phi} + J^\mu_\fl$, a {\it total non-Abelian current} as
$\mathcal{J}_a^\mu = \mathcal{J}^\mu_{\Phi, a} + \mathcal{J}^\mu_{\fl, a}$, and made use of the relations 
$\frac{\partial V_0}{\partial \varphi*} = \frac{\partial V_0}{\partial |\varphi|} \frac{\partial |\varphi|}{\partial \varphi*} = {1\over2}\frac{\partial V_0}{\partial |\varphi|} \frac{\varphi}{|\varphi|}$, and $\frac{\partial V_0}{\partial \Phi^\dag} = \frac{\partial V_0}{\partial |\Phi|}\frac{\partial |\Phi|}{\partial \Phi^\dag} = {1\over2}\frac{\partial V_0}{\partial |\Phi|}\frac{\Phi}{|\Phi|}$.

We note that we have assumed that ${\cal L}_\fl$ only depends on the fields $\{ f_i^{\rm IR} \} $ and their derivatives  $\{ \partial f^{\rm IR}_i \} $.
In general, \Eqs{gauge_fluid_couplings}{friction_Lagrangian} could be generalized to include dependence with higher-order derivatives, though these terms are likely to be suppressed.
Furthermore, since the variation of the action has to be gauge invariant, in light of
\Eqs{scalar_fluid_gauge_minkowski4}{scalar_fluid_gauge_minkowski5} the source terms $\Omega_\varphi$ and $\Omega_\Phi$ must transform, under a gauge transformation, as $\varphi$ and $\Phi$, respectively. 
Equivalently $J^{\nu}$ must be gauge invariant and ${\cal J}^{\nu}_a$ must transform as $G^{\mu \nu}_a$ (i.e., in the adjoint representation) under a gauge transformation. 

In order to solve the system, we still need EOM to solve for the fluid variables. While such equations are formally represented by $\frac{\partial \mathcal{L}_\fl}{\partial q_i} = 0$, in practice, these conditions do not fix them, 
unless we provide an exact form for $\mathcal{L}_\fl$.  As a fluid is an emergent macroscopic description of the dynamics of a many-body system, instead of using a Lagrangian description, we will rather describe the fluid dynamics in terms of some general properties of the system. As invariance under 
\textit{spacetime} translations is expected to hold on general grounds, conservation of the
total stress-energy tensor $T^{\mu \nu}$ of the system should be satisfied as $\partial_{\mu}T^{\mu\nu} = 0$. Denoting the flat spacetime metric as $\eta_{\mu \nu} = \eta^{\mu \nu} = {\rm diag} (-1, 1, 1, 1)$, the total stress-energy tensor of the system can be obtained as
\begin{eqnarray}
\label{eq:StressEnergyTensor}
    T^{\mu \nu} &\equiv& \eta^{\mu \nu} \, \mathcal{L} \, - 2 
    \frac{\partial \mathcal{L}}{\partial \eta_{\mu \nu}} \nonumber \\
    &=& T^{\mu\nu}_\fl +
    \eta^{\mu \nu} \, \mathcal{L}_{\rm Bos, IR}
    + F^{\mu \lambda}F^{\nu}_{ \, \, \, \lambda} +
    G_a^{\mu \lambda}G^{\nu}_{a \, \lambda}
    + \partial^{\mu} \phi \, \partial^{\nu} \phi 
    + \bigl[\bigl(D^{\mu}_A \varphi \bigr)^*
    \bigl(D^{\nu}_A \varphi \bigr) +
    \bigl(D^{\mu} \Phi\bigr)^{\dagger}
    \bigl(D^{\nu} \Phi \bigr) + c.c.\bigr] \,,
\end{eqnarray}
where we have defined the fluid contribution, formally, as
\begin{equation}
    T^{\mu\nu}_\fl 
    \equiv \eta^{\mu \nu} \, \mathcal{L}_\fl - 2 \frac{\partial  \mathcal{L}_\fl}{\partial \eta_{\mu \nu}}\,.
\end{equation}
Interestingly, we do not need to specify the 
explicit form of the stress-energy tensor of the fluid, $T^{\mu\nu}_\fl$, in order to obtain an equation describing its evolution. 

Plugging \Eq{eq:StressEnergyTensor}
into the conservation of the total stress-energy tensor,
$\partial_\mu T^{\mu \nu}=0$, leads to the following EOM for the fluid
(see \App{app:EOM} for details),
\begin{align}
    \label{Tmunu_cons_aux}
    \partial_\mu T^{\mu \nu}_\fl = &\,
    J^\mu_\fl F^{\nu}_{\ \, \mu} +
    {\cal J}^{\mu}_{\fl, a} G^{\nu}_{a \, \mu}
    +
    \Omega_\phi \, \partial^{\nu}\phi
    + \bigl(\Omega_\varphi^\ast \, D^\nu_A \varphi 
    + \Omega_\Phi^\dagger \, D^\nu \Phi  + c. c. \bigr)\,,
\end{align}
where $\{J^\mu_\fl, {\cal J}^{\mu}_{\fl, a}, \Omega_\phi, \Omega_\varphi, 
\Omega_\Phi\}$ are the same fluid currents and source terms that we already introduced previously in \Eqs{gauge_fluid_couplings}{friction_Lagrangian}.
Identifying the terms in the \rhs of \Eq{Tmunu_cons_aux} as forces upon the fluid, we distinguish the Abelian and non-Abelian Lorentz forces,
\begin{equation}
    {\cal F}_{U(1)}^\nu \equiv
    J^\mu_\fl
    F^{\nu}_{\, \ \mu}\,, \label{eq:F_U(1)}\qquad 
    {\cal F}_{SU(N)}^\nu \equiv {\cal J}^{\mu}_{\fl, a} G^{\nu}_{a \, \mu}\,,
\end{equation}
and define new forces due to the interaction with the scalar fields as
\begin{align}
    {\cal F}_\phi^\nu \equiv \Omega_\phi\, \partial^\nu \phi \,, \qquad
    {\cal F}_\varphi^\nu \equiv \Omega_\varphi^\ast\,
    D^\nu_A \varphi 
     + c.c. \,, \qquad
    {\cal F}_\Phi^\nu \equiv \Omega_\Phi^\dagger\,
    D^\nu \Phi + c. c.
    \,.
    \label{fluid_kernels}
\end{align}
The EOM for the fluid can then be expressed
in the following way,
\EQb
\label{cons_fluid}
\partial_\mu T^{\mu \nu}_\fl =
{\cal F}_{\rm int}^\nu \equiv
{\cal F}_{U(1)}^\nu + {\cal F}_{SU(N)}^\nu 
+ {\cal F}_{\phi}^\nu + {\cal F}_{\varphi}^\nu + 
{\cal F}_{\Phi}^\nu\,,
\end{mymathbox}
\noindent
where ${\cal F}_{\rm int}^\nu$ represents all the forces acting upon the fluid, due to its interactions with the gauge and scalar fields of the IR bosonic sector. 

\subsection{Dynamics in an expanding Universe}
\label{subsec:FLRW_Dynamics_Cont}

The background evolution is described by a homogeneous and isotropic 
Friedmann-Lema\^itre-Robertson-Walker (FLRW) metric, with line element 
\be\label{eq:FLRW}
\dd s^2 \equiv g_{\mu\nu}\dd x^\mu\dd x^\nu = - a^{2\alpha} (\eta)
\, \dd \eta^2 + a^2 (\eta) \, \delta_{ij} \dd x^i \dd x^j \,,
\ee
where the
spacetime coordinates are $x^\mu = (\eta, x^i)$, being
$\eta$ the $\alpha$-time and $x^i$ the comoving space coordinates.
The $\alpha$-time is defined such that $\dd t = a^{\alpha} \dd \eta$,
with $t$ the cosmic time and $\alpha$ a real constant parameter that can be conveniently chosen for a given problem.
By fixing $\alpha=0$, we recover cosmic time $t$, whereas $\alpha=1$ gives conformal time $\tau$.
The EOM in a curved background are obtained by the following replacements of the metric and derivatives, 
\begin{align} 
\eta_{\mu\nu} \longrightarrow g_{\mu\nu} \, ; \hspace{1.5cm}
\partial_{\mu} f^{\alpha\beta...}_{\sigma\rho...} \longrightarrow D_{\mu}^Gf^{\alpha\beta...}_{\sigma\rho...} = \partial_{\mu} f^{\alpha\beta...}_{\sigma\rho...} 
+\Gamma^{\alpha}_{\mu\lambda}
f^{\lambda\beta...}_{\sigma\rho...}-  \Gamma^\lambda_{\mu\sigma} f^{\alpha\beta...}_{\lambda\rho...}+...\, ,
\end{align}
where $D_{\mu}^G$ represents the gravitational covariant derivative, and $\Gamma^\lambda_{\alpha\beta} \equiv \frac{1}{2} g^{\lambda \nu}(\partial_\alpha g_{\beta \nu}+\partial_\beta g_{\nu \alpha} - \partial_\nu g_{\alpha \beta})$ the Christoffel symbols of the Levi-Civita connection.
For an FLRW metric, the only non-vanishing Christoffel symbols are
\begin{equation}
    \Gamma^0_{\ 00} = \alpha \HH\,, \qquad \Gamma^0_{\ ij} = a^{2 (1-\alpha)} \HH \delta_{ij}\,, \qquad
    \Gamma^{i}_{\ 0j} = \HH \delta_{ij}\,,
\end{equation}
where prime denotes time derivatives with respect to $\alpha$-time, $'\equiv d/d \eta$, and $\HH \equiv a'/a$ is the ($\alpha$-time) Hubble rate.
The gravitationally covariant divergence of a vector ($f^\mu$),
and of rank-2 symmetric ($T^{\mu\nu}$) and antisymmetric ($F^{\mu\nu}$) tensors, are given by
\begin{subequations}
\begin{align}  
    D_{\mu}^G f^\mu &= \frac{1}{\sqrt{-g}}
    \partial_\mu (\sqrt{-g} f^\mu) = \partial_\mu (g^{\mu\nu}f_\nu) + (3 + \alpha) \HH f^0
    \nonumber \\ &=a^{-2\alpha}\left[-\partial_0 f_0 + a^{-2(1-\alpha)}\partial_j f_j - (3-\alpha) \HH f_0\right]\,,
    \\
    D_{\mu}^G T^{\mu\nu}& = \frac{1}{\sqrt{-g}}
    \partial_\mu (\sqrt{-g}\, T^{\mu\nu}) + \Gamma_{\mu\sigma}^\nu T^{\mu\sigma} = \partial_\mu T^{\mu\nu} + (3 + \alpha) \HH T^{0\nu}
    + \Gamma_{\mu\sigma}^\nu T^{\mu\sigma}
    \nonumber \\
    & = \partial_0 T^{0\nu} +\partial_j T^{j \nu} +(3+\alpha)\HH T^{0 \nu} + \left(\alpha \HH 
    T^{00}
    + a^{2(1-\alpha)}\HH \delta_{jl} T^{jl}\right)\delta^{\nu}_{~\,0} + 2\HH
    T^{0i}\delta^{\nu}_{~\,i}\,,
    \\
    -D_{\mu}^G F^{\mu\nu} & = \frac{-1}{\sqrt{-g}}
    \partial_\mu (\sqrt{-g}\, F^\mn)=
    -\partial_\mu (g^{\mu\alpha}g^{\nu\beta}F_\ab) - (3 + \alpha)
    \HH F^{0i}\delta^{\nu}_{\,~i} \nonumber \\
    & = a^{-2(\alpha+1)}\left[\left(\partial_0 F_{0i} - a^{-2(1-\alpha)}\partial_j F_{ji} +
    (1 - \alpha)\HH
    F_{0i}\right)\delta^{\nu}_{\,~i} -
    (\partial_j F_{0j})\delta^{\nu}_{\,~0}\right]\,,
\end{align}
\end{subequations}
where $g\equiv {\rm det}(g_{\mu\nu}) = - a^{2(3 + \alpha)}$.
Applying these prescriptions to the EOM in flat spacetime 
[cf.~\Eqq{EOM_flat}],
leads to the following set of equations for an isotropic and homogeneous FLRW background,
\begin{subequations}
\label{generic_EOMs}
\begingroup
\allowdisplaybreaks
\EQalign
    \partial_0 F_{0i} - a^{-2(1 - \alpha )}\partial_j F_{ji}
    + (1 - \alpha) \mathcal{H} F_{0i} =&\,
    a^{2 \alpha} J_i \,, \label{eq:Maxwell} \\
    (\mathcal{D}_0 )_{a b}
    G_{0i}^b
    - a^{-2(1 - \alpha )} ( \mathcal{D}_j )_{a b}
    G_{ji}^b
    + (1 - \alpha) \mathcal{H} G_{0i}^a
    =&\, a^{2 \alpha} {\cal J}_i^a \,,  \label{eq:Maxwell_SUN}
    \\
    \partial_i F_{0i} =&\, a^2 J_0 \,, \label{eq:Gauss_Abel} \\
    (\mathcal{D}_i )_{a b} G_{0i}^b
    =&\, a^2 {\cal J}_0^a \,,   \label{eq:Gauss_nonAbel} \\ 
    \phi'' - a^{-2(1 - \alpha)} {\vv\nabla}^{\,2} \hspace{-1mm}\phi +
    (3 - \alpha)\mathcal{H} {\phi'} =&\, - a^{2 \alpha}
    \frac{\partial V_0}{\partial \phi}
    + a^{2 \alpha}
    \, \Omega_\phi
    \,, \label{eq:klein_gordon_singlet_FLRW}
    \\
    \varphi'' - a^{-2(1 - \alpha)}
    {\vv D}_{\hspace{-0.5mm}A}^{\,2}\varphi + (3 - \alpha) \mathcal{H}  {\varphi'} =&\, - \tfrac{1}{2} \,
    a^{2 \alpha} \,
    \frac{\partial V_0}{\partial |\varphi|} \frac{\varphi}{|\varphi |} 
    + a^{2 \alpha}
    \Omega_\varphi \,, \\
    \Phi'' - a^{-2(1 - \alpha)} {\vv D}^{\,2}\Phi +
    (3 - \alpha) \mathcal{H}  {\Phi'} =&\, - \tfrac{1}{2}
    \, a^{2 \alpha} \,
    \frac{\partial V_0}{\partial |\Phi|} \frac{\Phi}{|\Phi |}
    + a^{2 \alpha}
    \Omega_\Phi \,, \\
    \partial_0 T_\fl^{0 0} + \partial_i T_\fl^{i0} + (3+2\alpha)\mathcal{H}T_\fl^{00} 
    + a^{-2\alpha} \mathcal{H}\, T^{(3)}_\fl
    =&\,  {\cal F}^0_{\rm int}
    \,, \label{eq:EnergyConservation} \\
    \partial_0 T_\fl^{0i} + \partial_j T_\fl^{ji} + (5+\alpha)\mathcal{H}\,T_\fl^{0i}
    =&\, {\cal F}^i_{\rm int}
    \,, \label{eq:MomentumConservation} 
\end{mymathbox}
\endgroup
\label{equations_of_motion}
\end{subequations}
\noindent
where $T_\fl^{(3)} \equiv g_{ij} T^{ij}_\fl = a^2 \delta_{ij} T^{ij}_\fl$ is the 3D trace,
and the external forces on the fluid from the bosonic fields are
\begin{subequations}
\begin{align}
    {\cal F}^{0}_{\rm int} = &\,
    -a^{-2(\alpha+1)} \bigl( J_i^\fl F_{0i} +  {\cal J}_i^{\fl,a} G_{0i}^a \bigr) - a^{-2\alpha} \bigl[ \Omega_\phi\, \phi' + \bigl(\Omega_\varphi^\ast\, D_0^A \varphi
    +  \Omega_\Phi^\dagger\, D_0 \Phi
    + c.c. \bigr) \bigr] 
    \,, \\
    {\cal F}^{i}_{\rm int} =& \,
    a^{-2 (\alpha+1)} \bigl( J_0^\fl F_{0i} + {\cal J}_0^{\fl,a} G_{0i}^a \bigr) -  a^{-4} \bigl(J_j^\fl F_{ji} +  {\cal J}_j^{\fl, a} G_{ji}^a \bigr)
    \nonumber \\
    &\,
    + a^{-2} \bigl[ \Omega_\phi  \, \partial_i  \phi +
    \bigl(\Omega_\varphi^\ast \, D_i^A \varphi
    + \Omega_\Phi^\dagger\,  D_i \Phi
    + c.c. \bigr) \bigr] \,.
\end{align}
\end{subequations}

The evolution of the scale factor $a(\eta)$ is determined by the matter content of the system through the Friedmann equations, which in $\alpha$-time read~\cite{Figueroa:2020rrl}
\begin{eqnarray}
    \mathcal{H}^2 \equiv \left({a'\over a}\right)^2 = \frac{a^{2 \alpha}}{3 \mpl^2} 
    \langle{\rho}\rangle \label{eq:Friedmann-1st}
    \,,~~~~~~~
    {a''\over a} = \frac{a^{2 \alpha}}{6 \mpl^2}\bigl[ (2 \alpha - 1) \langle{\rho}\rangle - 3 \langle{ p}\rangle \bigr]\,, \hspace{0.7cm}
    {\rm with} \hspace{0.4cm}   \begin{cases}
    \langle{\rho}\rangle 
    \equiv - \langle g^{00} T_{00} \rangle 
    =  a^{-2 \alpha}\, \langle{T_{00}}\rangle \vspace{0.2cm}\\
    \langle{p}\rangle 
    \equiv \frac{1}{3} \langle{T}^{(3)} \rangle 
    = {1\over 3}
    a^{-2} \delta_{ij} \langle{{T}_{ij}}\rangle 
    \end{cases} 
    \!\!\!\!\!\!\!\,, \ \ \label{eq:Friedmann}
\end{eqnarray}
where $\mpl \equiv (8 \pi G)^{-1/2} \simeq 2.44 \times 10^{18}$ GeV is the reduced Planck mass, $\langle{\rho}\rangle$
and $\langle{p}\rangle$ are the background energy density
and pressure of the different
components of the Universe, and $T^{(3)} \equiv g_{ij} T^{ij} = a^2 \delta_{ij} T^{ij}$.
Here $\langle...\rangle$ denotes volume averaging over sufficiently large regions that encompass well all relevant scales of the excited fields, so that
a well-defined notion of a `homogeneous and isotropic' expanding background is recovered.

For the case where the matter content is completely determined by the field constituents, whose dynamics is described by \Eqq{equations_of_motion},
the stress-energy tensor is given by
\Eq{eq:StressEnergyTensor}
with the replacement $\eta_{\mu\nu}\rightarrow g_{\mu\nu}$,
\begin{align}
    T_{\mu\nu} = &\, T_\mn^\fl + g_{\mu \nu}
    \mathcal{L}_{\rm Bos, IR}
    + F_{\mu}^{\, \, \lambda} F_{\nu \lambda} +
    G^{a \, \lambda}_{\mu} G^a_{\nu \lambda}
    + \partial_{\mu} \phi \, \partial_{\nu} \phi 
    + \Bigl[\bigl(D_{\mu}^A \varphi \bigr)^*
    \bigl(D_{\nu}^A \varphi \bigr) +
    \bigl(D_{\mu} \Phi\bigr)^{\dagger}
    \bigl(D_{\nu} \Phi \bigr) + c.c.\Bigr]\,.
\end{align}
The resulting Friedmann equations become 
\begin{subequations}\label{Friedmann}
\EQalign
    \left({a'\over a}\right)^2 = &\, \frac{a^{2 \alpha}}{3 \mpl^2}\Bigr\langle 
    a^{-2\alpha}T^\fl_{00}
    + {K}_{\phi} + {K}_{\varphi} + {K}_{\Phi} + {G}_{\phi} + {G}_{\varphi} +
    {G}_{\Phi} +   {K}_{U(1)} + {G}_{U(1)} \nonumber \\
    &\, \hspace{77.5mm}
    + {K}_{SU(N)} + {G}_{SU(N)} +  V_0 \Bigr\rangle\,, \label{Friedmann1}\\
    {a''\over a} = &\,
    \frac{a^{2 \alpha}}{3 \mpl^2}\Big\langle {2 \alpha - 1 \over 2 a^{2\alpha}} 
    T^\fl_{00} -
    \frac{1}{2} T_\fl^{(3)} + (\alpha-2) ({K}_{\phi}  + {K}_{\varphi} + K_\Phi ) + \alpha \, ({G}_{\phi} + {G}_{\varphi} + {G}_{\Phi} ) \nonumber  \\
    &\, \hspace{23mm} + (\alpha + 1) V_0 + (\alpha-1)({K}_{U(1)} +  
    {G}_{U(1)} + {K}_{SU(N)} + {G}_{SU(N)} ) \Big\rangle\,. \label{Friedmann2}
\end{mymathbox}
\end{subequations}
\noindent
The different energy components of the scalar and gauge field constituents are given by
\bea
\begin{array}{l} \label{eq:energy-contrib}
    {K}_{\phi} = \frac{1}{2 a^{2\alpha}} {\phi'}^2 \vspace{0.1cm}\\
    {K}_{\varphi} = \frac{1}{a^{2\alpha}}
    |D_0^A \varphi|^2 
    \\
    {K}_{\Phi} = \frac{1}{a^{2\alpha}}
    |D_0 \Phi|^2
    \vspace{0.2cm}\\
\end{array}
\hspace{1cm};\hspace{1cm}
\begin{array}{l}
    {G}_{\phi} = \frac{1}{2 a^2}
    |{\vv\nabla}\phi|^2
    \vspace{0.1cm}\\
    {G}_{\varphi} = \frac{1}{a^2}
    |{\vv D}_A \phi|^2
    \vspace{0.2cm}\\
    {G}_{\Phi} = \frac{1}{a^2}
    |{\vv D}\Phi|^2
    \vspace{0.2cm}\\
\end{array}
\hspace{1cm};\hspace{1cm}
\begin{array}{l}
    {K}_{U(1)} = \frac{1}{2 a^{2(1 + \alpha)}}  \sum_{i} F_{0i}^2
    \vspace{0.1cm}\\
    {G}_{U(1)} = \frac{1}{2 a^4}  \sum_{i,j<i} F_{ij}^2
    \vspace{0.2cm}\\
    {K}_{SU(N)} = \frac{1}{2 a^{2(1 + \alpha)}}  \sum_{a, i} (G_{0i}^a)^2
    \vspace{0.1cm}\\
    {G}_{SU(N)} = \frac{1}{2 a^4}  \sum_{a,i,j<i} (G_{ij}^a)^2
    \vspace{0.2cm}\\
\end{array}
\hspace{0.1cm}.
\\
{\rm(Kinetic-Scalar)} \hspace*{2.4cm} {\rm(Gradient-Scalar)} \hspace*{2.2cm} {\rm (Electric ~\&~ Magnetic)} \hspace{1.5cm}
\nonumber
\eea

For completeness, we may also consider the case where the expansion is sourced by an external background component with equation of state (EOS) $\omega_{\rm ext} \equiv \langle p \rangle/\langle \rho \rangle$. For constant EOS, one obtains the following solutions to  \Eqq{eq:Friedmann} for the scale factor and Hubble parameter, in arbitrary $\alpha$-time~\cite{Figueroa:2020rrl},
\begin{eqnarray}\label{eq:sf_const}
    a(\eta)=a(\eta_i)\left(1+\frac{1}{{\iota}} \HH_i (\eta-\eta_i)\right)^{\iota} \, ,~~~~~~ \HH(\eta)= \frac{ \HH_i }{1+\frac{1}{{\iota}} \HH_i (\eta-\eta_i)}\,, ~~~~~~ {\iota} = \frac{2}{3(1+\omega_{\rm ext})-2\alpha}\, ,
\end{eqnarray}
where $\eta_i$ is some initial time and $\HH_i\equiv \HH(\eta_i)$.

\subsection{Fluid dynamics}
\label{subsec:fld_dynamics_cont}

In this section, we present the EOM describing the
dynamics of an isolated fluid (i.e., not coupled to other {\it dof}) in an expanding Universe.
We first discuss perfect
fluids in \Sec{perfect_fluids},
which we consider in full generality for relativistic velocities.
We then consider viscous fluids in \Sec{viscous_term},
where viscosity is presented only in its subrelativistic formulation.
This section closely follows the
description presented in ref.~\cite{RoperPol:2025lgc}, where more in-depth discussions can be found. The interactions of a fluid with gauge and scalar fields will be covered in \Sec{subsec:bosonic_inters}.

The \textit{perfect} fluid description applies to 
an ensemble of particles in local thermal equilibrium (LTE)~\cite{Rezzolla:2013dea}, i.e., when considering scales
that are much larger than the mean free path scale(s) of the particle ensemble, $\lambda \gg l_{\rm mfp}$ 
(recall the discussion at the beginning of \Sec{sec:cont}).
A perfect fluid is completely characterized in its rest frame by its energy density $\rho_f$ and isotropic pressure $p_\f$, which also determine its enthalpy $w_\f = p_\f +\rho_f$.
In order to describe it in a generic frame we need to introduce the fluid
four-velocity $U^\mu \equiv \dd x^\mu/\dd \tilde{s}$,
with $\tilde{s}$ being the proper time defined from $\dd \tilde{s}^2 \equiv - \dd s^2$, which implies $\dd \tilde{s}= a^{\alpha}\dd \eta/\gamma$. We then have
\begin{equation} \label{four-velocity}
    U^\mu = \gamma a^{-\alpha} (1, u_i)\,, \qquad \text{where \ \, }U^\mu U_\mu = -1\,,\qquad
    \gamma = \frac{1}{\sqrt{1 - a^{2(1 - \alpha )} |\vec{u}|^2}}\,, \quad \text{and} \quad 
    u_i \equiv \frac{\dd x^i}{\dd \eta}\,.
\end{equation}
We note that $\gamma$ in \Eq{four-velocity} is the physical relativistic gamma factor, despite the appearance of $a^{2(1-\alpha)}$ in the denominator. This is because one needs to distinguish carefully the three-velocity $u_i$ defined above with respect to a generic $\alpha$-time $\eta$, from the physical or peculiar velocity $u_i^{\rm p}$, defined with respect to conformal time $\tau$, 
\begin{eqnarray}
u_i \equiv \frac{\dd x^i}{\dd \eta}\,, \qquad 
u_i^{\rm p} \equiv \frac{\dd x^i}{\dd \tau} \qquad \Rightarrow
\qquad u_i^{\rm p} = a^{1 - \alpha} u_i \, .
\end{eqnarray}
The four-velocity in terms of the peculiar velocity is then given by $U^\mu = \gamma(a^{-\alpha}, u_i^{\rm p}/a)$, and the gamma factor becomes the more familiar expression $\gamma = 1/\sqrt{1-|\vec u^{\rm p}|^2}$. The normalization condition of the four-velocity, $U^\mu U_\mu = -1$,
is satisfied either way. In the following we are going to consider the four-velocity as defined in \Eq{four-velocity}.

The total stress-energy tensor of a \textit{perfect fluid} (pf) can then be 
written in a generic frame as 
\begin{eqnarray} \label{perf_fluid}
    T_{\rm pf}^{\mu \nu} = 
    w_\f 
    U^\mu U^\nu + 
    p_\f
    g^{\mu \nu}\, \qquad {\rm \Longrightarrow}\qquad
    \left\lbrace\begin{array}{l}
    T^{00}_\pf = a^{-2\alpha}(w_\f\gamma^2-p_f) \equiv a^{-2\alpha}(\rho_{\rm kin} + \rho_f)\,,  \vspace*{2mm}
    \\
    T^{0i}_\pf = a^{-2\alpha}(w_\f\gamma^2 u_i) \,,\vspace*{2mm}\\ 
    T^{ij}_\pf = a^{-2\alpha}(w_\f\gamma^2 u_i u_j + a^{2(\alpha - 1)} p_\f \delta_{ij})
    \,,
\end{array}\right.
\end{eqnarray}
where $\rho_{\rm kin} \equiv w_f \gamma^2 |\vec{u}_{\rm p}|^2$
is the {\it bulk kinetic} energy density of the fluid. 
To take into account deviations with respect to LTE (see \Sec{viscous_term}), we need to go
beyond the idealization of a perfect fluid, considering instead the dynamics of {\it imperfect} (ipf) fluids.
Indeed, due to interactions among the fluid constituents, or due to interactions with external fields as for example the IR bosonic sectors of scalar and gauge fields introduced at the beginning of \Sec{sec:cont},
the distribution of the particles that form the fluid may exhibit deviations with respect to LTE.
The total stress-energy tensor of an imperfect
fluid can be written as $T^{\mu \nu}_\fl = T^{\mu \nu}_{\rm pf} - \dev^\mn$,
where we refer to
$\dev^\mn \equiv T^\mn_{\rm pf} - T^\mn_\fl$
as the {\it deviatoric}\footnote{In cosmology, the distribution of the dominant field species in the Universe is homogeneous and isotropic, and hence its stress-energy tensor also takes the form of a perfect fluid, which at the background level must be at its rest frame, i.e., with vanishing background velocities $\vec u = 0$. In cosmological perturbation theory, up to linear order in small metric perturbations $|h_{\mu \nu}| \sim \mathcal{O}(\epsilon)$, with $\epsilon \ll 1$,
the perfect fluid form is maintained, though the fluid variables (pressure, energy density, and velocity) are perturbed to $\mathcal{O}(\epsilon)$.
If another contribution $\Pi_{\mn}$ appears in the stress-energy tensor, on top of the (now perturbed) perfect-fluid form, 
this is referred to as the {\it anisotropic} stress tensor, and it is expected to be also (at most) of $\mathcal{O}(\epsilon)$. Therefore, the anisotropic stress and the deviatoric tensor are, in general, not the same. For example, if there are some subdominant species in the energy budget of the Universe and they behave as a fluid, the perfect fluid part of their stress-energy tensor, given that in general it has nonzero transverse and traceless projection, will contribute to the anisotropic stress tensor of the Universe.
} 
stress tensor \cite{2008Kundu,RoperPol:2025lgc}.

The fluid EOM on an expanding FLRW background 
[cf.~\Eqs{eq:EnergyConservation}{eq:MomentumConservation}]
in general contain Hubble friction terms proportional to $\HH$.
However, as we describe in full detail in \Sec{rescaling_fluid_equations}, it is possible, for any $\alpha$-time choice, to partially get rid of these terms
by performing an appropriate rescaling of the fluid stress-energy tensor.
In the following,
we rescale the total stress-energy tensor as
\begin{equation}
    \conf{T}^\mn = a^{4 + 2\alpha} \, T^\mn\,. \label{components_Tmunu}
\end{equation}
The fluid equations for the rescaled stress-energy tensor then become
\begin{align}
    \partial_\mu {\conf T}^{\mu \nu}_\fl + a^{-2\alpha} {\cal H} {\conf T}_\fl  \, \delta^{\nu}_{~\,0} + (1-\alpha) {\cal H} {\conf T}^{0i}_\fl \, \delta^{\nu}_{~\,i}  = {\conf {\cal F}}_{\rm int}^{\nu}\,,
\label{eq:conformalEOM}
\end{align}
where we have rescaled the external forces as ${\conf {\cal F}}_{\rm int}^{\nu} = a^{4+2\alpha} {\cal F}_{\rm int}^{\nu}$, and
introduced the trace
\begin{equation}
    {\conf T}_\fl \equiv g_{\mu \nu} {\conf T}^{\mu \nu}_\fl
    = g_{00} \conf{T}^{00}_\fl + g_{ij} \conf{T}^{ij}_\pf
    = - a^{2\alpha} \conf{T}^{00}_\fl + \conf{T}_\fl^{(3)}\,.
\end{equation}
The rescaled components of the stress-energy tensor are
\begin{equation}
    \conf{T}^{00}_\fl
    = \conf{\rho}_{\rm kin} + \conf{\rho}_\f - \cdev^{00} \,, \qquad 
    \conf{T}^{0i}_\fl = \conf{w}_\f \gamma^2 u_i - \cdev^{0i}  \,, \qquad
    \conf{T}^{ij}_\fl = \conf{w}_\f \gamma^2 u_i u_j + a^{2(\alpha - 1)}
    \conf{p}_\f \delta_{ij} - \cdev^{ij}\,, \label{Tij_perf_fluid}
\end{equation}
with
$\conf{\rho}_\f = a^4 \rho_\f$,
$\conf{p}_\f = a^4 p_\f$, and $\conf{w}_\f = a^4 w_\f$
the comoving fluid energy density, pressure, and enthalpy, respectively, whereas
$\conf{\rho}_{\rm kin} = \conf{w}_\f \gamma^2 a^{2(1 - \alpha)}
|\vec{u}|^2 = \conf{w}_\f \gamma^2 |\vec{u}_{\rm p}|^2$ is
the comoving bulk kinetic energy density.
The factor $a^4$ in \Eq{components_Tmunu}
compensates the decay of the energy density in an expanding
Universe when the fluid is dominated by massless particles, and the factor $a^{2\alpha}$ compensates for the $a^{-2\alpha}$
term from the product of four-velocities in \Eq{perf_fluid},
such that $\conf{T}^\mn_{\rm pf} =
\conf{w}_\f \conf{U}^\mu \conf{U}^\nu + a^{2\alpha} \conf{p}_f g^\mn$, with
$\conf{U}^\mu = a^\alpha U^\mu = \gamma (1, u_i)$ the rescaled four-velocity. 
We note that in the rescaled fluid EOM we have a term including the trace of ${\conf T}^{\mu \nu}_\fl = {\conf T}^{\mu \nu}_{\rm pf} - \cdev^\mn$, 
which is given by
\begin{align}
    {\conf T}_\fl \equiv g_{\mu \nu} (\conf{T}^{\mu \nu}_{\rm pf} - \cdev^{\mu\nu}) = \conf{T}_{\rm pf} - \cdev =
    a^{2\alpha} (3 \conf{p}_\f - \conf{\rho}_\f) - \cdev\,.
\label{trace_fluid}
\end{align}
Inspecting \Eq{eq:conformalEOM},
we observe that if the trace of the stress-energy tensor vanishes, i.e., $\conf{T}_\fl = 0$, choosing conformal time ($\alpha = 1$) leads to {\it conformally flat} equations, i.e., $\partial_\mu \conf{T}^\mn_\fl = \conf{\cal F}^\nu_{\rm int}$. This is particularly relevant for perfect fluids composed
of relativistic particles, as for them $\dev^\mn = 0$
and ${\conf p}_\f = \conf{\rho}_\f/3$, so that the trace naturally vanishes. 
Furthermore, in such cases, the comoving components are given by $\conf{T}^\mn = a^6 T^\mn$, and their
dynamics in conformal time
becomes explicitly independent of the scale factor $a(\tau)$ \cite{Brandenburg:1996fc,Subramanian:1997gi,RoperPol:2025lgc}.

Plugging the decomposition $T_\fl^\mn  \equiv  T^{\mn}_{\rm pf} - \dev^{\mn}$ in 
\Eq{eq:conformalEOM}, and passing all Hubble friction terms to the \rhs of the equations allows to rewrite conveniently the EOM for the evolution of an imperfect fluid in an expanding Universe sourced by the forces due to the fluid interaction(s) with other $dof$, precisely as
\EQb
    \partial_\mu \conf{T}^\mn_{\rm pf} =
    \conf{\cal F}^\nu_{\rm int} + \conf{\cal F}^\nu_{\rm ipf} + \conf{\cal F}^\nu_H\,.
    \label{general_eqs}
\end{mymathbox}
\noindent
In this equation, the Hubble friction (for the perfect fluid part) is defined as
\begin{equation} \label{Hubble_friction}
    \conf{\cal F}^\nu_H 
    \equiv \partial_\mu \conf{T}^{\mu \nu}_{\rm pf} - a^{4+2\alpha} D_\mu^G T^{\mu \nu}_{\rm pf}
    = - a^{-2\alpha} \conf{T}_\pf\, \HH \delta^{\nu}_{~\,0} +
    (\alpha - 1) \HH \,\conf{T}_\pf^{0i} \delta^{\nu}_{~\,i}
    = 
    (\conf{\rho}_\f - 3 \conf{p}_\f)\, \HH \delta^{\nu}_{~\,0} +
    (\alpha - 1) \HH \,\conf{T}_\pf^{0i} \delta^{\nu}_{~\,i}
    \,.
\end{equation}
The force from the imperfect fluid term reads
\begin{equation} \label{ipf_force}
    \conf{\cal F}_{\rm ipf}^\nu 
    \equiv a^{4 + 2 \alpha} D_\mu^G {\dev}^\mn
    = \partial_\mu \cdev^\mn + a^{-2\alpha} \cdev\, \HH \delta^{\nu}_{~\,0} + (1 - \alpha)\HH\, \cdev^{0i} \delta^{\nu}_{~\,i}\,,
\end{equation}
which receives contributions of $\cdev_{\mn} \equiv a^{4+2\alpha} \dev_{\mu \nu}$, with trace $\cdev \equiv  g_{\mu \nu} \cdev^{\mu \nu}$, from both the divergence of the rescaled
deviatoric stress tensor and Hubble friction terms. 

\EEqss{general_eqs}{ipf_force}
are then equivalent to our previous \Eqs{eq:EnergyConservation}{eq:MomentumConservation}. The contributions of the fluid to the total energy density and pressure that enter the Friedmann equations in \Eq{Friedmann} are then
given by the following expressions 
\begin{subequations}
\begin{align}
    \langle\rho\rangle \supset &\, a^{-2 \alpha} \bra{T_{00}^\fl} = 
    a^{-4} \bra{\conf{\rho}_{\rm kin} + \conf{\rho}_\f -
    \cdev^{00}}\,, \\ 
    \langle p \rangle \supset &\, \frac{1}{3}
    \bra{T^{(3)}_\fl} = \frac{1}{3}  a^{-4} \bra{\conf{\rho}_{\rm kin} + 3 \conf{p}_\f -
    a^{-2 \alpha} \cdev^{(3)}}\,.
\end{align}
\end{subequations}
In the following sections, we consider specific
cases of \Eq{general_eqs}: {\em i}) dynamics of an isolated perfect fluid in an expanding background, neglecting interactions with bosonic fields and/or deviations with respect to LTE, i.e., with
${\cal F}_{\rm int}^\nu = {\cal F}_{\rm ipf}^\nu = 0$ (\Sec{perfect_fluids}), {\em ii)}
dynamics in an expanding background of imperfect (viscous) fluids due to deviations with respect to
LTE, i.e., with ${\cal F}_{\rm ipf}^\nu \neq 0$, thus allowing for a realistic increase of the entropy in the fluid (\Sec{viscous_term}),
{\em iii)} fluid dynamics in expanding backgrounds with electromagnetic (Lorentz) forces due to coupling of the fluid with Abelian gauge fields, i.e., with ${\cal F}_{\rm int}^\nu = {\cal F}_{U(1)}^\nu \neq 0$
(\Sec{subsec:gauge_fluid_dyns}), and {\em iv)}
fluid dynamics in an expanding background with external forces due to interactions of the fluid with a singlet scalar field, i.e., with ${\cal F}_{\rm int}^\nu = {\cal F}_{\phi}^\nu \neq 0$ (\Sec{subsec:scalar_fluid}).

In general, the number of fluid equations is insufficient to describe the dynamics of the fluid variables. This is the case already for perfect fluids and first-order imperfect fluids (see \Sec{viscous_term}), for which we have $4$ equations and $5$ fluid variables ($u_1, u_2, u_3, p_f, \rho_\f$).
Therefore, we need to introduce
an equation of state (EOS) that
relates the pressure and the energy density\footnote{For a general description of relativistic
imperfect fluids or non-Newtonian fluids (i.e., fluids for which Navier-Stokes
shear stress is not a valid description), one needs additional constitutive
relations or dynamical equations
to describe all components of $\dev^\mn$ \cite{2008Kundu,Romatschke:2009im,Rezzolla:2013dea}.
} 
as 
\begin{eqnarray}\label{eq:EOS}
  p_\f = \omega \, \rho_\f\,.   
\end{eqnarray}
It is sometimes useful to describe the relation between pressure and energy density in terms of the speed of sound,
\begin{eqnarray} \label{general_sound_speed}
    \cs^2 \equiv \frac{\partial p_f}{\partial \rho_f}\biggr|_{\rm s}\,,
\end{eqnarray}
where the subscript $s$ indicates derivation while keeping the entropy constant, which happens in LTE.
In general, $c_{\rm s}^2 \neq \omega$, however, when the EOS is constant, both quantities are equal, e.g., for a pure radiation perfect fluid,
$\omega = \cs^2 = 1/3$.
In the case of couplings of the fluid with a singlet scalar field (see \Sec{subsec:scalar_fluid}), instead, it is not possible in general to have a constant 
EOS, as both pressure and energy density depend on the scalar field and temperature.
However, in those cases it is still
possible to fully describe the fluid
pressure in terms of the scalar field and the energy density, allowing
us to close the system.

\subsubsection{Perfect fluid dynamics}
\label{perfect_fluids}

\subsubsubsection*{Conservation form of fluid dynamics}
\label{conservation_form}

In this section, we assume that the particle ensemble that forms the fluid can maintain LTE, so that we specialize the EOM [cf.~\Eq{general_eqs}]
for a perfect fluid in an expanding background, in the absence of external forces (${\cal F}_{\rm int}^\nu = {\cal F}_{\rm ipf}^\nu =  0$).
The EOM then reduce to
\begin{subequations}
\label{eqs_fluid}
\EQalign
    \partial_0 \conf{T}^{00}_\pf + \partial_i \conf{T}^{i0}_\pf  &= (\conf{\rho}_\f - 3 \conf{p}_\f) \HH = {\conf{\cal F}_H^0} \,, 
    \\
    \partial_0 \conf{T}^{0i}_\pf + \partial_j \conf{T}^{ji}_\pf &=
    (\alpha - 1) \HH 
    {\conf T}^{0i}_{\rm pf} = {\conf{\cal F}_H^i} \,.
    \end{mymathbox}
\end{subequations}
\noindent
A possible strategy to solve this system is the so-called
{\it conservation} form, in which $\lbrace\conf{T}^{0\mu}_\pf\rbrace$,
with $\mu = 0, 1, 2, 3$, are directly considered the dynamical variables to solve for.\footnote{In the {\it conservation} form, the variables $\conf{T}^{0\mu}_{\rm pf}$ evolve through external forces and the divergence terms $\partial_i \conf{T}^{i\mu}_\pf$. As a consequence, the volume averages of $\conf{T}^{0\mu}_\pf$ are conserved at the discrete level, in the absence of external forces and for periodic boundaries in the lattice, as in the continuum.}
Here we follow a similar approach to that described
in Section 3.4 of ref.~\cite{RoperPol:2025lgc}, by noting that in order to advance in time the temporal components $\conf{T}^{0\mu}_\pf$ according to \Eqq{eqs_fluid},
we need to find a closed expression for $\conf{T}^{ij}_\pf$
in terms of the former.

First of all, using \Eq{Tij_perf_fluid}
we can express the primitive fluid variables, i.e., the energy density and the velocity of the fluid,
in terms of the perfect fluid temporal components $\conf{T}^{0 \mu}_\pf$, as
\EQb
    \conf{\rho}_f = \frac{1}{z} \conf{T}^{00}_\pf
    \qquad \text{and \ } \qquad u_i =   \frac{z}{z + \omega}
    \frac{\conf{T}^{0i}_\pf}{\conf{T}^{00}_\pf}\,, \qquad
    \text{with \ }~~ z = (1 + \omega) \gamma^2 - \omega\,,
    \label{rho_u_fromT0mu}
\end{mymathbox} \noindent
where $\omega$ relates the pressure and the energy density as ${\conf p}_\f = \omega \conf{\rho}_\f$. Notice that, since $\gamma^2 \geq 1$, we have that $z \geq 1$ for any value of $\omega$.
This leads to construct $\conf{T}_{ij}^\pf$ from \Eq{Tij_perf_fluid} as
\EQb
    \label{Tij_gamma2}
    \conf{T}^{ij}_\pf =  \frac{z}{z + \omega}
    \frac{\conf{T}^{0i}_\pf \conf{T}^{0j}_\pf}{
    \conf{T}^{00}_\pf} + \frac{\omega}{z}  \, a^{2(\alpha - 1)}
    \conf{T}^{00}_\pf \delta_{ij}\,,
\end{mymathbox} \noindent
which is expressed in terms of $\conf{T}^{0\mu}_\pf$, but still
depends on $z$ (and, hence,
on the velocity through $\gamma^2$ and on ${\conf \rho}_f$ through $\omega$). To obtain $z$ in terms of our dynamical variables,
we first write
\begin{equation}
    r^2 \equiv \sum_i
    \Biggl[\frac{\conf{T}^{0i}_\pf}{\conf{T}^{00}_\pf}\Biggr]^2  a^{2(1 - \alpha)}
    = \frac{(z + \omega)^2}{z^2} |\vec{u}_{\rm p}|^2
    = \frac{(z + \omega)(z - 1)}{z^2}
    \label{T0i_sq}
    \,,
\end{equation}
which implies that $r^2 < 1$ for physical $|\omega| \leq 1$. 
Inverting this relation, we obtain
\EQb
    z = \frac{1}{2 (1 - r^2)}
    \Bigl(1 - \omega + \sqrt{(1 + \omega)^2 - 4 \omega r^2}\Bigr)
    \,, \label{gamma2_rel}
\end{mymathbox}
\noindent
where we took the positive value of the radicand to ensure that
$z \geq 1$. This then guarantees that the gamma factor, recovered from $\gamma^2 = (z + \omega)/(1 + \omega)$, actually verifies $\gamma \geq 1$
for physical $|\omega| \leq 1$.

Plugging \Eq{gamma2_rel} into \Eq{Tij_gamma2} allows to compute $\conf{T}^{ij}_\pf$ in terms of our dynamical variables. However, in practice, this trick is only useful if
$\omega$ is
independent of the fluid variables.
In that case, $z$ only depends on $r$ and $\omega$, 
and thus we can directly advance the system of \Eqq{eqs_fluid}
by plugging into them
\Eqss{rho_u_fromT0mu}{gamma2_rel}.
If we want to recover the primitive fluid variables $\lbrace
\conf{u}_1, \conf{u}_2, \conf{u}_3, \conf{p}_f, \conf{\rho}_f \rbrace$, we just need to use \Eqs{eq:EOS}{rho_u_fromT0mu}. Furthermore, if the EOS is constant (both in space and time), the sound speed also becomes constant and equal to $\omega$,
\begin{eqnarray} \label{sound_speed}
\cs^2 =\omega = {p_f\over\rho_f} \,, \qquad {\rm if}~~~\omega = const.
\end{eqnarray}
This is the case, for example, of a fluid formed by free relativistic species, for which $\omega = \cs^2 = 1/3$. On the other hand, when $w$ depends on the fluid primitive variables, this procedure needs to be generalized and adapted in a case-by-case basis.
We will see an example of this when we introduce interactions of the fluid with a scalar field in \Sec{subsec:scalar_fluid}.

\subsubsection*{Limit of subrelativistic fluid velocity}

In the limit of subrelativistic fluid bulk motion we take $|\vec{u}_{\rm p}|^2 \ll 1$.
Then, the variable $z = (1 + \omega) \gamma^2 - \omega$
reduces to $1$.
However, as shown in ref.~\cite{RoperPol:2025lgc} (see Sec.~3.4 there), taking $z=1$ 
directly in
the relation between the stress-energy tensor
components [cf.~\Eq{Tij_gamma2}]
leads to the wrong subrelativistic limit of the EOM.
This is due to the fact that the time derivative of the Lorentz factor, $\partial_0 \gamma^2$, contains terms that are non-negligible in the subrelativistic limit. 
To recover the correct limit, it is required to expand
$z$ up to first order in $r^2 \sim {\cal O} (|\vec{u}_{\rm p}|^2)$
using \Eq{gamma2_rel},
\begin{equation}
    z = 1 + \frac{r^2}{1 + \omega} + {\cal O} (|\vec{u}_{\rm p}|^4)\,.
\end{equation}
Then, the corrected expression to relate $\conf{T}^{ij}_\pf$ to the dynamical
variables $\conf{T}^{0\mu}_\pf$ becomes 
\EQb 
    \lim_{|\vec{u}_{\rm p}|^2 \ll 1}
    \conf{T}^{ij}_\pf = \frac{1}{1 + \omega} \frac{\conf{T}^{0i}_\pf \conf{T}^{0j}_\pf}
    {\conf{T}^{00}_\pf} \biggl(1 + \frac{r^2 \omega}{(1 + \omega)^2} \biggr) +
    \omega \, a^{2(\alpha - 1)}
    \conf{T}^{00}_\pf \biggl(1 - \frac{r^2}{1 + \omega} \biggr) \, \delta_{ij}\,. \label{eq:Tij_sub_rel}
\end{mymathbox} \noindent

\subsubsubsection*{Non-conservation form of fluid dynamics}
\label{nonconser_relativistic}

We now proceed to find EOM for the
primitive fluid variables $\conf{\rho}_f$ and $u_i$ in the so-called non-conservation form
of fluid dynamics.
In first place, we write the EOM, taking
the temporal and spatial components of \Eqq{eqs_fluid}, explicitly in
terms of $\conf{p}_f$, $\conf{w}_f$, and $u_i$,
\begin{subequations}
\label{fluid_noncons_equations}
\begin{align}
    \partial_0 (\conf{w}_f \gamma^2) \, - & \, \partial_0 \conf{p}_f +
    \partial_i
    (\conf{w}_f \gamma^2 u_i) = \conf{\cal F}_H^0
    \,, \label{continuity} \\
    \conf{w}_f \gamma^2  \partial_0 u_i + & \, u_i
    \partial_0 (\conf{w}_f \gamma^2) +
    \partial_j (\conf{w}_f \gamma^2 u_i u_j) 
    + a^{2(\alpha - 1)} \partial_i \conf{p}_f = \conf{\cal F}_H^i \,. \label{momentum}
\end{align}
\end{subequations}
Taking
\Eq{continuity} multiplied by $u_i$ and subtracting it from
\Eq{momentum}, we find
\begin{equation}\label{momentum2}
    \conf{w}_f  \gamma^2 \partial_0 u_i + u_i \partial_0 \conf{p}_f
    + 
    \conf{w}_f \gamma^2 u_j \partial_j u_i   + a^{2(\alpha - 1)}
    \partial_i \conf{p}_f = \conf{\cal F}_H^{i} - u_i
    \conf{\cal F}_H^{0}\,.
\end{equation}
Then, contracting this equation with $u_i$, and using the following identities,
\begin{subequations}
\begin{align}
    \partial_0 \ln \gamma &= a^{2(1 - \alpha)} \bigl[\gamma^2 u_i \partial_0 u_i
    -(\alpha - 1) \HH \gamma^2 {\vec u}^2\bigr]\,, \\
    u_j \partial_j \ln \gamma & = a^{2(1-\alpha)} \gamma^2 u_i u_j \partial_j u_i \,,
\end{align}
\end{subequations}
we find an evolution equation for $\gamma$,
\begin{equation}
    \conf{w}_f (\partial_0 + u_i \partial_i) \ln \gamma = - |\vec{u}_{\rm p}|^2
    \bigl(\partial_0 \conf{p}_f + \conf{\cal F}_H^0\bigr) - u_i \partial_i \conf{p}_f 
    +
    \underbrace{\biggl[
    a^{2(1-\alpha)} u_i \conf{\cal F}_H^i - \conf{w}_f (\alpha-1) {\cal H} \gamma^2 |\vec{u}_{\rm p}|^2
    \biggr]}_{=0}\,.
    \label{Dt_lngamma}
\end{equation}
Notice that using the explicit form of the Hubble friction force,
$\conf{\cal F}_H^i = (\alpha - 1) \HH \conf{w}_f \gamma^2 u_i$,
the term inside square brackets vanishes.
However, we write this term
explicitly in order to be able to trivially
generalize the equations to include external (interaction)
forces $\conf{\cal F}_{\rm int}^\nu$ in the following by replacing $\conf{\cal F}_H^\nu \to \conf{\cal F}_{\rm int}^\nu$.
We can now substitute \Eq{Dt_lngamma} into
\Eqs{continuity}{momentum2},
using $\partial_0 \gamma ^2 = 2 \gamma ^2 \partial_0 \ln \gamma$,
to find
\begin{subequations}
\label{equations_relativistic_general}
\begin{align}
    \partial_0 \conf{\rho}_f -
    |\vec{u}_{\rm p}|^2 \partial_0 \conf{p}_f  = & -
    \conf{w}_f \partial_i u_i
    - u_i \partial_i (\conf{\rho}_f - \conf{p}_f)
    + \conf{\cal F}_H^0
    (1 +
    |\vec{u}_{\rm p}|^2)
    -
    2 \underbrace{\biggl[ a^{2(1-\alpha)} u_i \conf{\cal F}_H^i - \conf{w}_\f (\alpha-1) \HH \gamma^2 |\vec{u}_{\rm p}|^2 \biggr]}_{=0}
    \,,
    \label{relativistic_density_general}
    \\
    \partial_0 u_i + u_j \partial_j u_i 
    = &  - \frac{u_i}{\conf{w}_\f \gamma^2}
    \bigl(\partial_0 \conf{p}_\f + \conf{\cal F}_H^0 \bigr)
    - a^{2(\alpha-1)} \frac{\partial_i \conf{p}_\f}{\conf{w}_\f \gamma^2} + \frac{\conf{\cal F}_H^i}{\conf{w}_\f \gamma^2}\,.
    \label{relativistic_momentum_general}
\end{align}
\end{subequations}
Taking into account that for the given form of the Hubble friction,
the values in the square brackets vanish,
the energy conservation equation
simplifies to 
\begin{align}
    \label{relativistic_eqs}
    \partial_0 \conf{\rho}_f -
    |\vec{u}_{\rm p}|^2 \partial_0 \conf{p}_f = &  -
    \conf{w}_f \partial_i u_i
    - u_i \partial_i (\conf{\rho}_f - \conf{p}_f)
    + \conf{\cal F}_H^0
    (1 +
    |\vec{u}_{\rm p}|^2)\,.
\end{align}
As in the conservation case, we need to provide an EOS to
close the system.
Assuming a constant EOS,
$\omega = \conf{p}_f/\conf{\rho}_f =  const.$, we find
the equations of conservation of energy and momentum in the relativistic regime of fluid bulk motion
\cite{RoperPol:2025lgc},
\begin{subequations}
\label{rho_u_eqs}
\EQalign
    & \, \hspace{-6mm}  (1 - \omega  |\vec{u}_{\rm p}|^2)\, \partial_0 \ln \conf{\rho}_f \, + 
    (1 + \omega) \partial_i u_i
    + (1 - \omega) u_i \partial_i \ln \conf{\rho}_f = (1 - 3 \omega) \HH ( 1+ 
    |\vec{u}_{\rm p}|^2)\,,  \label{rho_dens} \\ 
    &\,\hspace{-6mm}  \partial_0 u_i + u_j \partial_j u_i =  
    \frac{1 - |\vec{u}_{\rm p}|^2}{1 - \omega |\vec{u}_{\rm p}|^2}
    \biggl[ \omega \partial_j u_j +  \omega \frac{1 - \omega}{1 + \omega} u_j  \partial_j \ln \conf{\rho}_f
    + (3 \omega - 1) \HH
    \biggr] u_i \nonumber \\   &\, \hspace{17mm} -
    \frac{\omega}{1 + \omega}(1 - |\vec{u}_{\rm p}|^2) \,
    a^{2(\alpha - 1)} \, \partial_i \ln  \conf{\rho}_f + (\alpha - 1) \HH u_i \,, \qquad \text{for a constant 
    $\omega = p_f/\rho_f$} \,.  \label{u_vel}
\end{mymathbox}
\end{subequations}
\noindent
In terms of the peculiar velocity ${u}^{\rm p}_i = a^{1 - \alpha} u_i$,
the Hubble friction $\conf{\cal F}_H^i$
combines with the time derivative, $\partial_0 u_i = a^{\alpha - 1}
\partial_0 u_i^{\rm p} + a^{\alpha - 1}(\alpha - 1) \HH u_i^{\rm p}$, and
the momentum equation becomes
\begin{align}
    \partial_0 u_i^{\rm p} + a^{\alpha -1}
    u_j^{\rm p} \partial_j u_i^{\rm p} =  &\,
    \frac{1 - |\vec{u}_{{\rm p}}|^2}{1 - \omega |\vec{u}_{\rm p}|^2}
    \biggl[ a^{\alpha - 1} \omega \partial_j u_j^{\rm p} +
    a^{\alpha - 1} \omega \frac{1 - \omega}{1 + \omega}
    u_j^{\rm p}  \partial_j \ln \conf{\rho}_f
    + (3 \omega - 1) \HH
    \biggr] u_i^{\rm p}
    \nonumber \\   &\, - \frac{\omega}{1 + \omega} (1 - |\vec{u}_{\rm p}|^2) 
    \, a^{\alpha - 1} \partial_i \ln  \conf{\rho}_f
    \,, \qquad \text{for a constant 
    $\omega = p_f/\rho_f$}\,.
\end{align}

\subsubsection*{Limit of subrelativistic fluid velocity}

The continuity and momentum equations for the energy density
and peculiar velocity [cf.~\Eqq{rho_u_eqs}],
can be taken in the subrelativistic limit of fluid velocity,
$|\vec{u}_{\rm p}|^2 \ll 1$,
\begin{subequations}
\EQalign
    \lim_{|\vec{u}_{\rm p}|^2 \ll 1} \partial_0 \ln \conf{\rho}_f = &\,
    - (1 + \omega) \partial_i u_i
    - (1 - \omega) u_i \partial_i \ln \conf{\rho}_f +  (1 - 3 \omega) \HH\,,
    \label{rho_dens_sub} \\ 
    \lim_{|\vec{u}_{\rm p}|^2 \ll 1} \partial_0 u_i = &\, - u_j \partial_j u_i +
    u_i \, \omega \biggl(\partial_j u_j
    + \frac{1 - \omega}
    {1 + \omega} u_j  \partial_j \ln \conf{\rho}_f \biggr)
     + (3 \omega + \alpha - 2) \HH \, u_i 
    \nonumber \\ &\, - \frac{\omega}{1 + \omega} a^{2(\alpha - 1)} \partial_i \ln \conf{\rho}_f\,,  \qquad \text{for a constant
    $\omega = p_f/\rho_f$}\,.\label{u_vel_sub}
\end{mymathbox}
\end{subequations}
\noindent
From these equations, we find that a suitable choice to get rid of the Hubble friction
in the momentum equation [cf.~\Eq{u_vel_sub}] is\footnote{The value $\alpha = 2$ coincides with the so-called choice of {\it super-comoving} coordinates in cosmological simulations with fluids formed by non-relativistic species, for which $\omega \to 0$~\cite{Martel:1997hk,RoperPol:2025lgc}.} $\alpha = 2 - 3 \omega$.
As mentioned above, if one takes $\gamma^2 \to 1$ in the time derivatives, 
i.e., taking $\conf{T}^{00}_\pf \to \conf{\rho}_f$
and $\conf{T}_{0i} \to \conf{w}_f u_i$
in \Eqq{eqs_fluid} to get to \Eqq{fluid_noncons_equations},
the subrelativistic limit of the fluid EOM found is incorrect, see ref.~\cite{RoperPol:2025lgc}.
This is a consequence of neglecting the derivative of the Lorentz factor in the subrelativistic limit,
\begin{equation}
    \lim_{|\vec{u}_{\rm p}|^2 \ll 1} \partial_0 \ln \gamma = -
    \frac{\omega}{1 + \omega} u_i \partial_i \ln \conf{\rho}_f \,,
    \qquad \text{for a constant 
    $\omega = p_f/\rho_f$}\,,
\end{equation}
which is of the same order as other terms appearing in the subrelativistic EOM, and omitted
in previous work as, e.g., in ref.~\cite{Brandenburg:1996fc}.

\subsubsection*{Alternative non-conservation form for relativistic fluid variables}\label{Sec:AlternativeFormulation}

An equivalent form of the fluid EOM is sometimes used in relativistic fluid dynamics \cite{2003rnh..book.....W,Rezzolla:2013dea}, 
where the EOM
are expressed for the relativistic energy density $\conf{E} \equiv
\gamma \conf{\rho}_f$ and the
fluid momentum density $\conf{Z}_i \equiv \conf{T}^{0i}$.
This corresponds to a non-conservation form with an alternative choice of the primitive fluid
variables, $\conf{E}$ and $\conf{Z}_i$. To obtain the EOM for the comoving relativistic energy density,
$\conf{E}$, we project the conservation equation for the perfect fluid stress-energy tensor of \Eq{perf_fluid}
in the parallel direction to $U_\nu$, 
\begin{align}
    - U_\nu D_\mu^G T^\mn_\pf = U^\mu \partial_\mu \rho_f + w_f D_\mu^G U^\mu
    &= U^{\mu} \partial_\mu \rho_\f + w_\f [\partial_\mu U^\mu + (3+\alpha) \HH U^0] \nonumber \\
    & = a^{-(4+\alpha)} \bigl[\partial_\mu (\conf{\rho}_f \conf{U}^\mu ) + \conf{p}_f
    \partial_\mu \conf{U}^\mu
    +(3\conf{p}_\f - \conf{\rho}_\f) \HH \conf{U}^0
    \bigr] = 0 \,.\label{cons_energy}
\end{align}
Then, the EOM for $\conf{E}$ and for $\conf{Z}_i$, which can be
obtained from the space components of \Eq{general_eqs},
are
\begin{subequations}
\label{relativistic_EZeqs}
\EQalign
    &\,
    \partial_0 \conf{E}\, + 
    \partial_i (\conf{E} u_i) + \conf{p}_f \bigl[\partial_0 \gamma + \partial_i (\gamma u_i)\bigr]
    = (\conf{E} - 3 \conf{p}_f \gamma) \HH\,,
    \label{relativistic_Eeq} \\ 
     &\, \partial_0 \conf{Z}_i +
    \partial_j (\conf{Z}_i u_j) + a^{2(\alpha - 1)}
    \partial_i \conf{p}_f = \conf{\cal F}_H^i\,.
    \label{relativistic_Zeq}
\end{mymathbox}
\end{subequations}
\noindent
As before, we note that neglecting $\gamma$ in $\conf{E} = \gamma \conf{\rho}_f$ or in $\conf{Z}_i = \conf{w}_f \gamma^2 u_i$
would lead to the wrong subrelativistic limits due to the inconsistent
underlying assumption that $\partial_0 \gamma$ is negligible.

\subsubsection{First-order imperfect fluid dynamics}
\label{viscous_term}

The perfect fluid description applies for fluids that are exactly in LTE, characterized by an
equilibrium distribution function $f_{\rm eq}$, which can be the Bose-Einstein distribution for bosons, or the Fermi-Dirac distribution
for fermions \cite{2002rbet.book.....C,Rezzolla:2013dea}.
Both distributions reduce to
the Maxwell-Boltzmann (or Maxwell-J\"uttner 
for relativistic particle velocities)
in the classical limit.
Perfect fluids, however, cannot account for dissipative effects since, as we will show in this section, their entropy is conserved. Hence, the
description of any realistic fluid requires to consider deviations with respect to LTE.

Let us then consider the perturbative expansion around LTE for small
(but non-zero) Knudsen number, which is given by the
ratio between the particle mean free path and the characteristic fluid scales
$\lambda$, ${\rm Kn} = l_{\rm mfp}/\lambda$,
such that the distribution function is
\begin{equation}
    f = f_{\rm eq} +
\varepsilon f_1 + {\cal O} (\varepsilon^2)\,, \qquad \text{where \,}~~ 
f_1/f_{\rm eq} \sim {\cal O} (1), \quad \text{and}
    \quad \varepsilon \sim {\cal O} ({\rm Kn}) \ll 1\,,
    \label{eq:knudsen_expansion}
\end{equation}
with $f_{\rm eq}$ the equilibrium distribution function.
The perfect fluid description is recovered when $\varepsilon = 0$
(i.e., Kn = 0), leading to the perfect fluid stress-energy tensor
$T^\mn_\pf = w_f U^\mu U^\nu + p_\f \, g^{\mu \nu}$ [cf.~\Eq{perf_fluid}].
A small Knudsen number corresponds then to a highly collisional fluid.
The inclusion of $\delta f \equiv f - f_{\rm eq} \approx \varepsilon f_1$
leads to an anisotropic stress tensor that is
characterized by the deviatoric tensor $\dev^{\mu \nu}$,
already introduced above \Eq{general_eqs},
such that
$T^\mn_\fl = T^\mn_\pf - \dev^\mn$.

In order to deduce the form of $\dev^{\mu \nu}$, we need to understand which quantities characterizing the fluid are small in the limit ${\rm Kn} \ll 1$, such that we can perform a perturbative expansion over them. In particular, as we will show in the following, the transport coefficients describing dissipative effects are proportional to the mean free path of the particles, hence the expansion over ${\rm Kn} = l_{\rm mfp}/\lambda$ corresponds to an expansion over transport coefficients $\propto l_{\rm mfp}$ times gradients of the fluid variables $(\partial_\mu |\vec{u}|)/|\vec{u}|, \, 
(\partial_\mu \rho)/\rho
\propto 1/\lambda$.
Since for imperfect fluids
the definition of the
four-velocity is not unique \cite{PhysRev.58.919,Landau1987Fluid}
(see also discussion in refs.~\cite{Romatschke:2009im,Rezzolla:2013dea,RoperPol:2025lgc}),
we restrict to the Landau frame in the following.
In this frame, the four-velocity is defined as
an eigenvector of the stress-energy
tensor, $U_\mu T^\mn_\fl = - \rho_f U^\nu$.
Hence, the
deviatoric tensor in the Landau frame satisfies $U_\mu \dev^\mn = 0$.
This condition, together with a linear expansion around a small
Knudsen number,
leads to the Navier-Stokes stress-energy tensor,
corresponding to the most general rank-2 tensor that is linear in gradients
of the velocity field
\cite{Romatschke:2009im,Rezzolla:2013dea},
\EQb \label{Pi_visc}
    \dev^{\mu \nu} = \pi^\mn + \tfrac{1}{3} h^\mn \dev = 2 \, \nu \, w_f \, \sigma^{\mu \nu} + \xi \, w_f
    \, \theta \, h^{\mu \nu}\,,
\end{mymathbox}
\noindent
where $\nu \geq 0$ and $\xi \geq 0$ are the transport coefficients corresponding
to the kinematic
shear and bulk viscosities;
$\pi^\mn \equiv 2\, \nu\, w_f\, \sigma^\mn$ is the shear stress tensor with zero trace, $\pi = g_\mn \pi^\mn = 0$;
$h^{\mu \nu} = g^{\mu \nu} + U^\mu U^\nu$ is the projection tensor,
orthogonal to $U_\mu$, i.e., satisfying $U_\mu h^{\mu \nu} = 0$,
with trace $h = g_\mn h^\mn = 3$; $\theta = D_\mu^G U^\mu$
is the fluid expansion scalar,
and
$\sigma^{\mu \nu}$ is the traceless rate-of-strain tensor, defined as
\begin{equation} \label{sigma_munu}
    \sigma^{\mu \nu} = S^{\mu \nu} - {1 \over 3} \theta h^{\mu \nu}\,,  \qquad \text{where \ }
    S^{\mu \nu} =
    h^{(\mu \lambda} D_\lambda^G U^{\nu)} \quad
    \text{with \ } 
    S^\mu_{\ \mu} = \theta\,.
\end{equation}
The parentheses indicate symmetrization, in the free indices,
such that $S^\mn = \tfrac{1}{2}(h^{\mu \lambda} D_\lambda^G U^\nu + h^{\nu\lambda} D_\lambda^G U^\mu)$.
Notice that $\sigma^{\mu \nu}$ and $\theta$ are linear in the
gradients of the velocity, such that $\dev^{\mu \nu} \sim {\cal O}({\rm Kn}) \sim \delta f$.
Given that the trace of $\sigma^\mn$ vanishes,
only the bulk viscosity contributes to the trace of the deviatoric tensor $\dev \equiv g_{\mu \nu} \dev^{\mu \nu} =  3 \, \xi \, w_f\, \theta$.

The Navier-Stokes description with positive viscous coefficients
ensures that the entropy four-current
$s^\mu_f \equiv s_f U^\mu$
satisfies the second law of thermodynamics
$\partial_\mu s_f^\mu \geq 0$ (in flat spacetime).
To show this, we recast the equation of conservation of energy
for the fluid
in the following way [cf.~\Eq{cons_energy}],
\begin{equation}
    - U_\nu D_\mu^G T^\mn_\pf = U^\mu \partial_\mu \rho_f + w_f
    \, \theta
    = - U_\nu D_\mu^G \dev^\mn\,.
    \label{cons_energy2}
\end{equation}
The \rhs can be simplified using
the following relation,
\begin{equation} \label{div_Pi}
    U_\nu D_\mu^G \dev^\mn = D_\mu^G (U_\nu \dev^\mn) - 
    \dev^\mn D_{(\mu}^G U_{\nu)} =  - \dev^\mn S_\mn = - \pi^\mn S_\mn - \tfrac{1}{3} \, \dev \, \theta\,,
\end{equation}
where we have used the Landau frame condition
$U_\nu \dev^\mn = 0$ in the second equality.
We now introduce the thermodynamic relations from the first principle of thermodynamics for zero chemical potential,
$T s_f = w_f$ and $T \dd s_f = \dd \rho_f$,
in \Eq{cons_energy2},
to find the entropy evolution equation,
\begin{equation} \label{entropy}
    D_\mu^G s^\mu_f
    = \frac{1}{T} \dev^\mn S_\mn = 
    \frac{w_f}{T} \bigl(2 \,\nu \, \sigma^\mn \sigma_\mn + \xi\,
    \theta^2 \bigr) \geq 0\,.
\end{equation}
Notice that, as anticipated, for a perfect fluid the \rhs is zero, hence entropy is conserved, i.e., $D_\mu^G s_f^\mu = 0$.
For an imperfect fluid, instead, entropy increases, as it should from the second principle of thermodynamics. 

As mentioned above, the transport coefficients $\nu$ and $\xi$ can be related to the mean free path of the plasma $l_{\rm mfp}$ as
\cite{Weinberg:1971mx,Durrer:2013pga,RoperPol:2025lgc},
\begin{equation}
    \nu = \frac{4}{15} \frac{l_{\rm mfp}}{1 + \omega}\,, \qquad
    \xi = 4\, \bigl(\tfrac{1}{3} - \cs^2 \bigr)
    \frac{l_{\rm mfp}}{1 + \omega} \,.
    \label{transport_coefficients}
\end{equation}
In the early Universe, at temperatures larger than those of neutrino
decoupling 
$T_\nu \sim 1$ MeV and smaller than the $W$ boson's mass
$T_W \sim 80$ GeV, the largest mean free path in the primordial
plasma is that of
neutrinos
\cite{Heckler:1993nc,Arnold:2000dr,Caprini:2009yp,Durrer:2013pga,RoperPol:2025lgc}, providing
the dominant contribution to \Eq{transport_coefficients}.
For conformal theories with $\cs^2 = \omega = 1/3$, the bulk viscosity and,
hence, the trace of the deviatoric tensor, vanish.
At temperatures above $T_W$, neutrino interactions are no longer
suppressed and, hence, their mean free path is no longer the
dominant contribution.
The shear viscosity at these temperatures is estimated to be
$\nu \sim g^{-4} T^{-1}/\ln g^{-1}$ \cite{Arnold:2000dr}, where $g$ is the hypercharge
coupling at the electroweak scale.
The bulk viscosity is estimated for the primordial plasma
in ref.~\cite{Arnold:2006fz} to be $\xi \sim g^2 T^{-1}/\ln g^{-1}$,
such that, even if it is not exactly zero, it is suppressed by the
coupling constant $g$ with respect to the shear viscosity,
$\xi/\nu \sim g^6 \ll 1$.
In general, for temperatures above $T_{W}$ but well below the Planck mass, the shear and bulk viscosities are orders of magnitude smaller than
the Hubble time $1/H$. This is because,
barring aside coupling dependencies, in radiation domination
they scale as $\nu H \sim H/T \sim T/\mpl \ll 1$.
However, for temperatures below $T_W \sim 80$ GeV but above $T_\nu \sim 1$ MeV, again barring aside coupling dependencies, it holds that $\nu \propto T^{-5}$, and hence $\nu H$ starts approaching unity as the temperature decreases down to $T_\nu \sim 1$ MeV (see, e.g., figure 3 in ref.~\cite{RoperPol:2025lgc}).
This motivates to model the primordial plasma at temperatures above neutrino decoupling as an imperfect fluid close to LTE.
In particular, for high temperatures, and as we approach the electroweak scale, we get closer to the perfect fluid limit.
For a realistic description of the fluid perturbations
in the case of turbulence, however, dissipative effects need to be included
even if they are extremely small.
This is due to the fact that non-linear fluid systems
transport energy from large-scale eddies into
small-scale eddies.
Hence,
since viscous effects, at a given fluid length
scale $\lambda$, are proportional to $\nu/\lambda$, eventually they become relevant, leading to dissipation
of energy at sufficiently small scales, reached when the Reynolds
number is of order one, i.e., ${\rm Re} \equiv \lambda u_\lambda/\nu \sim {\cal O}(1)$, with $u_\lambda$ being a characteristic velocity of the flow
at the scale $\lambda$.
Note that the Reynolds number measures the ratio
of the non-linear
convective derivative {$u_j \partial_j u_i \sim u_\lambda^2/\lambda$
to the viscous Navier-Stokes
stresses $\nu \nabla^2 u_i \sim \nu \, u_\lambda/\lambda^2$ in the momentum equation \cite{2008Kundu}.

To incorporate the expansion of the Universe,
we now define the rescaled components of the deviatoric stress tensor in analogy to \Eq{components_Tmunu}, $\cdev^\mn = a^{4 + 2\alpha} \dev^\mn$.
Since $\nu$ and $\xi$ are proportional to $T^{-1}$
at large $T > T_W$,
we define the comoving viscous coefficients as
\EQb
    \conf{\nu} = \nu\, a^{-1}\,, \qquad \conf{\xi} = \xi \, a^{-1} \,,
    \label{viscosity_rescalings}
\end{mymathbox}
\noindent
such that the decrease of $a^{-1}$ in time, compensates the increase of $T^{-1}$. Notice that in the region $T_{\nu} < T< T_W$, where $\nu$ and $\xi$ scale as $T^{-5}$, the factor $a^{-1}$ does not decrease fast enough to compensate for the increase as $T^{-5}$, hence $\conf{\nu}$ and 
$\conf{\xi}$ increase in time.
Then, the comoving
rate-of-strain and the comoving projection tensors are defined such that
\begin{equation}
\label{rescaled_deviatoric_tensor}
    \cdev^{\mu \nu} \equiv \conf{\pi}^\mn + \conf{\xi}\, \conf{w}_f \, 
    \conf{\theta}\, \conf{h}^\mn\,, \qquad \text{with \ }       \conf{\pi}^\mn \equiv
    2
    \, \conf{\nu}\, \conf{w}_f \, \conf{\sigma}^\mn\,.
\end{equation}
This scaling implies the following ones for $S^\mn$ and $h^\mn$,
\begin{subequations}
    \label{comoving_stuff}
    \begin{align}
    \conf{S}^\mn = &\, a^{2 \alpha + 1} S^\mn = a^{1 - \alpha}\bigl[
    \conf{h}^{(\mu\lambda} D_\lambda^G \conf{U}^{\nu)}
    - \alpha \HH \, \conf{h}^{(\mu 0} \conf{U}^{\nu)}
    \bigr]\,, \qquad \text{with \ }
    \conf{\sigma}^\mn = \conf{S}^\mn - \frac{1}{3}
    \conf{\theta}\, \conf{h}^\mn\,, \\
    \conf{h}^\mn = &\, a^{2 \alpha} h^\mn = 
    \conf{U}^\mu \conf{U}^\nu
    + a^{2\alpha} g^\mn
    \,.
\end{align}
\end{subequations}
The comoving trace of the rate-of-strain tensor, i.e.,
the comoving fluid expansion scalar is  
\begin{equation}
\label{comoving_theta_visc}
    \conf{\theta} \equiv a \theta = 
    a^{-2\alpha} g_\mn 
    \conf{S}^\mn = - \conf{S}^{00} + a^{2(1- \alpha)}
    \delta_{ij}\conf{S}^{ij} = a^{1 - \alpha} \bigl[ \partial_0 \gamma + \partial_i (\gamma u_i) +
    3 \HH \gamma \bigr]
     \,.
\end{equation}
We note that we have decided to define $\conf{\theta}$
differing from the trace of $\conf{S}^{\mu \nu}$ by the factor $a^{-2\alpha}$,
such that $\cdev = g_\mn \cdev^\mn = 3 
\, \conf{\xi} \, \conf{w}_f \, \conf{\theta}\, a^{2\alpha}$.

\subsubsection*{Conservation form of first-order imperfect
fluid dynamics}

Including the viscous force of \Eq{ipf_force} to the fluid energy
and momentum equations in \Eq{general_eqs}, one finds
\EQb \label{conservation_viscous}
    \partial_{\mu} \conf{T}_\pf^\mn = \conf{\cal F}_{\rm ipf}^\nu +
    \conf{\cal F}_H^{\nu} \,, \qquad \text{where \ }
    \conf{\cal F}_{\rm ipf}^\nu = \partial_\mu \cdev^\mn + a^{-2\alpha} \cdev \,\HH 
    \delta^{\nu}_{~\,0} + (1 - \alpha) \HH \, \cdev^{0i}\delta^{\nu}_{~\,i}\,.
\end{mymathbox}
\noindent
Using the rescaled deviatoric stress tensor
of \Eq{rescaled_deviatoric_tensor}, the viscous four-force components
are
\begin{subequations}
\label{forces_ipf_expanding}
\begin{align}
    \conf{\cal F}^0_{\rm ipf} = &\,
    \partial_0 \conf{\pi}^{00} + \partial_i \conf{\pi}^{0i} + \gamma^2 \bigl(
    |\vec{u}_p|^2 \partial_0 + u_i \partial_i
    \bigr) \bigl(\conf{\xi} \, \conf{w}_f\, 
    \conf{\theta} \bigr) + \conf{\xi} \, \conf{w}_f
    \, \conf{\theta}\, \bigl[\partial_0 \gamma^2 
    + \partial_i (\gamma^2 u_i) \bigr]
    + 3 \, \conf{\xi} \, \conf{w}_f\,
    \conf{\theta} \, \HH
    \,, \\ 
    \conf{\cal F}^i_{\rm ipf} = &\,
    \partial_0 \conf{\pi}^{0i} + \partial_j 
    \conf{\pi}^{ij} + \bigl[\gamma^2 u_i
    (\partial_0 + u_j \partial_j)
    + a^{2(\alpha - 1)} \partial_i \bigr]
    \bigl( \conf{\xi} \, \conf{w}_f \, 
    \conf{\theta} \bigr)
    + \conf{\xi} \, \conf{w}_f \, \conf{\theta}
    \bigl[\partial_0 (\gamma^2 u_i)
    + \partial_j (\gamma^2 u_i u_j) \bigr]
    \nonumber \\ &\, + (1 - \alpha)\, 
    \bigl(\conf{\pi}^{0i} + \conf{\xi} \, \conf{w}_f \, \conf{\theta} \, \gamma^2 u_i \bigr) \HH \,,
\end{align}
\end{subequations}
where we have used $\partial_\mu (\conf{\xi}\, \conf{w}_f\,
\conf{\theta}\, \conf{h}^\mn) = \conf{h}^\mn \partial_\mu (\conf{\xi}\, \conf{w}_f\,
\conf{\theta}) + \conf{\xi}\, \conf{w}_f\,
\conf{\theta} \, \partial_\mu (\conf{U}^\mu \conf{U}^\nu)$.
Since these components
depend on the fluid variables $\conf{w}_f$ and $u_i$,
if we want to solve \Eq{conservation_viscous} in the conservation form,
we need to first reconstruct the primitive fluid variables
from the dynamical variables $\conf{T}^{0\mu}_\pf$
using \Eq{rho_u_fromT0mu}.
Then,
we can follow the same procedure as in \Sec{conservation_form}
to solve this system of equations for a constant EOS.
However, the viscous four-force components depend
on the fluid primitive variables and their spacetime
first and second derivatives, making the time advancement
of these equations cumbersome.
For this reason, we will restrict our description to the
subrelativistic limit of bulk velocity, for which 
$\cdev^{00} \sim {\cal O} (|\vec{u}_p|^2)$
and
$\cdev^{0i} \simeq 3\, a^{1 - \alpha}\, \conf{\xi} \, \HH  \,
\conf{w}_f \,  u_i \simeq 0$,
where the latter is neglected for small values
of the bulk viscosity
$\xi H = a^{1 - \alpha} \conf{\xi} \HH \ll 1$.
Hence, under these assumptions,
$\conf{\cal F}^\nu_{\rm ipf}$ only depends on spatial derivatives
of $u_i$ and $\conf{w}_f$
(see \Eqq{components_viscous_force} below).
Alternatively, when $\xi H$ is non-negligible, we can instead
advance $\conf{T}^{00}_\fl \simeq \conf{T}^{00}_\pf$
and $\conf{T}^{0i}_\fl = \conf{T}^{0i}_\pf - \cdev^{0i}
\simeq (\gamma^2 - {\cal A}) \, \conf{w}_f u_i$, where ${\cal A} = 3\,
a^{1 - \alpha}\, \conf{\xi}\, \HH$.
Then, the fluid primitive variables can be expressed in terms of
the dynamical variables in a similar way as done for perfect
fluid components in \Eq{rho_u_fromT0mu},
\begin{equation}
    \conf{\rho}_f = \frac{1}{z} \conf{T}^{00}_\fl \, \qquad
    \text{and \ } \qquad u_i = \frac{z}{z+ \omega_{\cal A}} \frac{\conf{T}^{0i}_\fl}{\conf{T}^{00}_\fl}\,, \qquad \omega_{\cal A} = \omega - {\cal A} ( 1+ \omega)\,.
\end{equation}
Then, $\conf{T}^{ij}_\pf$ can be reconstructed for a given value of $z$ as
\begin{equation}
    \conf{T}^{ij}_\pf = \conf{w}_f \gamma^2 u_i u_j + \conf{p}_f a^{2(\alpha - 1)}
    \delta_{ij} = \frac{z (z + \omega) }{(z + \omega_{\cal A})^2}
    \frac{\conf{T}^{0i}_\fl \conf{T}^{0j}_\fl}{\conf{T}^{00}_\fl}
    + \frac{\omega}{z}  a^{2(\alpha - 1)} \conf{T}^{00}_\fl \delta_{ij}\,.
\end{equation}
We can again write a polynomial equation to solve for $z$, by computing
the ratio $r^2$, as in \Eq{T0i_sq},
\begin{align}
    r^2 \equiv \sum_i
    \Biggl[\frac{\conf{T}^{0i}_\pf}{\conf{T}^{00}_\pf}\Biggr]^2  a^{2(1 - \alpha)}
    = &\, \frac{(z + \omega_{\cal A})^2}{z^2}
    |\vec{u}_p|^2 = \frac{(z + \omega_{\cal A})^2}{z + \omega} \frac{z - 1}{z^2}
    \nonumber \\ &\, \Rightarrow 
    z^3 (r^2-1)
    + z^2 (\omega r^2+1-2\omega_{\cal A}) 
     + z \, \omega_{\cal A} (2-\omega_{\cal A})
    + \omega_{\cal A}^2 = 0 \,.
\end{align}
Note that when $\omega_{\cal A} = \omega$, which happens for $\omega = -1$ or ${\cal A} = 0$, this polynomial
reduces to \Eq{T0i_sq} with the solution in \Eq{gamma2_rel} for $z$.
When ${\cal A} > 0$, we always have $\omega_{\cal A} < \omega$ for $\omega > -1$. Restricting to the physical values $|\omega| \leq 1$ we also have $r^2 < 1$ [see discussion after \Eq{T0i_sq}], and
using Cardano's cubic formula, the generic three solutions can be expressed as
\begin{subequations}
\begin{align}
\label{cardano_cubic_solutions}
    z_k
    = &\,
    e^{2k \pi i/3} \sqrt[3]{-\frac{q}{2}+\sqrt{\Delta}}
    +e^{-2k\pi i/3} \sqrt[3]{-\frac{q}{2}-\sqrt{\Delta}}
    - \frac{\omega r^2+1-2\omega_{\cal A}}{3(r^2 - 1)} \,,
    \qquad k=0,1,2\,,
\end{align}
where
\begin{align}
    p = &\, 
    \frac{
     3(r^2-1)\,\omega_{\cal A}(2-\omega_{\cal A})
    -\left(\omega r^2+1-2\omega_{\cal A}\right)^2
    }
    {3(r^2-1)^2}\,, \\ 
    q  = &\, 
    \frac{
    2\left(\omega r^2+1-2\omega_{\cal A}\right)^3
    - 9(r^2-1)\left(\omega r^2+1-2\omega_{\cal A}\right)
    \omega_{\cal A}(2-\omega_{\cal A})
    +27(r^2-1)^2\omega_{\cal A}^2
    }
    {27(r^2-1)^3}\,.
\end{align}
\end{subequations}
with the Cardano discriminant $\Delta=
\left(\frac{q}{2}\right)^2
+
\left(\frac{p}{3}\right)^3$.
Then, one needs to choose for $z$, on a case-by-case basis, a real root satisfying $z \geq 1$.
Since we are restricting to physical values of the EOS $|\omega| \leq 1$, which implies $r^2 <1$, such root always exists and is unique for $\omega_{\cal A} \geq -1$. We have, in fact, that, for $|\omega| \leq 1$ and since $\omega_{\cal A} \leq \omega$, the function $r^2(z) = \bigl[(z+\omega_{\cal A})^2(z-1)\bigr]/\bigl[(z+\omega)z^2\bigr]$ takes values in $[0, 1)$ and is monotonous for $z \geq 1$ iff $(1+\omega)(1+\omega_{\cal A}) \geq 0$, which is satisfied if both $\omega \geq -1$ and $\omega_{\cal A} \geq -1$, the latter requiring ${\cal A} = 3  \,  {\xi} \, H \leq 1$.

\subsubsection*{Non-conservation form of first-order imperfect
fluid dynamics}

To obtain the non-conservation form of the fluid
EOM, we follow the derivation in \Sec{nonconser_relativistic}, including the viscous four-force components, $\conf{\cal F}_{\rm ipf}^\nu$.
We find that the energy and momentum conservation equations are given by \Eqq{equations_relativistic_general} with the replacement $\conf{\cal F}_H^\nu \to \conf{\cal F}_H^\nu + \conf{\cal F}_{\rm ipf}^\nu$, 
\begin{subequations} 
\label{general_EOMs}
\begin{align}
    \partial_0 \conf{\rho}_f -
    |\vec{u}_{\rm p}|^2 \partial_0 \conf{p}_f  = &\,-
    \conf{w}_f \partial_i u_i
    - u_i \partial_i (\conf{\rho}_f - \conf{p}_f)
    + (\conf{\cal F}_H^0 + \conf{\cal F}_{\rm ipf}^0)
    (1 +
    |\vec{u}_{\rm p}|^2) - 2 \, a^{2(1 - \alpha)} u_i
    \conf{\cal F}_{\rm ipf}^i\,,
    \label{general_EOM_forces} \\
    \partial_0 u_i + u_j \partial_j u_i  =&\, - \frac{u_i}{\conf{w}_\f \gamma^2}
    \bigl( \partial_0 \conf{p}_\f  + \conf{\cal F}_H^0 + \conf{\cal F}_{\rm ipf}^0 \bigr) - a^{2(\alpha-1)} \frac{\partial_i \conf{p}_\f}{\conf{w}_\f \gamma^2} + \frac{\conf{\cal F}_H^i+\conf{\cal F}^i_{\rm ipf}}{\conf{w}_\f \gamma^2}\,.
    \label{general_EOM_forces2}
\end{align}
\end{subequations}
Taking a constant EOS, we find the EOM
for the fluid primitive
variables in their non-conservation form,
\begin{subequations}
\label{rel_eqs_visc}
\EQalign
    \hspace{-10mm} (1 - \omega |\vec{u}_{\rm p}|^2)
    \partial_0  \ln \conf{\rho}_f \, +   &\,
    (1 + \omega) \partial_i u_i +
    (1 - \omega) u_i \partial_i \ln \conf{\rho}_f \nonumber \\  = &\,
    \frac{1}{\conf{\rho}_f}\Bigl\{  \label{rel_energy_visc}
    \bigl[\conf{\cal F}_{\rm ipf}^0 +
    (1 - 3 \omega) \conf{\rho}_f \HH
    \bigr](1 +
    |\vec{u}_{\rm p}|^2) - 2 \,a^{2(1-\alpha)} u_i\conf{\cal F}_{\rm ipf}^i
    \Bigr\}\,, \\ 
    \hspace{-10mm} \partial_0 u_i + u_j \partial_j u_i = &\, \frac{1 - |\vec{u}_p|^2}
    {1 - \omega |\vec{u}_{\rm p}|^2} \biggl[ \omega \partial_j u_j  
    +\omega \frac{1 - \omega}{1 + \omega} u_j \partial_j \ln \conf{\rho}_f -
    \frac{1}{\conf{\rho}_f} \biggl(
    \conf{\cal F}_{\rm ipf}^0
    - \frac{2 \omega}{1 + \omega} a^{2(1 - \alpha)}
    u_j \conf{\cal F}_{\rm ipf}^j
    \biggr)
    \biggr] u_i \nonumber \\ 
    &\, - \frac{1 - |\vec{u}_{\rm p}|^2}
    {1 + \omega} \biggl(\omega a^{2(\alpha - 1)}  \partial_i \ln \conf{\rho}_f -
    \frac{1}{\conf{\rho}_f}
    \conf{\cal F}_{\rm ipf}^i \biggr)
    +
    \biggl[\frac{1 - |\vec{u}_{\rm p}|^2}
    {1 - \omega|\vec{u}_{\rm p}|^2}  (3 \omega  - 1)  +
    \alpha - 1\biggr] \HH u_i \,.
    \label{rel_mom_visc}
\end{mymathbox}
\end{subequations}
\noindent
As mentioned above when we presented
the conservation form of the EOM, the viscous force
depends, in general, on first and second-order time
derivatives of the fluid variables.
In the limit of subrelativistic fluid velocity and small bulk
viscosity, then $\conf{\cal F}^0_{\rm ipf} \simeq 0$
and $\conf{\cal F}^i_{\rm ipf}$ only depends on gradients
of the primitive fluid variables $\conf{w}_f$ and $u_i$ [cf.~\Eqq{components_viscous_force}].

\subsubsection*{Limit of subrelativistic fluid velocity}

\EEq{Pi_visc} corresponds to the deviatoric
stress tensor found after expanding the distribution function
up to first order for small Knudsen numbers,
following Chapman-Enskog theory \cite{ChapmanCowling1970}.
Therefore, the resulting expression is linear in gradients of $U^\mu$,
describing first-order hydrodynamics,
also known as
classical irreversible thermodynamics (CIT)
\cite{Landau1987Fluid,Romatschke:2009im}.
This fluid description is known to suffer from causality violations
due to the presence of diffusive operators like $\nu 
\vec{\nabla}^2 \vec{u}$ and, therefore, CIT needs to be
extended to Maxwell-Cattaneo equations to account for
finite non-zero relaxation times over which the deviatoric
tensor responds to changes in the velocity (see refs.~\cite{Israel:1976efz,Israel:1976tn}, the review
\cite{Romatschke:2009im}, and references therein).
Since the extension to a causal relativistic description goes beyond the scope of our current work,
we restrict our presentation
to Navier-Stokes viscosity to describe the deviatoric
stress tensor
and we defer its extension to future work.
We note that the introduction of the
dissipative force $\conf{\cal F}_{\rm ipf}^\nu$ in the \rhs
of \Eqs{conservation_viscous}{rel_eqs_visc}
allows to generalize our treatment to relativistic imperfect fluids
beyond first-order hydrodynamics.

At linear order in the velocity perturbations,
the comoving rate-of-strain tensor components
are $\conf{S}^{00} \simeq 0$, $\conf{S}^{0i} \simeq a^{1 - \alpha} \HH u_i$, and $\conf{S}^{ij} \simeq 
a^{\alpha - 1} \bigl[\partial_{(i} u_{j)} + \HH \delta_{ij} \bigr]$.
Hence, the comoving fluid expansion scalar is $\conf{\theta} \simeq a^{2(1 - \alpha)} \delta_{ij}\, \conf{S}^{ij} \simeq
a^{1 - \alpha} (\partial_i u_i + 3 \HH).$
In this limit, the components of the deviatoric tensor reduce to
\begin{subequations}
\begin{align}
    \lim_{|\vec{u}_p|^2 \ll 1}\cdev^{0i} = &\, 3 \, \HH \, \conf{\xi}\, 
    \conf{w}_f \, a^{1 - \alpha} \, u_i\,,  \\ 
    \label{Pi_ij_subrel}
    \lim_{|\vec{u}_p|^2 \ll 1}\cdev^{ij} = &\, a^{\alpha - 1} \Bigl[  2 \, \conf{\nu} \,
    \partial_{(i} u_{j)} + \bigl(\conf{\xi} - \tfrac{2}{3}
    \conf{\nu}\bigr) (\partial_l u_l)\,
    \delta_{ij} + 3 \, \conf{\xi} \, \HH \,
    \delta_{ij} \Bigr] \conf{w}_f \,,
\end{align}
\end{subequations}
with $\cdev^{00} \sim {\cal O} (|\vec{u}_p|^2)$.
Then,
the viscous four-force
components become
\begin{subequations}
\label{components_viscous_force}
\begin{align}
\label{components_viscous_force_zero}
    \lim_{|\vec{u}_{\rm p}|^2 \ll 1} &\, a^{\alpha - 1}
    \conf{\cal F}^{0}_{\rm ipf} =
    3 \, \HH \bigl[ u_i \partial_i (\conf{\xi} \, \conf{w}_f)
    + 2 \, \conf{\xi}\, \conf{w}_f \, \partial_i u_i + 3 \HH \,  \conf{\xi} \, \conf{w}_f \bigr] \,,
    \\
    \lim_{|\vec{u}_{\rm p}|^2 \ll 1} &\, a^{1- \alpha}
    \conf{\cal F}_{\rm ipf}^i = 
    3 \, 
    \partial_0 \bigl(a^{2(1 - \alpha)}
    \conf{\xi} \, \HH \, \conf{w}_f \, u_i \bigr)
    +
    \partial_j \bigl[2 \,\conf{\nu}\, \conf{w}_f\,
    \partial_{(i} u_{j)}\bigr] + \partial_i \bigl[ (\conf{\xi} - 
    \tfrac{2}{3} \conf{\nu}) \, \conf{w}_f \, \partial_j u_j\bigr]
    + 3\, \HH \,  \partial_i (\conf{\xi}\,
    \conf{w}_f)
    \,.
    \label{components_viscous_force_spatial}
\end{align}
\end{subequations}
Under the assumption that the
bulk viscosity is smaller than the Hubble time,
$\conf{\xi} \,\HH \ll 1$,
the deviatoric momentum density and, hence,
the viscous energy dissipation vanish,
i.e., $\cdev^{0i} \simeq 0 \Rightarrow \conf{\cal F}^{0i}_{\rm ipf} \simeq 0$.
On the other hand, when both
viscous coefficients and
the EOS are homogeneous, we can express
$\partial_i \conf{w}_f = \conf{w}_f \partial_i \ln \conf{\rho}_f$,
and
the viscous force becomes
\EQb
    \lim_{|\vec{u}_{\rm p}|^2 \ll 1,\,
    \xi H \ll 1}
    a^{1 - \alpha}
    \frac{\conf{\cal F}_{\rm ipf}^i}{\conf{w}_f}
    = \conf{\nu} \, \partial_j \partial_j u_i +
     \bigl(\tfrac{1}{3} \conf{\nu} + \conf{\xi}\bigr) \, 
    \partial_i \partial_j u_j
    + \Bigl[2\, \conf{\nu} \, \partial_{(i} u_{j)} + \bigl( \conf{\xi} -\tfrac{2}{3}
    \conf{\nu} \bigr) (\partial_l u_l) \delta_{ij} \Bigr]
     \partial_j \ln \conf{\rho}_f\,.
    \label{fviscmu}
\end{mymathbox} \noindent

In the limit of subrelativistic fluid velocity $|\vec{u}_{\rm p}|^2 \ll 1$,
the fluid EOM for a constant EOS, i.e., constant
$\omega = p_f/\rho_f$, are
\begin{subequations}
\label{subrelativistic_visc_all}
\EQalign
    \lim_{|\vec{u}_{\rm p}|^2 \ll 1} \partial_0  \ln \conf{\rho}_f\, = &\, - 
    (1 + \omega) \partial_i u_i - 
    (1 - \omega) u_i \partial_i \ln \conf{\rho}_f \nonumber \\ &\, +
    \frac{1}{\conf{\rho}_f}\bigl[
    \conf{\cal F}_{\rm ipf}^0 + (1 - 3 \omega)\conf{\rho}_f \HH -
    2 \, a^{-2(\alpha-1)} u_i\conf{\cal F}_{\rm ipf}^i\bigr]\,, \\ 
    \lim_{|\vec{u}_{\rm p}|^2 \ll 1}
    \partial_0 u_i = & \, - u_j \partial_j u_i + u_i \biggl[\omega \partial_j u_j 
    +\omega \frac{1 - \omega}{1 + \omega} u_j \partial_j \ln \conf{\rho}_f -
    \frac{1}{\conf{\rho}_f} \biggl( 
    \conf{\cal F}_{\rm ipf}^0
    - \frac{2 \omega}{1 + \omega} a^{2(1 - \alpha)}
    u_j \conf{\cal F}_{\rm ipf}^j \biggr)
    \biggr] \nonumber \\ &\, - \frac{1}{1 + \omega} \biggl(\omega
    \, a^{2(\alpha - 1)} \partial_i \ln \conf{\rho}_f 
    - \frac{1}{\conf{\rho}_f}
    \conf{\cal F}_{\rm ipf}^i \biggr) + (3 \omega + \alpha - 2) \HH u_i\,.
    \label{subrel_mom_visc}
\end{mymathbox}
\end{subequations}
\noindent
Since the viscous four-force components only
depend on the fluid primitive variables $\conf{\rho}_f$
and $u_i$, and their gradients in this limit when
the bulk viscosity is negligible, this
form of the EOM allows for a direct integration in time.

In order to express the 
work done
by the viscous forces, $u_i \conf{\cal F}_{\rm ipf}^i$,
we can use \Eqs{ipf_force}{div_Pi}, from which
\begin{align} \label{contraction_ipf}
    U_\nu \, a^{-(4+2\alpha)} \, \conf{\cal F}^{\nu}_{\rm ipf} = U_\nu D_\mu^G \dev^{\mu \nu} = - \dev^\mn S_\mn = - a^{-5}
    \cdev^{00} \conf{S}^{00}  + 2 a^{-(3 + 2 \alpha)}
    \cdev^{0i} \conf{S}^{0i}
    - a^{-(1 + 4 \alpha)} \cdev^{ij} \conf{S}^{ij}\,. 
\end{align}
From this property, we can then compute a generic expression
for $u_i \conf{\cal F}_{\rm ipf}^i$,
\begin{equation}
    \gamma u_i \conf{\cal F}_{\rm ipf}^i =
    \gamma a^{2(\alpha - 1)} \conf{\cal F}_{\rm ipf}^0 - a^{3(\alpha - 1)}
    \cdev^{00} \conf{S}^{00} + 2 a^{\alpha - 1}\cdev^{0i} \conf{S}^{0i}
    - a^{1 - \alpha} \cdev^{ij} \conf{S}^{ij}\,.
\end{equation}
Then, in the limit of subrelativistic 
bulk velocity, with 
$\cdev^{00} \simeq 0$, $\cdev^{0i} \conf{S}^{0i}
\simeq 0$,
$\conf{S}^{ij} \simeq a^{\alpha - 1} \bigl[
\partial_{(i} u_{j)} + \HH \delta_{ij} \bigr]$,
$\conf{h}^{ij} \simeq a^{2(\alpha-1)} \delta_{ij}$,
and
$\conf{\theta} \simeq a^{1 - \alpha} (\partial_i u_i + 3 \, \HH)$,
and with $\cdev^{ij}$ and $\conf{\cal F}_{\rm ipf}^0$ given in
\Eqs{Pi_ij_subrel}{components_viscous_force_zero}, respectively,
we find 
\begin{align}
    \hspace{-3mm}
    a^{1 - \alpha} u_i \conf{\cal F}_{\rm ipf}^i
    = &\,
    a^{\alpha - 1} \conf{\cal F}_{\rm ipf}^0 - a^{2(1 - \alpha)}
    \cdev^{ij} \conf{S}^{ij}
    + {\cal O} (|\vec{u}_p|^3)
    \nonumber \\ = &\,
    3 \, \HH \, u_i \partial_i (\conf{\xi} \, \conf{w}_f)
    -2 \, \conf{\nu} \, \conf{w}_f \, 
    (\partial_i u_j) \, \partial_{(i} u_{j)}
    - \bigl(\conf{\xi} - \tfrac{2}{3} \conf{\nu}\bigr)
    \conf{w}_f (\partial_i u_i)^2
    + {\cal O} (|\vec{u}_p|^3) \,.
\end{align}

\subsection{Fluid dynamics with bosonic interactions}
\label{subsec:bosonic_inters}

In the previous section, we have omitted the effect of gauge and scalar
bosonic fields on the fluid dynamics, assuming
that ${\cal F}_{\rm int}^\mu = 0$.
In the following, we first consider the dynamics of a fluid coupled to an Abelian gauge field, ${\cal F}_{\rm int}^\mu = {\cal F}_{U(1)}^\mu$ in \Sec{subsec:gauge_fluid_dyns}, and coupled to a singlet scalar field,
${\cal F}_{\rm int}^\mu = {\cal F}_\phi^\mu$ in \Sec{subsec:scalar_fluid}.
In the case of a fluid coupled to electromagnetism, i.e., to the $U(1)$ gauge field,
we follow the description of ref.~\cite{RoperPol:2025lgc}
for relativistic MHD in an expanding Universe.
We present the EOM for this system in a novel way,
suitable to apply
Runge-Kutta schemes in order to perform the numerical integration in time, as
described in \Sec{sec:RK}.

\subsubsection{Gauge field-fluid dynamics}
\label{subsec:gauge_fluid_dyns}

The dynamical system of a perfect
fluid that carries
charged particles coupled to
a $U(1)$ Abelian gauge field requires the inclusion of the Lorentz force
and the coupling to Maxwell
equations.
The latter are given in \Eqs{eq:Maxwell}{eq:Gauss_Abel},
\EQb
    \partial_0 F_{0i} - a^{-2(1 - \alpha )}\partial_j F_{ji} + (1 - \alpha)
    \mathcal{H} F_{0i} =
    a^{2 \alpha} J_i\,, \qquad \partial_i F_{0i} =
    a^2 J_0\,, \label{Maxwell2}
\end{mymathbox} \noindent
where $J_\mu = J^{\varphi}_{\mu} + J^{\Phi}_\mu + J^\fl_\mu$ corresponds to the four-current density from all charged particles,
including charged scalar fields $\varphi$ (singlet) and $\Phi$ (doublet).
The dynamical variables are the electric field ${\cal E}_i \equiv F_{0i}$
and the spatial components of the gauge field $A_i$,
\EQb
    \partial_0 A_i = {\cal E}_{i} + \partial_i A_0 \,.
    \label{gauge_eq}
\end{mymathbox}
\noindent
This equation can be solved within a chosen gauge,
e.g., temporal gauge ($A_0 = 0$).
The spatial components of the Faraday tensor in \Eq{Maxwell2}
are related to the
magnetic field ${\cal B}_i$,
\begin{equation}
    F_{ij} = \partial_i A_j - \partial_j A_i \equiv \varepsilon_{ijk} {\cal B}_k\,,
    \label{eq:B_field}
\end{equation}
such that $\vec{\cal B} = \vec{\nabla} \times \vec{A}$.

The effective
current induced by the charged particles in the fluid, $J^\mu_\fl$,
that appears in the \rhs  of \Eq{Maxwell2}, can be described
by the generalized Ohm's law  \cite{landau1984electrodynamics,Starke:2014tfa},
\begin{equation}
    J^\mu_\fl = \rho_e U^\mu + \sigma_f U_\nu F^{\mu \nu} \,, \label{Ohms_law}
\end{equation}
where $\rho_e$ and $\sigma_f$ are the electrical charge density and conductivity
of the fluid.
Note that the current induced from the fluid conductivity
corresponds to the linear response of the fluid
to the gauge fields due to
out-of-equilibrium interactions,
similar to the expansion described in \Sec{viscous_term}.
In general, we assume quasi-neutral fluids
with $\rho_e = 0$, but we will keep the charge density in
the equations for generality.
The rescaled
components of the current density and
the Faraday tensor are
$\conf{J}^\mu = \sqrt{-g} J^\mu = a^{3 + \alpha} J^\mu$ and
$\conf{F}^\mn = \sqrt{-g} F^\mn = a^{3 + \alpha} F^\mn$.
They are chosen
to make Maxwell equations conformally flat,
\begin{equation}
    - D_\mu^G F^\mn = - \frac{1}{\sqrt{-g}} \partial_\mu \bigl(
    \sqrt{-g} F^\mn\bigr) = J^\nu \Rightarrow
- \partial_\mu \conf{F}^{\mu \nu} = - \partial_\mu (
\sqrt{-g} \,F^\mn) 
= \sqrt{-g} \, J^\nu = \conf{J}^\nu\,,
\end{equation}
which is always possible due to the Faraday tensor being an antisymmetric tensor.
Note that the rescaled components of the Faraday tensor are
related to the comoving
electric and magnetic fields when $\alpha = 1$ 
\cite{Brandenburg:1996fc,RoperPol:2025lgc}, since
$\conf{F}^{0i} = - a^{1 - \alpha}
{\cal E}_i$ and $\conf{F}^{ij} = a^{\alpha - 1} \varepsilon_{ijk} {\cal B}_k$.
Based on this scaling, the current density components are
\begin{subequations}
\label{charge_current_density}
\EQalign
    \conf{J}^0_\fl = &\,
    \gamma(\conf{\rho}_e - a^{2 (1 -\alpha)}
    \conf{\sigma}_f u_i {\cal E}_i) \,,
    \label{charge_density}
    \\ \conf{J}^i_\fl = &\,
    \gamma \bigl(\conf{\rho}_e u_i +
    \conf{\sigma}_f \varepsilon_{ijk} u_j {\cal B}_k
    - \conf{\sigma}_f {\cal E}_i \bigr) \,, \label{current_density}
\end{mymathbox} \noindent
\end{subequations}
where $\conf{\rho}_e = a^3 \rho_e$ and
$\conf{\sigma}_f = a \sigma_f$
are the comoving charge
density and conductivity \cite{Brandenburg:1996fc,Subramanian:1997gi,Jedamzik:1996wp,Durrer:2013pga,RoperPol:2025lgc}.
    
The system of EOM is given by Maxwell equations
[cf.~\Eqq{Maxwell2}] and
the conservation of the
fluid stress-energy tensor [cf.~\Eq{general_eqs}].
The interaction force of the gauge field with the fluid (Lorentz force)
is given as ${\cal F}^{\nu}_{U(1)} = J_\mu^\fl \, F^{\nu \mu}$ [cf.~\Eq{eq:F_U(1)}]
with comoving components
$\conf{\cal F}_{U(1)}^\nu = a^{4 + 2\alpha}
{\cal F}_{U(1)}^\nu$, 
\begin{subequations}
\label{lorentz_forces}
\EQalign
    \hspace{-3mm}
    \conf{\cal F}^0_{U(1)} &= - a^{1 - \alpha} {\cal E}_i
    \conf{J}^i_\fl
    = -a^{1-\alpha} \gamma \Bigl[ \conf{\rho}_e \, \vec{u} \cdot \vec{\cal E}  - \conf{\sigma}_\f \, \vec{u} \cdot (\, \vec{\cal E} \times \vec{{\cal B}} \, ) - \conf{\sigma}_\f |\vec{\cal E}|^2 \Bigr]\,,
    \label{lorentz_0}\\ 
    \hspace{-3mm}
    \conf{\cal F}^i_{U(1)} &= -  a^{\alpha - 1}
    \bigl({\cal E}_i \conf{J}^0_\fl
    - \varepsilon_{ijk} \conf{J}^j_\fl {\cal B}_k\bigr) \nonumber \\
    \hspace{-3mm}
    & =  -a^{\alpha-1} \gamma \Bigl[(\, \conf{\rho}_e - a^{2(1-\alpha)} \conf{\sigma}_\f \, \vec{u} \cdot \vec{\cal E} \, ) \, {\cal E}_i - \conf{\rho}_e ( \, \vec{u} \times \vec{\cal B} \, )_i + \conf{\sigma}_\f ( \, \vec{\cal E} \times \vec{\cal B} \,)_i - \conf{\sigma}_\f (\, \vec{u} \cdot \vec{\cal B} \,) \, {\cal B}_i + \conf{\sigma}_\f |\vec{\cal B}|^2 u_i  \Bigr] \,.
    \label{lorentz_i}
\end{mymathbox}
\end{subequations}

Since the Ohm's law in \Eq{Ohms_law} introduces out-of-equilibrium dynamics,
it is expected to yield entropy production.
Including the Lorentz force in the fluid EOM, the entropy evolution
equation [cf.~\Eq{entropy}] becomes
\begin{equation} \label{entropy2}
    D_\mu^G s^\mu_f = - \frac{1}{T} U_\nu D_\mu^G T^\mn_{\rm pf}
    = \frac{1}{T} \dev^\mn S_\mn - \frac{1}{T} U_\nu {\cal F}^\nu_{U(1)} = 
    \frac{w_f}{T} \bigl(2 \,\nu \, \sigma^\mn \sigma_\mn + \xi\,
    \theta^2 \bigr) + 
    \frac{\eta_{\rm diff}}{T} J^\mu_{\rm cond} J_\mu^{\rm cond}
    \geq 0\,,
    \end{equation}
where the contribution from the equilibrium term in Ohm's law vanishes, as it is proportional to
$\rho_e U^\mu U^\nu F_\mn = 0$.
The entropy production from electromagnetic
dissipation is proportional to $\sigma_f U_\nu F^{\nu \mu} \, U^\lambda F_{\lambda \mu} = \eta_{\rm diff} J^\mu_{\rm cond} J_\mu^{\rm cond}$,
with $\eta_{\rm diff} = 1/\sigma_f$ the magnetic diffusivity, and indeed it only comes from the conduction component
of the fluid current density $J^\mu_{\rm cond} \equiv
\sigma_f U_\nu F^\mn$.

\subsubsection*{Conservation form of gauge-fluid dynamics} 
\label{conservation_mhd}

The resulting fluid conservation laws take the following
form [cf.~\Eq{general_eqs}],
\EQb \label{eq:conservation_mhd}
    \partial_\mu \conf{T}^\mn_\pf =
    \conf{\cal F}^\nu_{U(1)} + \conf{\cal F}^\nu_{\rm ipf} + 
    \conf{\cal F}_H^\nu \,,
\end{mymathbox} \noindent
which can then be solved evolving $\conf{T}^{0\mu}_\pf$ and
adding $\conf{\cal F}_{U(1)}^\nu$ as a force term on the \rhs\!.
The reconstruction of $\conf{T}^{ij}_\pf$ and of the fluid primitive variables
is equivalent to the one described in \Sec{subsec:fld_dynamics_cont}.

\subsubsection*{Alternative conservation form of gauge-fluid dynamics}

The stress-energy tensor associated to the Abelian gauge field is
\begin{equation}
    T^{\mu \nu}_{U(1)} =
    F^{\mu \sigma} F^{\nu}_{\ \, \sigma} - {1 \over 4}
    g^{\mu \nu} F^{\lambda \sigma} F_{\lambda \sigma}\,,
\end{equation}
which has vanishing trace,  $T_{U(1)} = g_\mn T_{U(1)}^\mn = 0$.
The comoving components of the electromagnetic stress-energy tensor, 
$\conf{T}^\mn_{U(1)} = a^{4 + 2\alpha} T^\mn_{U(1)}$,
become
\begin{subequations}
\label{Tmunu_EM_comps}
\EQalign
    &\,  \conf{T}^{00}_{U(1)}
    = {\frac{1}{2}} (a^{2(1 - \alpha)} |\vec{\cal E}|^2 +
    |\vec{\cal B}|^2)\,, \\ &\,  \conf{T}^{0i}_{U(1)}
    = -
    (\vec{\cal E} \times \vec{\cal B})_i \,, \\ 
    &\,  \conf{T}^{ij}_{U(1)}
    = - {\cal E}_i {\cal E}_j - a^{2(\alpha - 1)}
    {\cal B}_i {\cal B}_j +
    a^{2(\alpha - 1)} \conf{T}_{U(1)}^{00}
    \delta_{ij}
    \,.
\end{mymathbox}
\end{subequations}
Extending the computation that yields \Eq{Tmunu_cons_aux}
[see also
\App{app:EOM} and, in particular, \Eq{term_Abelian}]
to an expanding Universe,
one finds
$D_\mu^G {T}^\mn_{U(1)}
= - {\cal F}_{U(1)}^\nu$.
Therefore, since the comoving trace $\conf{T}_{U(1)} = g_\mn \conf{T}^\mn_{U(1)}$
vanishes, in analogy to \Eq{ipf_force}, one finds
\begin{equation}
    \partial_\mu \conf{T}^\mn_{U(1)}
    + (1 - \alpha) \HH \,
    \conf{T}^{0i}_{U(1)} \delta^{\nu}_{~\,i}
    = - \conf{\cal F}^\nu_{U(1)}\,.
\end{equation}
Substituting this relation into \Eq{eq:conservation_mhd},
we can express the fluid conservation laws
in the following way
\begin{equation} \label{cons_form_T0mu_tot}
    \partial_\mu \conf{T}^\mn_+
    =
    \conf{\cal F}_{\rm ipf}^\nu
    + \conf{\cal F}_H^\nu + (\alpha - 1)\,
    \conf{T}^{0i}_{U(1)} \,
    \HH\delta^{\nu}_{~\,i}\,,
\end{equation}
where 
$\conf{T}^{\mu \nu}_+ = \conf{T}^\mn_\pf + 
\conf{T}^\mn_{U(1)}$.
These equations remain conformally flat for a
radiation-dominated fluid with $p_f = \rho_f/3$
and zero bulk viscosity $\xi = 0$ \cite{Brandenburg:1996fc,RoperPol:2025lgc}.
The components of $\conf{T}^{0\mu}_+ =
\conf{T}^{0\mu}_\pf + \conf{T}^{0\mu}_{U(1)}$,
associated to the perfect
fluid and the electromagnetic fields, can be evolved using \Eq{cons_form_T0mu_tot}.
To reconstruct $\conf{T}^{ij}_+ \equiv \conf{T}^{ij}_\pf + \conf{T}^{ij}_{U(1)}$
we again define, as in \Eq{T0i_sq} (see \Sec{conservation_form}),
\begin{equation}
    r^2 \equiv \sum_i \Biggl[\frac{\conf{T}^{0i}_\pf}{\conf{T}^{00}_\pf}\Biggr]^2
    a^{2(1 - \alpha)}
    = \sum_i \Biggl[\frac{\conf{T}^{0i}_+ - 
    \conf{T}^{0i}_{U(1)}}{\conf{T}^{00}_+ - 
    \conf{T}^{00}_{U(1)}}  \Biggr]^2  a^{2( 1- \alpha)}
    = \frac{(z + \omega)(z - 1)}{z^2}\,,
\end{equation}
where the components
$\conf{T}^{\mu\nu}_{U(1)}$
are computed from the solution to
\Eqs{Maxwell2}{gauge_eq} using \Eqq{Tmunu_EM_comps}.
From $r^2$, one can compute $z$ and reconstruct $\conf{T}^{ij}_\pf$
using \Eq{Tij_gamma2} for an EOS that does
not depend on the fluid variables.
Then, $\conf{T}^{ij}_{+}$ is computed by adding $\conf{T}^{ij}_\pf + \conf{T}^{ij}_{U(1)}$,
allowing us to evolve the system of equations in 
\Eq{cons_form_T0mu_tot}.
The fluid primitive variables $\conf{\rho}_f$ and $u_i$ that are required to compute the Ohmic current components,
$\conf{J}_\mu^\fl$ [cf.~\Eqq{charge_current_density}],
and the imperfect fluid force [see \Eq{conservation_viscous}],
are finally reconstructed using \Eq{rho_u_fromT0mu}.
This is an alternative approach to the one presented above to solve the conservation form of the fluid equations.

\subsubsection*{Non-conservation form of gauge-fluid dynamics}
\label{nonconservation_mhd}

The non-conservation form of the gauge-fluid EOM
can be obtained following the same procedure as in
sections~\ref{nonconser_relativistic} and \ref{viscous_term}
(see also section~6.4 in ref.~\cite{RoperPol:2025lgc}).
In this approach, we evolve the primitive fluid
variables $\conf{\rho}_f$ and $u_i$ instead of the temporal components of the stress-energy tensor
$\conf{T}^{0\mu}_\pf$.
The fluid equations of conservation of energy and momentum with a constant EOS are equivalent to those in \Eqq{rel_eqs_visc}, substituting $\conf{\cal F}_{\rm ipf}^\nu$ by $\conf{\cal F}_{\rm tot}^\nu = \conf{\cal F}_{\rm ipf}^\nu + \conf{\cal F}_{U(1)}^\nu$,
\begin{subequations}
\label{equations_Lorentz}
\EQalign
    \hspace{-10mm} (1 - \omega |\vec{u}_{\rm p}|^2)
    \partial_0  \ln \conf{\rho}_f\, +  &\,
    (1 + \omega) \partial_i u_i + 
    (1 - \omega) u_i \partial_i \ln \conf{\rho}_f \nonumber \\  = &\,
    \frac{1}{\conf{\rho}_f}\Bigl\{  \label{rel_energy_gauge}
    \bigl[\conf{\cal F}_{\rm tot}^0 +
    (1 - 3 \omega) \conf{\rho}_f \HH
    \bigr](1 +
    |\vec{u}_{\rm p}|^2) - 2 \,a^{2(1-\alpha)} u_i\conf{\cal F}_{\rm tot}^i
    \Bigr\}\,, \\ 
    \hspace{-10mm} \partial_0 u_i + u_j \partial_j u_i = &\, \frac{1 - |\vec{u}_p|^2}
    {1 - \omega |\vec{u}_{\rm p}|^2} \biggl[\omega  \partial_j u_j 
    +\omega \frac{1 - \omega}{1 + \omega} u_j \partial_j \ln \conf{\rho}_f -
    \frac{1}{\conf{\rho}_f} \biggl( 
    \conf{\cal F}_{\rm tot}^0
    - \frac{2 \omega}{1 + \omega} a^{2(1 - \alpha)}
    u_j \conf{\cal F}_{\rm tot}^j
    \biggr)
    \biggr] u_i \nonumber \\ 
    &\, - \frac{1 - |\vec{u}_{\rm p}|^2}
    {1 + \omega} \biggl(\omega a^{2(\alpha - 1)} \partial_i \ln \conf{\rho}_f -
    \frac{1}{\conf{\rho}_f}
    \conf{\cal F}_{\rm tot}^i \biggr)
    +
    \biggl(\frac{1 - |\vec{u}_{\rm p}|^2}
    {1 - \omega|\vec{u}_{\rm p}|^2}  (3 \omega  - 1)  +
    \alpha - 1\biggr) \HH u_i \,.
\end{mymathbox}
\end{subequations}
\noindent
The components of the Lorentz four-force are computed from the gauge field and the fluid primitive variables
using \Eqq{lorentz_forces}, where
Ohm's law [cf.~\Eqs{Ohms_law}{charge_current_density}]
is also used, allowing us to close the system by
providing the relation between the peculiar velocity and the fluid
current density $J^\mu_\fl$.
This system of equations can be generalized by adding any external forces
due to interactions with other fields to $\conf{\cal F}_{\rm tot}^\nu$
as long as the EOS, $\omega = \conf{p}_f/\conf{\rho}_f$, is a constant.

\subsubsection*{Limit of subrelativistic fluid velocity}

In the limit of subrelativistic fluid velocity,
the fluid EOM reduce to the analogous version of \Eqq{subrelativistic_visc_all}
with the replacement $\conf{\cal F}_{\rm ipf}^{\nu} \to \conf{\cal F}_{\rm tot}^{\nu} = \conf{\cal F}_{\rm ipf}^{\nu} + \conf{\cal F}_{U(1)}^\nu$.
Furthermore, the components of Ohm's four-current
[cf.~\Eqq{charge_current_density}], up to second order in $|\vec{u}_p|$,
reduce to 
\begin{subequations}
\begin{align}
    \conf{J}^0_\fl &=
    \conf{\rho}_e \bigl(1 + \tfrac{1}{2} |\vec{u}_{\rm p}|^2
    \bigr) - a^{2 (1 - \alpha)}\conf{\sigma}_f\, u_i\, {\cal E}_i + {\cal O} (|\vec{u}_{\rm p}|^3)\,,
    \\ 
    \conf{J}^i_\fl &=
    \conf{\rho}_e u_i - \conf{\sigma}_f \,{\cal E}_i \bigl(
    1 + \tfrac{1}{2} |\vec{u}_{\rm p}|^2\bigr)
    + \conf{\sigma}_f
    \varepsilon_{ijk} u_j {\cal B}_k
    + {\cal O} (|\vec{u}_{\rm p}|^3)\,,
\end{align}
\end{subequations}
such that the Lorentz force components are [cf.~\Eqq{lorentz_forces}]
\begin{subequations}
\begin{align}
    a^{\alpha - 1} \conf{\cal F}_{U(1)}^0
    = &\, - \conf{\rho}_e u_i {\cal E}_i
    + \conf{\sigma}_f
    |\vec{\cal E}|^2 \bigl(1 + \tfrac{1}{2} |\vec{u}_{\rm p}|^2\bigr) - \conf{\sigma}_f \, u_i \,
    \conf{T}^{0i}_{U(1)}
    + {\cal O} (|\vec{u}_{\rm p}|^3)\,,  \\ 
    a^{1 - \alpha} 
    \conf{\cal F}_{U(1)}^i = &\,
    - \conf{\rho}_e \bigl(1 + \tfrac{1}{2} |\vec{u}_{\rm p}|^2\bigr) 
    {\cal E}_i + a^{2(1 - \alpha)}
    \conf{\sigma}_f (\, \vec{u} \cdot \vec{\cal E}\,) {\cal E}_i + 
    \conf{\rho}_e (\vec{u} \times \vec{\cal B})_i \nonumber \\ &\, +
    \conf{\sigma}_f \, 
    \conf{T}^{0i}_{U(1)}
    \bigl(1 + \tfrac{1}{2}|\vec{u}_{\rm p}^2|\bigr) - \conf{\sigma}_f u_i \, |\vec{\cal B}|^2 + \conf{\sigma}_f (\, \vec{u} \cdot \vec{\cal B}\,) {\cal B}_i +
    {\cal O} (|\vec{u}_{\rm p}|^3)\,.
\end{align}
\end{subequations}

\subsubsection*{Large conductivity limit: Ideal and resistive magnetohydrodynamics}
\label{nodispcurrent}

A common assumption in magnetohydrodynamics (MHD) is that of large conductivity, being the
limit of infinite conductivity known as ideal MHD.
Ideal MHD corresponds to the limit
when electromagnetic entropy production vanishes, as seen in \Eq{entropy2}, and
is applicable in
many astrophysical and cosmological scenarios \cite{Brandenburg:2004jv,2021amff.book.....S}.
In particular, the limit of large but finite conductivity is applicable during the radiation-dominated era, when the fluid is mostly composed 
by a thermal plasma of relativistic particles.
Indeed, the value
of the magnetic diffusivity $\eta_{\rm diff} = 1/\sigma_f$
in the radiation-dominated era
is several orders of magnitude smaller than one Hubble time
(see figure 4 in ref.~\cite{RoperPol:2025lgc}), justifying the assumption of ideal MHD for any relevant oscillations at large scales, i.e., $\lambda \gg \eta_{\rm diff}$.
In this limit, the displacement current $\partial_0 {\cal E}_i$
can be neglected and, hence,
instead of a dynamical equation for the electric field, we get a constraint equation from \Eq{eq:Maxwell},
directly relating the current density to the magnetic field,
\EQb
    \conf{J}^i_\fl
    = - \conf{J}^i_\varphi - \conf{J}^i_\Phi +
    a^{\alpha - 1}
    (\vec{\nabla} \times \vec{\cal B})_i
    + (1 -\alpha)\, a^{1 - \alpha}\,
    \HH \, {\cal E}_i\,.
    \label{J_curlB}
\end{mymathbox} \noindent
On the other hand, the evolution of the gauge field in \Eq{gauge_eq}
reduces to a first-order differential equation,
\EQb \label{induction}
    \partial_0 A_i = \frac{\conf{\eta}_{\rm diff} \conf{\rho}_e u_i
    + (\vec{u} \times \vec{\cal B})_i + (\conf{\eta}_{\rm diff}/\gamma)
    \bigl[
    \conf{J}_\varphi^i + \conf{J}^i_\Phi -
    a^{\alpha - 1}
    (\vec{\nabla} \times \vec{\cal B})_i\bigr]}{1 + (1 - \alpha) a^{1- \alpha}
    (\conf{\eta}_{\rm diff}/\gamma)
    \HH} + \partial_i A_0\,,
\end{mymathbox} \noindent
where we have used Ohm's law [cf.~\Eq{Ohms_law}]
to relate the electric field to the fluid variables and the magnetic field, 
and we have defined the comoving magnetic diffusivity $\conf{\eta}_{\rm diff}
= 1/\conf{\sigma}_f = a^{-1} \eta_{\rm diff}$.

\EEq{induction} is commonly known as the induction equation in MHD,
due to the induction term $\vec{u} \times \vec{\cal B}$ that
drives the gauge field evolution in the limit of ideal MHD,
$\eta_{\rm diff} = 0$.
Then, the induction equation is evolved together with the fluid 
EOM,
either in their conservation or non-conservation form.
The Lorentz four-force components can be computed using \Eqq{lorentz_forces}.
Finally, the Ohmic current density is substituted by \Eq{J_curlB} to close the system.
Similarly, the components of the electromagnetic stress-energy tensor in \Eqq{Tmunu_EM_comps} can be expressed in
terms of the magnetic field and the fluid variables in this limit, getting rid of one degree of freedom (the electric field)
in the dynamical equations.

\subsubsection*{Relativistic Alfv\'en speeds and Boris correction}

The relativistic MHD equations in the limit of large
conductivity discussed above present
Alfv\'en waves that propagate in the plasma and can be studied
in the linear regime (see, for example, Sec.~6.5 in ref.~\cite{RoperPol:2025lgc}).
The speed of propagation of transverse Alfv\'en waves becomes
\begin{equation}
    \vA = \frac{{\cal B}_0}{\sqrt{(1 + \omega) \conf{\rho}_0}}\,,
\end{equation}
where ${\cal B}_0$ and $\conf{\rho}_0$ are the  background
comoving homogeneous magnetic
and energy density fields
and $\omega = p_f/\rho_f$ is the constant EOS of the
fluid perturbations.
Therefore, the electric and magnetic field perturbations propagate as transverse
waves with angular
frequency $ k_\parallel \vA$,
where $k_\parallel$ is the projected wavenumber along the direction of propagation.
From this expression one finds that the Alfv\'en speed
is not bounded by the speed of light and it can become superluminal
in regions of small energy density or large magnetic fields,
independently of the
amplitude of the peculiar velocity $u_i^{\rm p}$.
The solution to avoid unphysical superluminal Alfv\'en waves
is to either drop the assumption of negligible displacement current
and evolve Maxwell equations for both the electric and
magnetic fields or, alternatively, to include the Boris correction
due to the
displacement current in the Lorentz force \cite{jay_p__boris_1970}
(see details in Sec.~6.6 of ref.~\cite{RoperPol:2025lgc},
where the Boris correction
is applied to relativistic MHD in an expanding Universe).
Applying this correction, the propagation speed
of transverse Alfv\'en waves becomes
\begin{equation}
    v_{\rm A} = \frac{{\cal B}_0}{\sqrt{(1 + \omega) \tilde \rho_0 +{\cal B}_0^2}} \leq 1\,.
\end{equation}

\subsubsection{Scalar field-fluid dynamics}
\label{subsec:scalar_fluid}

In this section, we deal with the dynamics of a fluid coupled to a singlet scalar field.
This system has been studied numerically to describe the dynamics
of first-order phase transitions in the pioneer work of
ref.~\cite{Hindmarsh:2013xza} and follow-up work \cite{Hindmarsh:2015qta,Hindmarsh:2017gnf,Cutting:2019zws,Correia:2025qif}, 
using the alternative non-conservation form for relativistic fluid variables in flat spacetime,
already generalized to an expanding background in \Eqq{relativistic_EZeqs},
and which we discuss at the end of this section.
Here we present our own approach for the fluid equations, both in conservation and non-conservation forms, and in an expanding background.
As we will show later, our formalism naturally allows for a lattice implementation
of arbitrary higher-order accuracy of both timestepping (see \Sec{sec:RK}) and spatial derivatives (see \Sec{lattice_fluid_scalar}).

The EOM of the scalar field-fluid system have already been presented in \Sec{subsec:FLRW_Dynamics_Cont}, precisely \Eq{eq:klein_gordon_singlet_FLRW} for the scalar field and \Eqs{eq:EnergyConservation}{eq:MomentumConservation} for the fluid, as
\begin{subequations}
\begin{align} \label{scalar_here}
    \phi'' - a^{-2(1 - \alpha)} {\vv\nabla}^{\,2} \hspace{-1mm}\phi +
    (3 - \alpha)\mathcal{H} {\phi'} =&\, - a^{2 \alpha}
    \frac{\partial V_0}{\partial \phi}
    + a^{2 \alpha}
    \, \Omega_\phi
    \,, \\
    \partial_0 T_\fl^{0 0} + \partial_i T_\fl^{i0} + (3+2\alpha)\mathcal{H}T_\fl^{00} 
    + a^{-2\alpha} \mathcal{H}\, T^{(3)}_\fl
    =&\,  {\cal F}^0_{\rm int}
    \,, \\
    \partial_0 T_\fl^{0i} + \partial_j T_\fl^{ji} + (5+\alpha)\mathcal{H}\,T_\fl^{0i}
    =&\, {\cal F}^i_{\rm int}
    \,,
\end{align}
\end{subequations}
where, in the absence of
gauge interactions, ${\cal F}_{\rm int}^\nu = {\cal F}_\phi^{\nu} = \Omega_\phi \, \partial^\nu \phi$, with $\Omega_\phi$ given in \Eq{friction_Lagrangian}. Hence, in order to solve the system we need to know the form of $\Omega_\phi$ in terms of scalar and fluid variables. For this purpose, we present the EOM for the stress-energy tensor
components, derived from Boltzmann equation.
For simplicity, we first present the computation
in flat Minkowski spacetime and later recover the expansion of the Universe in \Eq{equation_higgs_exp} by adding the Hubble friction terms (recall \Sec{subsec:fld_dynamics_cont}).

\subsubsubsection*{Scalar field-fluid dynamics in flat spacetime}

In kinetic theory, the conservation of $T^\mn_\fl$ is obtained taking the first
moment of Boltzmann equation.
In flat spacetime, in the absence of external forces, we find \cite{2002rbet.book.....C,Rezzolla:2013dea}
\begin{equation}
    \partial_\mu T^\mn_\fl = 0\,, \qquad \text{with \ } \quad
    T^\mn_\fl = \sum_i
    T^{\mu \nu}_{\fl, i}\,, \quad 
    T^{\mu \nu}_{\fl, i} \equiv  N_i
    \int \frac{\dd^3 \kk}{(2 \pi)^3}
    \frac{k^\mu k^\nu}{E_i} \, f_{i}\,,
\label{boltzmann_emt}
\end{equation}
where $N_i$ is the number of internal degrees of freedom for each particle species $i$, $f_i$ is their distribution function,
$k^\mu = (E, \vec{k})$ the relativistic four-momentum, and $E_i = \sqrt{|{\vec k}|^2 + m_i^2}$ the associated relativistic energy dispersion relation, i.e., $k^\mu k_\mu = -m^2_i$. In equilibrium $f_i = f_{{\rm eq},i}$, with $f_{{\rm eq},i}$ the Fermi-Dirac distribution for fermions and the Bose-Einstein distribution for bosons,
\begin{equation} \label{distr_function}
    f_{\rm eq} = \biggl( \exp\biggl[\frac{U_\mu k^\mu - \mu}{T}\biggr]
    - (-1)_F \biggr)^{-1}\,,
\end{equation}
where $\mu$ and $T$ are the local chemical potential and temperature,
and $(-1)_{\rm F}$ takes the value $1$ ($-1$) for bosons (fermions).

In LTE, it can be shown, introducing \Eq{distr_function} in \Eq{boltzmann_emt}
and integrating by parts,
that $T^\mn_\fl$ takes the perfect fluid form [cf.~\Eq{perf_fluid}],
\begin{equation}
\label{tmunu_perfect_fluid}
    T^\mn_{\rm pf} = w_\f U^\mu U^\nu + p_\f g^\mn\,,
\end{equation}
where $U^\mu$ is the (bulk) four-velocity of the particle ensemble, and
the energy density and pressure of the fluid in LTE are defined as
\begin{equation}
    \rho_\f =
    \sum_i N_i \int \frac{\dd^3 \kk}{(2 \pi)^3} E_i \, 
    f_{{\rm eq}, i}\,,
    \qquad p_\f = \frac{1}{3} \sum_i N_i \int \frac{\dd^3 \kk}{(2 \pi)^3} \frac{|\kk|^2}{E_i} f_{{\rm eq}, i}\,.
    \label{pressure_and_energy_density_def}
\end{equation}
The enthalpy is given by the sum of the energy density and pressure $w_\f \equiv \rho_\f + p_\f$.
Combining the energy density and the pressure defined in
\Eq{pressure_and_energy_density_def},
again assuming LTE and integrating by parts,
one finds the following thermodynamic
relation,
\begin{equation}
    w_\f = T \frac{\partial p_\f}{\partial T} \biggr|_{\mu} + \mu n_\f 
   \,, \qquad \text{where \ } \quad
   n_\f \equiv \sum_i N_i \int \frac{\dd^3 \kk}{(2 \pi)^3} f_{{\rm eq}, i}\,. 
\end{equation}
Based on the baryon-to-photon ratio measured at present time $\eta_B 
\equiv n_B/n_\gamma \simeq 6 \times 10^{-10}$ \cite{Burles:1999zt,OMeara:2000tmq}, the chemical potential can usually
be neglected at high $T$, since we expect $\mu/T \sim \eta_B$ at $T \gg m_e \simeq 0.5$ MeV \cite{Schwarz:2003du,Fromerth:2002wb}.
Therefore, from now on, we neglect $\mu/T \ll 1$,
such that the enthalpy can be expressed as
\begin{equation} \label{enthalpy_zero_mu}
    \lim_{\mu/T \ll 1} w_f = T \frac{\partial p_f}{\partial T}\,.
\end{equation}

We can reach a deeper insight on how the above description is affected by the presence of mass corrections
due to interactions of
the relativistic primordial plasma with the
scalar field by studying the relations between the fluid variables and the free energy density of the system, ${\frak F}$.
For a system composed by fermions and bosons, this is \cite{Dolan:1973qd,Kirzhnits:1976ts,Quiros:1999jp}
\begin{align}
    {\frak F} &\, \equiv 
    V_0 (\phi)
    + T \sum_i  \, (-1)_F  N_i \int \frac{\dd^3 \kk}{(2 \pi)^3}
    \log \bigl[1 -(-1)_F \,
    e^{-E_i/T}\bigr]
    \nonumber \\ 
    &\, = V_0(\phi) + T^4 \sum_{i \in B} J_{B,i} \biggl(\frac{m_i}{T} \biggr) + T^4 \sum_{i \in F} J_{F,i} \biggl(\frac{m_i}{T} \biggr)\,,
    \label{V_tot}
\end{align}
where the subscripts $B$ and $F$ stand for bosons and fermions. In the limit of high temperatures, i.e.,
$m_i/T \ll 1$ for each species,
the leading terms in the expansions of the functions $J_{B,F}(m_i/T)$ are 
respectively $J_B(m_i/T) = -\tfrac{\pi^2}{90} + {\cal O}(m_i^2/T^2)$
and $J_F(m_i/T) = -\tfrac{7}{8} \tfrac{\pi^2}{90} + {\cal O}(m_i^2/T^2)$.
Defining the effective potential as the
free energy density in this limit, we find
\begin{align} \label{large_T_limit}
    V_{\rm eff} (\phi, T) \equiv \lim_{\tfrac{m_i}{T} \ll 1}
    {\frak F}  = V_0 (\phi) - \frac{\pi^2}{90} \, g(\phi, T) \, T^4
    + V_T(\phi, T)\,,
\end{align}
where the thermal correction $V_T$ to the effective potential
is defined as the
contribution from the
higher-order terms in the expansion over $m/T$ of the $J_{B,F}$ functions,
identically zero only in the limit of infinite temperature or zero mass,
when the $T^4$ term dominates.
In the expression above,
the number of relativistic degrees of freedom,
$g(\phi, T) \equiv \sum_{i \in B} N_i + \frac{7}{8} \sum_{i \in F} N_i$,
and the thermal potential, $V_T(\phi, T)$,
are written as a function of $\phi$ and $T$, since we assume that the particles in the plasma may become massive 
through a Higgs-like mechanism characterized
by the scalar field $\phi$.

For a fluid composed of fermions and bosons in LTE,
integrating \Eq{V_tot} by parts, we find
the relation between the free
energy density and the fluid pressure,
\begin{align}
    {\frak F} - V_0(\phi) &= T \sum_i (-1)_F N_i \int \dd k \frac{k^2}{2 \pi^2} \log \bigl[ 1-(-1)_F \, e^{-E_i/T}\bigr] \nonumber \\
    & = - \frac{T}{3} \sum_i (-1)_F N_i \int \dd k \frac{k^3}{2\pi^2} \partial_k \log \bigl[ 1-(-1)_F \, e^{-E_i/T}\bigr]  = - \frac{1}{3} \sum_i N_i \int \frac{d^3 \vec{k}}{(2\pi)^3} \frac{k^2}{E_i} f_{\rm eq,i} \equiv - p_\f\,.
\label{proof_freeenergy_pressure}
\end{align}
Hence, in the high temperature expansion, we can write the fluid pressure as
\begin{equation}
    p_\f  = - V_{\rm eff} +
    V_0 = p_{\rm rad} - V_T 
    \,, \qquad \text{where \ }
    p_{\rm rad} = \frac{1}{3} \sigma_{\rm dof} T^4 \quad \text{and \ \ }
    \sigma_{\rm dof} \equiv \frac{\pi^2}{30} g(\phi, T)\,,
\end{equation}
with the radiation pressure $p_{\rm rad}$
representing the leading-order term in $m/T$.
Then, we find $\rho_{\rm rad} = w_{\rm rad} - p_{\rm rad} =
T (\dd p_{\rm rad}/\dd T)|_{\sigma_{\rm dof}} - p_{\rm rad}
= \sigma_{\rm dof} T^4$, which yields the EOS $\omega =
p_{\rm rad}/\rho_{\rm rad} = \tfrac{1}{3}$ considered in \Sec{perfect_fluids}
for a relativistic radiation fluid.
For a non-zero $V_T(\phi, T)$ or varying {\em dof}, this EOS changes,
and we can no longer treat $\omega = p_\f/\rho_\f$ as a constant.
The fluid pressure, energy density, and enthalpy become
\begin{subequations}
 \label{pressure_density}
\begin{align}
    &\, p_\f (\phi, T) = p_{\rm rad}
    - V_T\,,
    \\ 
    &\, \rho_\f (\phi, T) = \rho_{\rm rad} \biggl( 1  +
    \frac{1}{3} \frac{\partial \ln \sigma_{\rm dof}}
    {\partial \ln T}\biggr) + V_T
    - T \frac{\partial V_T}{\partial T}\,,
    \\ &\,
    w_\f (\phi, T) =
    w_{\rm rad} \biggl(1  + \frac{1}{4}
    \frac{\partial \ln \sigma_{\rm dof}}
    {\partial \ln T} \biggr) - T \frac{\partial V_T}{\partial T}\,.
\end{align}
\end{subequations}

When the particles in the plasma interact with a Higgs-like
scalar field,
their masses depend on the expectation value of
$\phi$.
To incorporate this interaction,
the conservation law of the stress-energy tensor [cf.~\Eq{boltzmann_emt}] is modified
due to the interaction force ${\cal F}^\nu_\phi$
\cite{Ignatius:1993qn,Moore:1995si,Moore:1995ua,Espinosa:2010hh,Hindmarsh:2020hop,Ekstedt:2025awx}, 
\begin{align} \label{cons_scalar} 
    \partial_\mu T^\mn_\fl = {\cal F}^\nu_\phi =  &\, -
    (\partial^\nu \phi) \sum_i N_i \frac{\dd m^2_i}{\dd \phi} \int
    \frac{\dd^3 \kk}{(2 \pi)^3} \frac{1}{2 E_i} f_i
    \nonumber \\ = &\, - (\partial^\nu \phi) \sum_i N_i \frac{\dd \ln m_i}{\dd \phi} \int
    \frac{\dd^3 \kk}{(2 \pi)^3} \frac{m_i^2}{E_i} f_i
    = (\partial^\nu \phi) \sum_i \frac{\dd 
    \ln m_i}{\dd \phi} T_{\fl, i}
    \,,
\end{align}
where
$T_{\fl,i} = g_{\mu \nu} T^{\mu \nu}_{\fl, i}$ is the trace of the
stress-energy tensor
associated to the species $i$.

We now consider
perturbations around LTE, such that $f_i = f_{\rm eq, i} + \delta f_i$.
The \rhs of the stress-energy tensor evolution equation can then be split as
\begin{align}
\label{fluid_eqs_scalar_split}
    {\cal F}_\phi^\nu =
    \Omega_\phi  \, \partial^\nu \phi\,, ~~~
    \text{with} ~~~~
    \Omega_\phi = - 
    \Biggl[ \,  \underbrace{\biggl( 
    \sum_i N_i \frac{\dd \ln m_i}{\dd \phi} \int \frac{d^3 \vec{k}}{(2\pi)^3} \frac{m_i^2}{E_i} f_{{\rm eq},i} \biggr)}_{\text{{\rm LTE} contribution}} + \underbrace{\biggl( \sum_i N_i \frac{\dd \ln m_i}{\dd \phi} \int \frac{d^3 \vec{k}}{(2\pi)^3} \frac{m_i^2}{E_i} \delta f_i \biggr)}_{\text{out-of-equilibrium contribution}}  \, \Biggr]
    \,.
\end{align}
We note that \Eq{fluid_eqs_scalar_split} is equivalent to the one found
in \Eqs{fluid_kernels}{cons_fluid},
which allows us to identify the term $\Omega_\phi$ directly
from kinetic theory.

The LTE contribution to the scalar interaction term $\Omega_\phi$
can be expressed as a derivative of the pressure $p_f$
with respect to $\phi$ \cite{Moore:1995ua}.
This can be shown using the fact that the only dependence on $\phi$ in $p_f$ is through the dependence of $E_i$ on $m_i(\phi)$ and of $f_{{\rm eq},i}$ on $E_i$, hence we can use the chain rule for deriving with respect to $\phi$, such that $\dd (f_{{\rm eq},i}/E_i)/\dd\phi = (\dd m_i^2/\dd \phi) (\dd E_i/\dd m_i^2) (\dd (f_{{\rm eq},i}/E_i)/\dd E_i)$,
\begin{equation}
    \frac{\partial p_f}{\partial \phi} = \frac{1}{6} \sum_i N_i 
    \frac{\dd m_i^2}{\dd \phi} \int \frac{\dd^3 \kk}{(2 \pi)^3}
    \frac{|\kk|^2}{E_i^2} \biggl(\frac{\partial f_{{\rm eq},i}}{\partial E_i}
    - \frac{f_{{\rm eq},i}}{E_i}\biggr)\,.
\end{equation}
Since the only dependence on $\kk$ of $f_{{\rm eq},i}$ is through its dependence on $E_i = \sqrt{|\kk|^2 + m_i^2(\phi)}$, we can write $\partial_{E_i} f_{{\rm eq},i} = 
(\partial_{E_i} k) \, (\partial_k f_{{\rm eq},i}) = (E_i/k)
\,\partial_k f_{{\rm eq},i}$. Then, spherical symmetry of the integrand, and integration by parts of the term including $\partial_k f_{{\rm eq},i}$, yield the following result,
\begin{equation} \label{pressure_trace}
    \frac{\partial p_f}{\partial \phi} = -
    \sum_i N_i \frac{\dd \ln m_i}{\dd \phi}
    \int \frac{\dd^3 \kk}{(2 \pi)^3} \frac{m_i^2}{E_i} f_{{\rm eq},i} =
    \sum_i \frac{\dd \ln m_i}{\dd \phi}
    \, T_{{\rm pf},i} \,,
\end{equation}
where $T_{{\rm pf}, i} = g_{\mu \nu} T^{\mu \nu}_{{\rm pf},i}$ is the perfect-fluid
contribution to the trace $T_{\fl, i}$.
Moreover, we define the integral corresponding to the out-of-equilibrium contribution in \Eq{fluid_eqs_scalar_split} as \cite{Ignatius:1993qn,Hindmarsh:2020hop,Ekstedt:2025awx}
\begin{align}
\label{delta_phi}
    \delta_{\phi} \equiv  \sum_i N_i \frac{\dd \ln m_i}{\dd \phi} \int \frac{d^3 \vec{k}}{(2\pi)^3} \frac{m_i^2}{E_i} \delta f_i 
    \,.
\end{align}
Hence, substituting the equilibrium and out-of-equilibrium contributions, respectively given
in \Eqs{pressure_trace}{delta_phi},
into \Eq{fluid_eqs_scalar_split}, we find
\begin{equation} \label{equation_higgs}
    \Omega_\phi = \frac{\partial p_f}{\partial \phi} 
    - \delta_\phi = - \frac{\partial V_T}{\partial \phi}
     + p_{\rm rad}
    \frac{\partial \ln \sigma_{\rm dof}}{\partial \phi} - \delta_\phi\,.
\end{equation}

In particular, in the case in which the masses are proportional to the Higgs field,
$m_i \propto \phi$ (hence $\dd \ln m_i/\dd \phi = 1/\phi$),
the LTE contribution to the scalar interaction term becomes
\begin{align}
    \label{Omega_phi_trace}
     \frac{\partial p_\f}{\partial\phi} = - \sum_i N_i \frac{\dd \ln m_i}{\dd \phi} \int \frac{d^3 \vec{k}}{(2\pi)^3} \frac{m_i^2}{E_i} f_{{\rm eq},i} = -\frac{1}{\phi} \sum_i N_i \int \frac{d^3 \vec{k}}{(2\pi)^3} \frac{m_i^2}{E_i} f_{{\rm eq},i}  = \frac{1}{\phi} T_{\rm pf}\,,
\end{align}
where 
$T_\pf = \sum_i T_{\pf, i}$ is the sum of the traces of each species $i$.
We note that  \Eq{Omega_phi_trace}
provides a consistency relation on the $T$ and $\phi$
dependence of the thermal corrections to the effective potential,
\begin{equation} \label{consistency_potential}
    \phi \frac{\partial p_f}{\partial \phi} = T_{\rm pf} = 3 p_f - \rho_f
    \Rightarrow \phi
    \frac{\partial V_T}{\partial \phi} + T \frac{\partial V_T}{\partial T}
    = 4 V_T + \frac{1}{3} \rho_{\rm rad} \biggl( \frac{\partial \ln \sigma_{\rm dof}}{\partial \ln \phi} + \frac{\partial \ln \sigma_{\rm dof}}{\partial \ln T} \biggr)\,.
\end{equation}
Regarding the out-of-equilibrium 
component,
following the pioneer work of ref.~\cite{Ignatius:1993qn},
the phenomenological description
\begin{eqnarray}
    \delta_\phi = \eta_\phi U^\mu \partial_\mu \phi\,,
    \label{eta_phi_fric}
\end{eqnarray}
based on the Lorentz invariant $\Theta \equiv U^\mu \partial_\mu \phi$, has been used
in the literature.
For example, numerical simulations of the scalar-fluid system
considered a constant $\eta_\phi$ 
\cite{Hindmarsh:2013xza,Hindmarsh:2015qta,Cutting:2019zws,Correia:2025qif}.
Recent work identified, in the limit of large amount of collisions, the
scalar damping coefficient as $\eta_\phi = A_\phi \, \phi^2/T$
\cite{Huber:2013kj,Konstandin:2014zta,Ekstedt:2025awx},
$A_\phi$ being a
constant that depends on the particle content of the plasma.
Hence, in the following, we allow $\eta_\phi$ to be a function of $\phi$
and $T$.

The entropy production that has been
studied for an imperfect fluid in \Sec{viscous_term} [cf.~\Eq{entropy}] and 
with fluid-gauge field interactions in \Sec{subsec:gauge_fluid_dyns} [cf.~\Eq{entropy2}],
can be considered for the scalar-fluid system by
including the external force $U_\mu
{\cal F}_\phi^\mu = {\Omega}_\phi \, U_\mu \partial^\mu \phi = \Omega_\phi \, \Theta$,
\begin{equation}
    D_\mu^G s_f^\mu = \frac{1}{T} \bigl(\pi^\mn S_\mn 
    + \tfrac{1}{3} \dev \, \theta
    -  (\Omega_\phi - \partial_\phi p_f) \, \Theta \bigr)
    = \frac{1}{T} \bigl(
    2 \, \nu \, w_f \, \sigma^\mn \sigma_\mn
    + \xi \, w_f \, \theta^2 +
    \eta_\phi \Theta^2
    \bigr) \geq 0\,,
\end{equation}
where we have again defined the fluid
entropy $s_f = w_f/T = \partial_T \, p_\f$, and used the definition of the deviatoric stress tensor given in \Eq{Pi_visc}. 

\subsubsubsection*{Scalar field-fluid dynamics in an expanding Universe}

The extension of \Eq{fluid_eqs_scalar_split} to an expanding Universe trivially follows from the derivation of \Sec{viscous_term}.
In fact, from \Eq{general_eqs}, we have
\begin{equation}
    \partial_\mu \conf{T}^\mn_\pf
    =
    \conf{\cal F}_\phi^\nu
    + \conf{\cal F}_{\rm ipf}^\nu + \conf{\cal F}_H^\nu
    \,,
    \label{equation_higgs_exp}
\end{equation}
where, as in previous sections, we introduce the rescaling $\conf{\cal F}^{\nu}_\phi = a^{4+2\alpha} {\cal F}^{\nu}_\phi$, and we have an analogous rescaling for the stress-energy tensor and all forces in the \rhs\!\!\,,
\begin{equation}
\label{components_fluidscalar_interaction}
    \conf{\cal F}_\phi^0 = a^{4 + 2 \alpha} {\cal F}_\phi^0 = - \conf{\Omega}_\phi \pi_\phi \,, \qquad
    \conf{\cal F}_\phi^i = a^{4 + 2\alpha} {\cal F}_\phi^i
    = a^{2(\alpha - 1)} \conf{\Omega}_\phi \, \partial_i \phi\,,
    \qquad \text{where} \quad \conf{\Omega}_\phi = a^4 \Omega_\phi\,.
\end{equation}
Note that we have defined the scalar field conjugate
momentum as $\pi_\phi = \phi'$.
Since we do not rescale
the scalar field $\phi$, the rescaling on $\Omega_\phi$ implies that
$\conf{V}_T = a^4 V_T$ and $\conf{\delta}_\phi = a^4 \delta_\phi$.
Moreover, the rescaled perfect fluid stress-energy tensor, defined as in previous sections $\conf{T}^\mn_\pf = a^{4 + 2\alpha} T_\pf^\mn$, is given by
\begin{align}
    \conf{T}^\mn_\pf =
    \conf{w}_f
    \conf{U}^\mu \conf{U}^\nu + \conf{p}_f \,a^{2\alpha} g^\mn\,,
\end{align}
where the comoving fluid enthalpy and pressure are given by
\begin{equation} \label{components_scalar_fluid}
    \conf{w}_f = \conf{w}_{\rm rad}
    \biggl(1 + \frac{1}{4} \frac{\partial \ln \sigma_{\rm dof}}{\partial \ln T}\biggr) 
    - \conf{T} \frac{\partial \conf{V}_T}{\partial \conf{T}}\,, 
    \qquad
    \conf{p}_f = \conf{p}_{\rm rad} - \conf{V}_T
    \,, \qquad \text{and} \qquad \conf{w}_{\rm rad} = 4 \, \conf{p}_{\rm rad} = \tfrac{4}{3} \sigma_{\rm dof} \conf{T}^4 \,,
\end{equation}
where $\tilde{T}=aT$ represents the rescaled temperature.

In summary, the EOM of the scalar-fluid system
can be written
explicitly as [cf.~\Eq{equation_higgs_exp}],
\begin{subequations}
    \label{fluid_eqs_scalar}
\EQalign
        \partial_0 \conf{T}^{00}_\pf +
        \partial_i \conf{T}^{0i}_\pf = &\, \conf{\cal F}^0_{\phi} + \conf{\cal F}^0_{\rm ipf} + 
        \conf{\cal F}^0_{H} = \pi_\phi
        \biggl(- \frac{\partial\conf{p}_f}{\partial \phi}
        + \conf{\delta}_\phi
        \biggr) + \conf{\cal F}^0_{\rm ipf}
        + (\conf{\rho}_f - 3 \conf{p}_f) \HH
        \,, \\
        \partial_0 \conf{T}^{0i}_\pf + \partial_j \conf{T}^{ij}_\pf = &\,
        \conf{\cal F}^i_{\phi} + \conf{\cal F}^i_{\rm ipf} + 
        \conf{\cal F}^i_{H}  = a^{2(\alpha - 1)} (\partial_i \phi) \,
        \biggl(\frac{\partial\conf{p}_f}{\partial \phi}
        - \conf{\delta}_\phi
        \biggr) + \conf{\cal F}^i_{\rm ipf} + (\alpha - 1) \conf{w}_f \gamma^2 \HH u_i\,,
\end{mymathbox}
\end{subequations}
\noindent
where the components of the Hubble friction and imperfect four-force are respectively given in \Eqs{Hubble_friction}{ipf_force}.
In particular, since the rescaled deviatoric tensor
can be expressed as in 
\Eq{rescaled_deviatoric_tensor}, the rescaled
imperfect four-force
is [cf.~\Eq{conservation_viscous}]
\begin{equation}
    \conf{\cal F}_{\rm ipf}^\nu = \partial_\mu
    \conf{\pi}^\mn + 
    \tfrac{1}{3} \partial_\mu \bigl(\conf{h}^\mn  a^{-2\alpha} \cdev 
    \bigr) + a^{-2\alpha} \cdev \HH \delta^\nu_{\ \, 0} + (1 - \alpha) 
    \HH \cdev^{0i} \delta^\nu_{\ \, i}\,,
\end{equation}
where the rescaled variables describing viscosity
and the rate-of-strain-tensor components
have been introduced in \Eqss{viscosity_rescalings}{comoving_theta_visc},
and the comoving trace of the deviatoric tensor is $a^{-2\alpha} \cdev = 3 \, \conf{\xi} \, \conf{w}_f \, \conf{\theta}$.
Finally, the rescaled scalar damping term is given by
\begin{align} \label{rescaled_frict}
    \conf{\delta}_\phi = a^4 \delta_{\phi} = \conf{\eta}_\phi \, \conf{\Theta}\,, \qquad \text{with \ } \conf{\eta}_\phi = a^3 \eta_\phi
    \quad \text{and \ \ } \conf{\Theta} = a \Theta = a^{1 - \alpha} \gamma (\pi_\phi + u_i \partial_i \phi)\,.
\end{align}

Once we have identified ${\Omega}_\phi \equiv \partial_\phi {p}_f - {\delta}_\phi$ from the scalar-fluid interactions, we can express the EOM of the scalar
field, given in \Eq{scalar_here}, as
\begin{mymathbox}
\begin{align}
    \label{equations_scalar}
    \phi'' - a^{-2(1 - \alpha)} {\vv\nabla}^{\,2} \hspace{-1mm}\phi +
    (3 - \alpha)\mathcal{H} {\phi'} =&\, - a^{2 \alpha}
    \frac{\partial V_{\rm eff}}
    {\partial \phi}
    - a^{2 \alpha}
    \, \delta_\phi
    \,, 
\end{align}
\end{mymathbox}
\noindent
where we recall that $V_{\rm eff} = V_0 - p_f$, defined in \Eq{large_T_limit}.

\subsubsubsection*{Conservation form}

As in previous sections, we can solve the system of \Eqq{fluid_eqs_scalar}
for the variables $\conf{T}^{0\mu}_\pf$.
However, in this case, we cannot set the EOS $\omega = p_f/\rho_f$
to a constant, since $\omega$ depends on the
values of the scalar field $\phi$ and the temperature $T$.
Therefore, solving the system of \Eqq{fluid_eqs_scalar} in the conservation form requires computing the temperature of the fluid
at each step in order to compute the
$\conf{T}^{ij}_\pf$ components, as well as the forces $\conf{\cal F}_{\rm tot}^\nu = \conf{\cal F}_\phi^\nu + \conf{\cal F}_{\rm ipf}^\nu$
and
$\conf{\cal F}_H^\nu$.
Since $\conf{\pi}^\mn$ and $\cdev$ are linear in first-order spacetime
derivatives of the velocity field, $\conf{\cal F}_{\rm ipf}^\nu$
contains up to second-order
time derivatives of the velocity field.
Therefore, the total four-force components
cannot
be directly reconstructed from the dynamical
variables (i.e., $\conf{w}_f$, $u_i$, $\phi$, $\pi_\phi$) and their spatial gradients.
Only in the limit of subrelativistic fluid velocity
and small bulk viscosity, $\xi H \ll 1$,
the 
force ${\conf{\cal F}}^{\nu}_{\rm tot} = \conf{\cal F}^{\nu}_{\rm ipf} + \conf{\cal F}^{\nu}_{\phi}$,
does not depend on time
derivatives of the dynamical variables [cf.~\Eq{fviscmu}].
We will restrict the description of the imperfect fluid forces in the fluid-scalar system to this limit when we describe
the conservation form of the equations in the lattice in 
\Secs{sec:fluidLattice}{sec:fluid_bosonic}, such that
$\conf{\cal F}_{\rm ipf}^\nu$ can be reconstructed after advancing the dynamical variables,
$\conf{T}^{0\mu}_\pf$.

For the perfect fluid stress-energy tensor
components, given in \Eq{perf_fluid}, the
following relations hold
\begin{align} \label{equation_to_solve}
    a^{2(1 - \alpha)}
    \conf{T}^{0i}_\pf \conf{T}^{0i}_\pf = 
    (\conf{T}^{00}_\pf + \conf{p}_f) (\conf{T}^{00}_\pf - \conf{\rho}_f) 
    = 
    (\conf{T}^{00}_\pf)^2 + \conf{T}^{00}_\pf (\conf{p}_f - \conf{\rho}_f)
    - \conf{p}_f \conf{\rho}_f\,.
\end{align}
Defining $r^2 \equiv a^{2(1 - \alpha)}
\sum_i (\conf{T}_\pf^{0i}/\conf{T}_\pf^{00})^2$
as in \Eq{T0i_sq}, this relation becomes
\begin{equation} \label{scalar_fluid_conservationform_relation}
    r^2 = 1 + \frac{\conf{p}_f - \conf{\rho}_f}{\conf{T}_\pf^{00}}
    - \frac{\conf{p}_f \conf{\rho}_f}{(\conf{T}_\pf^{00})^2}\,.
\end{equation}
Using the expressions of
$\conf{p}_f$ and $\conf{\rho}_f = \conf{w}_f - \conf{p}_f$
in terms of $\phi$ and $\conf{T}$ from \Eq{components_scalar_fluid},
this equation allows us to reconstruct the temperature from the
dynamical variables $\conf{T}^{00}_\pf$, $r^2$, and $\phi$.
Let us assume for the moment that we can invert
\Eq{scalar_fluid_conservationform_relation}
for a given $\conf{V}_T (\phi, \conf{T})$, such that we can
reconstruct $\conf{p}_f (\phi, \conf{T})$ and $\conf{w}_f (\phi, \conf{T})$.
In order to close the system, we still need to express
$\conf{T}^{ij}_\pf$, $u_i$, and $\gamma^2$
in terms of $\conf{T}^{0\mu}_\pf$, $\phi$, and $\conf{T}$,
\begin{equation} \label{reconstruction}
    \conf{T}^{ij}_\pf = \frac{\conf{T}^{0i}_\pf \conf{T}_\pf^{0j}}{\conf{T}_\pf^{00} +
    \conf{p}_f} + a^{2(\alpha - 1)}
    \conf{p}_f \delta_{ij}\,, \qquad u_i = \frac{\conf{T}^{0i}_\pf}{\conf{T}^{00}_\pf + \conf{p}_f}\,,
    \qquad \text{and \ } \quad \gamma^2 = \frac{\conf{T}^{00}_\pf + \conf{p}_f}
    {\conf{w}_f}\,.
\end{equation}

Let us now focus on inverting the relation in \Eq{scalar_fluid_conservationform_relation} in 
order to express the temperature in terms of the
fluid and scalar field variables.
If we restrict ourselves to renormalizable power law effective potentials, which are typically found
in the Standard Model and several extensions  of it at high
temperatures (see, e.g., refs.~\cite{Weinberg:1974hy,Kirzhnits:1976ts}, and 
ref.~\cite{Mustafa:2022got} for a recent review),
we can write  
\begin{align}
    V_T (\phi, T) =
    T^2 \, \phi^2 \, C_2 + T \, \phi^3 \, C_1
\label{model_effective_potential}
\end{align}
where 
$C_i$ for $i = 1, 2$ are dimensionless constants.
As an example, in the Standard Model's
electroweak theory, the next-to-leading order
term, proportional to $m_i^2/T^2$
in the effective potential, yields
$C_1=-(2m_W^3+m_Z^3)/(4 \pi \phi_b^3)$
and $C_2 =(2m_W^2+m_Z^2+2m_t^2)/(8 \phi_b^2)$,
with $m_W$, $m_Z$, and $m_t$ respectively being the W and Z bosons, and the top quark masses, and $\phi_b$ being the Higgs vacuum expectation value in the broken phase \cite{Dine:1992wr}. In extensions of the standard model one would have in general $C_1 = -\bigl[\sum_S m_S^3 (\phi) + \sum_V m_V^3(\phi)\bigr]/(4 \pi \phi^3)$ and $C_2 = \bigl[ \sum_S m_S^2(\phi)+3\sum_V m_V^2(\phi) + 6 \sum_F m_F^2(\phi)\bigr]/(24 \phi^2)$, with $m_S, m_V, m_F$ respectively being the $\phi$-dependent masses of the scalar, vector and fermionic fields in the theory \cite{Quiros:1999jp}.
Introducing
\Eq{model_effective_potential}
into \Eq{scalar_fluid_conservationform_relation},
we find that the solutions for the temperature become
the roots of an eigth-order polynomial $\sum_{j=0}^8 x_j \conf{T}^j \phi^{8 - j} = 0$,
with the following coefficients:
\begin{equation}
\begin{array}{lll}
    x_8 = \tfrac{1}{3} \sigma_{\rm dof}^2 \,, &
    x_7 = 0
    \,, &
    x_6 = - \tfrac{4}{3} a^2 \sigma_{\rm dof} C_2 \,, \\
    x_5 = - \sigma_{\rm dof} a^3 C_1 \,,
    &
    x_4 =
    \tfrac{2}{3} \sigma_{\rm dof} (\conf{T}_\pf^{00}/\phi^4)
    + a^4 C_2^2  \,, &
    x_3 =
    a^5 C_1 C_2 \,, \\
    x_2 = 0\,, &
    x_1 = a^3 C_1 (\conf{T}_\pf^{00}/\phi^4) \,, &
    x_0 = (\conf{T}_\pf^{00}/\phi^4)^2 (r^2 - 1)\,,
\end{array}
\end{equation}
where we have omitted the temperature dependence of
$\sigma_{\rm dof}$.
In general, from Abel-Ruffini theorem, since we have a polynomial of order $\geq 5$,
there is no algebraic solution for the roots, but they can be found 
numerically for each value of $\conf{T}_\pf^{00}/\phi^4$ and $r^2$,
and then one can select the relevant solutions (real and positive)
for $\conf{T}$.

A particular choice $C_1 = 0$, gives the following polynomial,
\begin{equation}
    V_T(\phi, T) =
    T^2 \phi^2 C_2 \quad \Rightarrow \quad
    x_8 \conf{T}^8 + x_6 \conf{T}^6 \phi^2
    + x_4 \conf{T}^4 \phi^4 + x_0 \phi^8 = 0\,.    
\end{equation}
This choice applies to theories predicting second-order or crossover phase transitions, which respectively correspond to the $C_1=0$ and $|C_1| \ll \lambda$ limits \cite{Senaha:2020mop}, with $\lambda$ being the coefficient of the quartic term in $V_{\rm eff} \supset \lambda \, \phi^4$. As an example, this approximately describes the Standard Model electroweak case at high temperatures, in which $|C_1| \ll \lambda$ \cite{Dine:1992wr}.
Then, the coefficients of the eigth-order polynomial become
\begin{equation}
\begin{array}{llll}
    x_8 = \tfrac{1}{3} \sigma_{\rm dof}^2 \,, &
    x_6 = - \tfrac{4}{3} a^2 \sigma_{\rm dof} C_2 \,, &
    x_4 = \tfrac{2}{3} \sigma_{\rm dof} 
    (\conf{T}_\pf^{00}/\phi^4) + a^4 C_2^2\,, &
    x_0 = (\conf{T}_\pf^{00}/\phi^4)^2 (r^2 - 1)\,.
\end{array}
\end{equation}
The roots of this polynomial can be found analytically using Ferrari's
quartic solution.

Another relevant case corresponds to the
toy model used in lattice simulations of the
scalar-fluid system \cite{Cutting:2019zws,Correia:2025qif}.
Within this toy model, again under the assumption that $\sigma_{\rm dof}$ does not depend on $T$ during the phase transition, the effective potential
is parameterized as \cite{Cutting:2019zws,Correia:2025qif,Bhusal:2026iuw},
\begin{equation}
\label{free_energy_toy_model}
    V_{\rm eff} = V_0 (\phi) - \tfrac{1}{3} \sigma_{\rm dof} (\phi) T^4
    = V_0 (\phi) - \tfrac{1}{3} \sigma_0 T^4 + \Delta V_0 (\phi_b)
    \frac{T^4}{T_c^4}
    \bigl[ 3 (\phi/\phi_b)^2 - 2 (\phi/\phi_b)^3 \bigr]
    \,,
\end{equation}
where $\phi_b$ is the vacuum expectation value
of the scalar field in the broken phase and $\Delta V_0 (\phi_b) =
V_0 (0) - V_0 (\phi_b)$.
In this model, the thermal corrections to the effective potential
are encapsulated in the
leading-order term $\tfrac{1}{3} \sigma_{\rm dof} (\phi)
T^4$
with $V_T = 0$.
Notice that in this case we have $\conf{p}_\f = \frac{1}{3} \sigma_{\rm dof}
\conf{T}^4$ and $\omega = \conf{p}_\f/\conf{\rho}_\f = 1/3$, as expected for a plasma composed of relativistic massless particles.
Therefore, the eigth-order polynomial becomes
\begin{equation} \label{equation_T4}
    x_8 \conf{T}^8 + x_4 \conf{T}^4 + x_0 = 0\,,
    \qquad
    \text{with \ } \left\{ \begin{array}{l}
        x_8 = \tfrac{1}{3} \sigma_{\rm dof}^2 
        (\phi)\,, \\
        x_4 = \tfrac{2}{3} \sigma_{\rm dof} (\phi)
        \conf{T}_\pf^{00} \,, \\
        x_0 = (\conf{T}_\pf^{00})^2 (r^2 - 1)\,.
    \end{array}  \right.
\end{equation}
Solving this polynomial, we find that the temperature is
\begin{equation}
    \sigma_{\rm dof} \conf{T}^4 = \conf{T}_\pf^{00} \bigl(\sqrt{4 - 3 r^2} - 1\bigr)\,,
\end{equation}
where we choose only the positive root.
We note that from this expression we can reconstruct $\gamma^2$ using \Eq{reconstruction}, which yields the same expression as what
we found in \Eqs{rho_u_fromT0mu}{gamma2_rel} taking $\omega = \tfrac{1}{3}$,
as expected,
\begin{equation}
    \gamma^2 = \frac{1}{4} \frac{2 + \sqrt{4 - 3 r^2}}{\sqrt{4 - 3 r^2} - 1}
    = \frac{1}{12} \frac{(2 + \sqrt{4 - 3 r^2})(\sqrt{4 - 3r^2} + 1)}
    {1 - r^2} = \frac{1}{2 (1 - r^2)}
    \Bigl(1 - \tfrac{1}{2} r^2 + \sqrt{1 - \tfrac{3}{4} r^2} \, \Bigr)
    \,.
\end{equation}
Therefore, the only difference with the system studied in \Sec{perfect_fluids}
with $\omega = 1/3$, is the appearance of the imperfect fluid viscous forces [cf.~\Eq{fviscmu}], and of the scalar-field
force in the EOM, 
\begin{equation}
    \conf{\cal F}_\phi^\nu = (\partial_\phi \conf{p}_f) \,
    \partial^\nu \phi = \frac{1}{3} T^4 (\partial_\phi \sigma_{\rm dof}) \, \partial^\nu \phi\,.
\end{equation}

\subsubsubsection*{Non-conservation form}

The energy and momentum conservation equations were found in terms of generic $\conf{p}_f$ and $\conf{\rho}_f$ for a perfect fluid
in \Eqq{equations_relativistic_general}, and trivially generalized to imperfect fluids
in 
\Eqq{general_EOMs}.
Extending the energy equation to also include
the scalar interaction force, i.e.,
$\conf{\cal F}^\nu_{\rm tot} =
\conf{\cal F}^\nu_{\phi} + \conf{\cal F}^\nu_{\rm ipf}$, we have
\begin{equation} \label{cont_T}
    \partial_0 \conf{\rho}_f - 
    |\vec{u}_{\rm p}|^2 \partial_0 \conf{p}_f+
    \conf{w}_f \partial_i u_i
    + u_i \partial_i (\conf{\rho}_f - \conf{p}_f) = 
    (\conf{\cal F}^0_{\rm tot} + {\cal F}_H^0) (1 + |\vec{u}_{\rm p}|^2)
     - 2 \, a^{2(1 - \alpha)}
    u_i
    \conf{\cal F}^i_{\rm tot}\,.
\end{equation}
In this case, the EOS depends on $\conf{T}$ and, hence, $\omega = \conf{p}_f/\conf{\rho}_f$ cannot be taken as a constant as we
have done in previous sections.
Taking into account that both $\conf{p}_f$
and $\conf{\rho}_f$
are functions of the temperature and of the scalar field,
we can rewrite \Eq{cont_T} as a dynamical equation for $\conf{T}$,
\EQalign
    \label{conservation_energy0}
    \partial_0 \conf{T} = -
    \bigl[\partial_{\conf{T}}(\conf{\rho}_f -
    |\vec{u}_{\rm p}|^2 \conf{p}_f)\bigr]^{-1}
    \Bigl\{ &
    \conf{w}_f \, \partial_i u_i + u_i \bigl[ (\partial_i \conf{T})\,
    \partial_{\conf{T}} + (\partial_i \phi)\, \partial_\phi \bigr] \bigl(\conf{\rho}_f - 
    \conf{p}_f\bigr) - \bigl(\conf{\cal F}^0_{\rm tot} + \conf{\cal F}^0_H\bigr)
    \bigl(1 + 
    |\vec{u}_{\rm p}|^2 \bigr) \nonumber \\
    & + 2 \, a^{2(1 - \alpha)} u_i {\cal F}_{\rm tot}^i
    + \pi_\phi \, \partial_\phi
    \bigl(\conf{\rho}_f - |\vec{u}_{\rm p}|^2 \conf{p}_f\bigr) \Bigr\}\,,
\end{mymathbox}
\noindent
where we have used the property 
$\partial_\mu X
= (\partial_\mu \conf{T})\,
\partial_{\conf{T}} X + (\partial_\mu \phi) \partial_\phi X$, which holds for both $X = \conf{\rho}_\f, \conf{p}_\f$.

The momentum equation was given in \Eq{general_EOM_forces2}
with external forces $\conf{\cal F}^\nu_{\rm ipf}$.
Extending this equation to include external  forces
$\conf{\cal F}_{\rm tot}^\nu
= \conf{\cal F}^\nu_{\phi} + \conf{\cal F}^\nu_{\rm ipf}$,
and substituting \Eq{conservation_energy0} in it, we find
\begin{mymathbox}
\begin{align}
    \label{conservation_momentum0}
    \partial_0 u_i + u_j \partial_j u_i  = \frac{u_i}{\conf{w}_\f \gamma^2}
    \Biggl\{
    \frac{\partial_{\conf{T}} \conf{p}_f}{\partial_{\conf{T}} \bigl(\conf{\rho}_f -
    |\vec{u}_{\rm p}|^2 \conf{p}_f\bigr)}
    \Bigl( &\,
    \conf{w}_f \, \partial_j u_j + u_j \bigl[ (\partial_j \conf{T})\,
    \partial_{\conf{T}} + (\partial_j \phi) \, \partial_\phi \bigr]
    \bigl(\conf{\rho}_f - 
    \conf{p}_f\bigr)
    + 2 \, a^{2(1 - \alpha)} u_j {\cal F}_{\rm tot}^j
    \nonumber \\ +   \pi_\phi \, &\, 
    \partial_\phi \bigl(\conf{\rho}_f - |\vec{u}_{\rm p}|^2 \conf{p}_f\bigr)  \Bigr) -
    \frac{\partial_{\conf{T}} \conf{w}_f}{\partial_{\conf{T}} \bigl(\conf{\rho}_f - 
    |\vec{u}_{\rm p}|^2 \conf{p}_f\bigr)} \bigl(\conf{\cal F}_{\rm tot}^0 + \conf{\cal F}_{H}^0 \bigr)
    - \pi_\phi 
    \partial_\phi \conf{p}_f\Biggr\}
     \nonumber \\
     - \, a^{2(\alpha-1)} \frac{1}{\conf{w}_\f \gamma^2} \bigl[(\partial_i \conf{T})\,\partial_{\conf{T}} + &\, (\partial_i \phi) \,   
     \partial_\phi \bigr] \conf{p}_f
     + \frac{1}{\conf{w}_\f \gamma^2}
     \bigl(\conf{\cal F}^i_{\rm tot} +   \conf{\cal F}_H^i\big)
     \,.
\end{align}
\end{mymathbox}
\noindent
As mentioned above, in the \rhs of the
dynamical equations for the fluid variables $\conf{T}$
and $u_i$, the imperfect four-force includes terms that
also depend on first and second-order time derivatives of the fluid variables.
We have shown in \Sec{viscous_term} that, at leading order in the fluid velocity
$|\vec{u}_p| \ll 1$, and neglecting the bulk viscosity,
$\xi \, H \ll 1$,
$\conf{\cal F}^0_{\rm ipf} \simeq 0$ [cf.~\Eq{components_viscous_force_zero}], and the resulting Navier-Stokes
force $\conf{\cal F}^i_{\rm ipf}$ only
depends on spatial gradients of the fluid variables.
Indeed, for homogeneous viscous coefficients, we find 
[cf.~\Eq{components_viscous_force_spatial}],
\begin{equation}
\label{subrel_ipf_in_fluidscalar}
    a^{1 - \alpha} {\cal F}_{\rm ipf}^i \simeq 2
    \, \conf{\nu} \partial_j \bigl[\conf{w}_f \partial_{(i} u_{j)}\bigr]
    + \bigl(\conf{\xi} - \tfrac{2}{3} \conf{\nu} \bigr) 
    \partial_i \bigl( \conf{w}_f \partial_j u_j \bigr)\,.
\end{equation}
Then, the system of \Eqs{conservation_energy0}{conservation_momentum0} can be directly
solved integrating the \lhs in time and then
evaluating the total forces $\conf{\cal F}_{\rm tot}^\nu = \conf{\cal F}_{\rm ipf}^\nu + \conf{\cal F}^{\nu}_{\phi}$ under these assumptions.

\subsubsubsection*{Non-conservation form in terms of a generalized sound speed}

Introducing the speed of sound ${\cs^2}_\mu = \partial_\mu \conf{p}_f/\partial_\mu \conf{\rho}_f$,
which generalizes the definition of \Eq{general_sound_speed}
along different directions $x^\mu$, as
\begin{align}
    \label{generic_cs2}
    {\cs^2}_\mu = \frac{\partial_\mu \conf{p}_f}{\partial_\mu \conf{\rho}_f}
    &= \frac{(\partial_{\conf{T}} \conf{p}_f) (\partial_\mu \conf{T})
    + (\partial_{\phi} \conf{p}_f) (\partial_\mu \phi)}
    {(\partial_{\conf{T}} \conf{\rho}_f) (\partial_\mu \conf{T}) + (\partial_{\phi} \conf{\rho}_f)(\partial_\mu \phi)}\,,
\end{align}
the equation of energy conservation [cf.~\Eq{cont_T}] becomes
\EQalign
\label{conservation_energy}
    (1 - {\cs^2}_0
    |\vec{u}_{\rm p}|^2)
    \partial_0  \ln \conf{\rho}_f \, +
    &\,
    (1 + \omega) \partial_i u_i +
    \sum_i (1 - {\cs^2}_i) u_i \partial_i \ln \conf{\rho}_f
    \nonumber \\  =
    &\,
    \frac{1}{\conf{\rho}_f}\Bigl[ 
    \bigl(\conf{\cal F}_{\rm tot}^0 +
    \conf{\cal F}_H^0)(1 +
    |\vec{u}_{\rm p}|^2) - 2 \,a^{2(1-\alpha)} u_i\conf{\cal F}_{\rm tot}^i
    \Bigr]\,.
\end{mymathbox}
\noindent
For the case studied in \Sec{perfect_fluids},
we have $V_T = 0$ and constant EOS, such that ${\cs^2}_\mu = \omega = {\rm constant}$. 
Hence, the equation of energy conservation
reduces to \Eq{rel_energy_visc}.
Note that for the generic case in which both $p_f$ and $\rho_f$
depend on $T$ and $\phi$, this equation cannot be directly advanced using explicit integrators,
since the speed of sound depends on the local variable $\partial_\mu T$,
which needs to be approximated from $p_f$ and $\rho_f$ at earlier times.
For this reason, \Eq{conservation_energy0}
is advantageous when applying explicit numerical time-stepping schemes.

In terms of the speed of
sound defined in \Eq{generic_cs2}, the momentum equation \Eq{general_EOM_forces2} becomes
\EQalign
    \partial_0 u_i + u_j \partial_j u_i  =&\, \frac{1 - |\vec{u}_p|^2}
    {1 - {\cs^2}_0 |\vec{u}_p|^2}
    \biggl[ {\cs^2}_0\partial_j u_j
    + \frac{{\cs^2}_0}{1 + \omega}
    \sum_j (1 - {\cs^2}_j)
    u_j \partial_j \ln \conf{\rho}_f \nonumber \\ &\, 
    \hspace{20mm} - \frac{1}{\conf{\rho}_f} \biggl(
    \frac{1 + {\cs^2}_0}{1 + \omega}
    \bigl(\conf{\cal F}_{\rm tot}^0 + \conf{\cal F}_H^0 \bigr)
    - \frac{2  {\cs^2}_0}{1 + \omega}
    a^{2(1 - \alpha)} u_j \conf{\cal F}_{\rm tot}^j \biggr) \biggr] u_i
    \nonumber \\ &\, - \frac{1 -
    |\vec{u}_p|^2}{1 + \omega}
    \biggl[
    {\cs^2}_i a^{2(\alpha-1)} 
    \partial_i \ln \conf{\rho}_\f - \frac{1}{\conf{\rho}_f} \bigl(  \conf{\cal F}^i_{\rm tot} + \conf{\cal F}_H^i \bigr) \biggr]\,.
\end{mymathbox}
\noindent
As before, if we take a constant EOS, we recover the
momentum equation given in \Eq{rel_mom_visc}.

\subsubsection*{Alternative non-conservation form for relativistic fluid variables}

The scalar-fluid system of equations can also be written expressing the fluid equations in the alternative non-conservation form for relativistic fluid variables, already discussed at the end of \Sec{perfect_fluids}
[cf.~\Eqq{relativistic_EZeqs}].
This formulation has been used
for simulations of phase transitions in flat spacetime
(see, e.g., 
refs.~\cite{Hindmarsh:2013xza,Hindmarsh:2015qta,Hindmarsh:2017gnf,Cutting:2019zws}),
where the EOM are expressed for
the relativistic energy density $\conf{E} \equiv
\gamma \conf{\rho}_f$ and fluid momentum density $\conf{Z}_i \equiv \conf{T}^{0i}$.

To obtain the dynamical equation for the comoving relativistic energy density,
$\conf{E}$, we project the fluid equations $D^G_\mu T^{\mu \nu}_{\rm pf} = {\cal F}^{\nu}_{\rm tot}$ [cf.~\Eqs{eq:EnergyConservation}{eq:MomentumConservation}] 
in the parallel direction to $U_\nu$.
The projection in the parallel direction of $D_\mu^G T^\mn_\pf$
is given in \Eq{cons_energy},
\begin{align}
    -U_\nu D_\nu^G T^\mn_\pf = a^{-(4+\alpha)} \bigl[\conf{U}^\mu \partial_\mu \conf{\rho}_f + \conf{w}_f
    \partial_\mu \conf{U}^\mu
    +(3\conf{p}_\f - \conf{\rho}_\f) \HH \conf{U}^0
    \bigr]\,,
\end{align}
while 
the projection of the total
four-force, ${\cal F}^\nu_{\rm tot} = {\cal F}_\phi^\nu + {\cal F}^\nu_{\rm ipf}$,
is
\begin{align}
    U_\nu {\cal F}_{\rm tot}^\nu = a^{-(4+3\alpha)} {\conf U}_\nu {\conf {\cal F}}_{\rm tot}^{\nu}
    \,.
\end{align}
Hence, the EOM for the rescaled variables become
\begin{align}
    \conf{U}^\mu \partial_\mu \conf{\rho}_f + \conf{w}_f
    \partial_\mu \conf{U}^\mu
    + \conf{T}_{\rm pf} \HH \conf{U}^0  =
    - a^{-2\alpha} {\conf U}_\mu
    \bigl(\conf{\cal F}_\phi^\mu + \conf{\cal F}_{\rm ipf}^\mu \bigr)\,.
    \label{cons_energy3}
\end{align}
The projection of the force due to the interaction of the
fluid with the scalar field, $\conf{\cal F}^\phi_\nu = \conf{\Omega}_\phi \, a^{2\alpha} \, \partial_\nu \phi$, with $\conf{\Omega}_\phi = \partial_\phi \conf{p}_f -
\conf{\delta}_\phi$, 
and $\conf{\delta}_\phi$ given in \Eq{rescaled_frict},
is
\begin{equation}
    - a^{-2\alpha} {\conf U}_\mu \conf{\cal F}_{\phi}^\mu = - \conf{\Omega}_\phi
    \conf{U}^\mu \partial_\mu
    \phi =
    \biggl(\conf{\delta}_\phi - \frac{\partial \conf{p}_f}{\partial \phi}
    \biggr) \conf{\Theta}
    = \conf{\eta}_\phi \conf{\Theta}^2 - 
    \frac{\partial \conf{p}_f}{\partial \phi} \conf{\Theta}
    \,.
\end{equation}
The imperfect fluid force can be expressed using \Eq{contraction_ipf} as,
\begin{equation}
    - a^{-2\alpha} {\conf U}_\mu \conf{\cal F}_{\rm ipf}^\mu = a^{-\alpha} \cdev^\mn S_\mn\,.
\end{equation}
Then, we find
\EQalign
    \partial_0 \conf{E} +
    \partial_i (\conf{E} u_i) + \conf{p}_f \bigl[\partial_0 \gamma + \partial_i (\gamma u_i)\bigr]
    = &\, (\conf{E} - 3 \conf{p}_f \gamma) \HH
    + a^{-\alpha} 
    \cdev^\mn
    \,S_\mn 
     - \frac{\partial \conf{p}_\f}{\partial \phi}
    \conf{\Theta} + a^{1 - \alpha} \conf{\eta}_\phi 
    \conf{\Theta}^2
    \,.
    \label{relativistic_Eeq_scalar}
\end{mymathbox}

The resulting equation for the fluid momentum density $\conf{Z}_i$ is obtained from the space components of \Eq{general_eqs},
\EQalign
    \partial_0 \conf{Z}_i +
    \partial_j (\conf{Z}_i u_j) + a^{2(\alpha - 1)}
    \partial_i \conf{p}_f =  &\,
    (\alpha - 1) \HH \conf{T}^{0i}_\pf 
    + \conf{\cal F}_{\rm ipf}^i
    + a^{2(\alpha - 1)}
    \biggl(
    \frac{\partial \conf{p}_\f}{\partial \phi}
    - \conf{\eta}_\phi
    a^{1 - \alpha} \gamma [\pi_\phi + u_j \partial_j \phi]\biggr) \partial_i \phi\,.
    \label{relativistic_Zeq_scalar}
\end{mymathbox}

These equations generalize the scalar-fluid system considered, for example,
in ref.~\cite{Hindmarsh:2013xza} (see their Eqs.~(5) and (6)), to first-order imperfect fluids and to an expanding Universe.
As before, we note that neglecting $\gamma$ in $\conf{E} = \gamma \conf{\rho}_f$ or in $\conf{Z}_i = \conf{w}_f \gamma^2 u_i$
would lead to the wrong subrelativistic limits due to the inconsistent
underlying assumption that $\partial_0 \gamma$ is negligible.

Unlike the previous formulations of the fluid equations
(conservation and non-conservation forms),
this system of equations, without further manipulations, cannot be directly solved following the
time integration schemes that we will present in
\Sec{sec:RK}. In fact, although one could relate $\gamma,u_i$ to ${\conf E}, {\conf Z}_i$ and have a system of four equations in four variables, the presence of time derivatives of both ${\conf E}$ and $\gamma$ in
\Eq{relativistic_Eeq_scalar} does not allow to apply explicit Runge-Kutta integrators to this system.
However, the system can be evolved using operator splitting methods,
where each term is advanced at
different intermediate time steps \cite{2003rnh..book.....W}.
In our numerical implementation, we focus on the systems of equations given by \Eqq{eqs_fluid}
or \Eqq{rho_u_eqs}, corresponding respectively
to the conservation and non-conservation forms of the equations of motion
for relativistic perfect fluids and their respective generalizations in the case of imperfect fluids and interactions with scalar and gauge fields.
These systems are readily available in a suitable form to apply Runge-Kutta time-advancing schemes, as we explain in \Sec{sec:RK}.

\subsubsubsection*{Non-conservation form in the limit of subrelativistic
fluid velocity}

In the subrelativistic limit $|\vec{u}_{\rm p}|^2 \ll 1$, the
fluid equations [cf.~\Eqs{conservation_energy0}{conservation_momentum0}] reduce to the following,
\begin{subequations}
\begin{align}
    \hspace{-1mm} \lim_{|\vec{u}_{\rm p}|^2 \ll 1} \partial_0 \conf{T} =  &\, -
    \frac{1}{\partial_{\conf{T}} \conf{\rho}_f}
    \biggl\{ 
    \conf{w}_f \, \partial_i u_i + u_i \Bigl[ (\partial_i \conf{T})\,
    \partial_{\conf{T}} + (\partial_i \phi) \partial_\phi \Bigr] \bigl(\conf{\rho}_f - 
    \conf{p}_f\bigr) - \conf{\cal F}^0_{\rm tot} - \conf{\cal F}^0_H \nonumber \\ &\, 
    \hspace{16mm} 
    + 2 \, a^{2(1 - \alpha)} u_i
    \conf{\cal F}^i_{\rm tot} + \pi_{\phi} \, \partial_\phi \conf{\rho}_f \biggr\}\,, \\
    \hspace{-1mm} \lim_{|\vec{u}_{\rm p}|^2 \ll 1} \partial_0 u_i =&\,-u_j \partial_j u_i + \frac{u_i}{\conf{w}_\f}
    \Biggl\{
    \frac{\partial_{\conf{T}} \conf{p}_f}{\partial_{\conf{T}} \conf{\rho}_f}
    \Bigl( \,
    \conf{w}_f \, \partial_i u_i + u_i \bigl[ (\partial_i \conf{T})\,
    \partial_{\conf{T}} + (\partial_i \phi) \, \partial_\phi \bigr]
    \bigl(\conf{\rho}_f - 
    \conf{p}_f\bigr)
    + 2 \, a^{2(1 - \alpha)} u_i {\cal F}_{\rm tot}^i +   \pi_\phi \,  
    \partial_\phi \conf{\rho}_f \Bigr)
    \nonumber \\ & \, - 
    \frac{\partial_{\conf{T}} \conf{w}_f}{\partial_{\conf{T}} \conf{\rho}_f} \bigl(\conf{\cal F}_{\rm tot}^0 + \conf{\cal F}_{H}^0 \bigr)
    - \pi_\phi 
    \partial_\phi \conf{p}_f\Biggr\} - \, a^{2(\alpha-1)} \frac{1}{\conf{w}_\f} \bigl[(\partial_i \conf{T})\,\partial_{\conf{T}} +  (\partial_i \phi) \,   
     \partial_\phi \bigr] \conf{p}_f
     +  \frac{1}{\conf{w}_\f}
     \bigl(\conf{\cal F}^i_{\rm tot} +   \conf{\cal F}_H^i\big)
     \,.
\end{align}
\end{subequations}

\subsection{Summary of the equations}
\label{equations_summary}

In this section, we provided the generic system
of equations of motion for a fluid that interacts
with an Abelian gauge field $A_\mu$ and with a singlet scalar fields
$\phi$ in a FLRW expanding Universe.

We consider a perfect fluid with small deviations with respect to
local thermal equilibrium (LTE), such that its stress-energy momentum
tensor can be expressed as
\begin{equation}
    T^\mn_\fl = T^\mn_\pf - \dev^\mn\,,
\end{equation}
with
\begin{equation}
    T^\mn_\pf =  w_f U^\mu U^\nu + p_f
    g^\mn
    \qquad \text{and \ } \qquad \dev^\mn = 2 \, \nu \, w_f \, 
    \sigma^\mn + \xi \, w_f \, \theta \, h^\mn 
\end{equation}
respectively corresponding to the perfect fluid stress-energy tensor, and to the deviatoric stress tensor for first-order imperfect fluids.
In the above expressions $w_f$ and $p_f$ are the perfect fluid enthalpy and pressure,
$U^\mu = \gamma a^{-\alpha}(1, u_i)$ is the four-velocity,
with $u_i \equiv \dd x^i/\dd \eta$, $\nu$ and $\xi$ are the shear
and bulk kinematic viscosity coefficients,
$\sigma^\mn = S^\mn - \tfrac{1}{3}
\theta h^\mn$ is the traceless rate-of-strain tensor,
with $S^\mn = h^{(\mu \lambda} D_\lambda^G U^{\nu)}$ and $\theta = S^\mu_{\ \mu}$,
and $h^\mn = U^\mu U^\nu + g^\mn$ is the projection tensor orthogonal to $U^\mu$.

The fluid equations of motion 
are obtained from the conservation
of the total stress-energy tensor,  
which, in an expanding
Universe, can be recast in the following way [cf.~\Eq{general_eqs}],
\EQb
\label{summary_fluid}
    \partial_\mu \conf{T}^\mn_\pf = \conf{\cal F}_{\rm tot}^\nu + \conf{\cal F}_H^\nu\,, \quad \text{with}\, \quad  \conf{\cal F}_{\rm tot}^\nu \equiv  \conf{\cal F}_{\rm int}^\nu
    + \conf{\cal F}_{\rm ipf}^\nu  \,,
\end{mymathbox}
\noindent
with the rescalings $\conf{T}^{\mu \nu}_{\rm pf}= a^{4 + 2\alpha} 
T^{\mu \nu}_{\rm pf}$
and $\conf{\cal{F}}^\nu_{\rm tot} = a^{4 + 2\alpha} {\cal F}^\nu_{\rm tot}$.
The effect of the Universe expansion is absorbed in the Hubble friction [cf.~\Eq{Hubble_friction}]
\begin{equation}
    \conf{\cal F}^\nu_H = 
    (\conf{\rho}_\f - 3 \conf{p}_\f)
    \HH \delta^{\nu}_{~\,0} + (\alpha - 1)
    \HH \, \conf{T}^{0i}_\pf \delta^{\nu}_{~\,i},
\end{equation}
while the viscous forces due to deviations with respect
to LTE are [cf.~\Eq{ipf_force}]
\begin{equation}
    \conf{\cal F}_{\rm ipf}^\nu = \partial_\mu \cdev^\mn + 
    3 \, \conf{\xi}
    \, \conf{w}_f \, \conf{\theta} 
    \,
    \HH 
    + (1 - \alpha) \HH \, \cdev^{0i}\delta^{\nu}_{~\,i}\,.
\end{equation}
In the expressions above, $\cdev^{\mu \nu} =a^{4+2\alpha} \dev^{\mu \nu} = 2 \, \conf{\nu} \,  \conf{w}_\f \,  \conf{\sigma}^{\mu \nu} + \conf{\xi} \, \conf{w}_\f \, \conf{\theta} \,  \conf{h}^{\mu \nu}$ with $\conf{\theta} = a\,  \theta$, $\conf{\xi} = a^{-1} \xi$, $\conf{\nu}=a^{-1} \nu$, while the fluid variables are rescaled as $\conf{\rho}_\f/\rho_\f = \conf{p}_\f/p_\f = \conf{w}_\f/w_\f = a^4$.

In \Sec{subsec:gauge_fluid_dyns}, we considered interactions of the fluid with an Abelian U(1) gauge field and
in \Sec{subsec:scalar_fluid}, with 
a singlet scalar field $\phi$.
The resulting interaction forces are 
\begin{equation}
    \conf{\cal F}^\nu_{\rm int} = 
    \conf{\cal F}^\nu_{U(1)} + \conf{\cal F}^\nu_\phi\,,
\end{equation}
where the components of $\conf{\cal F}^\nu_{U(1)}$ are given
in \Eqq{lorentz_forces},
\begin{subequations}
\begin{align}
    \conf{\cal F}^0_{U(1)} =  &\,  
    -a^{1-\alpha} \gamma \Bigl[ \conf{\rho}_e \, \vec{u} \cdot \vec{\cal E}  - \conf{\sigma}_\f \, \vec{u} \cdot (\, \vec{\cal E} \times \vec{{\cal B}} \, ) - \conf{\sigma}_\f |\vec{\cal E}|^2 \Bigr]\,, \\
    \conf{\cal F}^i_{U(1)} =  
    &\, -a^{\alpha-1} \gamma \Bigl[(\, \conf{\rho}_e - a^{2(1-\alpha)} \conf{\sigma}_\f \, \vec{u} \cdot \vec{\cal E} \, ) \, {\cal E}_i - \conf{\rho}_e ( \, \vec{u} \times \vec{\cal B} \, )_i + \conf{\sigma}_\f ( \, \vec{\cal E} \times \vec{\cal B} \,)_i - \conf{\sigma}_\f (\, \vec{u} \cdot \vec{\cal B} \,) \, {\cal B}_i + \conf{\sigma}_\f |\vec{\cal B}|^2 u_i  \Bigr]\,,
\end{align}
\end{subequations}
and the components of $\conf{\cal F}^\nu_\phi$
in \Eqq{fluid_eqs_scalar},
\begin{subequations}
\begin{align}
    \conf{\cal F}^0_\phi = &\,  \pi_\phi \biggl(-\frac{\partial  \conf{p}_\f}{\partial \phi} + \gamma \,  \conf{\eta}_\phi(
    \pi_\phi + u_i \partial_i \phi)\biggr)\,, \\
    \conf{\cal F}^i_\phi
    = &\,  - a^{2(\alpha - 1)}
    \partial_i \phi \biggl
    (-\frac{\partial \conf{p}_\f}{\partial \phi} 
    + \gamma \, \conf{\eta}_\phi(
    \pi_\phi + u_j \partial_j \phi)\biggr)\,.
\end{align}
\end{subequations}
They depend on the fluid velocity $u_i$, the Lorentz factor $\gamma$,
and, respectively,
also on the U(1) gauge field, $A_\mu$, the scalar field $\phi$,
and their time derivatives ${\cal E}_i \equiv F_{0i}
=\partial_0 A_i - \partial_i A_0$ and $\pi_\phi \equiv \partial_0 \phi$.
The fluid equations of motion are then coupled to Maxwell equations [cf.~\Eqq{Maxwell2}],
\EQb
    \partial_0 F_{0i} + a^{-2(1 - \alpha )}\partial_j F_{ij} +
    (1 - \alpha) \mathcal{H} F_{0i} =
    a^{2 \alpha} J^\fl_i\,, \qquad \partial_i F_{0i} = a^2 J_0^\fl\,, \qquad \partial_0 A_i = {\cal E}_{i} + \partial_i A_0 \,,
\end{mymathbox}
\noindent
and to the equations of motion of the scalar field [cf.~\Eqq{equations_scalar}],
\EQb
    \partial_0 \pi_{\phi} = a^{-2(1-\alpha)}
    {\vv\nabla}^{\,2} \hspace{-1mm}\phi -  (3 - \alpha)\mathcal{H} \, \pi_{\phi} - a^{2\alpha} \frac{\partial V_{\rm eff}}{\partial \phi}(\phi, T) -
    a^{2\alpha} 
    \eta_\phi \, \gamma \bigl(\pi_{\phi} + u_i \partial_i \phi \bigr)\,,
    \qquad \partial_0 \phi = \pi_\phi\,,
\end{mymathbox}
\noindent
with $V_{\rm eff} = V_0 - p_\f$.
We consider the large-temperature extension of the equilibrium distribution
function, such that the enthalpy and the pressure of the fluid are
\begin{equation}
    \conf{p}_f = \conf{p}_{\rm rad} (\conf{T}) - \conf{V}_T
    (\phi, \conf{T})\,, \qquad 
    \conf{w}_f = \conf{T} \frac{\partial \conf{p}_f}
    {\partial \conf{T}} = 4 \,\conf{p}_{\rm rad} (\conf{T}) +
    \conf{p}_{\rm rad} \frac{\partial \ln \sigma_{\rm dof}}{\partial \ln \conf{T}}
    - \conf{T} \frac{\partial \conf{V}_T}{\partial \conf{T}} (\phi, \conf{T})\,, 
\end{equation}
where $\conf{p}_{\rm rad} = \tfrac{1}{3} \sigma_{\rm dof} \conf{T}^4
= \tfrac{\pi^2}{90}\, g(\conf{T})\, \conf{T}^4$, with $\tilde{T}=aT$ being the rescaled temperature,
and we assume that the effective potential only depends on the temperature
and the
expectation value of the scalar field $\phi$. 
In the limit when the radiation term dominates ($\conf{V}_T \to 0$), we recover the
constant ultrarelativistic
equation of state (EOS),
$\conf{p}_f = \tfrac{1}{3} \conf{\rho}_f$, where $\conf{\rho}_f = \conf{w}_f - \conf{p}_f = 3 \conf{p}_{\rm rad}$ is the radiation energy density.

We have considered two main approaches to solve the equations of motion
of the fluid.
In the conservation form, we evolve the temporal components of the stress-energy
tensor of the perfect fluid $\conf{T}^{0\mu}_\pf$, while in the non-conservation
form, we evolve the fluid primitive variables $\conf{\rho}_f$ and $u_i$.

\subsubsection*{Conservation form of the fluid equations}

The system of equations in \Eq{summary_fluid} already provides
us with the fluid equations of motion in their conservation form.
The remaining step is to reconstruct the stress tensor $\conf{T}^{ij}_\pf$,
the velocity field, and the Lorentz factor that appears in the
interaction forces, as a function of the dynamical variables 
$\conf{T}^{0\mu}_\pf$.
This can be accomplished using \Eqq{reconstruction},
\begin{equation}
    \conf{T}^{ij}_\pf = \frac{\conf{T}^{0i}_\pf \conf{T}_\pf^{0j}}{\conf{T}_\pf^{00} +
    \conf{p}_f} + a^{2(\alpha - 1)}
    \conf{p}_f \delta^{ij}\,, \qquad u_i = \frac{\conf{T}^{0i}_\pf}{\conf{T}^{00}_\pf + \conf{p}_f}\,,
    \qquad \text{and \ } \quad \gamma^2 = \frac{\conf{T}^{00}_\pf + \conf{p}_f}
    {\conf{w}_f}\,.
\end{equation}
However, $\conf{p}_f$ and $\conf{w}_f$ still depend on the temperature
$\conf{T}$.
The temperature can be computed by substituting a particular
choice of the thermal corrections to the potential, $\conf{V}_T
(\phi, \conf{T})$, in the
following relation [cf.~\Eq{scalar_fluid_conservationform_relation}]
\begin{align}
    a^{2(1 - \alpha)} \conf{T}^{0i}_\pf \conf{T}^{0i}_\pf = 
    (\conf{T}^{00}_\pf + \conf{p}_f) (\conf{T}^{00}_\pf - \conf{\rho}_f) 
    = 
    (\conf{T}^{00}_\pf)^2 + \conf{T}^{00}_\pf (\conf{p}_f - \conf{\rho}_f)
    - \conf{p}_f \conf{\rho}_f\,,
\end{align}
and solving numerically for
the value of the temperature $\conf{T}$ as a function of
$\phi$, $g(T)$, $\conf{T}^{00}_\pf$, and $r^2 \equiv a^{2(1 - \alpha)}
\sum_i [\conf{T}^{0i}_\pf/\conf{T}^{00}_\pf]^2$.

In particular, in the limit $\conf{V}_T \to 0$, the EOS becomes
a constant $\omega \equiv \conf{p}_f/\conf{\rho}_f = 1/3$.
Then, defining the variable
$z = (1 + \omega)\gamma^2 - \omega$, it can be computed
as a unique function of the squared ratio $r^2$, as in \Eq{gamma2_rel},
\begin{equation}
    z = \frac{1}{2 (1 - r^2)}
    \Bigl(1 - \omega + \sqrt{(\omega + 1)^2 - 4 \omega r^2}\Bigr)
    \,,
\end{equation}
allowing us to reconstruct the fluid variables in the following way,
\begin{equation}
    \conf{T}^{ij}_\pf = \frac{z}{z + \omega}
    \frac{\conf{T}^{0i}_\pf \conf{T}^{0j}_\pf}{\conf{T}^{00}_\pf}
    + \frac{\omega}{z} a^{2(\alpha - 1)} \conf{T}^{00}_\pf \delta_{ij}\,, \qquad
    u_i = \frac{z}{z + \omega} \frac{\conf{T}^{0i}_\pf}{\conf{T}^{00}_\pf}\,,
    \qquad \gamma^2 = \frac{z + \omega}{1 + \omega}\,.
\end{equation}

\subsubsection*{Non-conservation form of the fluid equations}

On the other hand, the non-conservation form of the energy and momentum
equations are expressed in terms of the fluid primitive variables $\conf{\rho}_f$ and $u_i$, respectively.
In the general case, since the EOS $\omega = \conf{p}_f/\conf{\rho}_f$
is a function of the temperature,
we convert the equation of conservation of energy into
a dynamical equation for the temperature $\conf{T}$.
The fluid EOM are then given by 
\Eqs{conservation_energy0}{conservation_momentum0},
\begin{subequations}
\EQalign
    \partial_0 \conf{T} &\, = - 
    \bigl[\partial_{\conf{T}}(\conf{\rho}_f -
    |\vec{u}_{\rm p}|^2 \conf{p}_f)\bigr]^{-1}
    \Bigl\{
    \conf{w}_f \, \partial_i u_i + u_i \bigl[ (\partial_i \conf{T})\,
    \partial_{\conf{T}} + (\partial_i \phi)\, \partial_\phi \bigr] \bigl(\conf{\rho}_f - 
    \conf{p}_f\bigr) - \bigl(\conf{\cal F}^0_{\rm tot} + \conf{\cal F}^0_H\bigr)
    \bigl(1 + 
    |\vec{u}_{\rm p}|^2 \bigr) \nonumber \\
    & \hspace{44mm} + 2 \, a^{2(1 - \alpha)} u_i {\cal F}_{\rm tot}^i
    + \pi_\phi \, \partial_\phi
    \bigl(\conf{\rho}_f - |\vec{u}_{\rm p}|^2 \conf{p}_f\bigr) \Bigr\}\,, \\
\partial_0 u_i &\, + u_j \partial_j u_i  = \frac{u_i}{\conf{w}_\f \gamma^2}
    \Biggl\{
    \frac{\partial_{\conf{T}} \conf{p}_f}{\partial_{\conf{T}} \bigl(\conf{\rho}_f -
    |\vec{u}_{\rm p}|^2 \conf{p}_f\bigr)}
    \Bigl(
    \conf{w}_f \, \partial_j u_j + u_j \bigl[ (\partial_j \conf{T})\,
    \partial_{\conf{T}} + (\partial_j \phi) \, \partial_\phi \bigr]
    \bigl(\conf{\rho}_f - 
    \conf{p}_f\bigr)
    + 2 \, a^{2(1 - \alpha)} u_j {\cal F}_{\rm tot}^j
    \nonumber \\  &\, \hspace{33mm}  +   \pi_\phi \,
    \partial_\phi \bigl(\conf{\rho}_f - |\vec{u}_{\rm p}|^2 \conf{p}_f\bigr)  \Bigr) -
    \frac{\partial_{\conf{T}} \conf{w}_f}{\partial_{\conf{T}} \bigl(\conf{\rho}_f - 
    |\vec{u}_{\rm p}|^2 \conf{p}_f\bigr)} \bigl(\conf{\cal F}_{\rm tot}^0 + \conf{\cal F}_{H}^0 \bigr)
    - \pi_\phi  
    \partial_\phi \conf{p}_f\Biggr\}
     \nonumber \\  &\, \hspace{18mm}
     - \, a^{2(\alpha-1)} \frac{1}{\conf{w}_\f \gamma^2} \bigl[(\partial_i \conf{T})\,\partial_{\conf{T}} + (\partial_i \phi) \,   
     \partial_\phi \bigr] \conf{p}_f
     + \frac{1}{\conf{w}_\f \gamma^2}
     \bigl(\conf{\cal F}^i_{\rm tot} +   \conf{\cal F}_H^i\big)
     \,.
\end{mymathbox}
\end{subequations}
Again, in the limit $\conf{V}_T \to 0$, we recover the EOS $\omega = 1/3$, and the non-conservation
form of the fluid equations can be recast in terms of the energy density $\conf{\rho}_f$ and
the velocity $u_i$ [cf.~\Eqq{equations_Lorentz}]
\begin{subequations}
\EQalign
    (1 - &\,\omega |\vec{u}_{\rm p}|^2) \,
    \partial_0 \ln \conf{\rho}_f \, +
    (1 + \omega)\, \partial_i u_i +
    (1 - \omega) u_i 
    \partial_i \ln \conf{\rho}_f
    \nonumber \\ &\, \hspace{25mm}=
    \frac{1}{\conf{\rho}_f} \Bigl( \bigl[
    \conf{\cal F}^0_{\rm tot} + (1 - 3 \omega)\conf{\rho}_f \HH \bigr]
    (1 + 
    |\vec{u}_{\rm p}|^2)
    - 2 \, a^{2(1 - \alpha)} u_i
    \conf{\cal F}^i_{\rm tot} \Bigr)\,, \\ 
    \partial_0 u_i&\, + u_j \partial_j u_i =
    \frac{1 - |\vec{u}_{\rm p}|^2}{1 - \omega |\vec{u}_{\rm p}|^2}
    \biggl[ \omega \partial_j u_j 
    + \omega \frac{1 - \omega}{1 + \omega}
    u_j \partial_j \ln \conf{\rho}_f
      - \frac{1}{\conf{\rho}_f} \biggl(
    \conf{\cal F}^0_{\rm tot} - 
    \frac{2 \omega}{1 + \omega} a^{2(1 - \alpha)} u_j  \conf{\cal F}^j_{\rm tot} \biggr) \biggr] u_i
    \nonumber \\ &\, \hspace{16mm} - \frac{1 - |\vec{u}_{\rm p}|^2}{1 + \omega}
    \biggl( \omega \, a^{2(\alpha - 1)} \, \partial_i \ln \conf{\rho}_f
    - \frac{1}{\conf{\rho}_f} \conf{\cal F}^i_{\rm tot} \biggr) +
    \biggl( \frac{1 - |\vec{u}_{\rm p}|^2}{1 - \omega |\vec{u}_{\rm p}|^2}
    (3\omega - 1) + \alpha - 1\biggr) \HH u_i\,.
\end{mymathbox}
\end{subequations}


\newpage

\section{Introduction to the lattice}
\label{sec:generalLattice}

In this section, we provide a basic introduction to lattice techniques. The basic notions related to the lattice and definitions of discrete power spectra are given in \Sec{basic_lattice}. Lattice derivatives
and lattice momenta are covered in \Sec{Sect:LatticeDerivatives}. We then deal with more complex lattice derivative operators in \Sec{sec:LatticeOperators} and introduce, in \Sec{sec:RK},
higher-order accurate time evolution algorithms for a system of differential equations. Several of these concepts have been reviewed in {\tt The\,Art\,I} \cite{Figueroa:2020rrl} and {\tt II} \cite{Baeza-Ballesteros:2025tme}, which the reader may consult for more detailed discussions. Readers who are already familiar with lattice techniques, may directly jump to \Sec{sec:fluidLattice}, where we discuss the lattice implementation of the fluid EOM.

\subsection{The lattice}
\label{basic_lattice}

We consider a cubic lattice in three spatial dimensions that has {\it side length} $L$ and {\it number of grid points} $N$ (i.e., number of lattice sites) per dimension. The total number of points is then $N^3$ and each lattice site can be identified by the following vector,
\be
    {\bf n} = (n_1,n_2,n_3),~~~~ {\rm with}~~ n_i = 0,1,...,N-1 \,,~~~i = 1,2,3\,.
\ee
The distance between two lattice sites along one dimension, which represents the minimum distance that can be resolved on the lattice, is given by the \textit{lattice spacing},
\be
   \delta x \equiv \frac{L}{N} \ .
\ee
The value of any continuum function ${\tt f}({\bf x})$ can be represented by a discrete lattice equivalent $f({\bf n })$, such that they hold the same values at the discrete positions ${\bf x}={\bf n } \, \delta x$. We impose periodic boundary conditions, i.e., $f({\bf n }+\hat{\imath} N)=f({\bf n })$ for any $\hat{\imath}=\hat{1}, \hat{2}, \hat{3}$. Here, $\hat{\imath}$ represents the unit vector for each spatial direction,
\be
 \hat{1} \equiv (1,0,0) \ ,~~~~ \hat{2} \equiv (0,1,0) \ , ~~~~ \hat{3} \equiv (0,0,1) \ .
\ee
We can define a reciprocal lattice in momentum space, where we label each point by a vector $\tilde{\bf n}$,
\begin{eqnarray}
    \tilde{\bf n} = (\tilde n_1, \tilde n_2, \tilde n_3), ~~~~{\rm with}~~
    \tilde n_i = -\frac{N}{2}+1, -\frac{N}{2}+2, ... ,-1,0,1, ... , \frac{N}{2} - 1, \frac{N}{2}  \,,~~~ i  = 1,2,3\,.
\end{eqnarray}
The discrete Fourier transform (DFT) is defined as
\begin{eqnarray}
    f({\bf n}) \equiv {1\over N^3}\sum_{\tilde {\bf n}} e^{i{2\pi\over N} {\bf \tilde n n}} f({\bf \tilde n}) ~~~~ \Leftrightarrow ~~~~  f({\bf \tilde n}) \equiv \sum_{\bf n} e^{+i{2\pi\over N} {\bf n \tilde n} }f({\bf n})\, ,
\end{eqnarray}
and
\begin{eqnarray}
    \sum_{\tilde {\bf n}} e^{-i{2\pi\over N} {\bf \tilde n n}} \equiv N^3 \delta_{0, \tilde{n}}\, .
\end{eqnarray}
By imposing periodic boundary conditions in real space it follows that Fourier transformed functions are periodic in the reciprocal lattice, such that~$f({\bf \tilde n}+\hat{\imath} N)=f({\bf \tilde n})$.
Due to the discrete nature of the lattice we can only represent momenta between an
\textit{infrared} (IR) and an \textit{ultraviolet} (UV) cut-off. The smallest momentum that we can capture on the lattice is given by
\be
k_{\rm IR} \equiv \frac{2\pi}{L} = \frac{2\pi}{N\dx}\,,  \\
\ee
while the largest one along each spatial dimension by,
\be
k_{\rm UV} \equiv {N\over2}k_{\rm IR} = {\pi\over \dx} \ . 
\ee
The maximal momentum $k_{\rm max}$ that can be represented on the lattice lies in the 3D diagonal and is given by,
 \be
 k_{\rm max}  =\sqrt{3} \, k_{\rm UV}=\sqrt{3}{N \over 2}k_{\rm IR} = \sqrt{3}{\pi\over \dx} \ . 
\ee
The reciprocal lattice captures a range of discrete momenta, 
\begin{eqnarray}
{\bf k} = k_{\rm IR} (\tilde{n}_1,\tilde{n}_2,\tilde{n}_3)\,.
\end{eqnarray}
The modulus of the momentum will be denoted by $k=k(\tilde{n})\equiv k_{\rm IR} |\tilde{\bf{n}}|=k_{\rm IR}(\tilde{n}_1^2+\tilde{n}_2^2+\tilde{n}_3^2)^{1/2}$.

\subsubsection*{Power Spectrum}

The power spectrum $\Delta_{\tt f}(k)$ of a function ${\tt f}({\bf x})$ in the continuum is defined through its ensemble average $\langle{\tt f}^2\rangle$ by,
\be
\langle{\tt f}^2\rangle =\int {\rm d}\log k\, \Delta_{\tt f}(k)\ , \hspace{0.75cm} \Delta_{\tt f}(k)\equiv \frac{k^3}{2\pi^2}{\cal P}_{\tt f}(k)\ , \hspace{0.75cm} \langle {\tt f}_{\bf k}{\tt f}_{\bf k'}^*\rangle = (2\pi)^3\delta({\bf k-k'}){\cal P}_{\tt f}(k) \ ,
\ee
where $k=|\bf k |$ denotes the modulus of the momentum in the continuum. To obtain an equivalent lattice expression, the ensemble average is replaced by a volume average over the whole lattice,
\be
\langle f^2\rangle_V = \frac{\delta x^3}{V}\sum_{\bf n}f^2({\bf n})=\frac{1}{N^3} \sum_{\bf n}f^2({\bf n})\ .
\ee
Applying the discrete Fourier transform we obtain,
\be
\label{fourier_square_average}
\langle f^2\rangle_V = \frac{1}{2\pi}\sum_{|\tilde{{\bf n}}|}\Delta \log k(\tilde{{\bf n}}) \,  k(\tilde{{\bf n}}) \frac{\delta x}{N^5}\#_{R(\tilde{{\bf n}})}\langle |f(\tilde{{\bf n}}')|^2\rangle_{R(\tilde{{\bf n}})} \ ,
\ee
where $\Delta \log k(\tilde{{\bf n}})\equiv k_{\rm IR}/ k(\tilde{{\bf n}})$. Moreover, with $\langle (...) \rangle_{R(\tilde{{\bf n}})}\equiv (1/\#_{R(\tilde{{\bf n}})})\sum_{\tilde{{\bf n}}'\in {R(\tilde{{\bf n}})}} (...)$ we have introduced the angular average over spherical momentum shells $R(\tilde{{\bf n}})$, where each shell contains all reciprocal sites with radius $|\tilde{{\bf n}}'|\in [ |\tilde{{\bf n}}|,|\tilde{{\bf n}}|+\Delta \tilde{n}$), with $\Delta \tilde{n}$ being a chosen radial binning. The {\it multiplicity} $\#_{R(\tilde{{\bf n}})}$ gives the number of reciprocal lattice sites of each spherical shell. This leads to the following expression for the lattice power spectrum,
\be\label{eq:LatticePowerSpectrum}
\Delta_f  \bigl(k(|\tilde{{\bf n}}|) \bigr)= \frac{k(\tilde{{\bf n}}) }{2\pi} \frac{\delta x}{N^5}\#_{R(\tilde{{\bf n}})}\langle |f(\tilde{{\bf n}}')|^2\rangle_{R(\tilde{{\bf n}})}= \frac{k^3(\tilde{{\bf n}}) }{2\pi^2} \Upsilon_{| \tilde {\bf n}|} \left(\frac{\delta x}{N}\right)^3\langle |f(\tilde{{\bf n}}')|^2\rangle_{R(\tilde{{\bf n}})} \ ,
\ee
where we have adopted the notation of {\tt The Art\,-\,II} \cite{Baeza-Ballesteros:2025tme} and defined $ \Upsilon_{| \tilde {\bf n}|}\equiv \#_{R(\tilde{{\bf n}})}/4\pi |\tilde{{\bf n}}|^2$. A frequently used approximation is $\#_{R(\tilde{{\bf n}})}\simeq 4\pi |\tilde{{\bf n}}|^2$, such that $ \Upsilon_{| \tilde {\bf n}|}=1$. 
We note that in \href{https://cosmolattice.com}{\CLns}, the former is referred to as $\tt{Type-I}$, while the latter as $\tt{Type-II}$ power spectrum.

\subsection{Lattice derivatives and lattice momenta}\label{Sect:LatticeDerivatives}

The continuum derivatives that appear in the equations of motion
need to be replaced by lattice derivatives. These finite difference expressions are defined such that they reproduce the continuum
limit to a certain order $p$ in accuracy in the lattice spacing or time step, when expanded around a given location.
In the following, we discuss different types of lattice derivatives: the neutral (or central) $\nabla_\mu^{(0,p)}$ and a set of charged derivatives, which consists of the forward $\nabla_\mu^{(+,p)}$ and the backward $\nabla_\mu^{(-,p)}$ derivative. While in fundamental field dynamics it is usually
sufficient to consider lattice derivatives up to second order,
hydrodynamical systems often require higher-order expressions (i.e., more accurate for a given lattice spacing) \cite{VANLEER1974361, DUBEY2009512, 8153876, PencilCode:2020eyn}. 
Lattice derivatives can be easily constructed to any order of accuracy (see, e.g., \cite{abramowitz1964}  for the general expressions and \cite{DiscreteDerivatives} for tables of finite difference coefficients). In the following, we provide first the general expressions followed by examples up to sixth order in accuracy, the latter representing a good compromise between computational cost and effectiveness in capturing the properties of computationally demanding physical cases as, for example, shocks \cite{PencilCode:2020eyn}.

\subsubsection*{Neutral derivative}

The neutral (or central) derivative $\nabla_i^{(0, p)}$ accurate to order $p=2m$ for any $m=1,2,\dots$, is given by
\be
[\nabla_i^{(0,2m)} f ] =\frac{1}{\dx}\sum_{\ell=1}^{m} c_{\ell, m}^0\left[f({\bf n} + \ell \cdot \hat{\imath})- f({\bf n} - \ell \cdot \hat{\imath}) \right] ~~ \longrightarrow ~~ \partial_{i} {\tt f}({x})\big|_{{x}\,\equiv\, {\bf n}\dx} + \mathcal{O}(\dx^{2m}) \ .  \label{eq:CentralDerivativeOp}
\ee
with the coefficient $c_{\ell, m}^0$ being defined as,
\be
c_{\ell, m}^0\equiv (-1)^{\ell+1}\frac{(m!)^2}{\ell(m-\ell)!(m+\ell)!} \ , 
\ee
where $\delta x$ is the spatial spacing. On the \rhs of the arrow we denote around which location on the lattice the discrete expressions need to be expanded in order to recover the continuum limit to the respective order $p$ of accuracy. The first three cases that are accurate to order $p=2,4,6$ are
\begin{subequations}
\label{centralderivatives}
\begin{align}
{\rm 2}^{\rm nd}:\, [\nabla_i^{(0,2)}f]  =&  \frac{f({\bf n}+\hat{\imath})-f({\bf n}-\hat{\imath})}{2\delta x} ~~\longrightarrow ~~ \partial_{i} {\tt f}({x})\big|_{{x}\,\equiv\, {\bf n}\dx} + \mathcal{O}(\dx^2) \, , \label{eq:CentralDerivativeO2}\\
{\rm 4}^{\rm th}:\, [\nabla_i^{(0,4)}f]  =&  \frac{8f({\bf n}+\hat{\imath})-f({\bf n}+2\hat{\imath})+f({\bf n}-2\hat{\imath})-8f({\bf n}-\hat{\imath})}{12\dx} ~~ \longrightarrow ~~ \partial_i {\tt f}({x})\big|_{{x}\,\equiv\, {\bf n}\dx} + \mathcal{O}(\dx^4) \, ,\label{eq:CentralDerivativeO4}\\
{\rm 6}^{\rm th}:\, [\nabla_i^{(0,6)}f]  =& \frac{45f({\bf n}+\hat{\imath})-9f({\bf n}+2\hat{\imath})+f({\bf n}+3\hat{\imath})-f({\bf n}-3\hat{\imath})+9f({\bf n}-2\hat{\imath})-45f({\bf n}-\hat{\imath})}{60\dx} \nonumber\\
& \longrightarrow ~~ \partial_{i} {\tt f}({x})\big|_{{x}\,\equiv\, {\bf n}\dx} + \mathcal{O}(\dx^6)\, .\label{eq:CentralDerivativeO6}
\end{align}
\end{subequations}

\subsubsection*{Charged derivatives}

We can also define a set of \textit{charged} lattice derivatives, the so called forward $\nabla_i^{(+, p)}$ and backward $\nabla_i^{(-, p)}$ derivatives. Compared to the neutral derivative, the forward and backward derivatives are sensitive to $\dx$, i.e., the smallest possible distance captured by the lattice. These charged derivatives have the disadvantage that they are asymmetric, and when expanded around the lattice site $n$, they only recover the continuum limit to order ${\cal O}(\dx^p)$ for $p=1,2,3,\dots$, while $p+1$ stencils are needed (see, e.g., \cite{DiscreteDerivatives}). However, we can also define a set of charged derivatives living in between lattice sites (i.e., to recover the continuum limit we expand the derivative operator around the point $x = ({\bf n} \pm \hat{\imath}/2) \, \dx$ instead of $x = {\bf n} \, \dx$), which have the same accuracy as the neutral derivative. They are,
\be
[\nabla_i^{(\pm,2m)} f ] =\frac{1}{\dx}\sum_{\ell=1}^{m} c_{\ell,m}^{\pm}\left[f({\bf n} \pm \ell \cdot \hat{\imath})- f({\bf n} \mp \ell \cdot \hat{\imath} \pm \hat{\imath}) \right] ~~ \longrightarrow ~~ \partial_{i} {\tt f}({x})\big|_{{x}\,\equiv\, (n \pm \hat{\imath}/2)\dx}+ \mathcal{O}(\dx^{2m}) \ . \label{eq:FBDerivative05m}
\ee
with
\be
c_{\ell,m}^{\pm} =\pm(-1)^{\ell+1}\frac{(2m!)^2}{2^{4m-2}m!^2 (2\ell-1)^2(m-\ell)!(m+\ell-1)!} \ .
\ee
The first three example cases that are accurate up sixth order are then,
\begin{subequations}
\label{fbders2}
\begin{align}
{\rm 2}^{\rm nd}:\, [\nabla_i^{(\pm,2)} f] = & \frac{ f({\bf n}\pm\hat{\imath})- f({\bf n})}{\dx} ~~ \longrightarrow ~~\partial_{i} {\tt f}({x})\big|_{{x}\,\equiv\, ({\bf n} \pm \hat{\imath}/2)\dx} + \mathcal{O}(\dx^2) \, , \label{eq:FBDerivative05O2}\\
{\rm 4}^{\rm th}:\, [\nabla_i^{(\pm,4)} f] = & \frac{-f({\bf n}\pm 2\hat{\imath})+ 27 f({\bf n}\pm \hat{\imath}) - 27f({\bf n}) +  f({\bf n}\mp \hat{\imath})}{24\dx} \nonumber \\
& \longrightarrow ~~\partial_{i} {\tt f}({x})\big|_{{x}\,\equiv\, (n \pm \hat{\imath}/2)\dx} + \mathcal{O}(\dx^4) \, ,\label{eq:FBDerivative05O4} \\
{\rm 6}^{\rm th}:\, [\nabla_i^{(\pm,6)} f] = & \frac{9f({\bf n}\pm 3\hat{\imath}) - 125 f({\bf n}\pm 2\hat{\imath}) + 2250 f({\bf n}\pm \hat{\imath}) - 2250f({\bf n}) + 125f({\bf n}\mp \hat{\imath})- 9 f({\bf n}\mp 2\hat{\imath}) }{1920\dx} \nonumber \\ 
&~~ \longrightarrow ~~ \partial_{i} {\tt f}({x})\big|_{{x}\,\equiv\, ({\bf n} \pm \hat{\imath}/2)\dx} + \mathcal{O}(\dx^6) \, . \label{eq:FBDerivative05O6}
\end{align}
\end{subequations}
On the other hand, we can also assume that the field itself lives in between lattice sites at ${\bf n}\pm \hat{\imath}/2$. For example, in the discretization of gauge field theories it is typically assumed that the gauge field lives at $A_\mu(n+ \hat{\mu}/2)$, where $\hat{\mu}$ represents the unit four-vector (see, e.g., \cite{Figueroa:2020rrl}). Assuming such a vector field $f_{i}^\mp \equiv f_i^{\mp}({\bf n}')$ with ${\bf n}'\equiv {\bf n}\mp \hat{\imath}/2$, we can apply the above introduced charged derivative operators,
\be
[\nabla_i^{(\pm,2m)} f_j^{\mp}] =\frac{1}{\dx}\sum_{\ell=1}^{m} c_{\ell,m}^{\pm}\left[f_j^\mp({\bf n}' \pm \ell \cdot \hat{\imath})- f_j^\mp({\bf n}' \mp \ell \cdot \hat{\imath} \pm \hat{\imath}) \right] ~~ \longrightarrow ~~ \partial_{i} {\tt f}_j({x})\big|_{{x}\,\equiv\, ({\bf n}\mp \hat{\jmath}/2\pm\hat{\imath}/2) \dx}+ \mathcal{O}(\dx^{2m})  \ , \label{eq:FBDerivative0}
\ee
where we assume that the spacing is the same in all directions. The derivative operator now has to be expanded around the location $x=({\bf n}\mp \hat{\jmath}/2 \pm \hat{\imath}/2) \dx$ to recover the continuum limit to order ${\cal O}(\dx^{2m})$. The first three example cases that are accurate to $p=2,4$ and $6$ are,
\begin{subequations}
\label{eq:FBderivatives}
\begin{align}
{\rm 2}^{\rm nd}:\, [\nabla_i^{(\pm,2)} f_j^{\mp}] = & \frac{f_j^{\mp}({\bf n}' \pm\hat{\imath})- f_j^{\mp}({\bf n}')}{\dx} ~~ \longrightarrow ~~\partial_i {\tt f}_j({x})\big|_{{x}\,\equiv\, ({\bf n}\mp \hat{\jmath}/2 \pm \hat{\imath}/2) \dx} + \mathcal{O}(\dx^2) \, , \label{eq:FBDerivative05O2}\\
{\rm 4}^{\rm th}:\, [\nabla_i^{(\pm,4)} f_j^{\mp}] = & \frac{- f_j^{\mp}({\bf n}'\pm 2\hat{\imath})+ 27 f_j^{\mp}({\bf n}'\pm \hat{\imath}) - 27f_j^{\mp}({\bf n}') + f_j^{\mp}({\bf n}'\mp \hat{\imath})}{24\dx} \nonumber \\
& \longrightarrow ~~\partial_i {\tt f}_j({x})\big|_{{x}\,\equiv\, ({\bf n}\mp \hat{\jmath}/2 \pm \hat{\imath}/2) \dx} + \mathcal{O}(\dx^4) \, ,\label{eq:FBDerivative05O4} \\
{\rm 6}^{\rm th}:\, [\nabla_i^{(\pm,6)} f_j^{\mp}] = & \frac{9f_j^{\mp}({\bf n}'\pm 3\hat{\imath}) - 125 f_j^{\mp}({\bf n}'\pm 2\hat{\imath}) + 2250 f_j^{\mp}({\bf n}'\pm \hat{\imath}) - 2250f_j^{\mp}({\bf n}')}{1920\dx} \nonumber \\ 
&+\frac{125f_j({\bf n}' \mp \hat{\imath}) - 9 f_j^{\mp}({\bf n}' \mp 2\hat{\imath})}{1920\dx}~~ \longrightarrow ~~ \partial_i {\tt f}_j({x})\big|_{{x}\,\equiv\, ({\bf n}\mp \hat{\jmath}/2 \pm \hat{\imath}/2) \dx} + \mathcal{O}(\dx^6) \, . \label{eq:FBDerivative05O6}
\end{align}
\end{subequations}

\subsection*{Lattice momentum}

The discrete Fourier transform of the above mentioned spatial derivatives allows us to associate a so-called {\it lattice momentum} ${\bf k}_\LL$ to each of them,
\be
[\nabla_i f](\tilde{\bf n})=-i {\bf k}_\LL(\tilde{\bf n}) f(\tilde{\bf n}) \ , \label{eq:LMdef}
\ee
with each spatial component indicated as $({\bf k}_\LL)_i\equiv k_{\LL,i}$.

\subsubsection*{Lattice momentum for the neutral derivative}

The lattice momentum associated to the neutral derivative of order $p=2m$, denoted as ${\bf k}^{0}_\LL$, can be obtained by evaluating \Eq{eq:LMdef} with \Eq{eq:CentralDerivativeOp},
\be
k_{\LL,i}^0=\frac{1}{\dx}\sum_{\ell=1}^{m} (-1)^{\ell+1}\frac{2(m!)^2}{\ell(m-\ell)!(m+\ell)!} \sin\left(\frac{2\ell\pi \tilde{n}_i}{N}\right)  \ .
\ee
The explicit expressions for the lattice momenta associated to the neutral derivatives of order $p=2,4,6$ [cf. \Eqq{centralderivatives}], are given by
\begin{subequations}
\begin{align}
{\rm 2}^{\rm nd}:\,     k_{\LL,i}^0&=\frac{\sin(2\pi \tilde{n}_i/N)}{\dx} \ ,\\
{\rm 4}^{\rm th}:\,     k_{\LL,i}^0&=\frac{4}{3}\frac{\sin(2\pi \tilde{n}_i/N)}{\dx}-\frac{1}{6}\frac{\sin(4\pi \tilde{n}_i/N)}{\dx} \ ,\\
{\rm 6}^{\rm th}:\,     k_{\LL,i}^0&=\frac{3}{2}\frac{\sin(2\pi \tilde{n}_i/N)}{\dx} - \frac{3}{10}\frac{\sin(4\pi \tilde{n}_i/N)}{\dx} + \frac{1}{30}\frac{\sin(6\pi \tilde{n}_i/N)}{\dx} \,.
\end{align}
\end{subequations}

\subsubsection*{Lattice momenta for the forward and backward derivatives}

We can also associate a lattice momentum to the forward (+) and backward (--) derivatives, which we will label by ${\bf k}^{\pm}_\LL$. For the derivative operators that live at lattice site ${\bf n} \pm \hat{\imath}/2$ we use \Eq{eq:FBDerivative05m} to evaluate \Eq{eq:LMdef}, for which we obtain,
\be
k_{\LL,i}^+ = k_{\LL,i}^- = \frac{1}{\dx}\sum_{\ell=1}^{m}(-1)^{\ell+1}\frac{(2m!)^2}{2^{4m-3}m!^2 (2\ell-1)^2(m-\ell)!(m+\ell-1)!} \sin\left(\frac{(2\ell-1)\pi \tilde{n}_i}{N}\right) \ .
\ee
The explicit expressions for the lattice momenta corresponding to the charged derivatives of order $p=2,4,6$ [given by \Eqq{eq:FBderivatives}] are
\begin{subequations}
\begin{align}
{\rm 2}^{\rm nd}: k_{\LL,i}^+ &= k_{\LL,i}^- = 2\frac{\sin(\pi \tilde{n}_i/N)}{\dx}  \ , \\
{\rm 4}^{\rm th}: k_{\LL,i}^+ &= k_{\LL,i}^- = \frac{9}{4}\frac{\sin(\pi \tilde{n}_i/N)}{\dx}-\frac{1}{12}\frac{\sin(3\pi \tilde{n}_i/N)}{\delta x} \ , \\
{\rm 6}^{\rm th}:   k_{\LL,i}^+ &= k_{\LL,i}^- = \frac{75}{32}\frac{\sin(\pi \tilde{n}_i/N)}{\dx} -\frac{25}{192}\frac{\sin(3\pi \tilde{n}_i/N)}{\dx}+ \frac{3}{320}\frac{\sin(5\pi \tilde{n}_i/N)}{\dx} \, .
\end{align}
\end{subequations}

\subsection{Higher derivatives, semi-sums and clover averages}\label{sec:LatticeOperators}

In this section, we present higher-order finite-difference operators and semi-sums, first for the general cases and then for a few concrete examples up to sixth order in accuracy. We will see in \Secs{sec:fluidLattice}{sec:fluid_bosonic}, that higher derivative operators are in general required for the lattice implementation of the scalar field EOM and the viscous contributions to the
fluid equations. Semi-sums on the other hand appear frequently in
the staggered discretization scheme of the scalar-gauge-fluid system. In the following, we adopt, when dealing with long expressions, the short hand notation
$f_{a \mu}\equiv f({\bf n} + a \cdot \hat{\mu})$ and $f_{a \mu,b \nu}\equiv f({\bf n} + a \cdot \hat{\mu}+ b \cdot \hat{\nu})$.

\subsubsection*{Laplacian}

The discrete Laplace operator acting on a field $f(\bf n)$ can be constructed by consecutive application of forward and backward derivatives and summing over all spatial directions $d$, i.e., ${\Delta}\,f({\bf n})\equiv \sum_i^d{\nabla}_i^+\bigl[{\nabla}_i^-\, f({\bf n}) \,\bigr]$. For the case of $d=3$ dimensions the expressions are
\bea\label{eq:laplm}
[\Delta^{(2m)} f ] &=&\frac{1}{\dx^2}\sum_{\imath=1}^{3}\sum_{\ell=1}^{m}\left[\frac{2}{\ell}c_{\ell, m}^0\left(f({\bf n}+ \ell \cdot \hat{\imath})- f({\bf n} - \ell \cdot \hat{\imath}) \right)-\frac{2}{\ell^2}f({\bf n})\right] ~~ \longrightarrow ~~ \Delta {\tt f}({x})\big|_{{x}\,\equiv\, {\bf n}\dx} + \mathcal{O}(\dx^{2m}) \ .
\eea
The explicit expressions for the discrete Laplace operator of second, fourth, and sixth order in accuracy are,
\begin{subequations}\label{eq:lapl0}
\begin{align}
{\rm 2}^{\rm nd}:\, [\Delta^{(2)}f] = & \sum_{\imath=1}^3\frac{f({\bf n}+\hat{\imath})-2 f({\bf n})+f({\bf n}-\hat{\imath})}{\delta x^2} ~~ \longrightarrow ~~ \Delta {\tt f}({x})\big|_{{x}\,\equiv\, {\bf n}\dx} + \mathcal{O}(\dx^2) \, , \label{eq:lapl0-O2} \\
{\rm 4}^{\rm th}:\, [\Delta^{(4)}f] = & \sum_{\imath=1}^3\frac{-f({\bf n}+2\hat{\imath})+16f({\bf n}+\hat{\imath})-30 f({\bf n})+16f({\bf n}-\hat{\imath})-f({\bf {\bf n}}-2\hat{\imath})}{12\delta x^2}  \nonumber \\ &\longrightarrow ~~ \Delta {\tt f}({x})\big|_{{x}\,\equiv\, {\bf n}\dx} + \mathcal{O}(\dx^4) \,   ,\\
{\rm 6}^{\rm th}:\, [\Delta^{(6)}f] = & \sum_{\imath=1}^3\frac{2f({\bf n}+3\hat{\imath})-27f({\bf n}+2\hat{\imath})+270f(n+\hat{\imath})-490 f({\bf n})}{180\delta x^2}  \nonumber\\
&+\frac{270f({\bf n}-\hat{\imath})-27 f({\bf n}-2\hat{\imath})+2f({\bf n}-3\hat{\imath})}{180\delta x^2} ~~ \longrightarrow ~~ \Delta {\tt f}({x})\big|_{{x}\,\equiv\, {\bf n}\dx} + \mathcal{O}(\dx^6) \,  \label{eq:lapl0-O6},
\end{align}
\end{subequations}
where we have assumed that the lattice spacing $\delta x$ is the same in each direction.

\subsubsection*{Bidiagonal scheme for cross derivatives}

On the lattice, cross derivatives of the type $\partial^2 f/ \partial x^i \partial x^j$ with $i \neq j$ (that appear, e.g., in the viscous contributions to the fluid equations) can be evaluated by the consecutive application of lattice derivative operators in the directions $\hat{\imath}$ and $\hat{\jmath}$. However, these operations are typically computationally expensive to perform (in particular for higher-order accurate derivatives) and it is useful to consider more efficient schemes. One such scheme for fields that live at lattice site $\bf n$ is the so-called bidiagonal derivative scheme, 
\bea\label{eq:Bidiag2m}
[\Delta^{(2m)}_{ij} f ] &=&\frac{1}{\dx^2}\sum_{\ell=1}^{m}\frac{1}{2\ell}c_{\ell,m}^0\biggr[f({\bf n}+ \ell \, (\hat{\imath} + \hat{\jmath}) ) - f({\bf n} - \ell \, (\hat{\imath} - \hat{\jmath})) - f({\bf n} + \ell ( \hat{\imath} - \hat{\jmath})) + f({\bf n} - \ell ( \hat{\imath} + \hat{\jmath})) \biggr] \nonumber \\ &~~& \longrightarrow ~~ \partial_i\partial_j {\tt f}({x})\big|_{{x}\,\equiv\, {\bf n}\dx} + \mathcal{O}(\dx^{2m}) \ .
\eea
The cases of second, fourth and sixth order in accuracy are given by
\begin{subequations}\label{eq:bidiag}
\begin{align}
    [\Delta^{(2)}_{ij} f] =&  \frac{f_{i,j} - f_{i,-j} - f_{-i,j} + f_{-i, -j}}{4 \dx^2}
    ~~ \longrightarrow ~~ \partial_i \partial_j {\tt f}({x})\big|_{{x}\,\equiv\, {\bf n}\dx} + \mathcal{O}(\dx^2)\, , \\
    [\Delta^{(4)}_{ij} f] =& \frac{- f_{2i, 2j} + f_{-2i, 2j} + 16 f_{i, j} - 16 f_{-i, j} - 16 f_{i, -j} + 16 f_{-i, -j} + f_{2i, -2j} - f_{-2i, -2j}}{48 \dx^2} 
    \nonumber \\
    &~~ \longrightarrow ~~ \partial_i \partial_j {\tt f}({x})\big|_{{x}\,\equiv\, {\bf n}\dx} + \mathcal{O}(\dx^4)\, , \\
    [\Delta^{(6)}_{ij} f] =& \frac{2 f_{3i, 3j} - 2 f_{-3i,3j} - 27 f_{2i, 2j} + 27 f_{-2i, 2j} + 270 f_{i, j} - 270 f_{-i, j}}{720 \dx^2} \nonumber \\
    & + \frac{- 270 f_{i, -j} + 270 f_{-i, -j} + 27 f_{2i, -2j} - 27 f_{-2i, -2j} - 2 f_{3i, -3j} + 2 f_{-3i, -3j}}{720 \dx^2} \nonumber \\
    &~~ \longrightarrow ~~ \partial_i \partial_j {\tt f}({x})\big|_{{x}\,\equiv\, {\bf n}\dx} + \mathcal{O}(\dx^6)\, .
\end{align}
\end{subequations}

\subsubsection*{Semi-sums and clover averages}

Given a field that lives at lattice site ${\bf n}$, the value of the field in the middle point of the two lattice sites $f({\bf n}+\hat{\imath}/2)$ can be obtained to order $p$ in accuracy by taking the \textit{semi-sum} ${\mathcal S}_i^{(p)}$. It is defined as,
\be \label{eq:semisum}
{\cal S}_{i}^{(2m)} [f] \equiv \sum_{\ell=1}^m c_{m,\ell} \left[f({\bf n}+ \ell \cdot \hat{\imath}) + f({\bf n} - \ell \cdot \hat{\imath} + \hat{\imath}) \right] ~ \longrightarrow ~ {\tt f}(x)\big|_{{x}\,\equiv\, ({\bf n} + \hat\imath/2)\dx} + \mathcal{O}(\dx^{2m})\, ,
\ee
with the $c_{m,\ell}$ coefficients being,
\be
c_{m,\ell}\equiv (-1)^{\ell-1}\frac{(2m!)^2}{2^{4m-1}(2\ell-1)m!^2(m-\ell)!(m+\ell-1)!} \ . \label{eq:cml-coeff}
\ee
For the cases $p=2,4$ and $6$ the explicit expressions for semi-sum operators are,
\begin{subequations}
\begin{align} \label{eq:semisumO2}
{\rm 2}^{\rm nd}:\, {\mathcal S}_i^{(2)} [f] \equiv& \frac{f({\bf n}+\hat{\imath})+ f({\bf n})}{2} ~~ \longrightarrow ~~ {\tt f}(x)\big|_{{x}\,\equiv\, ({\bf n} + \hat\imath/2)\dx} + \mathcal{O}(\dx^2) \, , \\
{\rm 4}^{\rm th}:\, {\mathcal S}_i^{(4)} [f]  \equiv&  \frac{9f({\bf n} + 9f({\bf n}) + \hat{\imath}) - f({\bf n}+2\hat{\imath}) - f({\bf n}-\hat{\imath})}{16} ~~ \longrightarrow ~~ {\tt f}(x)\big|_{{x}\,\equiv\, ({\bf n} + \hat\imath/2)\dx} + \mathcal{O}(\dx^4) \, ,\\
{\rm 6}^{\rm th}:\, {\mathcal S}_i^{(6)} [f] \equiv&  \frac{150f({\bf n})+ 150f({\bf n}+\hat{\imath})-25f({\bf n}+ 2\hat{\imath})-25f({\bf n}-\hat{\imath})+3f({\bf n}+3\hat{\imath})+3f({\bf n}-2\hat{\imath})}{256} \nonumber \\ &~~ \longrightarrow ~~ {\tt f}(x)\big|_{{x}\,\equiv\, ({\bf n} + \hat\imath/2)\dx} + \mathcal{O}(\dx^6)\, , \label{eq:semisumO6}
\end{align}
\end{subequations}
where the subindex $i$ denotes the direction in which the semi-sum is taken. Conversely, for a field that lives in between lattice sites (i.e., at ${\bf n}'={\bf n} \pm \hat{\imath}/2$), the field value $f({\bf n})$ can be obtained to order $p$, by taking the semi-sum in the opposite direction, i.e., ${\mathcal S}_{\mp i}^{(p)} [f({\bf n}')]\to {\tt f}(x)|_{x\equiv {\bf n}\dx}+\mathcal{O}(\dx^p)$. 

For a field $f$ that lives at lattice site ${\bf n}$, we can obtain its field value at the position ${\bf n}+\hat{\imath}/2+\hat{\jmath}/2$ (sometimes referred to as the center of the plaquette) to order $p$ in accuracy, by applying the so-called \textit{clover} average. This average is given by the consecutive application of semi-sums in the $\hat{\imath}$ and $\hat{\jmath}$ directions ${\mathcal S}_{i,j}^{(p)}[f]\equiv{\mathcal S}_{j}^{(p)}[{\mathcal S}_{i}^{(p)}[f]]$. The clover average to arbitrary even order of accuracy is then given by
\bea
{\mathcal S}_{i,j}^{(2m)} [f] &\equiv& \sum_{k=1}^m\sum_{\ell=1}^m c_{m,k}c_{m,\ell}\biggr[f({\bf n}+ \ell \cdot \hat{\imath} + k \cdot \hat{\jmath})+ f({\bf n}+ \ell \cdot \hat{\imath}  - (k-1) \cdot \hat{\jmath}) + f({\bf n} - (\ell-1) \cdot \hat{\imath} + k \cdot \hat{\jmath}) \nonumber \\ 
&+& f({\bf n} - (\ell-1) \cdot \hat{\imath} - (k-1) \cdot \hat{\jmath} ) \biggr] ~~ \longrightarrow ~~ {\tt f}(x)\big|_{{x}\,\equiv\, ({\bf n} +  \hat\imath/2+\hat\jmath/2)\dx} + \mathcal{O}(\dx^{2m})\, ,
\eea
where the $c_{m,\ell}$ and $c_{m,k}$ coefficients have been defined above in \Eq{eq:cml-coeff}. The explicit expressions to second, fourth and sixth order in accuracy are,
\begin{subequations}
\begin{align}
{\rm 2}^{\rm nd}:\, {\mathcal S}_{i,j}^{(2)} [f] \equiv& \, \frac{1}{2}\left(\frac{f+ f_{+i}}{2} +\frac{f_{+j}+ f_{+i,+j}}{2}\right)  ~~ \longrightarrow ~~ {\tt f}(x)\big|_{{x}\,\equiv\,  ({\bf n} +  {\hat\imath}/{2} + {\hat\jmath}/{2})\dx} + \mathcal{O}(\dx^2)\, , \\
{\rm 4}^{\rm th}:\, {\mathcal S}_{i,j}^{(4)} [f] \equiv& \,  \frac{9}{16}\frac{9f+ 9f_{+i}-f_{+2i}-f_{-i}}{16}+\frac{9}{16}\frac{9f_{+j}+ 9f_{+i,+j}-f_{+2i+j}-f_{-i,+j}}{16}\, \nonumber \\ 
&-\frac{1}{16}\frac{9f_{+2j}+ 9f_{+i,+2j}-f_{+2i,+2j}-f_{-i,+2j}}{16}-\frac{1}{16}\frac{9f_{-j}+ 9f_{+i,-j}-f_{+2i,-j}-f_{-i,-j}}{16} 
\nonumber \\
& ~~ \longrightarrow ~~ {\tt f}(x)\big|_{{x}\,\equiv\,  ({\bf n} + {\hat\imath}/{2} + {\hat\jmath}/{2})\dx} + \mathcal{O}(\dx^4) \, ,\\
{\rm 6}^{\rm th}:\, {\mathcal S}_{i,j}^{(6)} [f] \equiv& \,  {\mathcal S}_{j}^{(6)}\left[\frac{150f+ 150f_{+i}-25 f_{+2i}-25f_{-i}+3f_{+3i}+3f_{-2i}}{256}\right] \nonumber \\
&~ \longrightarrow ~ {\tt f}(x)\big|_{{x}\,\equiv\, ({\bf n} +  {\hat\imath}/{2} + {\hat\jmath}/{2})\dx} + \mathcal{O}(\dx^6)\, .
\end{align}
\end{subequations}

\subsection{Evolution algorithms}\label{sec:RK}

In this section we present numerical algorithms for the time evolution of the equations governing the scalar-gauge-fluid dynamics in an expanding background.
The EOMs of the fluid and bosonic field variables [cf.~\Eqq{generic_EOMs} in terms of the physical variables, and \Sec{equations_summary} for a summary of the fluid EOMs written in  terms of the rescaled variables], together with the second Friedmann equation [cf.~\Eq{eq:Friedmann}], represent a system of coupled first and second-order partial differential equations. Furthermore,
in \Sec{sec:gravitational_waves}, we will discuss gravitational waves, which are governed by a wave-like equation [cf.~\Eq{eq:EOMGW}], and add them to the system of equations that we are interested to evolve on the lattice.
While we will focus on the spatial discretization of these equations in the following sections in great detail, here we address the issue of advancing them in time. In the following, we will denote the amplitudes of the fluid variables (i.e., the temporal components of the stress-energy tensor in the case of the conservation form, or the fluid velocity and either the energy density or the temperature in the non-conservation form) by $g_i$, whereas the different scalar and gauge field variables, as well as the gravitational wave amplitudes will be represented by $f_i$. The conjugate momenta associated to the bosonic fields $f_i$ and the scale factor $a$ will be denoted by $\pi_i$ and $b\equiv \pi_a$. Assuming that all derivative operators in the \rhs of the equations have been discretized (i.e., written in terms of field variables at different lattice points), our system of EOM can be brought into the following form of coupled first-order ordinary differential equations,
\begin{subequations}
\label{eq:SOPDEall}
\begin{eqnarray}
\label{eq:SOPDEDrift}
\pi_i({\bf x},\eta) &=& \mathcal{D}_i[f_i'({\bf x},\eta),a(\eta),\pi_a(\eta),\lbrace f_{j}({\bf x},\eta) \rbrace, \lbrace f'_{j\neq i}({\bf x},\eta) \rbrace]\,,\\
\label{eq:SOPDEKick}
\pi_i'({\bf x},\eta) &=& \mathcal{K}_i[f_i({\bf x},\eta),\pi_i({\bf x},\eta),a(\eta),\pi_a(\eta),\lbrace f_{j\neq i}({\bf x},\eta) \rbrace, \lbrace \pi_{j\neq i}({\bf x},\eta) \rbrace , \lbrace g_{l}({\bf x},\eta) \rbrace]\,,\\
\label{eq:FOPDE}
{g'_i}({\bf x},\eta) &=& \mathcal{G}_i[g_i({\bf x},\eta),a(\eta),\pi_a(\eta),\lbrace g_{j\neq i}({\bf x},\eta) \rbrace,\lbrace f_{l}({\bf x},\eta) \rbrace, \lbrace \pi_{l}({\bf x},\eta)\rbrace] \ ,\\
\label{eq:ScaleFactorDrift}
\pi_a(\eta) &=& a'(\eta)\,,\\
\label{eq:ScaleFactorKick}
\pi_a'(\eta) &=& \mathcal{K}_a[a(\eta), E_\fl^\rho(\eta), E_\fl^p(\eta), E_V(\eta), E_K(\eta), E_G(\eta)]\,,
\end{eqnarray}
\end{subequations}
where primes represents the time derivative with respect to $\alpha$-time. The \textit{drift} $\mathcal{D}_i$ defines the conjugate momentum of the respective scalar or gauge field, whereas the so-called \textit{kernels} $\mathcal{G}_i$ and $\mathcal{K}_i$ dictate the evolution and interaction of the fluid and bosonic field constituents. The kernel of the scale factor $\mathcal{K}_a$
is given by the \rhs of \Eq{eq:Friedmann} and only includes the volume averaged energy and pressure density components of the fluid, as well as the kinetic, gradient and potential energy of the remaining fields, defined by [cf.~\Eq{eq:Friedmann} and \Eqq{eq:energy-contrib}]
\be\label{eq:va_energy_densities}
E_\fl^\rho \equiv a^{2\alpha}\left\langle T_
\fl^{00} \right\rangle\,,~~~ E_\fl^p \equiv \frac{1}{3} \left\langle T_\fl^{(3)}\right\rangle\,,~~~ E_K \equiv \sum_{i}^{\phi, A_\mu,...} \left\langle K_i \right\rangle\,,~~~ E_G \equiv \sum_{i}^{\phi, A_\mu,...} \left\langle G_i \right\rangle\,, ~~~E_V \equiv \left\langle V_0 \right\rangle\,.  
\ee
We will see in \Sec{sec:gravitational_waves} that, by construction, the gravitational waves do not backreact onto the system, and therefore do not affect the evolution of the scale factor. At any time step, the Hubble constraint [cf.~\Eq{eq:Friedmann-1st}], as well as the Gauss constraints [cf.~\Eqs{eq:Gauss_Abel}{eq:Gauss_nonAbel}] for the gauge fields, need to be satisfied, which requires appropriate spatial discretization and time-integration schemes.

The kernels ${\cal K}_i$ and ${\cal G}_i$ of our system of \Eqq{eq:SOPDEall} depend on the conjugate momenta of the corresponding bosonic fields and fluid variables. A natural option to evolve such systems of equation in time, is represented by non-symplectic integrators, like the \textit{Runge--Kutta} methods (see {\tt The Art\,I} \cite{Figueroa:2020rrl} for a detailed review). These schemes typically involve intermediate sub-steps in time, and thus require
additional {\it auxiliary fields} to store the necessary information at each sub-step.
In the following, we will discuss the so-called \textit{low--storage Runge--Kutta} algorithms \cite{Williamson1980LowstorageRS,Carpenter:1994},
which are methods that require a minimal number of auxiliary fields. Among other Runge-Kutta algorithms with the same order of accuracy, these schemes represent, in fact, the most memory-efficient integrators.

We will see in \Sec{sec:gravitational_waves} that the gravitational waves are passive degrees of freedom. This allows us to either evolve their EOM with the same Runge-Kutta scheme as the remaining variables, or with a symplectic evolution algorithm up to second order in accuracy, with the time stepping appropriately synchronized to the Runge-Kutta scheme that is used to evolve the other fields. This procedure provides the possibility to evolve the gravitational waves in a more memory-efficient manner than with the Runge-Kutta method, since it requires no additional auxiliary fields, while the accuracy is kept to ${\cal O}(\delta \eta^2)$, which is typically sufficient for a wave-like equation \cite{Figueroa:2020rrl,Baeza-Ballesteros:2025tme} (see \Sec{sec:gravitational_waves} for a more detailed discussion). In the following we will also discuss the \textit{Leapfrog} method as an exemplary case of a symplectic integrator.

\subsubsection*{\textit{Runge--Kutta} methods} 

We first discuss the low-storage Runge-Kutta schemes that can be used to solve the EOM of the scalar-gauge-fluid system, together with the second Friedmann equation, as well as the gravitational wave equations. We assume that the following initial conditions at some initial time $\eta_0$ are given,
\be
IC  :  \lbrace f_i^{(0)},\pi_i^{(0)},g_i^{(0)},  a^{(0)},\pi_a^{(0)}\rbrace {\rm ~at~} \eta_0\,.\\ \nonumber
\ee
To evolve the system by one time step
$\eta \rightarrow \eta+\delta \eta$,
we introduce $s$ intermediate sub-steps. Each sub-step comes with a pair of coefficients $A_p$ and $B_p$, with $p=1,...,s$. With the help of the following set of auxiliary fields $\{ \Delta f_i^{(p)}, \Delta \pi_i, \Delta g_i^{(p)}, \Delta a_i^{(p)}, \Delta \pi_a^{(p)}\}$ the Runge-Kutta scheme can be written as, 
\begin{eqnarray}
\left\lbrace
\begin{array}{rcl}
\Delta f^{(p)}_i &=&A_p \Delta f^{(p-1)}_i + \delta  \eta \, \pi_i^{(p-1)}\, ,  \vspace{0.15cm}\\
 \Delta\pi_{i}^{(p)} &=&A_p\Delta \pi_{i}^{(p-1)}+  \delta \eta \,  \mathcal{K}^{(p-1)}_{ i}\, ,  \vspace{0.15cm}\\
 \Delta g_{i}^{(p)} &=&A_p\Delta g_{i}^{(p-1)}+  \delta \eta \,  \mathcal{G}^{(p-1)}_{i}\, ,  \vspace{0.15cm}\\
    \Delta a^{(p)} &=&A_p \Delta a^{(p-1)} + \delta \eta \, \pi_a^{(p-1)}\, ,  \vspace{0.15cm}\\
 \Delta \pi_a^{(p)} &=&A_p\Delta \pi_a^{(p-1)}+  \delta \eta \,  \mathcal{K}^{(p-1)}_{a}\, , 
 \end{array}
\quad \Longrightarrow \quad
 \begin{array}{rcl}
        f^{(p)}_i &=&  f^{(p-1)}_i + B_p  \Delta  f^{(p)}_i\,   \vspace{0.15cm}\\
        \pi_{i}^{(p)} &=&\pi_{i}^{(p-1)}+  B_p \Delta \pi_{i}^{(p)}\,    \vspace{0.15cm}\\
        g_{i}^{(p)} &=& g_{i}^{(p-1)}+  B_p \Delta g_{i}^{(p)}\,   \vspace{0.15cm}\\
        a^{(p)} &=&a^{(p-1)} +B_p  \Delta a^{(p)}\,  \vspace{0.15cm}\\
        \pi_a^{(p)} &=&\pi_a^{(p-1)} +B_p  \Delta \pi_a^{(p)}\, \vspace{0.15cm}
\end{array}
\right\rbrace_{p\,=\,1,\, ...,\, s} \Longrightarrow
\end{eqnarray}
\begin{eqnarray}
\quad\Longrightarrow \quad
\left\lbrace
\begin{array}{rcl}
f_{i,+0} &=&  f_i^{(s)}\,  \vspace{0.15cm}\\
\pi_{i,+0} &=& \pi_i^{(s)}\,  \vspace{0.15cm}\\
g_{i,+0} &=&  g_i^{(s)}\,  \vspace{0.15cm}\\
a_{+0} &=& a^{(s)}\, \vspace{0.15cm}\\
\pi_{a,+0} &=& \pi_a^{(s)}\, \\
\end{array}\right\rbrace \, ,
\end{eqnarray}
where $+0$ refers to the advancement of the fields by one full time step, i.e., $\eta\to \eta + \delta \eta$. A specific Runge-Kutta scheme is completely characterized by the coefficients $A_p, B_p$, and the number of intermediate sub-steps $s$. For example, the coefficients $A_p = \{0,-1\}$ and $B_p = \{1,1/2\}$ characterize a second order scheme, which reproduces the fields at the next timestep with an error of ${\cal O}(\delta \eta^2)$, and is known as the modified Euler method. For fluid dynamics, a frequently used low-storage scheme, is the third order accurate Williamson scheme \cite{Williamson1980LowstorageRS}. It has one auxiliary field per \textit{dof}, and is characterized by the coefficients $A_p = \{0,-5/9,-153/128\}$ and $B_p = \{1/3,15/16,8/15\}$. The coefficients for higher-order low-storage algorithms are provided, e.g., in~\cite{Williamson1980LowstorageRS,Carpenter:1994}.

\subsubsection*{\textit{Leapfrog} method}

As mentioned above the gravitational wave degrees of freedom can also be evolved with a symplectic evolution scheme, while the fluid and bosonic degrees of freedom are evolved by a non-symplectic algorithm. In the following we present, as an example algorithm, the Leapfrog method. For a broader overview on the topic of symplectic evolution algorithms with a range of further examples, such as the Verlet schemes, we refer to section 3 and 2 of {\tt The Art\,I} \cite{Figueroa:2020rrl} and {\tt II} \cite{Baeza-Ballesteros:2025tme} respectively.

In the following, we distinguish between the field amplitude and conjugate momenta that are evolved by the leapfrog method (e.g., the gravitational waves), labeled by $h_{lm}$ and $(\pi_h)_{lm}$, and the ones of the remaining fields, which we continue to name $f_i$ and $\pi_i$. Since these remaining fields and the fluid are evolved by the Runge-Kutta schemes, the $f_i$ and $g_i$, as well as $\pi_i$ live at integer times. In case of the Leapfrog algorithm the field amplitudes $h_{lm}$ and their conjugate momenta $(\pi_h)_{lm}$ are separated by a half time step. Thus, the gravitational waves can only be evolved with the Leapfrog method if the two algorithms are synchronized correctly. We set the field amplitude $h_{lm}$ to live at integer times and the conjugate momentum $(\pi_h)_{lm}$ at semi-integer times, which we denote by the subscript $+0/2$. The initial conditions to evolve the gravitational waves by one time step are: the field amplitudes $h_{lm}, f_i^{(0)}, g_i^{(0)}$ and the scale factor $a^{(0)}$ at the initial time step $\eta_0$ (where the latter three are the same initial conditions as for the Runge-Kutta scheme that has been explained above), the conjugate momenta $(\pi_h)_{lm}$ at minus half a time step $\eta_{-0/2}$, as well as the scale factor $a_{+0}$ at the advanced time step $\eta_{+0}$. 
\be
IC  : \quad  \lbrace  h_{lm}, f_i^{(0)}, g_i^{(0)}, a^{(0)} \rbrace {\rm ~at~} \eta_0\,, \quad\lbrace (\pi_h)_{lm,-0/2} \rbrace {\rm ~at~} \eta_{-0/2}=\eta_0-\frac{1}{2}\delta\eta,\quad {\rm and} \quad \lbrace a_{+0}\rbrace{\rm ~at~} \eta_{+0}=\eta_0+\delta\eta \, . \nonumber
\ee
By choosing the canonical momentum as $(\pi_h)_{lm} \equiv a^{\beta} \partial_0 h_{lm}$ the Leapfrog scheme can then be written in the following form,
\begin{align}
\begin{cases}
\ \ \ (\pi_h)_{lm,+0/2} &= \ (\pi_h)_{lm,-0/2}+\delta\eta \ {\cal K}_{lm}[h_{lm}, f_i^{(0)}, g_i^{(0)}, a^{(0)}]\, , \vspace{0.15cm}\\
 \ \ \ a_{+0/2} &= \ (a^{(0)}+a_{+0})/2\, ,  \vspace{0.15cm}\\
 \ \ \ h_{lm,+0} &= \ h_{lm}+a_{+0/2}^{-\beta}\delta \eta \ (\pi_h)_{lm,+0/2} \, , \\
\end{cases}
\end{align}
where $\beta=3-\alpha$ for gravitational waves (see \Sec{sec:gravitational_waves} of this manuscript, or section 8 of {\tt The Art\,-\,II} \cite{Baeza-Ballesteros:2025tme} for more details).


\newpage
\section{Lattice formulation of fluid dynamics}\label{sec:fluidLattice}

In this section we discuss the lattice formulation of the EOM of pure fluid dynamics in an expanding FLRW background, as discussed in \Sec{subsec:fld_dynamics_cont}, including first-order imperfect fluid effects (i.e., viscosity). In \Sec{sec:fluid_bosonic} we will then discuss the interactions with gauge or scalar fields. In the following, we will first introduce a set of so-called program variables associated to the fluid variables. We then deal in \Sec{discrete_fluid_cons_form} with the conservation form, where the dynamical degrees of freedom of the fluid sector are the components $\conf{T}^{00}_{\rm pf}$ and $\conf{T}^{0i}_{\rm pf}$ of the stress-energy tensor. In \Sec{sec:LatticeFluidDynamicsNC} we consider the non-conservation form, where we solve for the perfect fluid energy density $\conf{\rho}_f$ and the velocity field $u_i$. In both cases we will assume that the fluid has a \textit{constant} EOS. We will first present the case of a perfect fluid, as described in \Sec{perfect_fluids}. Then, we turn to imperfect fluids that have been discussed in \Sec{viscous_term}, restricting our treatment to the \textit{subrelativistic} description of the viscous contributions. In all the cases, the lattice equations will be presented using two different discretization procedures: the collocated and staggered formulations. Moreover, when describing the perfect fluid EOM, we will, for clarity, explicitly show the convergence to the continuum of the lattice expressions. All the discrete versions of the kernels will be presented with derivatives of generic order $p$ accuracy, however, these will always refer to the even order $p=2m$ lattice operators introduced in \Sec{sec:generalLattice}.

\subsubsection*{Program variables}

In the following we will work with rescaled dimensionless fluid and spacetime variables, which we refer to as the \textit{program variables} and label by a bar. We follow the same prescription as in {\tt The\,Art\,I} \cite{Figueroa:2020rrl} to rescale the spacetime variables $\dx^i$ and $\delta\eta$. Depending on whether we work in the conservation or non-conservation form, we have to rescale the components of the stress-energy tensor $ T^{\mu\nu}_{\rm pf}$, or the energy density $\rho_f$ and velocity field $u_i$. In both cases we are expressing the \textit{conformal fluid variables} in terms of program variables, such that,
\be
\delta\prgv\eta \equiv \omega_*\delta\eta= a^{-\alpha}\omega_* \delta t\, , \quad \delta\prgv x^i \equiv\omega_* \delta x^i \, , \qquad {\rm and} \qquad \prgv{T}^{\mu\nu}_{\rm pf} \equiv \frac{{\conf T}^{\mu\nu}_{\rm pf}}{T_*^4} \ , \qquad {\rm or} \qquad  \prgv{\rho} \equiv \frac{{\conf \rho}_f}{T_*^4} \ ,\quad  u_i = u_i  \ ,\label{eq:FluidProgramVariables}
\ee
where $\omega_*$ and $T_*$ are two mass scales with dimension $[\omega_*]=[T_*]=+1$. The values of $\alpha$, $\omega_*$, and $T_*$ are chosen such that the dynamics of the system under consideration is represented well on the lattice (for a more in depth discussion, see the box on pages 32/33 of {\tt The\,Art\,I} \cite{Figueroa:2020rrl}).
Note that the velocity field $u_i$ does not require any rescaling, since it is a dimensionless quantity. We will see in the following that, in the non-conservation form, we typically evolve the logarithm of the fluid energy density $\ln \prgv{\rho}$ as a dynamical variable. Moreover, both the pressure and the enthalpy will be rescaled as $\prgv{p} = p_\f/T_*^4, \prgv{w}=w_\f / T_*^4$. From the rescaling of the spacetime variables, it follows that the derivatives are rescaled as $\prgv{D}_\mu\equiv D_\mu/\omega_*$ and the Hubble parameter becomes $\prgv{\HH}\equiv \HH/\omega_*$. 

The forcing terms that appear in the conservation equations and stem from the Hubble friction and the out-of-equilibrium effects are rescaled accordingly,
\be
\prgv{{\mathcal F}}_{{\HH}}^\mu  \equiv  \frac{\conf{{\mathcal F}}_{H} ^\mu}{\omega_* T_*^4 }\, , \qquad {\rm and} \qquad \prgv{{\mathcal F}}_{\rm ipf}^\mu  \equiv  \frac{\conf{{\mathcal F}}_{\rm ipf} ^\mu}{\omega_* T_*^4} \ . 
\ee
The deviatoric tensor $\cdev^{\mu \nu}$ that provides the imperfect fluid contributions has to be rescaled in the same way as $T^{\mu\nu}_{\pf}$ and is given in terms of program variables by,
\be
\pdev^{\mu \nu} \equiv \frac{\cdev^{\mu \nu}}{T_*^4}  = 2 \, \prgv{\nu} \, \prgv{w} \, \prgv{\sigma}^{\mu \nu} + \prgv{\xi} \, \prgv{w}  \, \prgv{\theta} h^{\mu \nu} \, , \label{eq:PrgvPiMuNu}
\ee
where the rescaled enthalpy $\prgv{w}$, the traceless rate-of-strain tensor $\prgv{\sigma}^{\mu \nu}$, the  rate-of-strain tensor $\prgv{S}^{\mu \nu}$ and its trace $\prgv{\theta}$ are,
\be
\prgv{w} \equiv \frac{\conf{w}_f}{T_*^4} \ , \hspace{1cm} \prgv{\sigma}^{\mu \nu} \equiv \frac{\conf{\sigma}^{\mu \nu}}{\omega_*} \ , \hspace{1cm} \prgv{S}^{\mu \nu} \equiv  \frac{\conf{S}^{\mu \nu}}{\omega_*}\ , \hspace{1cm} {\rm and} \hspace{1cm}  \prgv{\theta} \equiv  \frac{\conf{\theta}}{\omega_*}\ .
\ee
Consequently, the shear and bulk viscosity are rescaled as,
\be
\prgv{\nu} \equiv \omega_*\conf{\nu} \, , \qquad {\rm and} \qquad \prgv{\xi} \equiv \omega_*\conf{\xi} \, .
\ee
From the above rescaling prescriptions it follows that the traces of the conformal stress energy tensor $\conf{T}_\pf \equiv g_{\mu \nu} \conf{T}^{\mu \nu}_\pf$ as well as the deviatoric stress tensor $\cdev \equiv g_{\mu \nu} \cdev^{\mu \nu}$ are rescaled in the same fashion as the full tensors, 
\be
\label{emt_rescalings_prgv_fluid}
\prgv{T}_{\pf} = \frac{\conf{T}_{\pf}}{T_*^4} \, , \qquad {\rm and} \qquad \pdev = \frac{\cdev}{T_*^4} \, .
\ee

\subsection{Conservation form}
\label{discrete_fluid_cons_form}

In the conservation form the dynamics of the system is described by the energy and momentum conservation equations of the fluid introduced in \Eq{general_eqs}. The background expansion on the other hand is governed by the Friedmann \Eqq{eq:Friedmann}, where we use the first one as a constraint equation and the second one to evolve the scale factor. The degrees of freedom we are solving for are the rescaled conformal program variables $\prgv{T}_{\rm pf}^{00}$ and $\prgv{T}_{\rm pf}^{0i}$, defined in \Eq{eq:FluidProgramVariables}, as well as the scale factor $a$ and its conjugate momentum $b\equiv a'$. The conservation equations together with the second Friedmann equation can be brought into the following form of coupled first-order differential equations,
\begin{subequations}
\begin{align}
    (\prgv{T}_{\rm pf}^{00})' & = {\cal G}^0
    [\prgv{T}_{\rm pf}^{00}, \prgv{T}_{\rm pf}^{0i}, \pdev^{\mu\nu}, a, b] ={\cal G}_\pf^0[\prgv{T}_{\rm pf}^{0i}]+\prgv{{\cal F}}^0_{\HH}[\prgv{T}_{\pf},\prgv{\HH}]+\prgv{{\cal F}}^0_{\rm ipf}[\pdev^{\mu0},\pdev,\HH]\, , \label{eq:EOMsetCF1}\\
    (\prgv{T}_{\rm pf}^{0i})' & = {\cal G}^i
    [\prgv{T}_{\rm pf}^{ij}, \prgv{T}_{\rm pf}^{0i}, \pdev^{\mu\nu}, a, b] = {\cal G}_\pf^i[\prgv{T}_{\rm pf}^{ij}] +\prgv{{\cal F}}^i_\HH[\prgv{T}_{\rm pf}^{0i},\prgv{\HH}]+\prgv{{\cal F}}^i_\ipf[\pdev^{\mu j},\prgv{\HH}]\, , \label{eq:EOMsetCF2} \\
    b' & = {\mathcal{K}}_{a}[a,\prgv{E}_\fl^\rho,\prgv{E}_\fl^p] \, , \label{eq:EOMsetCF3}\\
    a' & = b \, . \label{eq:EOMsetCF4}
\end{align}
\end{subequations}
where $\prgv{T}_{\pf} = \conf{T}_{\pf}/T_*^4$ and $\pdev = \cdev/T_*^4$ , with $\conf{T}_\pf \equiv g_{\mu \nu} \conf{T}^{\mu \nu}_\pf$ and $\cdev \equiv g_{\mu \nu} \cdev^{\mu \nu}$ respectively being the perfect fluid and deviatoric stress energy tensor traces, $\prgv{\HH}\equiv \HH/\omega_*= a'/a$ is the Hubble parameter in terms of program variables, and primes denote derivatives with respect to $\alpha$-time in program units. The above set of equations is written for a system with self-consistent expansion, which means that the volume average of the fluid energy and pressure density, as given in \Eq{eq:va_energy_densities}, source the evolution of the scale factor. In case the expansion of the background is sourced by other sectors as well,
the kernel ${\mathcal{K}}_{a}$ will depend on additional contributions. Of particular interest is the case of an external background fluid with constant equation of state dictating the expansion of the Universe. In that case the evolution of the scale factor is described by \Eq{eq:sf_const} and we do not need to explicitly evolve \Eqs{eq:EOMsetCF3}{eq:EOMsetCF4}.

The fluid kernels $\mathcal{G}^\mu$ depend on the different components of the stress-energy tensor and on the choice of constant equation of state $\omega$ [cf.~\Eq{sound_speed}]. In the second equality we have split the total fluid kernels into core kernels $\mathcal{G}_\pf^\mu$ associated to the perfect fluid, that do not depend on the Hubble parameter $\HH$ (and fully describe, for $\alpha=1$, the conformal dynamics of a pure radiation perfect fluid), and additional forces describing the Hubble friction $\prgv{{\mathcal F}}^\mu_{\mathcal{H}}$ and the viscous effects $\prgv{{\cal F}}^\mu_\ipf$. This separation allows us to discuss first the case of a perfect fluid in an expanding background, and then add the imperfect fluid contributions. Moreover, in \Sec{sec:fluid_bosonic} we will be able to discuss the fluid-gauge and fluid-scalar interactions by just adding additional force terms. The fluid equations written as in \Eqs{eq:EOMsetCF1}{eq:EOMsetCF2} can in principle be solved by using non-symplectic integrators, an example of such algorithms being the Runge-Kutta method, which we have discussed in \Sec{sec:RK}.

\subsubsection{Perfect fluid}
\label{discrete_perfect_fluids}

We first discuss the case of a perfect fluid, which corresponds to setting ${\cal F}^\mu_\ipf=0$ in \Eqs{eq:EOMsetCF1}{eq:EOMsetCF2}. The discretized versions of the kernels of the energy and momentum conservation equations, as well as the Hubble friction terms are given by [cf.~\Eqq{eqs_fluid}]
\begin{align}
    & \mathcal{G}^{0}_\pf[\prgv{T}_{\rm pf}^{0i}] = -\sum_i\prgv{\nabla}_i^\LL\prgv{T}_{\rm pf}^{0i} \ , \hspace{1cm} \mathcal{G}_\pf^i[\prgv{T}_{\rm pf}^{ij}] =-\sum_j\prgv{\nabla}_j^\LL\prgv{T}_{\rm pf}^{ij}\, , \label{eq:latticeGT}\\
    &  \prgv{{\mathcal F}}^0_{\HH}[\prgv{T}_{\rm pf}^{00},\prgv{\HH}]= \frac{(1-3\omega)}{z_\LL} \prgv{\HH}\prgv{T}_{\rm pf}^{00}\, , \hspace{1cm}  \prgv{{\mathcal F}}^i_{\HH}[\prgv{T}_{\rm pf}^{0i},\prgv{\HH}]=(\alpha-1)\prgv{\HH}\prgv{T}_{\rm pf}^{0i} \label{eq:latticeFH}
\end{align}
where $\prgv{\nabla}_j^\LL$ and $z_\LL$ denote the lattice versions of the derivative and the $z$-function introduced in \Eq{rho_u_fromT0mu}, which have different forms when using different discretization schemes. The purely spatial part of the stress-energy tensor $\prgv{T}_{\rm pf}^{ij}$, as well as $z_\LL$ can be expressed in terms of $\prgv{T}_{\rm pf}^{00}$ and $ \prgv{T}_{\rm pf}^{0i}$ [cf.~\Eqs{Tij_gamma2}{gamma2_rel}]
\begin{align}
    \prgv{T}_{\rm pf}^{ij}&=\left(\frac{z_\LL}{z_\LL+\omega}\right)\frac{\prgv{T}_{\rm pf}^{0i}\prgv{T}_{\rm pf}^{0j}}{\prgv{T}_{\rm pf}^{00}}+a^{2(\alpha-1)}\frac{\omega}{z_\LL} \prgv{T}_{\rm pf}^{00}\delta_{ij} \,,  \label{eq:TijL} \\
    z_\LL &= \frac{1}{2(1-[r^2]_\LL)} \left( \, 1-\omega  + \sqrt{(1+\omega)^2-4\omega [r^2]_\LL} \, \right) \, , \label{eq:zL}
\end{align}
where $[r^2]_\LL$ denotes the lattice version of the ratio $r^2 \equiv a^{2(1-\alpha)}\sum_i(\prgv{T}_{\rm pf}^{0i}/\prgv{T}_{\rm pf}^{00})^2$ [cf.~\Eq{T0i_sq}]. The kernel of the scale factor only depends on the volume averaged energy and pressure density of the perfect fluid itself, and is given by,
\be
{\mathcal{K}}_{a}[a,\prgv{E}_{\rm pf}^\rho,\prgv{E}_{\rm pf}^\rho]=\frac{1}{3} \left( \frac{ T_*^2}{\omega_*\mpl} \right)^2 a^{2\alpha+1} \left[ \frac{2\alpha - 1}{2}{\prgv E}_{\rm pf}^\rho-\frac{3}{2}{\prgv E}_{\rm pf}^p\right] \ , \label{eq:ScaleFactorPF}
\ee
with the energy density and pressure contributions defined as
\be\label{fluid_terms_in_friedmann}
{\prgv E}_\pf^\rho \equiv \frac{1}{a^4}\left\langle {\prgv T}_{\rm pf}^{00}\right\rangle \qquad {\rm and} \qquad {\prgv E}_\pf^p \equiv \frac{1}{3a^{2+2\alpha}}\left\langle \sum_{i}{\prgv T}_{\rm pf}^{ii} \right\rangle \ ,
\ee
where $\langle \dots\rangle$ denotes a volume average over the whole lattice.  When we evolve the above set of \Eqss{eq:EOMsetCF1}{eq:EOMsetCF4}, the Hubble constraint, given by 
\be
b^2 = \frac{a^{2(\alpha+1)}}{3} \left( \frac{ T_*^2}{\omega_*\mpl} \right)^2  {\prgv E}_\pf^\rho   \, ,
\ee
has to be preserved up to the accuracy of the chosen integrator.

\subsubsection*{Collocated formulation}

In the collocated formulation the fluid variables $\prgv{T}_{\rm pf}^{00}$ and $\prgv{T}_{\rm pf}^{0i}$ (as well as $\prgv{T}_{\rm pf}^{ij}$) are assumed to live at lattice site $\bf n$. The discretized versions of the energy and momentum conservation equations are then obtained by replacing the continuum derivatives by the neutral derivative $\prgv{\nabla}_i^{(0,p)}$, introduced in \Eqss{eq:CentralDerivativeO2}{eq:CentralDerivativeO6}. The spatial-spatial part of the stress-energy tensor $\prgv{T}_{\rm pf}^{ij}$ and $z_\LL$ are obtained by simply substituting the lattice field components $\prgv{T}_{\rm pf}^{00}({\bf n})$ and $\prgv{T}_{\rm pf}^{0i}({\bf n})$ into \Eqs{eq:TijL}{eq:zL}.\\

\subsubsection*{Energy conservation equation}

The core kernel $\mathcal{G}^0_\pf$ and the Hubble friction term $\prgv{{\mathcal F}}_\HH^0$ of the energy conservation equation are then,
\begin{align}
    \mathcal{G}^{0}_\pf[\prgv{T}_{\rm pf}^{0i}] &= - \sum_i\prgv{\nabla}_i^{(0,p)} \prgv{T}_{\rm pf}^{i0} ~~ \longrightarrow ~~  - \prgv{\partial}_i \prgv{{\tt T}}_{\rm pf}^{i0}|_{\prgv{x}\equiv {\bf n}\delta \prgv{x}} + {\mathcal O}(\delta \prgv{x}^p)\,, \label{eq:latticeG0Tcoll} \\
   \prgv{{\mathcal F}}^0_{\HH}[\prgv{T}_{\rm pf}^{i0},\prgv{{\mathcal H}}] &= \frac{(1-3\omega)}{z} \prgv{\HH}\prgv{T}_{\rm pf}^{00} \, , \label{eq:latticeF0Tcoll}
\end{align}
where the $z$-function is obtained from \Eq{eq:zL} by simply replacing $\prgv{T}_{\rm pf}^{00}$ and $\prgv{T}_{\rm pf}^{0i}$ by their lattice counterparts. The expressions on the \rhs of the arrows indicate to which order the continuum expressions are recovered.

\subsubsection*{Momentum conservation equation}

The core kernels $\mathcal{G}^i_\pf$ and the Hubble friction term $\prgv{{\mathcal F}}_\HH^i$ of the momentum conservation equation are given by
\begin{align}
    \mathcal{G}^{i}_\pf[\prgv{T}^{ij}_{\rm pf}] & = - \sum_j\prgv{\nabla}_j^{(0,p)} \prgv{T}_{\rm pf}^{ji}  ~~ \longrightarrow ~~  - \prgv{\partial}_j \prgv{{\tt T}}_{\rm pf}^{ji}|_{\prgv{x}\equiv {\bf n}\delta \prgv{x}} + {\mathcal O}(\delta \prgv{x}^p) \,, \label{eq:latticeGiTcoll} \\
    \prgv{{\mathcal F}}^i_{\HH}[\prgv{T}_{\rm pf}^{0i},\prgv{{\HH}}] &=(\alpha-1)\prgv{\HH}\prgv{T}_{\rm pf}^{0i}  \ , \label{eq:latticeFiTcoll}
\end{align}
where in \Eq{eq:latticeGiTcoll} $\prgv{T}^{ij}_\pf$ has to be replaced with its expression in \Eq{eq:TijL}. Since we are applying the neutral derivatives to order $p$, the terms including the spatial derivatives, $\prgv{\partial}_j\prgv{{\tt T}}_{\rm pf}^{j\mu}$, are recovered up to order ${\mathcal{O}(\delta \prgv{x}^p)}$ in accuracy at lattice site $\bf n$.

\subsubsection*{Staggered formulation}

In the staggered formulation, we follow a similar approach as for the discretization of gauge theories presented in {\tt The Art\,I} \cite{Figueroa:2020rrl} and {\tt II} \cite{Baeza-Ballesteros:2025tme}, and we assume that scalar and vector fields live in different locations on the lattice. The (scalar) temporal component of the stress-energy tensor $\prgv{T}_{\rm pf}^{00}({\bf n})$ lives at lattice sites, while the (vector) temporal-spatial component $\prgv{T}_{\rm pf}^{0i}({\bf n}+\hat{\imath}/2)$ lives in the middle of two lattice points. As a consequence, the spatial-spatial part of the stress-energy tensor naturally lives in the middle of the plaquette,
hence it is defined as $\prgv{T}_{\rm pf}^{ij}({\bf n}+\hat{\imath}/2+\hat{\jmath}/2)$. In the staggered formulation a particular difficulty arises from the fact that one has to ensure that all terms appearing in the same \rhs of the EOM live at the same location on the lattice -- at lattice site in the case of the energy conservation equation and in between lattice sites in the case of the momentum conservation equation. This makes the discretized kernels more involved in comparison to the collocated formulation. The advantage of this formulation, however, is that we can use forward/backward derivatives in our discretization procedure, which (contrary to the neutral derivatives), are sensitive to $\delta x$, i.e., the smallest resolvable distance on the lattice. Note that the consecutive application of forward/backward derivatives, semi-sums and clover averages can typically be done in more than one way. Thus, the discretization scheme of the fluid kernels we provide in the following is not the only possible realization.

\subsubsection*{Energy conservation equation}

The core kernel and the Hubble friction term $\prgv{{\mathcal F}}_\HH^0$ of the energy conservation equation are discretized as follows,

\begin{align}
    \mathcal{G}^{0}_\pf[\prgv{T}_{\rm pf}^{0i}] &= -\sum_i\prgv{\nabla}_i^{(-,p)}\prgv{T}_{\rm pf}^{0i} ~~ \longrightarrow ~~ -\prgv{\partial}_i\prgv{{\tt T}}_{\rm pf}^{0i} |_{\prgv{x}\equiv{\bf n}\delta \prgv{x}}+{\mathcal O}(\delta \prgv{x}^p) \ , \label{eq:latticeG0Tstagg}\\
    \prgv{{\mathcal F}}^0_{\HH}[\prgv{T}_{\rm pf}^{0i},\prgv{{\HH}}] &= \frac{(1-3\omega)}{z_{\rm e}} \prgv{\HH}\prgv{T}_{\rm pf}^{00} ~~ \longrightarrow ~~ \frac{(1-3\omega)}{z} \prgv{\HH}\prgv{{\tt T}}_{\rm pf}^{00}\biggr|_{\prgv{x}\equiv{\bf n}\delta \prgv{x}}+{\mathcal O}(\delta \prgv{x}^p)\, , \label{eq:latticeF0Tstagg}
\end{align}
where $\prgv{\nabla}_i^{(-,p)}$ denotes the backward derivative accurate to order $p$ [see \Eq{eq:FBDerivative0}]. The $z$-function $z_{\rm e}\equiv z(r^2_{\rm e})$ is evaluated by just substituting the $r^2$-ratio into \Eq{eq:zL}. However, the $\prgv{T}^{0i}_\pf$ components that appear in the $r^2$-ratio need to be shifted by a half lattice spacing,
\be
r^2_{\rm e}\equiv a^{2(1-\alpha)}\sum_{k=1}^{3}\frac{{\mathcal S}_{-k}^{(p)}[\prgv{T}_{\rm pf}^{0k}\prgv{T}_{\rm pf}^{0k}]}{[\prgv{T}_{\rm pf}^{00}]^2} ~~ \longrightarrow ~~ a^{2(1-\alpha)}\sum_{k=1}^{3}\left[\frac{\prgv{{\tt T}_{\rm pf}^{0k}}}{\prgv{{\tt T}}_{\rm pf}^{00}}\right]^2 \Biggr|_{\prgv{x}\equiv{\bf n}\delta \prgv{x}}+{\mathcal O}(\delta \prgv{x}^p) \, ,
\ee
where ${\mathcal S}^{(p)}$ denotes the semi-sum to order $p$
[given in \Eq{eq:semisum}] applied to the components $\prgv{T}_{\rm pf}^{0k}$.\\

\subsubsection*{Momentum conservation equation}

The core kernel of the momentum conservation equation and the Hubble friction term $\prgv{{\mathcal F}}_\HH^i$ are discretized as follows,
\begin{align}
    \mathcal{G}^{i}_\pf[\prgv{T}_{\rm pf}^{ij}] &= -\sum_j\prgv{\nabla}_j^{(-,p)}\prgv{T}_{\rm pf}^{ij} ~~ \longrightarrow ~~ -\prgv{\partial}_j\prgv{{\tt T}}_{\rm pf}^{ij} \, \biggr|_{\prgv{x}\equiv({\bf n}+\hat{\imath}/2)\delta \prgv{x}}+{\mathcal O}(\delta \prgv{x}^p)\, , \label{eq:latticeGiTstagg} \\
    \prgv{{\mathcal F}}^i_{\mathcal{H}}[\prgv{T}_{\rm pf}^{0i}] &=(\alpha-1)\mathcal{\prgv{H}}\prgv{T}_{\rm pf}^{0i} ~~ \longrightarrow ~~ (\alpha-1)\mathcal{\prgv{H}}\prgv{{\tt T}}_{\rm pf}^{0i}|_{\prgv{x}\equiv({\bf n}+\hat{\imath}/2)\delta \prgv{x}} \ , \label{eq:latticeFiTstagg}
\end{align}
where, again, $\prgv{\nabla}_j^{(-,p)}$ denotes the backward derivative to order $p$. The expression of $\prgv{T}_{\rm pf}^{ij}$ requires special care in the staggered formulation and is discretized in the following way,
\begin{align}
    \prgv{T}_{\rm pf}^{ij}&= {\mathcal S}_{i,j}^{(p)}\left[\left(\frac{z_{\rm m}}{z_{\rm m}+\omega}\right)\frac{{\mathcal S}_{-i}^{(p)}[\prgv{T}_{\rm pf}^{0i}]{\mathcal S}_{-j}^{(p)}[\prgv{T}_{\rm pf}^{0j}]}{\prgv{T}_{\rm pf}^{00}}+a^{2(\alpha-1)}\frac{\omega}{z_{\rm m}}\prgv{T}_{\rm pf}^{00} \delta_{ij}\right] ~~ \longrightarrow ~~ \prgv{{\tt T}}_{\rm pf}^{ij}|_{\prgv{x}\equiv({\bf n}+\hat{\imath}/2+\hat{\jmath}/2)\delta \prgv{x}} + {\mathcal O}(\delta \prgv{x}^p) \,,
\end{align}
such that $\prgv{T}_{\rm pf}^{ij}$ itself lives at ${\bf n} + \hat{\imath}/2+\hat{\jmath}/2$. The $z$-function, here defined as $z_{\rm m}\equiv z(r^2_{\rm m})$, can be evaluated according to \Eq{eq:zL}. However, the $r^2$-ratio
requires the application of a semi-sum on the $\prgv{T}_{\rm pf}^{0k}$ components,
\be
r^2_{\rm m}\equiv a^{2(1-\alpha)}\sum_{k=1}\frac{{\mathcal S}_{-k}^{(p)}[\prgv{T}_{\rm pf}^{0k}\prgv{T}_{\rm pf}^{0k}]}{(\prgv{T}_{\rm pf}^{00})^2} ~~ \longrightarrow ~~ \prgv{{\tt r}}|_{\prgv{x}\equiv{\bf n}\delta \prgv{x}} +{\mathcal O}(\delta \prgv{x}^p) \, . \label{eq:r2_m}
\ee

\subsubsection*{Pure radiation fluid in the subrelativistic limit}

A case of particular interest represents a pure radiation perfect fluid in the subrelativistic limit, i.e., a fluid that has constant equation of state $\omega=1/3$ and $|\vec{u}_{\rm p}|^2\ll 1$. With the convenient choice of $\alpha=1$, the full kernels of the fluid EOM then reduce to ${\mathcal{G}}^0_\pf$ and ${\mathcal{G}}^i_\pf$. The spatial-spatial part of the stress-energy tensor can be evaluated in the collocated formulation by just substituting $\prgv{T}_\pf^{00}$, $\prgv{T}_\pf^{0i}$ into \Eq{eq:Tij_sub_rel} and $r^2$. In the staggered formulation the lattice expression for $\prgv{T}_\pf^{ij}$ becomes,
\be 
    \prgv{T}^{ij}_\pf = {\mathcal S}_{i,j}^{(p)}\left[ \frac{3}{4} \frac{{\mathcal S}_{-i}^{(p)}[\prgv{T}_{\rm pf}^{0i}]{\mathcal S}_{-j}^{(p)}[\prgv{T}_{\rm pf}^{0j}]}{\prgv{T}_{\rm pf}^{00}} \biggl(1 + \frac{3}{16} r^2_{\rm m} \biggr) + a^{2(\alpha - 1)}
    \prgv{T}^{00}_\pf \biggl(\frac{1}{3} - \frac{r^2_{\rm m}}{4} \biggr) \, \delta_{ij}\right]\, , 
\ee
with $r^2_{\rm m}$ given by \Eq{eq:r2_m}.

\subsubsection{Imperfect fluids}

In the conservation form imperfect fluid effects can be incorporated by simply adding the forces $\prgv{{\mathcal F}}_\ipf^{\nu}$ to the perfect fluid equations, such that we end up with the system of equations given by \Eqss{eq:EOMsetCF1}{eq:EOMsetCF4}. 
In the following we will only focus on the case of imperfect fluids in the \textit{subrelativistic limit}, since it allows us to describe the deviatoric stress tensor through Navier-Stokes viscosity. In the limit of small fluid velocities, i.e., $|\vec{u}_{\rm p}|^2 \ll 1$, and small bulk viscosities compared to the Hubble parameter, i.e., $ a^{1-\alpha} \,  \prgv{\xi} \,  \prgv{\HH} = \xi \, H \ll 1$, the discretized imperfect fluid friction terms become [cf.~\Eq{fviscmu}],
\bea
\prgv{{\cal F}}_\ipf^0 &\simeq& 0 \ , \\
    \prgv{{\cal F}}_\ipf^i
    &=& a^{-2(1-\alpha)} \, (1+\omega) \,  \sum_{j}\Biggl\{
    \prgv{\nu} \, \frac{\prgv{T}^{00}_{\pf}}{z} \biggl[\prgv{\nabla}_j \prgv{\nabla}_j \biggl(\frac{z}{z+\omega}\prgv{r}_i \biggr)\biggr]_\LL 
    +
     \bigl(\tfrac{1}{3} \prgv{\nu} + \prgv{\xi}\bigr) \, \frac{\prgv{T}^{00}_{\pf}}{z} \,
    \biggl[\prgv{\nabla}_i \prgv{\nabla}_j \biggl(\frac{z}{z+\omega}\prgv{r}_j\biggr)\biggr]_\LL 
    \nonumber \\
    &+&
    \prgv{\nu} \,   \left[\prgv{\nabla}_i \left( \frac{z}{z+\omega}\prgv{r}_j\right) \prgv{\nabla}_j \left(\frac{\prgv{T}^{00}_\pf}{z}\right)\right]_\LL +
    \prgv{\nu} \,   \left[\prgv{\nabla}_j \left(  \frac{z}{z+\omega}\prgv{r}_i\right) \prgv{\nabla}_j \left(\frac{\prgv{T}^{00}_\pf}{z}\right)\right]_\LL \nonumber \\
    &+&
    \bigl( \prgv{\xi} - \tfrac{2}{3}
    \prgv{\nu} \bigr) \,  \left[\prgv{\nabla}_j \left( \frac{z}{z+\omega}\prgv{r}_j\right) \prgv{\nabla}_i \left(\frac{\prgv{T}^{00}_\pf}{z}\right)\right]_\LL \Biggr\}
    \,,
\eea
where the label `$\rm L$' denotes a discretized version of the respective object that has to be separately specified for each discretization scheme. For convenience we have defined the ratio $\prgv{r}_i=a^{1-\alpha}\prgv{T}^{0i}/\prgv{T}^{00}$. In the above equations we can see that in the subrelativistic limit the imperfect fluid forces do not depend on time derivatives of the dynamical variables. This circumstance allows us to solve the imperfect fluid equations in the form given by \Eqs{eq:EOMsetCF1}{eq:EOMsetCF2} by applying non-symplectic algorithms, as discussed in \Sec{sec:RK}.

In the following we discuss the discretization of $\prgv{{\cal F}}_{\rm ipf}^i$ in the collocated and staggered formulations. The remaining parts of the total fluid kernels, i.e., ${\cal G}_\pf^\mu$ and ${\cal F}_\HH^\mu$, are discretized in the same way as presented in \Sec{discrete_perfect_fluids}. The scale factor kernel obtains additional contributions from the out-of-equilibrium effects as well. The evolution of the scale factor is now dictated by,
\be
{\mathcal{K}}_{a}[a,\prgv{E}_\fl^\rho,\prgv{E}_\fl^\rho]=\frac{1}{3} \left( \frac{ T_*^2}{\omega_*\mpl} \right)^2 a^{2\alpha+1} \left[ \frac{2\alpha - 1}{2}{\prgv E}_\fl^\rho-\frac{3}{2}{\prgv E}_\fl^p\right] \ , \label{eq:ScaleFactorIPF}
\ee
with the total fluid energy density and pressure components given by
\be
{\prgv E}_\fl^\rho \equiv \frac{1}{a^4}\left\langle {\prgv T}_{\rm pf}^{00}\right\rangle \qquad {\rm and} \qquad {\prgv E}_\fl^p \equiv \frac{1}{3a^{2+2\alpha}}\left\langle \sum_{i}({\prgv T}_{\rm pf}^{ii}-\pdev^{ii}) \right\rangle \ ,
\ee
where $\langle \dots\rangle$ denotes a volume average over the entire lattice. In the subrelativistic limit $\pdev^{00}\simeq 0$, thus ${\prgv E}_\fl^\rho$ does not obtain any additional contribution.

\subsubsection*{Collocated formulation}

In the collocated formulation, all field components live at lattice site $\bf n$ and the additional imperfect fluid forces $\prgv{{\cal F}}_\ipf^i$ are obtained by replacing the relevant dynamical variables with their lattice counterparts and the derivative operators with neutral derivatives, such that,
\bea
    \prgv{{\cal F}}_{\rm ipf}^i
    &=& a^{-2(1-\alpha)} \, (1+\omega) \, \Biggl[\prgv{\nu} \,  \frac{\prgv{T}^{00}_{\pf}}{z}\prgv{\Delta}^{(p)}\left(\frac{z}{z+\omega}\prgv{r}_i\right)
    +
     \bigl(\tfrac{1}{3} \prgv{\nu} + \prgv{\xi}\bigr) \, \frac{\prgv{T}^{00}_{\pf}}{z} \,
    \sum_{j}\prgv{\Delta}^{(p)}_{ij}\left(\frac{z}{z+\omega}\prgv{r}_j\right) \nonumber \\
    &+&
    \prgv{\nu} \, \sum_{j} \prgv{\nabla}_i^{(0,p)}\left(\frac{z}{z+\omega}\prgv{r}_j\right) \prgv{\nabla}_j^{(0,p)}\left(\frac{\prgv{T}^{00}_\pf}{z}\right)
    +
    \prgv{\nu} \, \sum_{j} \prgv{\nabla}_j^{(0,p)} \left(\frac{z}{z+\omega}\prgv{r}_i\right) \prgv{\nabla}_j^{(0,p)}\left(\frac{\prgv{T}^{00}_\pf}{z}\right)  \nonumber \\
    &+& 
    \bigl( \prgv{\xi} - \tfrac{2}{3}
    \prgv{\nu} \bigr) \,  \prgv{\nabla}_i^{(0,p)}\left(\frac{\prgv{T}^{00}_\pf}{z}\right)\sum_{j}\prgv{\nabla}_j^{(0,p)} \left(\frac{z}{z+\omega}\prgv{r}_j\right)  \Biggr] \, ,
\eea
where $\prgv{\Delta}^{(p)}$ denotes the discrete Laplace operator to order $p$ in accuracy introduced in \Eq{eq:laplm}, and $\prgv{\Delta}^{(p)}_{ij}$ represents the bidiagonal scheme for cross derivatives introduced in \Eq{eq:Bidiag2m}. The 3D trace of the deviatoric stress tensor that appears in the volume averaged pressure density is given by,
\be
\sum_{i}\pdev^{ii} = 3 \, a^{2(\alpha-1)} \prgv{\xi}(1+\omega) \frac{\prgv{T}^{00}_\pf}{z} \sum_{i}\left[\prgv{\nabla}_i^{(0,p)}\frac{z}{z+\omega}\prgv{r}_i \right] \ ,
\ee
The $z$-function can be evaluated according to \Eq{eq:zL}. The $\prgv{T}^{00}_\pf$ and $\prgv{T}^{0i}_\pf$ that appear in the $r^2$-ratio, as well as $r_j$ are just replaced by their lattice counterparts at lattice site $\bf n$.

\subsubsection*{Staggered formulation}

In the staggered formulation $\prgv{T}^{00}_\pf$ lives at lattice sites, while vectors, such as $\prgv{T}^{0i}_\pf$ and $\prgv{{\cal F}}_{\rm ipf}^i$ live in between lattice sites at ${\bf n} + \hat{\imath} /2$. Thus, to incorporate the imperfect fluid contributions in the EOM [cf.~\Eqs{eq:EOMsetCF1}{eq:EOMsetCF2}] correctly, we have to ensure that the different terms in each equation are evaluated at the same location on the lattice. In the staggered formulation, the discretized versions of the imperfect fluid forces can be written as
\bea
    \prgv{{\cal F}}_\ipf^i
    &=& a^{2(1-\alpha)} \, (1+\omega) \, \Biggl[ \prgv{\nu} \, {\cal S}_{i}^{(p)} \Biggl[ \frac{\prgv{T}^{00}_{\pf}}{z}  \prgv{\Delta}^{(p)}\left(\frac{z}{z+\omega}\prgv{r}_i\right) \Biggr]
    \nonumber \\
    &+&
     \bigl(\tfrac{1}{3} \prgv{\nu} + \prgv{\xi}\bigr) \, \sum_{j} {\cal S}_i^{(p)} \biggl[\frac{\prgv{T}^{00}_{\pf}}{z}
    \prgv{\Delta}^{(p)}_{ij}
    \left(\frac{z}{z+\omega}
    \prgv{r}_j \right) \biggr] \nonumber \\
    &+&
    \prgv{\nu} \, \sum_{j} \prgv{\nabla}_i^{(+,p)} \left(\frac{z}{z+\omega}\prgv{r}_j\right) {\cal S}_{i}^{(p)}\left[\prgv{\nabla}_j^{(0,p)} \left(\frac{\prgv{T}^{00}_\pf}{z}\right)\right]
    +
    \prgv{\nu} \, \sum_{j} {\cal S}_{i}^{(p)}\left[\prgv{\nabla}_j^{(0,p)}\left(\frac{z}{z+\omega}\prgv{r}_i\right) \prgv{\nabla}_j^{(0,p)} \left(\frac{\prgv{T}^{00}_\pf}{z} \right) \right]  \nonumber \\
    &+& 
    \bigl( \prgv{\xi} - \tfrac{2}{3}
    \prgv{\nu} \bigr) \prgv{\nabla}_i^{(+,p)}\left(\frac{\prgv{T}^{00}_\pf}{z}\right) \sum_{j} {\cal S}_i^{(p)} \left[\prgv{\nabla}_j^{(0,p)}\left( \frac{z}{z+\omega} \prgv{r}_j\right) \right]  \Biggr]
    \,,
\label{ipf_forces_staggered_cons}
\eea
where $\prgv{\Delta}^{(p)}$ denotes the discrete Laplacian to order $p$ in accuracy introduced in \Eq{eq:laplm}, and $\prgv{\Delta}^{(p)}_{ij}$ the bidiagonal derivatives. The trace of the spatial part of the deviatoric stress tensor that appears in the volume averaged pressure density is given by,
\be
\sum_{i}\pdev^{ii} =3 \, a^{2(\alpha-1)}\prgv{\xi} \,  (1+\omega) \frac{\prgv{T}^{00}_{\rm pf}}{z} \sum_{i} 
\prgv{\nabla}_i^{(0,p)} \left(\frac{z}{z+\omega}\prgv{r}_i\right)  \,.
\label{trace_pi_ipf_staggered_cons}
\ee
In \Eqs{ipf_forces_staggered_cons}{trace_pi_ipf_staggered_cons}, the $z$-function can be evaluated according to \Eq{eq:zL}. However, the $r^2$-ratio that appears in it, as well as $\prgv{r}_j$ requires the application of a semi-sum on the $\prgv{T}_{\rm pf}^{0k}$ components,
\be
r^2\equiv a^{2(1-\alpha)}\sum_{k=1}^{3}\frac{{\mathcal S}_{-k}^{(p)}[\prgv{T}_{\rm pf}^{0k}\prgv{T}_{\rm pf}^{0k}]}{(\prgv{T}_{\rm pf}^{00})^2} \ , \qquad {\rm and} \qquad
     \prgv{r}_k = a^{1-\alpha} \frac{{\mathcal S}_{-k}^{(p)}[\prgv{T}_{\rm pf}^{0k}]}{\prgv{T}_{\rm pf}^{00}}  \ .
\ee

\subsection{Non-conservation form}\label{sec:LatticeFluidDynamicsNC}

In the non-conservation form the energy density and velocity field are the fundamental fluid variables. The dynamics of the fluid system is then described by \Eqs{rho_dens}{u_vel} for the perfect fluid case, and \Eqs{rel_energy_visc}{rel_mom_visc} for imperfect fluids. The background expansion on the other hand is governed by the Friedmann equations. We use \Eq{Friedmann2} to evolve the scale factor, whereas \Eq{Friedmann1} represents a constraint equation. The dynamical variables we solve for are then the logarithm of the perfect fluid energy density $\ln \prgv{\rho}$ and the velocity field $u_i$, introduced in \Eq{eq:FluidProgramVariables}, as well as the scale factor $a$ and its conjugate momentum $b\equiv a'$. The above mentioned equations can be brought into the following form of coupled first-order partial differential equations,
\begin{subequations}
\label{discretized_non_cons}
\begin{align}
(\ln\prgv{\rho})' &= {\cal G}^{\rho}[\ln \prgv{\rho},u_i,a,b]={\cal G}_0^\pf[\ln \prgv{\rho},u_i]+{\cal G}^{\HH}_0[u^2,\prgv{\HH}]+{\cal G}_0^{\rm ipf}[\ln \prgv{\rho},u_i,{\cal F}^\mu_{\rm ipf}]  \, , \label{eq:EOMsetNCF1}\\
(u_i)'&={\cal G}_i^u[\ln \prgv{\rho},u_j,a,b]={\mathcal{G}}_i^{\pf}[\ln \prgv{\rho},u_j] +{\mathcal{G}}^{\HH}_i[u_i,u^2,\prgv{\HH}]+{\mathcal G}_i^{\rm ipf}[\ln \prgv{\rho},u_j,{\mathcal{F}}^\mu_{\rm ipf}]\, , \label{eq:EOMsetNCF2}\\
b'&={\mathcal{K}}_{a}[a,{\prgv E}_\fl^\rho,{\prgv E}_\fl^p] \, , \label{eq:EOMsetNCF3}\\
a'&=b \, , \label{eq:EOMsetNCF4}
\end{align}
\end{subequations}
where primes denote derivatives with respect to $\alpha$-time in program units and $\mathcal{\prgv{H}}\equiv a'/a$. Similar to the fluid equations in the conservation form, the above equations are written for a system with self-consistent expansion, where the volume average of the fluid energy and pressure density given in \Eq{eq:va_energy_densities} source the evolution of the scale factor. In case other sources contribute to the expansion of the background as well, their volume averaged energy density components would need to be added to the kernel ${\cal K}_{a}$. We could also consider the scenario where the evolution of the scale factor is driven by an external background fluid with constant equation of state. In this case, the evolution of the scale factor is described by \Eq{eq:sf_const} and we do not need to explicitly solve for \Eqs{eq:EOMsetNCF3}{eq:EOMsetNCF4}.

The fluid kernels $\mathcal{G}^\rho$ and $\mathcal{G}^u_i$ depend, in addition to the dynamical variables, also on the choice of constant equation of state $\omega$ [cf.~\Eq{sound_speed}]. In the second equalities of \Eqs{eq:EOMsetCF1}{eq:EOMsetCF2}, we have split the fluid kernels into a sum of terms that can be associated with perfect fluid dynamics $\mathcal{G}_0^\pf$ and $\mathcal{G}_i^\pf$, terms that contain the Hubble parameter $\mathcal{G}_\mu^{\mathcal{H}}$, as well as the terms $\mathcal{G}_\mu^{\rm ipf}$ that incorporate possible first-order imperfect fluid effects, as discussed in \Sec{viscous_term}. In the case $\alpha=1$ any explicit dependence on the scale factor vanishes in the first two equations. Moreover, for $\omega=1/3$ also the dependence on the Hubble parameter vanishes and we are left with ${\mathcal{G}}_\mu^\pf$ and $\mathcal{G}_\mu^{\rm ipf}$. The above set of first-order differential equations can be solved with a non-symplectic integrator, such as the Runge-Kutta scheme introduced in \Sec{sec:RK}. In the following, we will discuss different discretization procedures of the different terms. First, we discuss the EOM for a perfect fluid in \Sec{discrete_non_cons_perfect} and then incorporate first-order imperfect fluid effects in the subrelativistic limit in \Sec{discrete_non_cons_imperfect}.

\subsubsection{Perfect fluid}
\label{discrete_non_cons_perfect}

We first discuss the case of a perfect fluid with self-consistent background expansion in the absence of out-of-equilibrium effects, i.e., setting  $\mathcal{G}_\mu^{\rm ipf}=0$ and $\dev^{\mu \nu}=0$ in \Eqq{discretized_non_cons}. From \Eqs{rho_dens}{u_vel}, the discretized version of the kernels for the perfect fluid energy density and velocity field are then,
\begin{align}
{\mathcal{G}}_0^\pf[\ln \prgv{\rho},u_i,a]&=-\underbrace{\frac{1+\omega}{1-\omega a^{2(1-\alpha)} u^2_\LL}\sum_{j=1}^3\prgv{\nabla}_j^\LL u_j}_{\rm (E.1)} -\underbrace{\frac{1-\omega}{1-\omega a^{2(1-\alpha)} u^2_\LL}\sum_{j=1}^3\left[u_j\prgv{\nabla}_j^\LL \ln \prgv{\rho}\right]_\LL}_{\rm (E.2)} \, , \label{eq:GKenrelRho}\\
{\mathcal{G}}_i^\pf[\ln \prgv{\rho},u_j,a]&= \underbrace{\frac{1-a^{2(1-\alpha)}u^2_\LL}{1-\omega a^{2(1-\alpha)} u^2_\LL}}_{\rm (M.0)} \Big [\omega\underbrace{\sum_{j=1}^3\prgv{\nabla}_j^\LL u_j}_{\rm (M.1)} +\omega\frac{1-\omega}{1+\omega}\underbrace{\sum_{j=1}^3\left[u_j\prgv{\nabla}_j^\LL \ln \prgv{\rho}\right]_\LL}_{\rm (M.2)}\Big]u_i -\underbrace{\sum_{j=1}^3\left[u_j\prgv{\nabla}_j^\LL u_i\right]_\LL}_{\rm (M.3)}\,   \nonumber\\
&-\omega\underbrace{\frac{1-a^{2(1-\alpha)}u^2_\LL}{1+\omega}a^{2(\alpha-1)}\prgv{\nabla}_i^\LL \ln\prgv{\rho}}_{\rm (M.4)} \, , \label{eq:GKenrelU} \\
{\mathcal{G}}^{\HH}_0[u^2,\prgv{\HH}] &=\underbrace{\frac{1-3\omega}{1-\omega a^{2(1-\alpha)} u^2}(1+a^{2(1-\alpha)}u^2_\LL)\prgv{\HH}}_{\rm (H.1)}\ , \\ 
{\mathcal{G}}^{\HH}_i[u_i,u^2,\prgv{\HH}] & = \underbrace{\left[\frac{1-a^{2(1-\alpha)}u^2_\LL}{1-\omega a^{2(1-\alpha)} u^2_\LL}(3\omega-1)+\alpha-1\right]\prgv{\HH}u_i}_{\rm (H.2)} \ , \label{eq:GKenrelH}
\end{align}
where the label '$\rm L$' denotes some lattice representation of the respective object. In the following section we will specify the discretization of these terms (E.1)--(E.2), (M.0)--(M.4) and (H.1)--(H.2), which can have different realizations in the collocated and staggered formulations. The kernel of the scale factor $a$ is given by,
\be
{\mathcal{K}}_{a}[a,{\prgv E}_\pf^\rho, {\prgv E}_\pf^p] = \frac{a^{2\alpha+1}}{3} \left( \frac{ T_*^2}{\omega_*\mpl} \right)^2 \left[ \frac{2\alpha - 1}{2}{\prgv E}_\pf^\rho-\frac{3}{2} {\prgv E}_\pf^p \right]  \, ,
\ee
with the averaged energy density and pressure components
\be\label{eq:AvFluidEnergiesNC}
{\prgv E}_\pf^\rho \equiv \frac{1}{a^4}\left\langle \left(1 + (1+\omega)\gamma_\LL^2a^{2(1-\alpha)}u_\LL^2\right){\prgv \rho}_\LL \right\rangle \ ,  \quad {\rm and} \quad {\prgv E}_\pf^p \equiv \frac{1}{3a^{2+2\alpha}}\left\langle \left(3\omega + (1+\omega)\gamma_\LL^2a^{2(1-\alpha)}u_\LL^2\right){\prgv \rho}_\LL  \right\rangle \ ,
\ee
where $\langle\dots\rangle$ denotes a volume average over the whole lattice. When we evolve the above set of \Eqss{eq:EOMsetNCF1}{eq:EOMsetNCF4}, the Hubble constraint, given by 
\be
b^2 = \frac{a^{2(\alpha+1)}}{3} \left( \frac{ T_*^2}{\omega_*\mpl} \right)^2 {\prgv E}_\pf^\rho   \, ,
\ee
has to be preserved up to the accuracy of the chosen integrator.

\subsubsection*{Collocated formulation}

In the collocated formulation all dynamical variables, i.e., energy density and velocity field, live at lattice site $\bf n$.
Discretizing the different terms of the energy density equation kernel (\ref{eq:GKenrelRho}) yields,
\begin{subequations}
\begin{align}
{\rm (E.1)}: \hspace{0.5cm} &\, \frac{1+\omega}{1-\omega a^{2(1-\alpha)}u^2}\sum_{j=1}^3\prgv{\nabla}_j^{(0,p)}u_j  \quad \longrightarrow \quad \frac{1+\omega}{1-\omega a^{2(1-\alpha)}{\tt u}^2}\sum_{j=1}^3\prgv{\partial}_j{\tt u}_j \biggr|_{\prgv{x}\equiv{\bf n}\delta \prgv{x}}+ {\mathcal{O}(\delta \prgv{x}^p)}\, , \label{eq:E1coll}\\
{\rm (E.2)}: \hspace{0.5cm} &\, \frac{1-\omega }{1-\omega a^{2(1-\alpha)} u^2}\sum_{j=1}^3u_j\prgv{\nabla}_j^{(0,p)} \ln\prgv{\rho} \quad \longrightarrow \quad \frac{1-\omega}{1-\omega a^{2(1-\alpha)}{\tt u}^2}\sum_{j=1}^3{\tt u}_j\prgv{\partial}_j \ln\prgv{{\ttrho}}\,  \biggr|_{\prgv{x}\equiv{\bf n}\delta \prgv{x}} + {\mathcal{O}(\delta \prgv{x}^p)} \, , \label{eq:E2coll}
\end{align}
\end{subequations}
while the discretized terms of the kernels of the momentum conservation equation (\ref{eq:GKenrelU}) are given by,
\begin{subequations}
\begin{align}
{\rm (M.0)}: \hspace{0.5cm} &\, \frac{1 - a^{2(1-\alpha)} u^2}{1-\omega a^{2(1-\alpha)} u^2} \, , \\
{\rm (M.1)}: \hspace{0.5cm} &\, \sum_{j=1}^3\prgv{\nabla}_j^{(0,p)}u_j ~~ \longrightarrow ~~ \sum_{j=1}^3\prgv{\partial}_j{\tt u}_j\, \biggr|_{\prgv{x}\equiv{\bf n}\delta \prgv{x}} + {\mathcal{O}(\delta \prgv{x}^p)} \, , \\
{\rm (M.2)}: \hspace{0.5cm} &\,  \sum_{j=1}^3u_j\prgv{\nabla}_j^{(0,p)} \ln\prgv{\rho} ~~ \longrightarrow ~~ \sum_{j=1}^3{\tt u}_j\prgv{\partial}_j \ln\prgv{{\ttrho}}\, \biggr|_{\prgv{x}\equiv{\bf n}\delta \prgv{x}}  + {\mathcal{O}(\delta \prgv{x}^p)} \,,\\
{\rm (M.3)}: \hspace{0.5cm} &\, \sum_{j=1}^3u_j\prgv{\nabla}_j^{(0,p)} u_i ~~ \longrightarrow ~~ \sum_{j=1}^3{\tt u}_j\prgv{\partial}_j {\tt u}_i\, \biggr|_{\prgv{x}\equiv{\bf n}\delta \prgv{x}}  + {\mathcal{O}(\delta \prgv{x}^p)}\, , \\
{\rm (M.4)}: \hspace{0.5cm} &\, \frac{1-a^{2(1-\alpha)} u^2}{1+\omega}a^{2(\alpha-1)}\prgv{\nabla}_i^{(0,p)} \ln\prgv{\rho} ~~ \longrightarrow ~~ \frac{1-a^{2(1-\alpha)}{\tt u}^2}{1+\omega}a^{2(\alpha-1)}\prgv{\partial}_i \ln\prgv{{\ttrho}}\,\biggr|_{\prgv{x}\equiv{\bf n}\delta \prgv{x}}  + {\mathcal{O}(\delta \prgv{x}^p)}\,.
\end{align}
\end{subequations}
where $u^2= \sum_{j}u_ju_j$. The expressions $\prgv{\nabla}_j^{(0,p)}$ represent the neutral derivative to order $p$, as introduced in \Sec{Sect:LatticeDerivatives}.  
The \rhs of the arrows indicates to which order ${\mathcal{O}(\delta \prgv{x}^p)}$ the continuum expressions are recovered when expanding around lattice sites $\bf n$. The lattice versions of the terms ${\mathcal{G}}^{\mathcal{H}}_0$ and ${\mathcal{G}}^{\mathcal{H}}_i$, stemming from the Hubble friction, are given by,
\begin{subequations}
\begin{align}
{\rm (H.1)}: \hspace{0.5cm} &\, \frac{1-3\omega}{1-\omega a^{2(1-\alpha)} u^2}(1+a^{2(1-\alpha)}u^2)\prgv{\HH} \, , \\
{\rm (H.2)}: \hspace{0.5cm} &\, \left[\frac{1-a^{2(1-\alpha)}u^2}{1-\omega a^{2(1-\alpha)} u^2}(3\omega-1)+\alpha-1\right]\prgv{\HH}u_i \,. \label{eq:H2coll}
\end{align}
\end{subequations}
The volume averaged energy and pressure densities, given in \Eq{eq:AvFluidEnergiesNC},
are obtained by simply replacing the expressions of the fluid energy density and the fluid velocity field by their discrete counterparts at lattice site $\bf n$.

\subsubsection*{Staggered formulation}

In the staggered discretization procedure the fluid degrees of freedom are assumed to live in different locations on the lattice. While the logarithm of the energy density $\ln \prgv{\rho}$ lives at lattice site $\bf n$, the velocity field $u_i$ lives in the middle of two lattice points, at ${\bf n}+\hat{\imath}/2$. The different terms indicated above in the kernel $\mathcal{G}^\rho$ may then be discretized as follows,
\begin{subequations}
\begin{align}
{\rm (E.1)}: \hspace{0.5cm} &\,\frac{1+\omega}{1-\omega a^{2(1-\alpha)}{\mathcal S}^{(p)}[u^2]}\sum_{j=1}^3\prgv{\nabla}_j^{(-,p)}u_j ~~\longrightarrow~~ \frac{1+\omega}{1-\omega a^{2(1-\alpha)}{\tt u}^2}\sum_{j=1}^3\prgv{\partial}_j {\tt u}_j\, \biggr|_{\prgv{x}\equiv{\bf n}\delta \prgv{x}} + {\mathcal{O}(\delta \prgv{x}^p)} \, , \label{eq:E1stagg} \\
{\rm (E.2)}: \hspace{0.5cm} &\, \frac{1-\omega}{1-\omega a^{2(1-\alpha)}{\mathcal S}^{(p)}[u^2]}\sum_{j=1}^3{\mathcal S}_{-j}^{(p)}[u_j]\prgv{\nabla}_j^{(0,p)} \ln\prgv{\rho} ~~\longrightarrow~~ \frac{1-\omega}{1-\omega a^{2(1-\alpha)}{\tt u}^2}\sum_{j=1}^3{\tt u}_j\prgv{\partial}_j \ln\prgv{{\ttrho}}\, \biggr|_{\prgv{x}\equiv{\bf n}\delta \prgv{x}} + {\mathcal{O}(\delta \prgv{x}^p)} \,,
\end{align}
\end{subequations}
where we have defined ${\mathcal S}^{(p)}[u^2]\equiv \sum_{j}{\mathcal S}_{-j}^{(p)}[u_ju_j]$.\footnote{Alternatively one could also discretize the expression by defining ${\mathcal S}^{(p)}[u^2] \equiv \sum_{j}\left({\mathcal S}_{-j}^{(p)}[u_j]\right)^2$.}
The different contributions to $\mathcal{G}^u_i$ on the other hand are given by,
\begin{subequations}
\begin{align}
{\rm (M.0)}: \hspace{0.5cm} &\, \frac{1-a^{2(1-\alpha)}{\mathcal S}_i^{(p)}[{\mathcal S}^{(p)}[u^2]]}{1-\omega a^{2(1-\alpha)} {\mathcal S}_i^{(p)}[{\mathcal S}^{(p)}[u^2]]} ~~\longrightarrow~~  \frac{1-a^{2(1-\alpha)}{\tt u}^2}{1-\omega a^{2(1-\alpha)} {\tt u}^2} \, \biggr|_{\prgv{x}\equiv({\bf n}+\hat{\imath}/2)\delta \prgv{x}} + {\mathcal{O}(\delta \prgv{x}^p)}\, , 
\\
{\rm (M.1)}: \hspace{0.5cm} &\, \sum_{j=1}^3{\mathcal S}_{i}^{(p)}[\prgv{\nabla}_j^{(-,p)}u_j]  ~~\longrightarrow~~ \sum_{j=1}^3\prgv{\partial}_j {\tt u}_j  \, \biggr|_{\prgv{x}\equiv({\bf n}+\hat{\imath}/2)\delta \prgv{x}} + {\mathcal{O}(\delta \prgv{x}^p)}\, , 
\\
{\rm (M.2)}: \hspace{0.5cm} &\, \sum_{j=1}^3 {\mathcal S}_{i,-j}^{(p)}[u_j\prgv{\nabla}_j^{(+,p)} \ln\prgv{\rho}] ~~ \longrightarrow ~~ \sum_{j=1}^3{\tt u}_j\prgv{\partial}_j \ln\prgv{{\ttrho}}  \, \biggr|_{\prgv{x}\equiv({\bf n}+\hat{\imath}/2)\delta \prgv{x}} + {\mathcal{O}(\delta \prgv{x}^p)}  \, ,
\\
{\rm (M.3)}: \hspace{0.5cm} &\, \sum_{j=1}^3{\mathcal S}_{i,-j}^{(p)}[u_j]\prgv{\nabla}_j^{(0,p)} u_i  ~~\longrightarrow~~ \sum_{j=1}^3{\tt u}_j\prgv{\partial}_j {\tt u}_i  \, \biggr|_{\prgv{x}\equiv({\bf n}+\hat{\imath}/2)\delta \prgv{x}} + {\mathcal{O}(\delta \prgv{x}^p)} \, , 
\\
{\rm (M.4)}: \hspace{0.5cm} &\,  \omega\frac{1-a^{2(1-\alpha)}{\mathcal S}_i^{(p)}[{\mathcal S}^{(p)}[u^2]]}{1+\omega}a^{2(\alpha-1)}\prgv{\nabla}_i^{(+,0)} \ln\prgv{\rho} \nonumber 
\\ 
&\longrightarrow~~ \omega\frac{1-a^{2(1-\alpha)}{\tt u}^2}{1+\omega}a^{2(\alpha-1)}\prgv{\partial}_i \ln\prgv{{\ttrho}}   \, \biggr|_{\prgv{x}\equiv({\bf n}+\hat{\imath}/2)\delta \prgv{x}} + {\mathcal{O}(\delta \prgv{x}^p)}\, . 
\end{align}
\end{subequations}
The discretized versions of the terms stemming from the Hubble friction are given by,
\begin{subequations}
\begin{align}
{\rm (H.1)}: \hspace{0.5cm} &\, \frac{1-3\omega}{1-\omega a^{2(1-\alpha)} {\mathcal S}^{(p)}[u^2]}(1+a^{2(1-\alpha)}{\mathcal S}^{(p)}[u^2])\prgv{\HH} \nonumber \\
& \longrightarrow ~~ \frac{1-3\omega}{1-\omega a^{2(1-\alpha)}{\tt u}^2}(1+a^{2(1-\alpha)}{\tt u}^2)\prgv{\HH}\, \biggr|_{\prgv{x}\equiv{\bf n}\delta \prgv{x}} + {\mathcal{O}(\delta \prgv{x}^p)} \, , \\
{\rm (H.2)}: \hspace{0.5cm} &\, \left[\frac{1-a^{2(1-\alpha)}{\mathcal S}^{(p)}_i[{\mathcal S}^{(p)}[u^2]]}{1-\omega a^{2(1-\alpha)} {\mathcal S}^{(p)}_i[{\mathcal S}^{(p)}[u^2]]}(3\omega-1)+\alpha-1\right]\prgv{\HH}u_i \nonumber \\
&\longrightarrow~~\left[\frac{1-a^{2(1-\alpha)}{\tt u}^2}{1-\omega a^{2(1-\alpha)} {\tt u}^2}(3\omega-1)+\alpha-1\right]\prgv{\HH} {\tt u}_i\, \biggr|_{\prgv{x}\equiv({\bf n}+\hat{\imath}/2)\delta \prgv{x}} + {\mathcal{O}(\delta \prgv{x}^p)}\, . \label{eq:H2stagg}
\end{align}
\end{subequations}
In the expressions above we have used the neutral $\prgv{\nabla}_j^{(0,p)}$ and charged derivatives $\prgv{\nabla}_j^{(\pm,p)}$ to order $p$ in accuracy, introduced in \Sec{sec:generalLattice}. The staggered formulation requires several semi-sums ${\mathcal S}_i^{(p)}$ and clover averages ${\mathcal S}_{i,j}^{(p)}$ to ensure that all terms live at the required position on the lattice, such that we recover the continuum expressions to the desired accuracy. On the \rhs of the arrows we show to which order the continuum expressions are recovered when expanding around the indicated location. The volume averaged energy and pressure densities that appear in the kernel of the scale factor are discretized as follows,
\be
{\prgv E}_\pf^\rho \equiv \frac{1}{a^4}\left\langle \left(1 + (1+\omega)\gamma^2a^{2(1-\alpha)}{\cal S}^{(p)}[u^2]\right){\prgv \rho} \right\rangle \qquad {\rm and} \qquad {\prgv E}_\pf^p \equiv \frac{1}{3a^4}\left\langle \left(3\omega + (1+\omega)\gamma^2a^{2(1-\alpha)}{\cal S}^{(p)}[u^2]\right){\prgv \rho} \right\rangle \ ,
\ee
with the lattice version of the $\gamma$-factor being defined via,
\be
\gamma^2=\frac{1}{1-a^{2(1-\alpha)}{\cal S}^{(p)}[u^2]} \ .
\ee

\subsubsection*{Subrelativistic limit}

In the subrelativistic limit, i.e., when $|\vec{u}_{\rm p}|^2\ll 1$, the kernels of the perfect fluid ${\cal G}^\rho$ and ${\cal G}^u_i$ reduce to the simpler forms given in \Eqs{rho_dens_sub}{u_vel_sub}. In the collocated formulation we have,
\begin{subequations}
\begin{align}
{\cal G}^\rho[\ln \prgv{\rho},u_i,\prgv{\HH}]=&-(1+\omega)\sum_{j=1}^3\prgv{\nabla}_j^{(0,p)}u_j +(\omega-1)\sum_{j=1}^3u_j\prgv{\nabla}_j^{(0,p)} \ln\prgv{\rho}+(1-3\omega)\prgv{\HH}\, , \\
{\cal G}^u_i[\ln \prgv{\rho},u_j,\prgv{\HH}]=&-\sum_{j=1}^3u_j\prgv{\nabla}_j^{(0,p)}u_i +\left [\omega\sum_{j=1}^3\prgv{\nabla}_j^{(0,p)}u_j +\omega\frac{1-\omega}{1+\omega}\sum_{j=1}^3u_j\prgv{\nabla}_j^{(0,p)} \ln\prgv{\rho}\right]u_i \nonumber \\
& - \frac{\omega}{1+\omega}a^{2(\alpha-1)}\prgv{\nabla}_i^{(0,p)} \ln\prgv{\rho} + (3\omega+\alpha-2)\prgv{\HH}u_i \, ,
\end{align}
\end{subequations}
while in the staggered formulation we have,
\begin{subequations}
\begin{align}
{\mathcal{G}}^\rho[\ln \prgv{\rho},u_i,\prgv{\HH}]=&-(1+\omega)\sum_{j=1}^3\prgv{\nabla}_j^{(-,p)}u_j +(\omega-1)\sum_{j=1}^3{\cal S}_{-j}^{(p)}[u_j\prgv{\nabla}_j^{(+,p)} \ln\prgv{\rho}] +(1-3\omega)\prgv{\HH} \, , \\
{\mathcal{G}}^u_i[\ln \prgv{\rho},u_j,\prgv{\HH}]=&-\sum_{j=1}^3{\mathcal S}_{i,-j}^{(p)}[u_j]\prgv{\nabla}_j^{(0,p)} u_i + \left[\omega\sum_{j=1}^3{\cal S}_{i}^{(p)}[\prgv{\nabla}_j^{(-,p)}u_j] +\omega\frac{1-\omega}{1+\omega}\sum_{j=1}^3{\cal S}_{i,-j}^{(p)}[u_j\prgv{\nabla}_j^{(+,p)} \ln\prgv{\rho}]\right]u_i \nonumber\\
& - \frac{\omega}{1+\omega}a^{2(\alpha-1)}\prgv{\nabla}_i^{(+,p)} \ln\prgv{\rho} + (3\omega+\alpha-2)\prgv{\HH}u_i 
\, .
\end{align}
\end{subequations}
In case of a relativistic fluid with $\omega=1/3$, the particular choice of $\alpha=1$ leads to conservation laws that have no explicit dependence on the scale factor $a$ and the Hubble parameter $\mathcal{\prgv{H}}$.

\subsubsection{Imperfect fluids}
\label{discrete_non_cons_imperfect}

We now discuss how to discretize the fluid EOM when incorporating first-order imperfect fluid effects. In the non-conservation form the fluid equations \Eqs{eq:EOMsetNCF1}{eq:EOMsetNCF2} now hold additional contribution on the \textit{rhs}, given by ${\mathcal{G}}_{\rm ipf}^\mu$. On the lattice these have the following form [cf.~\Eqs{rel_energy_visc}{rel_mom_visc}], 
\begin{subequations}
\begin{align}
{\mathcal{G}}_{\rm ipf}^0[\ln\prgv{\rho},u_i,{\mathcal{F}}^\mu_{\rm ipf}] &= \frac{1}{(1-\omega a^{2(1-\alpha)}u^2_\LL)\prgv{\rho}_\LL}\left[(1+a^{2(1-\alpha)}u^2_\LL) \left[\prgv{{\mathcal F}}^0_{\rm ipf}\right]_\LL- 2 \,  a^{2(1-\alpha)} \sum_j \left[u_j\prgv{{\mathcal F}}^j_{\rm ipf}\right]_\LL\right] \, , \\
{\mathcal{G}}_{\rm ipf}^i[\ln\prgv{\rho},u_j,{\mathcal{F}}^\mu_{\rm ipf}] &= \frac{1-a^{2(1-\alpha)}u^2_\LL}{(1-wa^{2(1-\alpha)}u^2_\LL)\prgv{\rho}_\LL}\left[ \frac{2\omega a^{2(1-\alpha)}}{1+\omega}\sum_j \left[u_j\prgv{{\mathcal F}}^j_{\rm ipf}\right]_\LL-\left[\prgv{{\mathcal F}}^0_{\rm ipf} \right]_\LL\right] u_i + \frac{1-a^{2(1-\alpha)} u^2_\LL}{(1+\omega)\prgv{\rho}_\LL} \left[\prgv{{\mathcal F}}^i_{\rm ipf}\right]_\LL \, ,
\end{align}
\end{subequations}
where the label `L' denotes some discretized version of the corresponding object. In the following we will specify the lattice versions of the terms in ${\mathcal{G}}_{\rm ipf}^\mu$ for the collocated and staggered formulations, and will also discuss the discretization of the imperfect fluid forces ${\mathcal{F}}_{\rm ipf}^i$ in the limit of subrelativistic velocity. The scale factor kernel obtains additional contributions from the out-of-equilibrium effects as well. The evolution of the scale factor is always dictated by,
\be
{\mathcal{K}}_{a}[a,\prgv{E}_\fl^\rho,\prgv{E}_\fl^\rho]=\frac{1}{3} \left( \frac{ T_*^2}{\omega_*\mpl} \right)^2 a^{2\alpha+1} \left[ \frac{2\alpha - 1}{2}{\prgv E}_\fl^\rho-\frac{3}{2}{\prgv E}_\fl^p\right] \ . \label{eq:ScaleFactorIPF}
\ee
However, the total fluid pressure component receives additional contributions from the deviatoric stress tensor [cf.~\Eq{Pi_visc}], and we have
\begin{subequations}
\bea\label{eq:Eav_ipf}
{\prgv E}_\pf^\rho &\equiv& \frac{1}{a^4}\left\langle \left(1 + (1+\omega)\gamma_\LL^2a^{2(1-\alpha)}u_\LL^2\right){\prgv \rho}_\LL \right\rangle \ , \\
{\prgv E}_\pf^p &\equiv& \frac{1}{3a^{2+2\alpha}}\left\langle \left(3\omega + (1+\omega)\gamma_\LL^2a^{2(1-\alpha)}u_\LL^2\right){\prgv \rho}_\LL  -a^{2(1-\alpha)} \sum_i\pdev_\LL^{ii}\right\rangle \ .
\eea
\end{subequations}
In the expressions above $\langle \dots\rangle$ again denotes a volume average over the whole lattice.

\subsubsection*{Collocated formulation}

In the collocated formulation the field components $\prgv{\rho}$ and $u_i$, as well as the friction terms $\prgv{{\mathcal F}}^\mu_{\rm ipf}$, live at lattice site $\bf n$. Thus, the friction terms that address the imperfect fluid contributions are just given by,
\begin{subequations}
\begin{align}
{\mathcal{G}}^{\rm ipf}_0[\ln\prgv{\rho},u_i,{\mathcal{F}}^\mu_{\rm ipf}] &= \frac{1}{(1-\omega a^{2(1-\alpha)}u^2)\prgv{\rho}}\left[(1+a^{2(1-\alpha)}u^2)\prgv{{\mathcal F}}^0_{\rm ipf}- 2 \, a^{2(1-\alpha)} \sum_j u_j\prgv{{\mathcal F}}^j_{\rm ipf}\right] \ ,  \\
{\mathcal{G}}^{\rm ipf}_i[\ln\prgv{\rho},u_j,{\mathcal{F}}^\mu_{\rm ipf}] &= \frac{1-a^{2(1-\alpha)}u^2}{(1-wa^{2(1-\alpha)}u^2)\prgv{\rho}}\left[ \frac{2\omega a^{2(1-\alpha)}}{1+\omega}\sum_j u_j\prgv{{\mathcal F}}^j_{\rm ipf}-\prgv{{\mathcal F}}^0_{\rm ipf}\right] u_i + \frac{1-a^{2(1-\alpha)} u^2}{(1+\omega)\prgv{\rho}}\prgv{{\mathcal F}}^i_{\rm ipf} \ ,
\end{align}
\end{subequations}
where $u^2= \sum_j u_j u_j$. The trace of the spatial part of the deviatoric stress tensor gives an additional contribution to the volume averaged pressure density. In the limit of subrelativistic velocities and small bulk viscosity, the discretized version of the energy and pressure density are now given by
\begin{subequations}
\bea
{\prgv E}_\pf^\rho &\equiv& \frac{1}{a^4}\left\langle \left(1 + (1+\omega)\gamma^2a^{2(1-\alpha)}u^2\right){\prgv \rho} \right\rangle \ , \\ 
{\prgv E}_\pf^p &\equiv& \frac{1}{3a^{2+2\alpha}}\left\langle \left(3\omega + (1+\omega)\gamma^2a^{2(1-\alpha)}u^2 + 3 \, a^{1-\alpha} \prgv{\xi}(1+\omega)\sum_{i}\prgv{\nabla}_i^{(0,p)}u_i\right){\prgv \rho} \right\rangle \ ,
\eea
\end{subequations}
with the lattice version of the gamma factor being,
\be
\gamma^2=\frac{1}{1-a^{2(1-\alpha)}u^2} \ .
\ee

\subsubsection*{Staggered formulation}

In the staggered formulation it is assumed that scalar fields, such as $\prgv{\rho}$ and $\mathcal{\prgv{F}}^0_{\rm ipf}$, live at lattice site $\bf n$, while vector fields, such as $u_i$ and $\mathcal{\prgv{F}}^i_{\rm ipf}$, live in between lattice sites, at ${\bf n}+\hat{\imath}/2$. The discretized imperfect fluid kernel contributions can then be written as,
\begin{subequations}
\begin{align}
\label{g0_ipf_stagg_noncons}
{\mathcal{G}}^{\rm ipf}_0[\ln\prgv{\rho},u_i,{\mathcal{F}}^\mu_{\rm ipf}] = \,& \frac{1}{(1-\omega a^{2(1-\alpha)}{\mathcal S}^{(p)}[u^2])\prgv{\rho}}\left[(1+a^{2(1-\alpha)}{\mathcal S}^{(p)}[u^2]) \prgv{{\mathcal F}}^0_{\rm ipf}- 2 \, a^{2(1-\alpha)} \sum_j {\mathcal S}_{-j}^{(p)}[u_j\prgv{{\mathcal F}}^j_{\rm ipf}]\right] \ ,  \\
{\mathcal{G}}^{\rm ipf}_i[\ln\prgv{\rho},u_j,{\mathcal{F}}^\mu_{\rm ipf}] = \, & {\mathcal S}_{i}^{(p)} \left[\frac{1-a^{2(1-\alpha)}{\mathcal S}^{(p)}[u^2]}{(1-wa^{2(1-\alpha)}{\mathcal S}^{(p)}[u^2])\prgv{\rho}}\left( \frac{2\omega a^{2(1-\alpha)}}{1+\omega}\sum_j {\mathcal S}_{-j}^{(p)}[u_j\prgv{{\mathcal F}}^j_{\rm ipf}]-\prgv{{\mathcal F}}^0_{\rm ipf}\right)\right] u_i \nonumber\\
&+ {\mathcal{S}}^{(p)}_{i}\left[\frac{1-a^{2(1-\alpha)} {\mathcal S}^{(p)}[u^2]}{(1+\omega)\prgv{\rho}}\right]\prgv{{\mathcal F}}^i_{\rm ipf} \ , 
\label{gi_ipf_stagg_noncons}
\end{align}
\end{subequations}
where again we have defined ${\mathcal S}^{(p)}[u^2]\equiv \sum_{j}{\mathcal S}_{-j}^{(p)}[u_ju_j]$ and the semi-sums ${\mathcal S}^{(p)}_i$ that are applied have been introduced in \Sec{sec:LatticeOperators}. The trace of the spatial part of the deviatoric stress tensor gives again an additional contribution to the volume averaged pressure density. In the limit of subrelativistic velocities and small bulk viscosity, we now have
\begin{subequations}
\bea
{\prgv E}_\pf^\rho &\equiv& \frac{1}{a^4}\left\langle \left(1 + (1+\omega)\gamma^2a^{2(1-\alpha)}{\cal S}^{(p)}[u^2]\right){\prgv \rho} \right\rangle \ , \\ 
{\prgv E}_\pf^p &\equiv& \frac{1}{3a^{2+2\alpha}}\left\langle \left(3\omega + (1+\omega)\gamma^2a^{2(1-\alpha)}{\cal S}^{(p)}[u^2] + 3 \, a^{1-\alpha} \prgv{\xi}(1+\omega)\sum_{i}\prgv{\nabla}_i^{(-,p)}u_i\right){\prgv \rho} \right\rangle \ ,
\eea
\end{subequations}
with the lattice version of the gamma factor being,
\be
\gamma^2=\frac{1}{1-a^{2(1-\alpha)}{\cal S}^{(p)}[u^2]} \ .
\ee

\subsubsection*{Subrelativistic limit}

In the subrelativistic regime, where $|\vec{u}_{\rm p}^2| \ll 1$, and considering small bulk viscosity compared to the Hubble time, i.e. $\xi H = a^{1-\alpha} \conf{\xi} {\cal H} \ll 1$, the imperfect fluid forces contributions appearing in \Eqs{g0_ipf_stagg_noncons}{gi_ipf_stagg_noncons} are given, in the collocated formulation, by the following expressions [cf.~\Eq{fviscmu}] 
\begin{subequations}
\begin{align}
\lim_{|\vec{u}_{\rm p}|^2 \ll 1,\xi H \ll 1} \prgv{{\mathcal F}}^0_{\rm ipf} = & \, 0 \, , \\
\lim_{|\vec{u}_{\rm p}|^2 \ll 1, \xi H \ll 1} \prgv{{\mathcal F}}^i_{\rm ipf} = & \ a^{\alpha-1} (1+\omega) \prgv{\rho} \, \biggr(\prgv{\nu} \, \prgv{\Delta}^{(p)} u_i + \, \prgv{\nu} \, \sum_j (\prgv{\nabla}^{(0,p)}_i u_j + \prgv{\nabla}^{(0,p)}_j u_i) \left[ \prgv{\nabla}^{(0,p)}_j \ln \prgv{\rho} \right]\nonumber \\  
&  + \bigl(\tfrac{1}{3} \prgv{\nu} + \prgv{\xi}\bigr) \,  \sum_j\prgv{\Delta}^{(p)}_{ij} u_j 
 + \left( \prgv{\xi}-\frac{2}{3}\prgv{\nu} \right) \left[ \prgv{\nabla}^{(0,p)}_i \ln \prgv{\rho} \right] \sum_j\left[ \prgv{\nabla}^{(0,p)}_j u_j \right] \biggr) \,, \\
\lim_{|\vec{u}_{\rm p}|^2 \ll 1,\xi H \ll 1} u_i \prgv{{\mathcal F}}^i_{\rm ipf} = & -a^{\alpha-1} (1+\omega) \prgv{\rho} \, \biggr(\prgv{\nu} \, \sum_{i,j} \left[\prgv{\nabla}^{(0,p)}_i u_j\right]^2+ \prgv{\nu} \, \sum_{i,j} \prgv{\nabla}^{(0,p)}_i u_j  \prgv{\nabla}^{(0,p)}_j u_i \nonumber \\
& + \bigl(\prgv{\xi}- \tfrac{2}{3} \prgv{\nu}\bigr) \, \sum_i \left[\prgv{\nabla}^{(0,p)}_i u_i\right]^2 \biggr)  \, ,
\end{align}
\end{subequations}
where $\prgv{\Delta}^{(p)}_{ij}$ represents the bidiagonal scheme for cross derivatives introduced in \Eq{eq:Bidiag2m}, and the lattice expression for the Laplacian $\prgv{\Delta}^{(p)}$ has been introduced in \Eqq{eq:laplm}. Here, the rescaled program enthalpy is defined as $\prgv{w}\equiv \prgv{\rho}+ \prgv{p} = (1+\omega) \prgv{\rho}$. In the staggered formulation the various contributions from the imperfect fluid forces are discretized as follows,
\begin{subequations}
\begin{align}
\lim_{|\vec{u}_{\rm p}|^2 \ll 1,\xi H \ll 1} \prgv{{\mathcal F}}^0_{\rm ipf} = & \, 0 \, , \\
\lim_{|\vec{u}_{\rm p}|^2 \ll 1,\xi H \ll 1} \prgv{{\mathcal F}}^i_{\rm ipf} = & \ a^{\alpha-1} (1+\omega) {\cal S}_{i}^{(p)}[\prgv{\rho}] \biggr(\prgv{\nu} \, \prgv{\Delta}^{(p)} u_i + \prgv{\nu} \, \sum_j{\cal S}_{-j}^{(p)}[\prgv{\nabla}^{(+,p)}_i u_j + \prgv{\nabla}^{(+,p)}_j u_i] {\mathcal S}_{i}^{(p)}[\prgv{\nabla}^{(0,p)}_j \ln \prgv{\rho}]\nonumber \\ 
& + \bigl(\tfrac{1}{3} \prgv{\nu} + \prgv{\xi}\bigr) \,
\prgv{\nabla}^{(+,p)}_i \sum_j \prgv{\nabla}^{(-,p)}_j u_j + \left( \prgv{\xi}-\frac{2}{3}\prgv{\nu} \right) [\prgv{\nabla}^{(+,p)}_i \ln \prgv{\rho} ] \sum_j {\mathcal S}_{i}^{(p)} [\prgv{\nabla}^{(-,p)}_j u_j] \biggr) \, , \\
\lim_{|\vec{u}_{\rm p}|^2 \ll 1,\xi H \ll 1} u_i \prgv{{\mathcal F}}^i_{\rm ipf} = & -a^{\alpha-1} (1+\omega) \prgv{\rho} \biggr(\prgv{\nu} \,   \, \sum_{i,j} {\cal S}_{-i,-j}^{(p)}[(\prgv{\nabla}^{(+,p)}_i u_j)^2]+ \prgv{\nu} \, \sum_{i,j}  {\mathcal S}_{-i,-j}^{(p)}[ [\prgv{\nabla}^{(+,p)}_i u_j ] [\prgv{\nabla}^{(+,p)}_j u_i] ] \nonumber \\
& + \bigl(\prgv{\xi}- \tfrac{2}{3} \prgv{\nu}\bigr) \,  \sum_i (\prgv{\nabla}^{(-,p)}_i u_i)^2 \biggr) \,.
\end{align}
\end{subequations}

\newpage
\section{Fluid dynamics with bosonic interactions on the lattice}
\label{sec:fluid_bosonic}

In this section we turn to the case of relativistic fluid dynamics with interactions to IR bosonic degrees of freedom on the lattice. We begin the discussion in \Sec{sec:fluid_gauge} with fluid-gauge field dynamics, and then turn to scalar-fluid dynamics in \Sec{lattice_fluid_scalar}. The corresponding equations in the continuum have been presented in \Sec{subsec:bosonic_inters}. Here we focus on the lattice formulations of such systems of equations. All the discrete versions of the kernels will be presented with derivatives of generic order $p$ accuracy, however, these will always refer to the even order $p=2m$ lattice operators introduced in \Sec{sec:generalLattice}.

\subsubsection*{Program variables}

In the previous section we have already introduced so-called dimensionless program variables, which we have used to express the discretized versions of the relevant EOM. Besides the spacetime and fluid program variables introduced in \Eqss{eq:FluidProgramVariables}{emt_rescalings_prgv_fluid}, we now have additional field variables that belong to the bosonic field sectors. We follow a similar prescription as in {\tt The Art I} \cite{Figueroa:2020rrl} and {II} \cite{Baeza-Ballesteros:2025tme}, such that the program variables associated to the scalar field $\phi$, the temperature $T$, as well as the $U(1)$ gauge field $A_\mu$ and the field strength tensor $F_{\mu\nu}$ become,
\begin{align}
   \prgv\phi \equiv \frac{\phi}{f_*} \ , 
   \hspace{0.5cm} 
   \prgv{T} \equiv \frac{\conf{T}}{T_*} \ ,
   \hspace{0.5cm} 
   \prgv{A}_\mu \equiv \frac{A_\mu }{\omega_*} \ , \hspace{0.5cm} {\rm and}  \hspace{0.5cm} \prgv F_{\mu\nu} \equiv \frac{F_{\mu\nu}}{\omega_*^2} \ ,\label{eq:FieldProgramVariables}
\end{align}
where the values of the three mass scales $T_*$, $f_*$, and $\omega_*$ (which all have dimensions $[T_*]=[f_*]=[\omega_*]=+1$) can be chosen such that the physics of the processes under consideration is well captured within the numerical simulation. 
The current $J^{\mu}_\fl$, the source term $\Omega_\phi$ and the Lorentz and fluid-scalar forces ${\cal F}^\mu_{U(1)}$ and ${\cal F}^{\mu}_{\phi}$ stemming from the fluid-gauge and fluid-scalar interactions are given, in terms of program variables, by
\be
\prgv{J}^{\mu}_\fl  \equiv \frac{\omega_*}{T_*^4 }\conf{J}^{\mu}_\fl \ , \qquad
\prgv{\Omega}_\phi \equiv \frac{f_*}{T_*^4}\conf{\Omega}_{\phi}\, , \qquad 
\prgv{{\mathcal F}}_{U(1)}^\mu  \equiv  \frac{\conf{{\mathcal F}}_{U(1)}^\mu}{\omega_*T_*^4}\, , \qquad {\rm and} \qquad \prgv{{\mathcal F}}_\phi^\mu  \equiv  \frac{\conf{{\mathcal F}}_{\phi}^\mu}{\omega_*T_*^4} \ ,
\ee
such that the relations $\prgv{{\mathcal F}}_{U(1)}^\nu=a^{1-\alpha}\prgv{J}^\mu_{\fl}\prgv{F}^{\nu}_{\ \mu}$ and $\prgv{{\mathcal F}}_{\phi}^0=\prgv{\Omega}_{\phi}\prgv{\partial}_0\prgv{\phi}, \prgv{{\mathcal F}}_{\phi}^i=a^{2(\alpha-1)}\prgv{\Omega}_{\phi}\prgv{\partial}_i\prgv{\phi} $ also hold [cf.~\Eq{components_fluidscalar_interaction}]. The fluid and gauge sectors are coupled to each other through Ohm's law, which was introduced in \Eq{Ohms_law}. In terms of rescaled program variables it reads,
\be\label{eq:Ohmprgv}
\prgv{J}^\mu_{\fl} = a^{3+\alpha}\left( \frac{\prgv{\rho}_e}{a^3} U^\mu +  \frac{\prgv{\sigma}_f}{a} \prgv{F}^{\mu \nu} U_\nu\ \right) ,
\ee
where the rescaled electrical charge density $\prgv{\rho}_e$ and conductivity $\prgv{\sigma}_f$ are now defined as,
\be
\prgv{\rho}_e\equiv  \left(\frac{\omega_*}{T_*}\right)^4 \frac{\conf{\rho}_e}{\omega_*^3} \ , \hspace{0.5cm} {\rm and} \hspace{0.5cm}  \prgv{\sigma}_f \equiv  \left(\frac{\omega_*}{T_*}\right)^3 \frac{\conf{\sigma}_{f}}{T_*}\ .
\ee
Note that we will occasionally use the diffusivity $\conf{\eta}_{\rm diff}$ instead of the conductivity as a parameter in the EOM. Its associated program variable is defined as $\prgv{\eta}_{\rm diff}\equiv 1/\prgv{\sigma}_f$. The scalar-fluid interaction term is rescaled as follows,
\be
\prgv{\Omega}_{\phi} = \frac{\partial \prgv{p}}{\partial \prgv{\phi}} - \prgv{\delta}_\phi \, , \qquad {\rm with} \qquad \prgv{\delta}_\phi=\prgv{\eta}_\phi U^\mu \prgv{\partial}_\mu \prgv{\phi} \, , 
\ee
where,
\be
\prgv{\delta}_\phi \equiv \frac{f_*}{T_*} \frac{\conf{\delta}_\phi}{T_*^3} \, , \qquad {\rm and} \qquad \prgv{\eta}_\phi \equiv \left(\frac{f_*\omega_*}{T_*^2}\right)^2 \frac{\eta_\phi}{\omega_*}  \, , 
\ee
such that the rescaled scalar damping term in the limit of large amounts of collisions can be identified as $\prgv{\eta}_\phi=\prgv{A}\prgv{\phi}^2/\prgv{T}$ with $\prgv{A}\equiv (f_*^4\omega_*/T_*^5) A$ [cf.~\Eq{rescaled_frict}].
Moreover, from the scalar field interactions with a fluid an effective scalar potential $\conf{V}_{\rm eff} \equiv -\conf{p}_{\rm rad} + \conf{V}_0(\phi) + \conf{V}_T(\phi, T)$ emerges [cf.~\Eq{large_T_limit}], which is also rescaled as 
\begin{align}
    \prgv{V}_{\rm eff}(\prgv{\phi}, \prgv{T}) \equiv \left. \frac{\conf{V}_{\rm eff}(\phi, \conf{T})}{T_*^4} \right|_{\phi = f_* \prgv{\phi}, \, \conf{T} = T_* \prgv{T}}\, , \qquad {\rm and} \qquad \prgv{V}_{T}(\prgv{\phi}, \prgv{T}) \equiv \left. \frac{\conf{V}_{T}(\phi, \conf{T})}{T_*^4} \right|_{\phi = f_* \prgv{\phi}, \, \conf{T} = T_* \prgv{T}} \, .
\end{align}
In the fluid-scalar system case, since the scalar kinetic term is rescaled as $f_*^2 \omega_*^2$ a typical convenient choice is $T_*^4 = f_*^2 \omega_*^2$, such that all terms in the energy momentum tensor have the same rescaling.

\subsection{Fluid-gauge dynamics on the lattice}
\label{sec:fluid_gauge}

We first turn to the case of relativistic fluid-gauge dynamics on the lattice. We will treat the gauge sector in the non-compact formulation, i.e., we will solve \Eq{eq:Maxwell} explicitly for the gauge field $A_\mu$ (see, e.g., \cite{Figueroa:2020rrl} for a detailed discussion on the alternative compact formulation). In \Sec{fluid_gauge_cons_form} we will deal with the conservation form, which we have discussed for the case of a perfect fluid in \Sec{discrete_fluid_cons_form}. Then we will turn to the non-conservation form in \Sec{fluid_gauge_non_cons_form}}, which we have already discussed for the perfect fluid case in \Sec{sec:LatticeFluidDynamicsNC}. We will cover different discretization procedures, including the so-called collocated, semi-collocated and staggered formulations. Throughout this section we will work in the temporal gauge, i.e., we set $\prgv{A}_0=0$. For simplicity, we will discuss the gauge-fluid interaction in the absence of any imperfect fluid contributions. However, they can be added, when needed, following the discretization techniques outlined in \Sec{sec:fluidLattice}.

\subsubsection{Conservation form}
\label{fluid_gauge_cons_form}
 
In the conservation form the set of equations that describe the fluid-gauge system with self-consistent expansion of the Universe are given, in the continuum, by 
\Eqs{eq:conservation_mhd}{Maxwell2} for the fluid-gauge degrees of freedom and by \Eqq{Friedmann} for the evolution of the background. The interaction between the fluid and the gauge field is described through Ohm's law [cf.~\Eq{Ohms_law} in terms of physical, and \Eq{eq:Ohmprgv} in terms of program variables], as we have seen in \Sec{subsec:gauge_fluid_dyns}. The degrees of freedom we solve for are, the components of the stress-energy tensor $\prgv{T}^{00}_\pf$ and $\prgv{T}^{0i}_\pf$, the gauge field $\prgv{A}_i$ and the scale factor $a$, together with the conjugate momenta $\prgv{\pi}^A_i \equiv a^{1-\alpha} \prgv{F}_{0i}$ and $b \equiv a'$. This choice of the gauge field conjugate momentum, allows us to write the EOM as the following set of coupled first-order differential equations,
\begin{subequations}\label{eq:MHDConservationForm}
\begin{align}
\prgv{\pi}^A_i &= a^{1-\alpha}\prgv{F}_{0i} \ , \\
(\prgv{\pi}^A_i)' &= {\mathcal{K}}_i^A[\prgv{A}_j,\prgv{\pi}^A_i,\prgv{T}_\pf^{00},\prgv{T}_\pf^{0j},a] \ , \\
(\prgv{T}_{\rm pf}^{00})' &= {\cal G}^T_0 [\prgv{T}_\pf^{00}, \prgv{T}_\pf^{0i}, \prgv{A}_i, \prgv{\pi}^A_i, a, b] = {\cal G}^0_\pf[\prgv{T}_\pf^{0i}]+\prgv{{\mathcal F}}^0_{\mathcal{H}}[\prgv{T}_\pf,\mathcal{\prgv{H}}] + \prgv{{\mathcal F}}^0_A[\prgv{T}_\pf^{00},\prgv{T}_\pf^{0i},\prgv{A}_i,\prgv{\pi}^A_i,a] \, , \\
(\prgv{T}_\pf^{0i})' &= {\cal G}^T_i[\prgv{T}_{\rm pf}^{ij},\prgv{T}_\pf^{0i},\prgv{A}_j,\prgv{\pi}^A_j,a,b]= {\cal G}^i_\pf[\prgv{T}_\pf^{ij}] +  \prgv{{\mathcal F}}^i_{\mathcal{H}}[\prgv{T}_\pf^{0i},\mathcal{\prgv{H}}] + \prgv{{\mathcal F}}^i_A[\prgv{T}_\pf^{00},\prgv{T}_\pf^{0j},\prgv{A}_j,\prgv{\pi}^A_j,a] \, ,\\
b' &= {\mathcal{K}}_{a}[{a,\prgv E}_\fl^\rho,{\prgv E}_\fl^p,{\prgv E}^A_K, {\prgv E}^A_G] \, ,\\
b &= a' \, ,
\end{align}
\end{subequations}
where $\mathcal{\prgv{H}}\equiv b/a$ and primes denote derivatives with respect to $\alpha$-time in program units. In the following we consider a constant EOS $\omega$, such that the value of $\prgv{T}_\pf^{ij}$ can be directly reconstructed from the dynamical variables $\prgv{T}_\pf^{00}$ and $\prgv{T}_\pf^{0i}$, by using \Eq{eq:TijL}. The discretization of the core kernels ${\cal G}^\mu_\pf$, as well as the Hubble friction $\prgv{{\mathcal F}}_H^\mu$, have been discussed in \Sec{sec:fluidLattice}. The new fluid-gauge interaction term $\prgv{{\mathcal F}}_A^\mu\equiv \prgv{J}^\nu\prgv{F}^\mu_{\ \nu}$ has been introduced in the continuum in \Eq{eq:F_U(1)}. The discretized versions of the gauge field kernel and the new terms $\prgv{{\mathcal F}}_A^\mu$ are [cf.~\Eqq{current_density} and \Eqq{lorentz_forces}],
\begin{align}
{\mathcal{K}}_i^A[\prgv{A}_j,\prgv{\pi}^A_i,\prgv{r}_j,a] & =  a^{\alpha-1} \sum_j \prgv{\nabla}_j^\LL \prgv{F}_{ji}^\LL+ a^{-2} \prgv{J}_i^\fl = a^{\alpha-1}\sum_j \prgv{\nabla}_j^\LL \prgv{F}_{ji}^\LL\nonumber 
\\
&+a^{\alpha-1} \left(\frac{T_*}{\omega_*} \right)^4 \sqrt{\frac{z_\LL+\omega}{\omega+1}}\left[ \frac{z_\LL}{z_\LL+\omega}\prgv{\rho}_{e}[\prgv{r}_i]_\LL + \prgv{\sigma}_f \biggl(\frac{z_\LL}{z_\LL+\omega}\sum_{j}[\prgv{F}_{ij}\prgv{r}_j]_\LL -  \prgv{\sigma}_f[\prgv{\pi}_i^A]_\LL \biggr)\right] \ ,    
\\
\prgv{{\mathcal F}}^0_A[\prgv{r}_i,\prgv{A}_i,\prgv{\pi}^A_i,a] &=  a^{\alpha-1} \prgv{\rho}_e\sqrt{\frac{z_\LL+\omega}{\omega+1}} \frac{z_\LL}{z_\LL+\omega} \sum_{j}[\prgv{\pi}_{j}^A\prgv{r}_j]_\LL \nonumber 
\\
&+ a^{\alpha-1} \prgv{\sigma}_f\sqrt{\frac{z_\LL+\omega}{\omega+1}}\biggl(\frac{z_\LL}{z_\LL+\omega}\sum_{j,k} [\prgv{\pi}_{j}^A\prgv{F}_{jk}\prgv{r}_k]_\LL -  \sum_{j}[\prgv{\pi}_{j}^A]_\LL^2 \biggr) \ ,  
\\
\prgv{{\mathcal F}}^i_A[\prgv{r}_j,\prgv{A}_j,\prgv{\pi}^A_j,a] & = a^{2(\alpha-1)} \prgv{\rho}_e \sqrt{\frac{z_\LL+\omega}{\omega+1}} \left[\frac{z_\LL}{z_\LL+\omega} \sum_{j} [\prgv{F}_{ij} \prgv{r}_j]_\LL -[\prgv{\pi}_{i}^A]_\LL \right]  \nonumber 
\\ 
&+a^{2(\alpha-1)} \prgv{\sigma}_f \sqrt{\frac{z_\LL+\omega}{\omega+1}}\sum_{j,k} \Biggl[\frac{z_\LL}{z_\LL+\omega}
[\prgv{F}_{ij}\prgv{F}_{jk} \prgv{r}_k]_\LL +\frac{z_\LL}{z_\LL+\omega}[\prgv{\pi}_i^A\prgv{\pi}_j^A\prgv{r}_j]_\LL -[\prgv{F}_{ij}\prgv{\pi}_j^A]_\LL \Biggr] \ ,
\end{align}
where we have defined for convenience the ratio $\prgv{r}_i\equiv a^{1-\alpha}\prgv{T}^{0i}_\pf/\prgv{T}^{00}_\pf$. In the second equality we have directly introduced Ohm's law [cf.~\Eq{eq:Ohmprgv}], written $\prgv{F}_{0i}$ in terms of the gauge field conjugate momentum $\prgv{\pi}_i^A$, and expressed the velocity field through \Eq{rho_u_fromT0mu}. The label `$\rm L$', such as in $[\prgv{\pi}_i^A]_\LL$, $\prgv{r}_i^\LL$, $z_\LL$, and $\prgv{F}_{\mu\nu}^\LL\equiv \prgv{\nabla}_\mu^\LL \prgv{A}_\nu - \prgv{\nabla}_\nu^\LL \prgv{A}_\mu$ denotes the lattice version of the corresponding quantity. These terms may have different realizations in different discretization schemes.

The kernel of the scale factor holds, besides the energy density and pressure of the fluid, additional contributions from the kinetic and gradient energy density of the gauge field and is given by,
\begin{align}
{\mathcal{K}}_{a}[a,{\prgv E}_\fl^\rho, {\prgv E}_\fl^p,{\prgv E}^A_K,{\prgv E}^A_G] = \frac{a^{2\alpha+1}}{3} \left( \frac{ T_*^2}{\omega_*\mpl} \right)^2 \left[ \frac{2\alpha - 1}{2}{\prgv E}_\fl^\rho-\frac{3}{2} {\prgv E}_\fl^p + (\alpha -1) ({\prgv E}^A_K+{\prgv E}^A_G) \right] \ .
\end{align}
The volume averaged fluid energy ${\prgv E}_\fl^\rho$ and pressure ${\prgv E}_\fl^p$ density contributions have been introduced in \Eq{eq:va_energy_densities}. On the other hand, the lattice expressions of the volume averaged kinetic and gradient energy density associated to the gauge field are given, in terms of program variables, by
\begin{align}
& {\prgv E}_\fl^\rho \equiv  \frac{1}{a^4}\left\langle {\prgv T}_{\rm pf}^{00}\right\rangle  \ , & \qquad & {\prgv E}^A_K = \frac{1}{2a^4} \frac{\omega_*^4}{T_*^4} \sum_{i}^3 \left\langle\left( {\prgv \pi_A} \right)_i^2\right\rangle &  &&\\ 
&{\prgv E}_\fl^p \equiv \frac{1}{3a^{2+2\alpha}}\Big\langle \sum_{i}{\prgv T}_{\rm pf}^{ii} \Big\rangle \ , & \qquad &{\prgv E}^A_G = \frac{1}{2a^4} \frac{\omega_*^4}{T_*^4} \sum_{i,j<i} \left\langle\left(\prgv{\nabla}_i^\LL A_j-  \prgv{\nabla}_i^\LL A_j \right)^2\right\rangle & &&
\end{align}
where $\langle \dots\rangle$ denotes a volume average over the whole lattice. The lattice derivative $\prgv{\nabla}_i^\LL$ in the expression for the gradient energy density ${\prgv E}^A_G $ needs to be replaced by the neutral derivative $\prgv{\nabla}_i^{(0,p)}$ in case we work in the collocated formulation or by the forward derivative $\prgv{\nabla}_i^{(+,p)}$ if we work in the semi-collocated or staggered formulation, in which the gauge field lives in between lattice sites.

For the case of the fluid-gauge field system the Hubble and Gauss [cf.~\Eq{Maxwell2}] constraints are given by,
\begin{align}
&b^2 = \frac{1}{3} \left(\frac{ T_*^2}{\omega_*\mpl} \right)^2a^{2(\alpha+1)} \Big({\prgv E}_\fl^\rho+ {\prgv E}^A_K+{\prgv E}^A_G\Big)\,, \\
&\sum_j \prgv{\nabla}_j^{\rm L} \prgv{\pi}_j^A = - \left(\frac{T_*}{\omega_*} \right)^4 \sqrt{\frac{z_\LL+\omega}{\omega+1}}\left(\prgv{\rho}_e - \prgv{\sigma}_f\frac{z_\LL}{z_\LL+\omega}\sum_j [\prgv{\pi}_j^A\prgv{r}_j]_\LL \right)\ .
\end{align}
In the following we present the discretized kernels of the gauge field in the non-compact formulation, as well as the force terms $\prgv{{\mathcal F}}_A^0$ and $\prgv{{\mathcal F}}_A^i$, for the collocated, semi-collocated and staggered formulations. The fluid kernels ${\mathcal{G}}_\mu$, as well as $\prgv{{\mathcal F}}_\HH^\mu$ that have been introduced in the previous section remain unchanged. We use \Eqss{eq:latticeG0Tcoll}{eq:latticeFiTcoll} for the collocated and semi-collocated discretization scheme, and \Eqss{eq:latticeG0Tstagg}{eq:r2_m} for the staggered formulation.

\subsubsection*{All-collocated implementation}

In the all-collocated implementation we assume that all fluid variables (i.e. $\prgv{T}^{00}_{\rm pf}$ and $\prgv{T}^{0i}_{\rm pf}$), as well as the gauge field and the corresponding conjugate momentum live at lattice sites $\bf n$. The discretized version of the gauge field kernel and the force terms are then,
\begin{align}
{\mathcal{K}}_i^A &= a^{\alpha-1} \sum_j \prgv{\nabla}_j^{(0,p)}(\prgv{\nabla}_j^{(0,p)}\prgv{A}_i-\prgv{\nabla}_i^{(0,p)}\prgv{A}_j) \nonumber \\
&+ a^{\alpha-1}\left(\frac{T_*}{\omega_*} \right)^4\sqrt{\frac{z+\omega}{\omega+1}} \left[ \prgv{\rho}_e \frac{z}{z+\omega} \prgv{r_i} + \prgv{\sigma}_f\biggl(\frac{z}{z+\omega}\sum_{j}(\prgv{\nabla}_i^{(0,p)}\prgv{A}_j- \prgv{\nabla}_j^{(0,p)}\prgv{A}_i)\prgv{r}_j -  \prgv{\pi}_i^A \biggr) \right]   \ , 
\\
\prgv{{\mathcal F}}^0_A &= a^{\alpha-1} \prgv{\rho}_e \sqrt{\frac{z+\omega}{\omega+1}} \frac{z}{z+\omega}\sum_{j}\prgv{\pi}_j^Ar_j \nonumber \\
& + a^{\alpha-1} \prgv{\sigma}_f \sqrt{\frac{z+\omega}{\omega+1}}  \left[\frac{z}{z+\omega} \sum_{j,k}(\prgv{\nabla}_j^{(0,p)}\prgv{A}_k- \prgv{\nabla}_k^{(0,p)}\prgv{A}_j) \prgv{\pi}_{j}^A r_k   - \sum_{j} (\prgv{\pi}_j^A)^2\right] \ , 
\\
 \prgv{{\mathcal F}}^i_A &=  a^{2\alpha-2} \prgv{\rho}_e\sqrt{\frac{z+\omega}{\omega+1}} \left[  \frac{z}{z+\omega} \sum_{j} (\prgv{\nabla}_i^{(0,p)}\prgv{A}_j - \prgv{\nabla}_j^{(0,p)}\prgv{A}_i )\prgv{r}_j -\prgv{\pi}_i^A\right]  \nonumber \\
&+a^{2\alpha-2} \prgv{\sigma}_f \sqrt{\frac{z+\omega}{\omega+1}}\frac{z}{z+\omega} \sum_{j,k} \Biggl[ (\prgv{\nabla}_i^{(0,p)}\prgv{A}_j- \prgv{\nabla}_j^{(0,p)}\prgv{A}_i)(\prgv{\nabla}_j^{(0,p)}\prgv{A}_k- \prgv{\nabla}_k^{(0,p)}\prgv{A}_j)r_k \Biggr] \nonumber \\
&+a^{2\alpha-2} \prgv{\sigma}_f \sqrt{\frac{z+\omega}{\omega+1}} \sum_j\left[\frac{z}{z+\omega} \prgv{\pi}_i^A\prgv{\pi}_j^A\prgv{r}_j -(\prgv{\nabla}_i^{(0,p)}\prgv{A}_j - \prgv{\nabla}_j^{(0,p)}\prgv{A}_i)\prgv{\pi}_{j}^A \right]\ ,
\end{align}
and the Gauss law is discretized as,
\be
\sum_j \prgv{\nabla}_j^{(0,p)} \prgv{\pi}_j^A = -\left(\frac{T_*}{\omega_*} \right)^4 \sqrt{\frac{z+\omega}{\omega+1}}\left(\prgv{\rho}_e - \prgv{\sigma}_f\frac{z}{z+\omega}\sum_j \prgv{\pi}_{j}^A\prgv{r}_j\right)\ .
\ee
Here $\prgv{\nabla}_j^{(0,p)}$ represents the neutral derivative introduced in 
\Eqq{centralderivatives}. The discretized version of the $z$-function [cf.~\Eq{eq:zL}] is a function of the ratio $\prgv{r}^2\equiv a^{2(1-\alpha)}\sum_i(\prgv{T}_{\rm pf}^{0i}/\prgv{T}_{\rm pf}^{00})^2$.

\subsubsection*{Semi-collocated implementation}

In the semi-collocated formulation we assume that the gauge field and its conjugate momentum live in between lattice sites, as $\prgv{A}_i({\bf n}+\hat{\imath}/2)$ and $\prgv{\pi}^A_i({\bf n}+\hat{\imath}/2)$, while the components of the stress-energy tensor live at lattice sites, as $\prgv{T}^{0\mu}(\bf n)$. The discretized versions of the gauge field kernel and the force terms are then,
\begin{align}
{\mathcal{K}}_i^A &= a^{\alpha-1} \sum_j \prgv{\nabla}_j^{(-,p)}(\prgv{\nabla}_j^{(+,p)}\prgv{A}_i-\prgv{\nabla}_i^{(+,p)}\prgv{A}_j) + a^{\alpha-1} \left(\frac{T_*}{\omega_*} \right)^4 \prgv{\rho}_e \sqrt{\frac{z_{(+i)} +\omega}{\omega+1}}\frac{z_{(+i)}}{z_{(+i)} +\omega}  \prgv{r}_{i,+i} \nonumber \\
& + a^{\alpha-1} \left(\frac{T_*}{\omega_*} \right)^4 \prgv{\sigma}_f\sqrt{\frac{z_{(+i)} +\omega}{\omega+1}} \left[ \frac{z_{(+i)}}{z_{(+i)} +\omega}  \sum_{j}{\mathcal S}_{-j}^{(p)}[\prgv{\nabla}_i^{(+,p)}\prgv{A}_j - \prgv{\nabla}_j^{(+,p)}\prgv{A}_i] \prgv{r}_{j, +i} - \prgv{\pi}_i^A \right]   \ , 
\end{align}
\begin{align}
\prgv{{\mathcal F}}^0_A &= a^{\alpha-1} \prgv{\rho}_e \sqrt{\frac{z+\omega}{\omega+1}} \sum_{j}\left[ \frac{z}{z+\omega}  {\mathcal S}_{-j}^{(p)}[\prgv{\pi}_{j}^A] \prgv{r}_j \right] \nonumber 
\\
& + a^{\alpha-1} \prgv{\sigma}_f\sqrt{\frac{z+\omega}{\omega+1}} \left[\frac{z}{z+\omega} \sum_{j,k} {\mathcal S}_{-j,-k}^{(p)}[\prgv{\nabla}_j^{(+,p)}\prgv{A}_k - \prgv{\nabla}_k^{(+,p)}\prgv{A}_j] {\mathcal S}_{-j}^{(p)}[\prgv{\pi}_{j}^A] \prgv{r}_k -   \sum_{j} ({\mathcal S}_{-j}^{(p)}[\prgv{\pi}_{j}^A])^2 \right] \ , 
\end{align}
\begin{align}
 \prgv{{\mathcal F}}^i_A &= a^{2\alpha-2} \prgv{\rho}_e\sqrt{\frac{z+\omega}{\omega+1}} \left[ \frac{z}{z+\omega} \sum_{j} {\mathcal S}_{-i,-j}^{(p)}[\prgv{\nabla}_i^{(+,p)}\prgv{A}_j - \prgv{\nabla}_j^{(+,p)}\prgv{A}_i ]\prgv{r}_j - {\mathcal S}_{-i}^{(p)}[\prgv{\pi}_i^A] \right] \nonumber 
\\
&+a^{2\alpha-2} \prgv{\sigma}_f \sqrt{\frac{z + \omega}{\omega + 1}}\frac{z}{z + \omega} \sum_{j,k}\Biggl[{\mathcal S}_{-i,-j}^{(p)}[\prgv{\nabla}_i^{(+,p)}\prgv{A}_j - \prgv{\nabla}_j^{(+,p)}\prgv{A}_i]{\mathcal S}_{-j,-k}^{(p)}[\prgv{\nabla}_j^{(+,p)}\prgv{A}_k - \prgv{\nabla}_k^{(+,p)}\prgv{A}_j]\prgv{r}_k \Biggr] \nonumber 
\\
&+a^{2\alpha-2} \prgv{\sigma}_f\sqrt{\frac{z+\omega}{\omega+1}}  \sum_{j}\left[\frac{z}{z+\omega}{\cal S}_{-i}^{(p)} [\prgv{\pi}_i^A] {\cal S}_{-j}^{(p)} [\prgv{\pi}_j^A]\prgv{r}_j - {\mathcal S}_{-j}^{(p)}[{\mathcal S}_{-i}^{(p)}[\prgv{\nabla}_i^{(+,p)}\prgv{A}_j- \prgv{\nabla}_j^{(+,p)}\prgv{A}_i]\prgv{\pi}_{j}^A]\right]\ ,
\end{align}
and the Gauss law is given by,
\be
\sum_j \prgv{\nabla}_j^{(-,p)} \prgv{\pi}_j^A = - \left(\frac{T_*}{\omega_*} \right)^4 \sqrt{\frac{z+\omega}{\omega+1}}\left(\prgv{\rho}_e -  \prgv{\sigma}_f\frac{z}{z+\omega}\sum_j \prgv{r}_j {\cal S}_{-j}^{(p)}[\prgv{\pi}_{j}^A]\right)\ .
\ee
The continuum derivative operators are replaced by $\nabla_j^{(\pm,p)}$, which represent forward ($+$) and backward ($-$) derivatives that have been introduced in \Eqq{fbders2}. The function $z\equiv z(\prgv{r}^2)$, defined in \Eq{eq:zL}, is obtained by replacing $\prgv{T}^{00}_\pf$ and $\prgv{T}^{0i}_\pf$ inside the $r^2$-ratio by their lattice version at lattice site $\bf n$. Moreover, we have introduced the two displaced ratios $\prgv{r}_{j,+i}$ and $\prgv{r}^2_{+i}$ (which appears in $z_{(+i)}\equiv z(r^2_{+i})$), that are defined as,
\be
\prgv{r}^2_{+i} \equiv a^{2(1-\alpha)}{\mathcal S}_i^{(p)}\left[\frac{\sum_{j=1}^{3}\prgv{T}^{0j}\prgv{T}^{0j}}{[\prgv{T}^{00}]^2}\right] \, , \qquad {\rm and} \qquad \prgv{r}_{j, +i} \equiv a^{1-\alpha}{\mathcal S}_i^{(p)}\left[\frac{\prgv{T}^{0j}}{\prgv{T}^{00}}\right] \ . 
\ee

\subsubsection*{Staggered implementation}

In the staggered formulation we assume that all vectors, i.e. $\prgv{A}_i$, $\prgv{\pi}^A_i$ and $\prgv{T}^{0i}$, live in between lattice sites, i.e., at ${\bf n}+\hat{\imath}/2$, while the scalar components as $\prgv{T}^{00}$, live at lattice sites $\bf n$. The discretized versions of the gauge field kernel and the force terms are then,
\begin{align}
{\mathcal{K}}_i^A &= a^{\alpha-1} \sum_j \prgv{\nabla}_j^{(-,p)}(\prgv{\nabla}_j^{(+,p)}\prgv{A}_i-\prgv{\nabla}_i^{(+,p)}\prgv{A}_j) + a^{\alpha-1} \left(\frac{T_*}{\omega_*} \right)^4 \prgv{\rho}_e \sqrt{\frac{z_{(+i)} +\omega}{\omega+1}} \frac{z_{(+i)}}{z_{(+i)}+\omega} \prgv{r}_{i,+i} \nonumber \\
&+ a^{\alpha-1} \left(\frac{T_*}{\omega_*} \right)^4 \prgv{\sigma}_f\sqrt{\frac{z_{(+i)} +\omega}{\omega+1}}  \left[\frac{z_{(+i)}}{z_{(+i)}+\omega} \sum_{j}{\mathcal S}_{-j}^{(p)}[\prgv{\nabla}_i^{(+,p)}\prgv{A}_j - \prgv{\nabla}_j^{(+,p)}\prgv{A}_i]\prgv{r}_{i,+i} - \prgv{\pi}_i^A \right]   \ , 
\\
\prgv{{\mathcal F}}^0_A &= a^{\alpha-1} \prgv{\rho}_e \sqrt{\frac{z+\omega}{\omega+1}}  \frac{z}{z+\omega} \sum_{j}{\mathcal S}_{-j}^{(p)}[\prgv{\pi}_{j}^A] \prgv{r}_j  \nonumber 
\\
& + a^{\alpha-1} \prgv{\sigma}_f \sqrt{\frac{z+\omega}{\omega+1}} \left[ \frac{z}{z+\omega}\sum_{j,k}{\mathcal S}_{-j}^{(p)}[\prgv{\pi}_{j}^A]{\mathcal S}_{-j,-k}^{(p)}[\prgv{\nabla}_j^{(+,p)}\prgv{A}_k - \prgv{\nabla}_k^{(+,p)}\prgv{A}_j] \prgv{r}_k - \sum_j({\mathcal S}_{-j}^{(p)}[\prgv{\pi}_{j}^A])^2 \right] \ ,  
\\
\prgv{{\mathcal F}}^i_A &= a^{2\alpha-2} \prgv{\rho}_e \sqrt{\frac{z_{(+i)}+\omega}{\omega+1}} \left[\frac{z_{(+i)}}{z_{(+i)} +\omega} \sum_{j} {\mathcal S}_{-j}^{(p)}[\prgv{\nabla}_i^{(+,p)}\prgv{A}_j - \prgv{\nabla}_j^{(+,p)}\prgv{A}_i ]\prgv{r}_{j,+i} - \prgv{\pi}_{i}^A \right] \nonumber 
\\
&+a^{2\alpha-2} \prgv{\sigma}_f \sqrt{\frac{z_{(+i)}+\omega}{\omega+1}} \Biggl[ \sum_j \left[\frac{z_{(+i)}}{z_{(+i)} +\omega}\prgv{r}_{j,+i}{\mathcal S}_{i,-j}^{(p)}[\prgv{\pi}_j^A]\prgv{\pi}_i^A -  {\mathcal S}_{-j}^{(p)}[\prgv{\nabla}_i^{(+,p)}\prgv{A}_j- \prgv{\nabla}_j^{(+,p)}\prgv{A}_i]{\mathcal S}_{i,-j}^{(p)}[\prgv{\pi}_{j}^A] \right] \nonumber \\
&+\frac{z_{(+i)}}{z_{(+i)} + \omega} \sum_{j,k} {\mathcal S}_{-j}^{(p)}[\prgv{\nabla}_i^{(+,p)}\prgv{A}_j - \prgv{\nabla}_j^{(+,p)}\prgv{A}_i]{\mathcal S}_{i}^{(p)}[{\mathcal S}_{-k,-j}^{(p)}[\prgv{\nabla}_j^{(+,p)}\prgv{A}_k - \prgv{\nabla}_k^{(+,p)}\prgv{A}_j]]\prgv{r}_{k,+i}  \Biggr] \ ,
\end{align}
and the Gauss law is given by,
\be
\sum_j \prgv{\nabla}_j^{(-,p)} \prgv{\pi}_j^A = - \left(\frac{T_*}{\omega_*} \right)^4 \sqrt{\frac{z+\omega}{\omega+1}}\left(\prgv{\rho}_e - \prgv{\sigma}_f\frac{z}{z+\omega}\sum_j {\cal S}_{-j}^{(p)}[\prgv{\pi}_{j}^A]\prgv{r}_j^{(0)}\right)\ .
\ee
Again, here $\nabla_j^{(\pm,p)}$ represents the forward ($+$) and backward ($-$) derivatives that have been introduced in \Eqq{fbders2}. The $r^2$-ratio that appears inside the function $z\equiv z(\prgv{r}^2)$ [defined in \Eq{eq:zL}], as well as the ration $\prgv{r}_j$ are given by,
\be
\prgv{r}^2 =  a^{2(1-\alpha)}\frac{\sum_{j=1}^{3}{\mathcal S}_{-j}^{(p)}[\prgv{T}^{0j}\prgv{T}^{0j}]}{[\prgv{T}^{00}]^2} \, , \hspace{1cm} {\rm and} \hspace{1cm}\prgv{r}_j =  a^{1-\alpha}\frac{{\mathcal S}_{-j}^{(p)}[\prgv{T}^{j0}]}{\prgv{T}^{00}} \ , 
\ee
where $\prgv{T}^{00}_\pf$ and $\prgv{T}^{0i}_\pf$ inside the ratios are replaced by their lattice version at lattice site $\bf n$. 
Moreover, we have introduced the two displaced ratios $\prgv{r}_{j,+i}$ and $\prgv{r}^2_{+i}$ (which appears in $z_{(+i)}\equiv z(r^2_{+i})$), that are defined as,
\be
\prgv{r}_{j,+i}= a^{1-\alpha}{\mathcal S}_i^{(p)}\left[\frac{{\mathcal S}_{-j}^{(p)}[\prgv{T}^{0j}]}{\prgv{T}^{00}}\right] \, , \hspace{1cm} {\rm and} \hspace{1cm} \prgv{r}^2_{+i}=\prgv{r}^2_{\rm i}= a^{2(1-\alpha)}{\mathcal S}_i^{(p)}\left[\frac{\sum_{j=1}^{3}{\mathcal S}_{-j}^{(p)}[\prgv{T}^{0j}\prgv{T}^{0j}]}{[\prgv{T}^{00}]^2}\right] \, . 
\ee

\subsubsection{Non-conservation form}
\label{fluid_gauge_non_cons_form}

In the non-conservation form the fluid equations are expressed in terms of the rescaled primitive variables $\prgv{\rho}$ and $u_i$ [cf.~\Eq{eq:FluidProgramVariables}].
The dynamics of the system is governed by the Maxwell equation [cf.~\Eq{eq:Maxwell}], the energy density and velocity field equations [cf.~\Eqs{rho_dens}{u_vel}], and the second Friedmann equation as introduced in \Eq{Friedmann}. With the dynamical variables being the gauge field $\prgv{A}_i$, its conjugate momentum $\prgv{\pi}_A$, the logarithm of the energy density $\ln \prgv{\rho}$ the velocity field $u_i$, as well as the scale factor $a$ and its conjugate momentum $b=a'$, we can bring the corresponding EOM into the following set of first-order partial differential equations,
\begin{subequations}
\begin{align}
\prgv{\pi}_i^A&=a^{1-\alpha}F_{0i} \\
(\prgv{\pi}_i^A)'&={\mathcal{K}}^A_i[\prgv{A}_j, \prgv{\pi}_i^A, u_j, a, b]\\
(\ln\prgv{\rho})' &= {\mathcal{G}}^{\rho}[\ln \prgv{\rho}, u_i, \prgv{A}_i, \prgv{\pi}_i^A, a, b]={\mathcal{G}}_0^\pf[\ln \prgv{\rho}, u_i]+{\mathcal{G}}^{\HH}_0[\ln \prgv{\rho}, u_i, \prgv{\HH}]+{\mathcal G}_0^{A}[\ln \prgv{\rho}, u_i, {\mathcal{F}}^\mu_A]  \, , \\
(u_i)'&={\mathcal{G}}_i^u[\ln \prgv{\rho}, u_j, \prgv{A}_j, \prgv{\pi}_j^A, a, b]={\mathcal{G}}_i^\pf[\ln \prgv{\rho}, u_j] +{\mathcal{G}}^{\HH}_i[\ln \prgv{\rho}, u_i, \prgv{\HH}]+{\mathcal G}_i^A[\ln \prgv{\rho}, u_j, {\mathcal{F}}^\mu_A]\, , \\
b'&={\mathcal{K}}_{a}[{a, \prgv E}_\fl^\rho, {\prgv E}_\fl^p, {\prgv E}^A_K,  {\prgv E}^A_G] \, ,\\
b&=a' \, ,
\end{align}
\end{subequations}
where $\mathcal{\prgv{H}}\equiv a'/a$ and primes denote derivatives with respect to $\alpha$-time in program units. The core kernels ${\mathcal{G}}_0^\pf$ and ${\mathcal{G}}_i^\pf$, as well as the terms ${\mathcal G}_\mu^\HH$ that explicitly depend on the Hubble parameter have been introduced in the previous section in \Eqss{eq:GKenrelRho}{eq:GKenrelH}. The gauge field kernel $\mathcal{K}_i^A$, as well as the additional contributions ${\mathcal{G}}_0^A$ and ${\mathcal{G}}_i^A$ to the fluid kernels, are given by, 
\begin{align}
{\mathcal{K}}_i^A[\prgv{A}_j,\prgv{\pi}_i^A,u_j,a] &= a^{\alpha-1} \sum_j \prgv{\nabla}_j^\LL \prgv{F}_{ji}^\LL+a^{-2} \prgv{J}_i^{A} \nonumber \\
&= a^{\alpha-1} \sum_j \prgv{\nabla}_j^\LL \prgv{F}_{ji}^\LL+\gamma_\LL \left(\frac{T_*}{\omega_*} \right)^4 \left[\prgv{\rho}_e [u_i]_\LL+\prgv{\sigma}_f \sum_{j}[\prgv{F}_{ij}u_j]_\LL- a^{\alpha-1}\prgv{\sigma}_f[\prgv{\pi}_i^A]_\LL \right] \ ,    \\
{\mathcal{G}}_0^A[\prgv{\rho},u_j,{\mathcal{F}}^\mu_A]&=\frac{1}{(1-\omega a^{2(1-\alpha)}u^2_\LL)\prgv{\rho}_\LL}\left[(1+a^{2(1-\alpha)}u^2_\LL)\left[\prgv{{\mathcal F}}^0_A\right]_\LL- 2a^{2(1-\alpha)} \sum_j \left[u_j\prgv{{\mathcal F}}^j_A \right]_\LL\right]  \ ,  \\
{\mathcal{G}}_i^A[\prgv{\rho},u_i,{\mathcal{F}}^\mu_A]&=\frac{1-a^{2(1-\alpha)}u^2_\LL}{(1-wa^{2(1-\alpha)}u^2_\LL)\prgv{\rho}_\LL}\left[ \frac{2\omega}{1+\omega}a^{2(1-\alpha)}\sum_j [u_j\prgv{{\mathcal F}}^j_A u_i]_\LL-[\prgv{{\mathcal F}}^0_A u_i]_\LL \right] + \frac{1-a^{2(1-\alpha)}u^2_\LL}{(1+\omega)\prgv{\rho}_\LL} [\prgv{{\mathcal F}}^i_A]_\LL \ ,
\end{align}
with the force terms being,
\begin{align}
\prgv{{\mathcal F}}^0_A = & \,\,\gamma_\LL  \prgv{\rho}_e \sum_{j}[\prgv{\pi}_j^Au_j]_\LL + \gamma_\LL \prgv{\sigma}_f\left[\sum_{j,k} [\prgv{\pi}_j^A\prgv{F}_{jk}u_k]_\LL - a^{\alpha-1}\sum_{j} [\prgv{\pi}_j^A]_\LL^2 \right] \ , 
\\
\prgv{{\mathcal F}}^i_A = & \,\,a^{\alpha-1}\gamma_\LL  \biggl[  \prgv{\rho}_e \biggl(\sum_{j} [\prgv{F}_{ij}u_j]_\LL - a^{\alpha-1} [\prgv{\pi}_i^A]_\LL \biggr) \nonumber
\\
&+ \prgv{\sigma}_f \biggl(\sum_{j,k}[u_k\prgv{F}_{ij}\prgv{F}_{jk}]_\LL+\sum_j[u_j \prgv{\pi}_j^A \prgv{\pi}_i^A]_\LL - a^{\alpha-1}\sum_j[\prgv{F}_{ij}\prgv{\pi}_j^A]_\LL  \biggr)\biggr]  \ ,
\end{align}
where the label `L' denotes some lattice version of the corresponding expressions, which we will specify for the collocated, semi-collocated and staggered formulation in the following sections. Note that we have directly substituted Ohm's law [cf.~\Eq{Ohms_law}] for the current $\prgv{J}_\mu^{\fl}$ in the gauge field kernel. The background expansion is sourced by both the gauge field and the fluid, thus the kernel of the scale factor holds contributions from the fluid as well as from the kinetic and gradient energy density of the gauge field,
\begin{align}
{\mathcal{K}}_{a}[a,{\prgv E}_\fl^\rho, {\prgv E}_\fl^p,{\prgv E}^A_K,{\prgv E}^A_G] = \frac{a^{2\alpha+1}}{3} \left( \frac{ T_*^2}{\omega_*\mpl} \right)^2 \left[ \frac{2\alpha - 1}{2}{\prgv E}_\fl^\rho-\frac{3}{2} {\prgv E}_\fl^p + (\alpha -1) ({\prgv E}^A_K+{\prgv E}^A_G) \right] \ .
\end{align}
The volume averaged energy ${\prgv E}_\fl^\rho$ and pressure ${\prgv E}_\fl^p$ density of the fluid, as well as the averaged kinetic and gradient energy density of the gauge field, in terms of program variables, are given by,
\begin{align}
& {\prgv E}_\fl^\rho \equiv \frac{1}{a^4}\left\langle \left(1 + (1+\omega)\gamma_\LL^2a^{2(1-\alpha)}u_\LL^2\right){\prgv \rho}_\LL \right\rangle  \ , & \qquad & {\prgv E}^A_K = \frac{1}{2a^4} \frac{\omega_*^4}{T_*^4} \sum_{i}^3 \left\langle\left( {\prgv \pi_A} \right)_i^2\right\rangle \ , &  &&\\ 
&{\prgv E}_\fl^p \equiv \frac{1}{3a^{2+2\alpha}}\left\langle \left(3\omega + (1+\omega)\gamma_\LL^2a^{2(1-\alpha)}u_\LL^2\right){\prgv \rho}_\LL  \right\rangle \ , & \qquad &{\prgv E}^A_G = \frac{1}{2a^4} \frac{\omega_*^4}{T_*^4} \sum_{i,j<i} \left\langle\left( \prgv{\nabla}_i^\LL A_j- \prgv{\nabla}_i^\LL A_j \right)^2\right\rangle \ , & &&
\end{align}
where $\langle \dots\rangle$ denotes a volume average over the whole lattice. The lattice derivative $\prgv{\nabla}_i^\LL$ in the expression for the gradient energy density ${\prgv E}^A_G $ needs to be replaced by the neutral derivative $\prgv{\nabla}_i^{(0,p)}$ in the case where we work in the collocated formulation or by the forward derivative $\prgv{\nabla}_i^{(+,p)}$ if we work in the semi-collocated or staggered formulation. For the fluid-gauge system the Hubble constraint and the Gauss law are,
\begin{align}
&b^2 = \frac{1}{3} \left(\frac{ T_*^2}{\omega_*\mpl} \right)^2 a^{2(\alpha+1)} \Big({\prgv E}_\fl^\rho+ {\prgv E}^A_K+{\prgv E}^A_G\Big)\,, \\
&\sum_j \prgv{\nabla}^\LL_j \prgv{\pi}_j^A =- \gamma_\LL \left(\frac{T_*}{\omega_*} \right)^4 \left(\prgv{\rho}_e - \prgv{\sigma}_f a^{1-\alpha}\sum_j [u_j \prgv{\pi}_j^A]_\LL \right)\ .
\end{align}
In the following we present the discretized kernels of the gauge field in the non-compact formulation, as well as the forces $\prgv{{\mathcal F}}_A^0$ and $\prgv{{\mathcal F}}_A^i$, in the collocated, semi-collocated, and staggered formulations. The fluid kernels ${\mathcal{G}}_0^\pf$ and ${\mathcal{G}}^\pf_i$, as well as $\prgv{{\mathcal F}}_\HH^\mu$ remain unchanged with respect to the ones we have introduced in \Sec{sec:fluidLattice}. We use \Eqss{eq:E1coll}{eq:H2coll} for the collocated and semi-collocated discretization scheme, and \Eqss{eq:E1stagg}{eq:H2stagg} for the staggered formulation.

\subsubsection*{All-collocated implementation}

In the all-collocated formulation we assume that all fluid components, as well as the gauge field, live at lattice sites $\bf n$. The discretized version of the gauge field kernel and of the additional contributions to the fluid kernels are then,
\begin{align}
{\mathcal{K}}^A_i[\prgv{A}_i, u_i,a] = & \,\, a^{\alpha-1} \sum_j \prgv{\nabla}_j^{(0,p)}(\prgv{\nabla}_j^{(0,p)}\prgv{A}_i-\prgv{\nabla}_i^{(0,p)}\prgv{A}_j) \nonumber \\
& + \gamma \left(\frac{T_*}{\omega_*} \right)^4 \left[\prgv{\rho}_e u_i+\prgv{\sigma}_f \biggl( \prgv{\sigma}_f\sum_{j}(\prgv{\nabla}_i^{(0,p)}\prgv{A}_j- \prgv{\nabla}_j^{(0,p)}\prgv{A}_i)u_j- a^{\alpha-1}\prgv{\pi}_i^A \biggr)\right]   \ , \\
{\mathcal{G}}_0^A[\prgv{\rho},u_i,{\mathcal{F}}^\mu_A] = & \,\, \frac{1}{(1-\omega a^{2(1-\alpha)}u^2)\prgv{\rho}}\left[(1+a^{2(1-\alpha)}u^2)\prgv{{\mathcal F}}^0_A - 2 a^{2(1-\alpha)} \sum_j u_j\prgv{{\mathcal F}}^j_A\right]  \ ,  \\
{\mathcal{G}}_i^A[\prgv{\rho},u_i,{\mathcal{F}}^\mu_A] = & \,\,\frac{1-a^{2(1-\alpha)}u^2}{(1-wa^{2(1-\alpha)}u^2)\prgv{\rho}}\left[ \frac{2\omega}{1+\omega}a^{2(1-\alpha)}\sum_j u_j\prgv{{\mathcal F}}^j_A u_i-\prgv{{\mathcal F}}^0_A u_i \right] + \frac{1-a^{2-2\alpha}u^2}{(1+\omega)\prgv{\rho}}\prgv{{\mathcal F}}^i_A \ ,
\end{align}
where the squared modulus of the velocity field is just defined by $u^2=\sum_i u_iu_i$. The discretized versions of the two force terms are given by,
\begin{align}
\prgv{{\mathcal F}}^0_A =& \,\, \gamma \left[ \prgv{\rho}_e \sum_j \prgv{\pi}_j^A u_j + \prgv{\sigma}_f \biggl[\sum_{j,k} \prgv{\pi}_j^A(\prgv{\nabla}_j^{(0,p)}\prgv{A}_k- \prgv{\nabla}_k^{(0,p)}\prgv{A}_j) u_k - a^{\alpha-1} \sum_j (\prgv{\pi}_j^A)^2 \biggr]\right] \ , 
\\
\prgv{{\mathcal F}}^i_A =& \,\, a^{\alpha-1} \gamma \Biggr[  \prgv{\rho}_e \sum_{j}\biggr( (\prgv{\nabla}_i^{(0,p)}\prgv{A}_j - \prgv{\nabla}_j^{(0,p)}\prgv{A}_i ) u_j - a^{\alpha-1} \prgv{\pi}_i^A\biggr) \nonumber
\\
&+  \prgv{\sigma}_\f \biggl[\sum_{j,k} (\prgv{\nabla}_i^{(0,p)}\prgv{A}_j- \prgv{\nabla}_j^{(0,p)}\prgv{A}_i) (\prgv{\nabla}_j^{(0,p)}\prgv{A}_k - \prgv{\nabla}_k^{(0,p)}\prgv{A}_k)u_k + \sum_{j} u_j \prgv{\pi}_j^A \prgv{\pi}_i^A \nonumber
\\
& - a^{\alpha-1} \sum_{j} \prgv{\pi}_j^A (\prgv{\nabla}_i^{(0,p)}\prgv{A}_j - \prgv{\nabla}_j^{(0,p)}\prgv{A}_i)\biggl]\Biggr] \ ,
\end{align}
where $\prgv{\nabla}_j^{(0,p)}$ represents the neutral derivative introduced in \Sec{Sect:LatticeDerivatives}. The gamma factor $\gamma$ is obtained by inserting the velocity field at lattice sites into the corresponding expression. The Gauss law, which has to be satisfied at every time step is discretized as follows,
\be
\sum_j \prgv{\nabla}^{(0,p)}_j \prgv{\pi}_j^A =- \gamma \left(\frac{T_*}{\omega_*} \right)^4 \left(\prgv{\rho}_e -  \prgv{\sigma}_f a^{1-\alpha}\sum_j u_j \prgv{\pi}_j^A \right)\ .
\ee

\subsubsection*{Semi-collocated implementation}

In the semi-collocated formulation we assume that the gauge field lives at semi-integer values ${\bf n}+\hat{\imath}/2 $, while $\ln \prgv{\rho}$ and $u_i$ live at lattice sites $\bf n$. The discretized versions of the gauge field kernel and of the additional terms in the fluid kernels are,
\begin{align}
{\mathcal{K}}^A_i[\prgv{A}_i, \prgv{\pi}_i^A, u_i,a] = & \,\, a^{\alpha-1} \sum_j \prgv{\nabla}_j^{(-,p)}(\prgv{\nabla}_j^{(+,p)}\prgv{A}_i-\prgv{\nabla}_i^{(+,p)}\prgv{A}_j)  \nonumber \\
& + \gamma_{+i} \left(\frac{T_*}{\omega_*} \right)^4  \left[ \prgv{\rho}_e \, {\mathcal S}_{i}^{(p)} [u_i] + 
\prgv{\sigma}_f\biggl(\sum_{j}{\mathcal S}_{-j}^{(p)} [(\prgv{\nabla}_i^{(+,p)}\prgv{A}_j- \prgv{\nabla}_j^{(+,p)}\prgv{A}_i)] {\mathcal S}_{i}^{(p)}[u_j] - a^{\alpha-1} \prgv{\pi}_i^A \biggr)\right]   \ , \\
{\mathcal{G}}_0^A[\prgv{\rho},u_i,{\mathcal{F}}^\mu_A] = & \,\, \frac{1}{(1-\omega a^{2(1-\alpha)}u^2)\prgv{\rho}}\left[(1+a^{2(1-\alpha)}u^2)\prgv{{\mathcal F}}^0_A- 2 a^{2(1-\alpha)} \sum_j u_j\prgv{{\mathcal F}}^j_A\right]  \ ,  \\
{\mathcal{G}}_i^A[\prgv{\rho},u_i,{\mathcal{F}}^\mu_A] = & \,\,\frac{1-a^{2(1-\alpha)}u^2}{(1-wa^{2(1-\alpha)}u^2)\prgv{\rho}}\left[ \frac{2\omega}{1+\omega}a^{2(1-\alpha)}\sum_j u_j\prgv{{\mathcal F}}^j_A u_i-\prgv{{\mathcal F}}^0_A u_i \right] + \frac{1-a^{2(1-\alpha)}u^2}{(1+\omega)\prgv{\rho}}\prgv{{\mathcal F}}^i_A \ ,
\end{align}
where the squared modulus of the velocity field is just defined by $u^2=\sum_i u_iu_i$. The forces that appear in the fluid kernels are,
\begin{align}
\prgv{{\mathcal F}}^0_A =& \,\, \gamma \biggr[\prgv{\rho}_e \sum_{j} {\mathcal S}_{-j}^{(p)}[\prgv{\pi}_j^A]u_j + \prgv{\sigma}_f\biggl[ \sum_{j,k}  {\mathcal S}_{-j}^{(p)}[ \prgv{\pi}_j^A] {\mathcal S}_{-j,-k}^{(p)}[\prgv{\nabla}_j^{(+,p)}\prgv{A}_k- \prgv{\nabla}_k^{(+,p)}\prgv{A}_j] u_k \nonumber 
\\
&- \sum_{j} a^{\alpha-1}{\mathcal S}_{-j}^{(p)}[(\prgv{\pi}_j^A)^2]  \biggr] \biggr]  \ , 
\\
\prgv{{\mathcal F}}^i_A=& \,\,  a^{\alpha-1}\gamma  \Biggr[ \prgv{\rho}_e  \left( \sum_{j}{\mathcal S}_{-i,-j}^{(p)}[\prgv{\nabla}_i^{(+,p)}\prgv{A}_j - \prgv{\nabla}_j^{(+,p)}\prgv{A}_i] u_j-a^{\alpha-1}{\mathcal S}_{-i}^{(p)}[\prgv{\pi}_i^A]\right) \nonumber 
\\ 
&+ \prgv{\sigma}_f \Biggl(\sum_{j,k} {\mathcal S}_{-i,-j}^{(p)}[\prgv{\nabla}_i^{(+,p)}\prgv{A}_j- \prgv{\nabla}_j^{(+,p)}\prgv{A}_i]{\mathcal S}_{-j,-k}^{(p)} [\prgv{\nabla}_j^{(+,p)}\prgv{A}_k- \prgv{\nabla}_k^{(+,p)}\prgv{A}_j]u_k \nonumber
\\
& +  \sum_j  u_j \, {\cal S}_{-j}^{(p)}[\prgv{\pi}_j^A] {\cal S}_{-i}^{(p)}[\prgv{\pi}_i^A]  - a^{\alpha-1} \sum_j {\mathcal S}_{-i,-j}^{(p)}[\prgv{\nabla}_i^{(+,p)}\prgv{A}_j- \prgv{\nabla}_j^{(+,p)}\prgv{A}_i]{\mathcal S}_{-j}^{(p)}[\prgv{\pi}_j^A]\Biggr)   \Biggr] \ .
\end{align}
The Gauss law is discretized as follows,
\be
\sum_j \prgv{\nabla}^{(-,p)}_j\prgv{\pi}_j^A =- \left(\frac{T_*}{\omega_*} \right)^4 \gamma \left(\prgv{\rho}_e - \prgv{\sigma}_f a^{1-\alpha}\sum_j u_j {\cal S}_{-j}^{(p)}[\prgv{\pi}_j^A] \right)\ .
\ee
The forward/backward discrete derivative $\prgv{\nabla}_j^{(\pm,p)}$, as well as the semi-sums ${\mathcal S}_{i}^{(p)}$ and clover averages ${\mathcal S}_{i,j}^{(p)}$ to even order $p$ have been introduced in \Sec{Sect:LatticeDerivatives}. The gamma factors that appear in the above equations, i.e., $\gamma_{+i}$ and $\gamma$ are defined as,
\be
\gamma_{+i}^2\equiv\frac{1}{1- a^{2(1 - \alpha )}{\mathcal{S}}_i^{(p)}[\sum_j u_ju_j]} \ , \hspace{1cm} {\rm and} \hspace{1cm} \gamma^2\equiv\frac{1}{1- a^{2(1 - \alpha )}\sum_j u_ju_j} \ .
\ee

\subsubsection*{Staggered implementation}

In the staggered formulation we assume that all vectors, i.e. $\prgv{A}_i$ and $u_i$, live at semi-integer lattice sites ${\bf n}+\hat{\imath}/2$, while $\ln \prgv{\rho}$ lives at lattice site $\bf n$. The discretized versions of the gauge field kernel and of the additional contributions to the fluid kernels are,
\begin{align}
{\mathcal{K}}_{A_i}[\prgv{A}_i, \prgv{\pi}_i^A, u_i,a] = & \,\, a^{\alpha-1} \sum_j \prgv{\nabla}_j^{(-,p)}(\prgv{\nabla}_j^{(+,p)}\prgv{A}_i-\prgv{\nabla}_i^{(+,p)}\prgv{A}_j) \nonumber \\
& + \gamma_{+1} \left(\frac{T_*}{\omega_*} \right)^4  \left[\prgv{\rho}_e  u_i + \prgv{\sigma}_f \biggl(\sum_{j}{\mathcal S}_{-j}^{(p)}[(\prgv{\nabla}_i^{(+,p)}\prgv{A}_j- \prgv{\nabla}_j^{(+,p)}\prgv{A}_i) {\mathcal S}_{i}^{(p)}[ u_j]] - a^{\alpha-1} \prgv{\pi}_i^A \biggr)\right]   \ , 
\\
{\mathcal{G}}_0^A[\prgv{\rho},u_i,{\mathcal{F}}^\mu_A] = & \,\,\frac{1}{(1-\omega a^{2(1-\alpha)}{\mathcal S}^{(p)}[u^2])\prgv{\rho}}\left[(1+a^{2(1-\alpha)}{\mathcal S}^{(p)}[u^2]) \prgv{{\mathcal F}}^0_A- 2 a^{2(1-\alpha)} \sum_j {\mathcal S}^{(p)}_{-j}[u_j\prgv{{\mathcal F}}^j_A]\right]  \ ,  \\
{\mathcal{G}}_i^A[\prgv{\rho},u_i,{\mathcal{F}}^\mu_A] = & \,\, {\mathcal S}^{(p)}_i\left[\frac{1-a^{2(1-\alpha)}{\mathcal S}^{(p)}[u^2]}{(1-wa^{2(1-\alpha)}{\mathcal S}^{(p)}[u^2])\prgv{\rho}}\left( \frac{2\omega}{1+\omega}a^{2(1-\alpha)}\sum_j {\mathcal S}^{(p)}_{-j}[u_j\prgv{{\mathcal F}}^j_A] -\prgv{{\mathcal F}}^0_A \right)\right] u_i \nonumber \\
& + {\mathcal S}^{(p)}_i\left[\frac{1-a^{2(1-\alpha)}{\mathcal S}^{(p)}[u^2]}{(1+\omega)\prgv{\rho}}\right]\prgv{{\mathcal F}}^i_A \ ,
\end{align}
where ${\mathcal S}^{(p)}[u^2]\equiv \sum_{j}{\mathcal S}_{-j}^{(p)}[u_ju_j]$. The Lorentz force terms in the staggered formulation are given by,
\begin{align}
\prgv{{\mathcal F}}^0_A =& \,\, \gamma  \biggr[\prgv{\rho}_e \sum_j {\mathcal S}_{-j}^{(p)}[\prgv{\pi}_j^A u_j] + \prgv{\sigma}_f \biggl[\sum_{j,k} {\mathcal S}_{-j}^{(p)}[\prgv{\pi}_j^A] {\mathcal S}_{-j,-k}^{(p)}[\prgv{\nabla}_j^{(+,p)}\prgv{A}_k - \prgv{\nabla}_k^{(+,p)}\prgv{A}_j] {\mathcal S}_{-k}^{(p)}[u_k] \nonumber \\
& - a^{\alpha-1}\sum_j {\mathcal S}_{-j}^{(p)}[(\prgv{\pi}_j^A)^2] \biggr] \biggr] \ ,  
\\
\prgv{{\mathcal F}}^i_A =& \,\, a^{\alpha-1}\gamma_{+i} \Biggr[ \prgv{\rho}_e  \biggr( \sum_{j} {\mathcal S}_{-j}^{(p)}[(\prgv{\nabla}_i^{(+,p)}\prgv{A}_j - \prgv{\nabla}_j^{(+,p)}\prgv{A}_i ) {\mathcal S}_{i}^{(p)}[u_j]]-\prgv{\pi}_i^A\bigg) \nonumber 
\\
& + \prgv{\sigma}_f \Biggr(\sum_{j,k}{\mathcal S}_{-j}^{(p)}[\prgv{\nabla}_i^{(+,p)}\prgv{A}_j- \prgv{\nabla}_j^{(+,p)}\prgv{A}_i]{\mathcal S}_{i, -k}^{(p)}[ {\mathcal S}_{-j}^{(p)} [\prgv{\nabla}_j^{(+,p)}\prgv{A}_k- \prgv{\nabla}_k^{(+,p)}\prgv{A}_j]u_k] \nonumber 
\\ 
& + \sum_j{\mathcal S}_{i,-j}^{(p)}[u_j \prgv{\pi}_j^A]\prgv{\pi}_i^A 
- a^{\alpha-1} \sum_j {\mathcal S}_{-j}^{(p)}[\prgv{\nabla}_i^{(+,p)}\prgv{A}_j- \prgv{\nabla}_j^{(+,p)}\prgv{A}_i]   {\mathcal S}_{i,-j}^{(p)}[\prgv{\pi}_j^A] \Biggr) \Biggr]\ .
\end{align}
The Gauss law is discretized as follows,
\be
\sum_j \prgv{\nabla}^{(-,p)}_j \prgv{\pi}_j^A =-\gamma \left(\frac{T_*}{\omega_*} \right)^4 \left(\prgv{\rho}_e - \prgv{\sigma}_fa^{1-\alpha}\sum_j {\cal S}_{-j}^{(p)}[u_j \prgv{\pi}_j^A] \right)\ .
\ee
The forward/backward discrete derivatives $\nabla_j^{(\pm,p)}$, as well as the semi-sums ${\mathcal S}_{i}^{(p)}$ and clover averages ${\mathcal S}_{i,j}^{(p)}$ to generic even order $p$ in accuracy have been introduced in \Sec{Sect:LatticeDerivatives}. The gamma factors, i.e., $\gamma_{+i}$ and $\gamma$ are defined as,
\be
\gamma_{+i}^2\equiv\frac{1}{1- a^{2(1 - \alpha )}\mathcal{S}_i^{(p)}[{\mathcal{S}}^{(p)}[u^2]]} \ , \hspace{1cm} {\rm and} \hspace{1cm} \gamma^2\equiv\frac{1}{1- a^{2(1 - \alpha )}\mathcal{S}^{(p)}[u^2]} \ .
\label{gamma_factors_def}
\ee

\subsubsection{Large conductivity limit on the lattice}

Towards the end of \Sec{nodispcurrent} we have discussed the large conductivity limit, also known as \textit{ideal MHD}, where the second-order differential equation for the gauge field [cf.~\Eq{eq:Maxwell}] can be replaced by the induction equation [cf.~\Eq{induction}], which is a first-order differential equation. In the following, we discuss the discretization of the induction equation in both the conservation and the non-conservation form.

\subsubsection*{Conservation form}

In the large conductivity limit, the set of equations describing the fluid-gauge dynamics [cf.~\Eq{eq:MHDConservationForm}] are replaced by the following set of first-order differential equations [cf.~\Eqs{J_curlB}{induction}],
\begin{subequations}
\begin{align}
A_i'&={\mathcal{K}}_i^A[A_i,a,\prgv{T}_\pf^{00},\prgv{T}_\pf^{0i}] \ , \\
(\prgv{T}_{\rm pf}^{00})' &= \mathcal{G}^{0}_T[\prgv{T}_{\rm pf}^{00}, \prgv{T}_{\rm pf}^{0i}, \prgv{T}_{\rm pf}, a, b] = \mathcal{G}^0_\pf[\prgv{T}_{\rm pf}^{0i}]+\prgv{{\mathcal F}}^0_{\mathcal{H}}[\prgv{T}_{\rm pf}^{00},\mathcal{\prgv{H}}] + \prgv{{\mathcal F}}^0_A[\prgv{T}_\pf^{00},\prgv{T}_\pf^{0i},\prgv{A}_i,\prgv{\pi}^A_i,a] \, , \\
(\prgv{T}_{\rm pf}^{i0})'&= {\mathcal{G}}^i_T[\prgv{T}_{\rm pf}^{0i},\prgv{T}_{\rm pf}^{ij},\prgv{A}_j,a,b]= {\mathcal{G}}^i_\pf[\prgv{T}_{\rm pf}^{ij}] +  \prgv{{\mathcal F}}^i_{\mathcal{H}}[\prgv{T}_{\rm pf}^{0i},\mathcal{\prgv{H}}] + \prgv{{\mathcal F}}^i_A[\prgv{T}_\pf^{00},\prgv{T}_\pf^{0j},\prgv{A}_j,\prgv{\pi}^A_j,a] \, ,\\
b'&={\mathcal{K}}_{a}[{a,\prgv E}_\fl^\rho,{\prgv E}_\fl^p,{\prgv E}^A_K, {\prgv E}^A_G] \, ,\\
b&=a' \, , 
\end{align}
\end{subequations}
with the lattice version of the constraint equation that relates the current density to the magnetic field being [cf. \Eq{J_curlB}]
\be
\prgv{J}^i_\fl = -\left( \frac{\omega_*}{T_*}\right)^4 \left[a^{\alpha-1}\sum_j[\nabla_j F_{ji}]_\LL - (1-\alpha)a^{1-\alpha}\prgv{\HH}[A_i']_L\right] \ .
\ee
The kernels ${\cal G}^\mu_\pf$, the force term ${\cal F}_A^\mu$, the Hubble friction terms ${\cal F}_\HH^\mu$, as well as the kernel of the scale factor remain the same as discussed in the previous and current \Secs{sec:fluidLattice}{sec:fluid_bosonic}. The new gauge field kernel as well as the constraint equation are discretized in the collocated, semi-collocated and staggered formulation as we present in the following.

\subsubsection*{Collocated formulation}

In the collocated formulation we have,
\begin{align}
{\mathcal{K}}^A_{i}[A_i,a,\prgv{T}_\pf^{00},\prgv{T}_\pf^{0i}]=&{\cal A}[z_{\rm c}] \biggr[\prgv{\rho}_e\prgv{\eta}_{\rm diff} \frac{z_{\rm c}}{z_{\rm c} + \omega}
\frac{\prgv{T}^{0i}_\pf}{\prgv{T}^{00}_\pf}
+ \frac{z_{\rm c}}{z_{\rm c} + \omega} \sum_j\left(\nabla_i^{(0,p)}A_j-\nabla_j^{(0,p)}A_i\right) \frac{\prgv{T}^{0j}_\pf}{\prgv{T}^{00}_\pf} \nonumber 
\\ 
&+ a^{\alpha-1}\left(\frac{\omega_*}{T_*}\right)^4 \prgv{\eta}_{\rm diff}\sqrt{\frac{1+\omega}{z_{\rm c}+\omega}}\sum_j\nabla_j^{(0,p)}(\nabla_j^{(0,p)}A_i-\nabla_i^{(0,p)}A_j) \biggr] \, , 
\\
\prgv{J}^i_\fl =& -\left( \frac{\omega_*}{T_*}\right)^4 \left[a^{\alpha-1}\sum_j \nabla_j^{(0,p)}(\nabla_j^{(0,p)}A_i-\nabla_i^{(0,p)}A_j) - (1-\alpha)a^{1-\alpha}\prgv{\HH}[A_i']_L\right] \ ,
\end{align}
where the (comoving) program magnetic diffusivity is defined by $\prgv{\eta}_{\rm diff} \equiv 1/\prgv{\sigma}_f$. The lattice version of the $z$-function $z_{\rm c}\equiv z(r_{\rm c}^2)$ can be obtained by substituting the appropriately discretized expression of the $r^2$-ratio into \Eq{eq:zL}. The prefactor ${\cal A}[z_{\rm c}]$ and the $r^2$-ratio are here defined as,
\be
{\cal A}[z_{\rm c}]=\left[1 + (1 - \alpha)\left(\frac{\omega_*}{T_*}\right)^4 a^{1- \alpha}\prgv{\eta}_{\rm diff}\sqrt{\frac{1+\omega}{z_{\rm c}+\omega}} \prgv{\HH} \right]^{-1} \, , \qquad {\rm and} \qquad r_{\rm c}^2\equiv a^{2(1-\alpha)}\sum_i\frac{\prgv{T}^{0i}_\pf \prgv{T}^{0i}_\pf}{[\prgv{T}^{00}_\pf]^2} \,.
\ee

\subsubsection*{Semi-collocated formulation}

For the semi-collocated case, the lattice expressions can be written as,
\begin{align}
{\mathcal{K}}^A_{i}[A_i,a,\prgv{T}_\pf^{00},\prgv{T}_\pf^{0i}]=&{\cal A}[z_{\rm sc}] \biggr[\prgv{\rho}_e\prgv{\eta}_{\rm diff}\frac{z_{\rm sc}}{z_{\rm sc} + \omega} {\mathcal S}_{i}^{(p)}\left[\frac{\prgv{T}^{0i}_\pf}{\prgv{T}^{00}_\pf}\right] \nonumber
\\
& + \frac{z_{\rm sc, +i}}{z_{\rm sc, +i} + \omega}\sum_j{\mathcal S}_{-j}^{(p)}[(\nabla_i^{(+,p)}A_j-\nabla_j^{(+,p)}A_i)]{\mathcal S}_{i}^{(p)}\left[\frac{\prgv{T}^{0j}_\pf}{\prgv{T}^{00}_\pf}\right] \nonumber 
\\ 
&+ a^{\alpha-1}\left(\frac{\omega_*}{T_*}\right)^4\prgv{\eta}_{\rm diff}\sqrt{\frac{1+\omega}{z_{\rm sc}+\omega}}\nabla_j^{(-,p)}(\nabla_j^{(+,p)}A_i-\nabla_i^{(+,p)}A_j)\biggr]\, ,
\\
\prgv{J}^i_\fl =& -\left( \frac{\omega_*}{T_*}\right)^4 \left[a^{\alpha-1}\sum_j \nabla_j^{(-,p)}(\nabla_j^{(+,p)}A_i-\nabla_i^{(+,p)}A_j) - (1-\alpha)a^{1-\alpha}\prgv{\HH}[A_i']_L\right] \ ,
\end{align}
where the (comoving) program magnetic diffusivity is defined by $\prgv{\eta}_{\rm diff} \equiv 1/\prgv{\sigma}_f$. The lattice version of the $z$-function $z_{{\rm sc},+i}\equiv z(r_{{\rm sc},+i}^2)$ can be obtained by substituting the appropriately discretized expression of the $r^2$-ratio into \Eq{eq:zL}. The prefactor ${\cal A}[z_{\rm sc}]$ and the $r^2$-ratio are here defined as,
\be
{\cal A}[z_{{\rm sc}, +i}]=\left[1 + (1 - \alpha)a^{1- \alpha}\left(\frac{\omega_*}{T_*}\right)^4 \prgv{\eta}_{\rm diff}\sqrt{\frac{1+\omega}{z_{{\rm sc}, +i}+\omega}} \prgv{\HH} \right]^{-1} \, , \qquad {\rm and} \qquad r_{{\rm sc}, +i}^2\equiv a^{2(1-\alpha)} {\mathcal S}_{i}^{(p)}  \left[\frac{\sum_j\prgv{T}^{0j}_\pf\prgv{T}^{0j}_\pf}{(\prgv{T}^{00}_\pf)^2}\right] \, .
\ee

\subsubsection*{Staggered formulation}

Finally, in the staggered formulation we have
\begin{align}
{\mathcal{K}}^A_{i}[A_i,a,\prgv{T}_\pf^{00},\prgv{T}_\pf^{0j}]=&{\cal A}[z_{{\rm s}, +i}]
\biggr[\frac{z_{{\rm s},+i}}{z_{{\rm s},+i} + \omega}\prgv{\rho}_e\prgv{\eta}_{\rm diff}\frac{\prgv{T}^{0i}_\pf}{{\mathcal S}_{i}^{(p)}[\prgv{T}^{00}_\pf]} \nonumber
\\
&+ \frac{z_{{\rm s}, +i}}{z_{{\rm s}, +i} + \omega}\sum_j{\mathcal S}_{-j}^{(p)}\biggl[(\nabla_i^{(+,p)}A_j-\nabla_j^{(+,p)}A_i)\frac{\prgv{T}^{0j}_\pf}{{\mathcal S}_{j}^{(p)}[\prgv{T}^{00}_\pf]}\biggr]\nonumber 
\\
& + a^{\alpha-1}\left(\frac{\omega_*}{T_*}\right)^4\prgv{\eta}_{\rm diff} \sqrt{\frac{1+\omega}{z_{{\rm s}, +i}+\omega}}\sum_j \nabla_j^{(-,p)}(\nabla_j^{(+,p)}A_i-\nabla_i^{(+,p)}A_j)\biggr]\, ,  
\\
\prgv{J}^i_\fl =& -\left( \frac{\omega_*}{T_*}\right)^4 \left[a^{\alpha-1}\sum_j \nabla_j^{(-,p)}(\nabla_j^{(+,p)}A_i-\nabla_i^{(+,p)}A_j) - (1-\alpha)a^{1-\alpha}\prgv{\HH}[A_i']_L\right] \ ,
\end{align}
where the (comoving) program magnetic diffusivity is defined by $\prgv{\eta}_{\rm diff} \equiv 1/\prgv{\sigma}_f$. The lattice version of the $z$-function $z_{{\rm s},+i}\equiv z(r_{{\rm s},+i}^2)$ can be obtained by substituting the appropriately discretized expression of the $r^2$-ratio into \Eq{eq:zL}. The prefactor ${\cal A}[z_{\rm s}]$ and the $r^2$-ratio are here defined as,
\be
{\cal A}[z_{{\rm s}, +i}]=\left[1 + (1 - \alpha)a^{1- \alpha}\left(\frac{\omega_*}{T_*}\right)^4\prgv{\eta}_{\rm diff}\sqrt{\frac{1+\omega}{z_{{\rm s}, +i}+\omega}} \prgv{\HH} \right]^{-1} \,  , \quad {\rm and} \quad r_{{\rm s}, +i}^2\equiv a^{2(1-\alpha)} {\cal S}_{i, -j}^{(p)}\left[ \frac{\sum_{j}\prgv{T}^{0j}_\pf\prgv{T}^{0j}_\pf}{{\mathcal S}_{j}^{(p)}[(\prgv{T}^{00}_\pf)^2]}\right]\, .
\ee

\subsubsection*{Non-conservation form}

Similarly as in the conservation form, in the non-conservation form the two first-order differential equations describing the dynamics of the gauge field are replaced by the induction equation. The resulting set of equations can be brought into the following form,
\begin{subequations}
\begin{align}
A_i'&={\mathcal{K}}^A_{i}[\prgv{A}_j,u_j,a]\\
(\ln\prgv{\rho})' &= {\mathcal{G}}^{\rho}[\ln \prgv{\rho},u_i,A_i,a,b]={\mathcal{G}}_0^\pf[\ln \prgv{\rho},u_i]+{\mathcal{G}}^{\HH}_0[u^2,\prgv{\HH}]+{\mathcal G}_0^{A}[\ln \prgv{\rho},u_i,{\mathcal{F}}^\mu_A]  \, , \\
(u_i)'&={\mathcal{G}}_i^u[\ln \prgv{\rho},u_j,A_j,a,b]={\mathcal{G}}_i^\pf u[\ln \prgv{\rho},u_j] +{\mathcal{G}}^{\HH}_i[u_i,u^2,\prgv{\HH}]+{\mathcal G}_i^A[\ln \prgv{\rho},u_j,{\mathcal{F}}^\mu_A]\, , \\
b'&={\mathcal{K}}_{a}[{a,\prgv E}_\fl^\rho,{\prgv E}_\fl^p,{\prgv E}^A_K, {\prgv E}^A_G] \, ,\\
b&=a' \, .
\end{align}
\end{subequations}
with the lattice version of the constraint equation being,
\be
\prgv{J}^i_\fl = -\left( \frac{\omega_*}{T_*}\right)^4 \left[a^{\alpha-1}\sum_j[\nabla_j F_{ji}]_\LL - (1-\alpha)a^{1-\alpha}\prgv{\HH}[A_i']_L\right] \ .
\ee
The kernels of the fluid as well as the scale factor remain the same as discussed in the previous and present section. The new gauge field kernel as well as the constraint equation are discretized in the collocated, semi-collocated and staggered formulation as in the following.

\subsubsection*{Collocated formulation}

In the collocated formulation we have,
\begin{align}
{\mathcal{K}}^A_{i}[\prgv{A}_j,u_j,a]=& {\cal B}[\gamma_{\rm c} ]\biggr[\prgv{\rho}_e\prgv{\eta}_{\rm diff} u_i
+ \sum_j(\nabla_i^{(0,p)}A_j-\nabla_j^{(0,p)}A_i)u_j \nonumber 
\\ 
& +  a^{\alpha-1}\left(\frac{\omega_*}{T_*}\right)^4\frac{\prgv{\eta}_{\rm diff}}{\gamma_{\rm c}}\sum_j\nabla_j^{(0,p)}(\nabla_j^{(0,p)}A_i-\nabla_i^{(0,p)}A_j)\biggr] \, , 
\\
\prgv{J}^i_\fl =& -\left( \frac{\omega_*}{T_*}\right)^4 \left[a^{\alpha-1}\sum_j \nabla_j^{(0,p)}(\nabla_j^{(0,p)}A_i-\nabla_i^{(0,p)}A_j) - (1-\alpha)a^{1-\alpha}\prgv{\HH}[A_i']_L\right] \ ,
\end{align}
where the (comoving) program magnetic diffusivity is defined by $\prgv{\eta}_{\rm diff} \equiv 1/\prgv{\sigma}_f$. The prefactor ${\cal B}[\gamma_{\rm c}]$ and the gamma-factor are defined as,
\be
{\cal B}[\gamma_{\rm c}]=\left[1 + (1 - \alpha) a^{1- \alpha} \left(\frac{\omega_*}{T_*}\right)^4 \frac{\prgv{\eta}_{\rm diff}}{\gamma_{\rm c}}\prgv{\HH} \right]^{-1} \, , \qquad {\rm and} \qquad \gamma_{\rm c}^2\equiv\frac{1}{1- a^{2(1 - \alpha )}\sum_j u_ju_j}\, .
\ee

\subsubsection*{Semi-collocated formulation}

For the semi-collocated case, the lattice expressions can be written as,
\begin{align}
{\mathcal{K}}^A_{i}[\prgv{A}_j,u_j,a]=&{\cal B}[\gamma_{{\rm sc},+i}] \biggr[\prgv{\rho}_e\prgv{\eta}_{\rm diff} {\mathcal S}_{i}^{(p)}[u_i]
+ \sum_j{\mathcal S}_{-j}^{(p)}[(\nabla_i^{(+,p)}A_j-\nabla_j^{(+,p)}A_i)]{\mathcal S}_{i}^{(p)}[u_j] \nonumber 
\\
&+ \left(\frac{\omega_*}{T_*}\right)^4 a^{\alpha-1}\frac{\prgv{\eta}_{\rm diff}}{\gamma_{\rm sc}} \sum_j\nabla_j^{(-,p)}(\nabla_j^{(+,p)}A_i-\nabla_i^{(+,p)}A_j)\biggr] \, ,
\\
\prgv{J}^i_\fl =& -\left( \frac{\omega_*}{T_*}\right)^4 \left[a^{\alpha-1}\sum_j \nabla_j^{(-,p)}(\nabla_j^{(+,p)}A_i-\nabla_i^{(+,p)}A_j) - (1-\alpha)a^{1-\alpha}\prgv{\HH}[A_i']_L\right] \ ,
\end{align}
where the (comoving) program magnetic diffusivity is defined by $\prgv{\eta}_{\rm diff} \equiv 1/\prgv{\sigma}_f$. The prefactor ${\cal B}[\gamma_{\rm sc}]$ and the gamma-factor being,
\be
{\cal B}[\gamma_{{\rm sc}, +i}]=\left[1 + (1 - \alpha) a^{1- \alpha} \left(\frac{\omega_*}{T_*}\right)^4 \frac{\prgv{\eta}_{\rm diff}}{\gamma_{{\rm sc}, +i}}\prgv{\HH} \right]^{-1} \, , \qquad {\rm and} \qquad \gamma_{{\rm sc}, +i}^2\equiv \frac{1}{1- a^{2(1 - \alpha )}\mathcal{S}_i^{(p)}[\sum_j u_ju_j]}\, .
\ee

\subsubsection*{Staggered formulation}

Finally, in the staggered formulation we have
\begin{align}
{\mathcal{K}}^A_{i}[\prgv{A}_j,u_j,a]=&{\cal B}[\gamma_{{\rm sc},+i}] \biggr[\prgv{\rho}_e\prgv{\eta}_{\rm diff}u_i
+ \sum_j{\mathcal S}_{-j}^{(p)}[\nabla_i^{(+,p)}A_j-\nabla_j^{(+,p)}A_i]{\mathcal S}_{i,-j}^{(p)}[u_j] \nonumber 
\\
&+ \left(\frac{\omega_*}{T_*}\right)^4 a^{\alpha-1}\frac{\prgv{\eta}_{\rm diff}}{\gamma_{\rm s}}\sum_j\nabla_j^{(-,p)}(\nabla_j^{(+,p)}A_i-\nabla_i^{(+,p)}A_j)\biggr] \, , 
\\
\prgv{J}^i_\fl =& -\left( \frac{\omega_*}{T_*}\right)^4 \left[a^{\alpha-1}\sum_j \nabla_j^{(-,p)}(\nabla_j^{(+,p)}A_i-\nabla_i^{(+,p)}A_j) - (1-\alpha)a^{1-\alpha}\prgv{\HH}[A_i']_L\right] \ ,
\end{align}
where the (comoving) program magnetic diffusivity is defined by $\prgv{\eta}_{\rm diff} \equiv 1/\prgv{\sigma}_f$. The prefactor ${\cal B}[\gamma_{\rm s}]$ and the gamma-factor being,
\be
{\cal B}[\gamma_{{\rm s}, +i}]=\left[1 + (1 - \alpha)a^{1- \alpha} \left(\frac{\omega_*}{T_*}\right)^4 \frac{\prgv{\eta}_{\rm diff}}{\gamma_{{\rm s}, +i}}\prgv{\HH} \right]^{-1} \, , \qquad {\rm and} \qquad \gamma_{{\rm s}, +i}^2\equiv \frac{1}{1- a^{2(1 - \alpha )}\mathcal{S}_i^{(p)}[\sum_j {\mathcal{S}}_{-j}^{(p)}[u_ju_j]]}\, .
\ee

\subsection{Scalar-fluid dynamics on the lattice}
\label{lattice_fluid_scalar}

We now turn to the case of relativistic scalar field-fluid dynamics on the lattice in an expanding background, whose continuum formulation has been discussed in \Sec{subsec:scalar_fluid}. For the scalar field, the equation of motion is the Klein-Gordon equation in an expanding background sourced by the fluid, as given in \Eq{equations_scalar}. For the fluid we have both the conservation form, given in \Eqq{fluid_eqs_scalar}, where the fluid degrees of freedom are $\prgv{T}^{00}_{\rm pf}, \prgv{T}^{0i}_{\rm pf}$, and the non-conservation form, expressed in \Eqs{conservation_energy0}{conservation_momentum0}, where the fluid variables are chosen to be the temperature $\prgv{T}$ and the fluid velocity $u_i$. The fundamental variables are then the scalar field $\prgv{\phi}$, its conjugate momentum $\prgv{\pi}_{\phi} = \prgv{\partial}_0 \prgv{\phi}$ and either $\prgv{T}^{0\mu}_{\rm pf}$ or $u_i$ and $\prgv{T}$.

\subsubsection{Conservation form}

In the conservation form the fluid equations are the same as described in \Sec{discrete_fluid_cons_form} with additional source terms due to the interaction between the fluid and the scalar field.  
The generic lattice version of the EOM of a scalar field-fluid system in an expanding background can be written as 
\begin{subequations}
\begin{align}
    \prgv{\phi}' & = \prgv{{\pi}_{\phi}} \,, \\
    \prgv{{\pi}_{\phi}}' & = {\cal K}_{\phi} [\prgv{\phi}, \prgv{\pi_{\phi}}, \prgv{T^{00}_\pf}, \prgv{T^{0i}_\pf},a,b]\,, \\
    (\prgv{T^{00}_\pf})' &= {\cal G}^{0}_T[\prgv{T^{00}_\pf}, \prgv{T^{0i}_\pf}, \prgv{\phi}, \prgv{\pi_{\phi}},a,b] =\mathcal{G}^{0}_\pf[\prgv{T}^{0i}_\pf]+\prgv{{\mathcal F}}_{\HH}^0[\prgv{T}_{\rm pf},\mathcal{\prgv{H}}]  + \mathcal{\prgv F}^0_\phi[\prgv{\phi}, \prgv{\pi}_\phi, \prgv{T}] \,, \\
    (\prgv{T^{0i}_\pf})' &= {\cal G}^{i}_T [\prgv{T^{00}_\pf}, \prgv{T^{0j}_\pf},\prgv{\phi}, \prgv{\pi_{\phi}},a,b] = {\cal G}_\pf^i[\prgv{T}_{\rm pf}^{ij}] +\prgv{{\cal F}}^i_\HH[\prgv{T}_{\rm pf}^{0i},\prgv{\HH}] +  \mathcal{\prgv F}^i_\phi[\prgv{\phi}, \prgv{\pi}_\phi, \prgv{T}] \,, \\
    a' & = b\,, \\
    b' & = {\cal K}_a[a, \prgv{E^{\rho}_\fl}, \prgv{E^p_\fl}, \prgv{E}_K^{\phi}, \prgv{E}_G^{\phi}, \prgv{E}_V] \,, 
\end{align}
\end{subequations}
with
\begin{align}
     \mathcal{\prgv F}^0_\phi &= \,  \prgv{\pi_\phi} \biggl[\frac{\partial {\prgv V_T}}{\partial {\prgv \phi}} ({\prgv \phi}, \prgv{T}) + \prgv{\delta_{\phi}} \biggr]\,, \\
   \mathcal{\prgv F}^i_\phi
    &= \,  - a^{2(\alpha - 1)}
    \prgv{\nabla}^\LL_i {\prgv \phi} \biggl[\frac{\partial \prgv{V}_T}{\partial {\prgv \phi}} ({\prgv \phi}, \prgv{T}) +
    \prgv{\delta_{\phi}} \biggr]\, , \\
    \prgv{\delta_{\phi}} &= a^{1 - \alpha} \gamma_\LL \prgv{\eta}_\phi \left({\prgv \pi_{\phi}}+ \left[\frac{\prgv{T}^{0j}_\pf}{\prgv{T}^{00}_\pf + \prgv{p}}\right]_\LL \prgv{\nabla}_j^\LL {\prgv \phi}\right)\,,
\end{align}
where the label '$\rm L$' denotes some lattice representation of the corresponding object. The core kernels of the perfect fluid $ \mathcal{G}_\mu^{\rm pf}$ and $\mathcal{\prgv F}_\mu^{\HH}$ are the same as defined in \Sec{sec:LatticeFluidDynamicsNC}.\footnote{Another typical choice of the conjugate momentum associated to the scalar field is $\pi_\phi = a^{3-\alpha} \partial_0 \phi$, such that the Hubble friction term in the Klein-Gordon equation disappears. In the absence of scalar-fluid friction (i.e., $\eta_\phi=0$), this choice leads to a scalar field equation that can be solved using a symplectic integrator.} The temperature $\prgv{T}$ can be reconstructed as a function of $\prgv{\phi}$ and $\prgv{r}^2 \equiv a^{2(1-\alpha)} \sum_j \prgv{T}_\pf^{0i} \prgv{T}_\pf^{0i}/(\prgv{T}_\pf^{00})^2$ by inverting \Eq{scalar_fluid_conservationform_relation}. Moreover, the lattice version of the $\gamma$-factor is given by $\gamma^2 = (\prgv{T}^{00}_\pf + \prgv{p})/\prgv{w}$, where $\prgv{p}$ and $\prgv{w}$ are expressed in terms of $\prgv{\phi}$ and the reconstructed $\prgv{T}$ using \Eqq{pressure_density}. The discretized version of the kernel of the scalar field is given by
\be
\label{kernel_k_phi_fluid}
{\cal K}_{\phi} [\prgv{\phi}, \prgv{\pi_{\phi}}, a, b, \prgv{\delta}_\phi] \equiv a^{-2(1-\alpha)} \prgv{\Delta}^{(p)} \prgv{\phi} - (3-\alpha) \frac{b}{a} \prgv{\pi}_{\phi} - a^{2\alpha-4} \left(\frac{T_*^4}{\omega_*^2f_*^2}\right) \left(\prgv{V}_{{\rm eff}, \prgv{\phi}} + \prgv{\delta}_{\phi}\right) \,,
\ee
where $\prgv{\Delta}^{(p)}$ represents the Laplace operator introduced in \Eqq{eq:laplm}, while the function $\prgv{\delta}_\phi$ has a different form in the collocated and staggered formulations, as we will show below.

The kernel of the scale factor $a$ is given by,
\be
\label{scale_factor_kernel_fluidscalar}
{\cal K}_a[a, \prgv{E^{\rho}_\fl}, \prgv{E^p_\fl}, \prgv{E}_K^{\phi}, \prgv{E}_G^{\phi}, \prgv{E}_V] \equiv \frac{a^{2\alpha+1}}{3}  \left( \frac{T_*^2}{\omega_*\mpl} \right)^2 \left[\frac{2\alpha-1}{2} \prgv{E}^{\rho}_\fl - \frac{3}{2} \prgv{E}^p_\fl + (\alpha-2) \prgv E_K^{\phi} + \alpha \prgv{E}_G^{\phi} + (\alpha+1) \prgv{E}_V \right] \,.
\ee
The volume averaged fluid energy ${\prgv E}_\fl^\rho$ and pressure ${\prgv E}_\fl^p$ density contributions, the lattice expressions of the volume averaged kinetic $\prgv{E}_K^{\phi}$ and gradient $\prgv{E}_G^{\phi}$ energy density associated to the scalar field, as well as the potential energy density $\prgv{E}_V$ are given by,
\begin{align}
& {\prgv E}_\fl^\rho \equiv  \frac{1}{a^4}\left\langle {\prgv T}_\fl^{00}\right\rangle  \ , & \quad &{\prgv E}_\fl^p \equiv \frac{1}{3a^{2+2\alpha}}\Big\langle \sum_{i}{\prgv T}_\fl^{ii} \Big\rangle \ ,  
\label{fluid_contr} &&\\ 
& \prgv{E}_K^{\phi} \equiv \frac{1}{2 a^{2\alpha}} \left(\frac{\omega_*f_*}{T_*^2}\right)^2 \left\langle \prgv{\pi}_{\phi}^2 \right\rangle \,, & \quad &\prgv{E}_G^{\phi} \equiv \frac{1}{2 a^2} \left(\frac{\omega_*f_*}{T_*^2}\right)^2 \sum_{k} \left\langle (\prgv{\nabla}^\LL_k \prgv{\phi})^2 \right\rangle \,, & & \prgv{E}_V \equiv \left\langle \prgv{V}_0(\prgv{\phi}, \prgv{T}) \right\rangle \,, 
\end{align}
where $\langle \dots\rangle$ denotes a volume average over the whole lattice. In the collocated discretization scheme the $\prgv{\nabla}^\LL_k$ needs to be replaced  by the neutral derivative $\prgv{\nabla}^{(0,p)}_k$, while in the staggered formulation we use the forward derivative $\prgv{\nabla}^{(+,p)}_k$. The Hubble constraint for the scalar-fluid system is given by 
\be
b^2 = \frac{a^{2(\alpha+1)}}{3} \left( \frac{ T_*^2}{\omega_*\mpl} \right)^2 \Big[ {\prgv E}_\pf^\rho + \prgv{E}_K^{\phi} + \prgv{E}_G^{\phi} + \prgv{E}_V \Big]  \, .
\ee

\subsubsubsection*{Collocated implementation}

In the implementation in which all fields $\prgv{T}^{0\mu},\prgv{\phi}, \prgv{\pi}_{\phi}$ live at lattice sites, we have already shown in \Sec{sec:fluidLattice} how to discretize the fluid equations in the absence of a scalar field. Therefore, we only need to specify the kernel of the scalar field and the additional scalar-fluid interaction terms that appear in the fluid equations. The additional scalar-fluid interaction terms can then be expressed in the discrete version as
\begin{align}
    \prgv{\delta_{\phi}} &= a^{1 - \alpha} \gamma \, \prgv{\eta}_\phi \left({\prgv \pi_{\phi}}+
    \sum_{j=1}^3 \frac{\prgv{T}^{0j}_\pf}{\prgv{T}^{00}_\pf + \prgv{p}}
    \prgv{\nabla}_j^{(0,p)} {\prgv \phi}\right) \,, \\
     \mathcal{\prgv F}^0_\phi &= \,  \prgv{\pi_\phi} \biggl[\frac{\partial {\prgv V_T}}{\partial {\prgv \phi}} ({\prgv \phi}, \prgv{T}) + \prgv{\delta_{\phi}}\biggr]\,, \\
   \mathcal{\prgv F}^i_\phi
    &= \,  - a^{2(\alpha - 1)}
     \left[\prgv{\nabla}_i^{(+,p)} {\prgv \phi}\right] \biggl[\frac{\partial \prgv{V}_T}{\partial {\prgv \phi}} ({\prgv \phi}, \prgv{T}) +
     \prgv{\delta_{\phi}} \biggr]\,,
\end{align}
where $\gamma^2 = (\prgv{T}^{00}_\pf + \prgv{p})/\prgv{w}$.

\subsubsubsection*{Staggered implementation}

In the staggered formulation we have ${\prgv T}^{00}_\pf, \prgv{\phi}, \prgv{\pi_{\phi}}$ living at lattice sites ${\bf n}$, while $\prgv{T}^{0i}_\pf({\bf n}+\hat{\imath}/2)$ lives in between lattice sites. The pressure $\prgv{p}$ is written in terms of $\prgv{\phi}, \prgv{T}$, where $\prgv{T}$ can be reconstructed by inverting the relation [cf.~\Eq{scalar_fluid_conservationform_relation}]
\begin{align}
    a^{2(1-\alpha)} \frac{ \sum_{k=1}^3 {\cal S}_{-k} [\prgv{T}^{0k}_\pf \prgv{T}^{0k}_\pf]}{(\prgv{T}^{00}_\pf)^2} \equiv \prgv{r}^2 = 1 + \frac{\prgv{p} - \prgv{\rho}}{\prgv{T}_\pf^{00}}
    - \frac{\prgv{p} \prgv{\rho}}{(\prgv{T}_\pf^{00})^2}\,.
\end{align}
The gamma factor is given by $\gamma^2 = (\prgv{T}^{00}_\pf + \prgv{p})/\prgv{w}$ [cf.~\Eq{reconstruction}]. We then have
\begin{align}
    \prgv{\delta_{\phi}} &= a^{1 - \alpha} \gamma \,  \prgv{\eta}_\phi \left[
    \prgv{\pi_\phi} + \sum_{j=1}^3
    \frac{{\cal S}_{-j} [\prgv{T}^{0j}_\pf]}{\prgv{T}^{00}_\pf + \prgv{p}}
    \prgv{\nabla}_j^{(0,p)}  \prgv{\phi} \right]\,, \\
     \mathcal{\prgv F}^0_\phi &= \,  \prgv{\pi_\phi} \biggl[\frac{\partial {\prgv V_T}}{\partial {\prgv \phi}} ({\prgv \phi}, \prgv{T}) +  \prgv{\delta_{\phi}} \biggr]\,, \\
   \mathcal{\prgv F}^i_\phi
    &= \,  - a^{2(\alpha - 1)}
    \left[\prgv{\nabla}_i^{(+,p)} {\prgv \phi}\right] \, {\cal S}_{i}^{(p)} \biggl[\frac{\partial \prgv{V}_T}{\partial {\prgv \phi}} ({\prgv \phi}, \prgv{T}) +
    \prgv{\delta_{\phi}} \biggr]\,.
\end{align}

\subsubsection{Non-conservation form}

In the non-conservation form the fluid variables are the velocity $u_i$ and temperature $\prgv{T}$, with the continuum fluid equations given in \Eqs{conservation_energy0}{conservation_momentum0}. The generic lattice version of the EOM of a scalar field-fluid system in an expanding background can be written in terms of program variables as 
\begin{subequations}
\begin{align}
    \prgv{\phi}' & = \prgv{\pi}_{\phi} \,, \\
    \prgv{\pi}_{\phi}' & = {\cal K}_{\phi} [\prgv{\phi}, \prgv{\pi}_{\phi}, \prgv{T}, u_i, a, b]\,, \\
    \prgv{T}' &= {\cal G}^T[\prgv{T}, u_i, \prgv{\phi}, \prgv{\pi}_{\phi}, a, b] \,, \\
    u_i' &= {\cal G}_{i}^u [\prgv{T}, u_j, \prgv{\phi}, \prgv{\pi_{\phi}}, a, b] \,, \\
    a' & = b\,, \\
    b' & = {\cal K}_a[a, \prgv{E^{\rho}_\fl}, \prgv{E^p_\fl}, \prgv{E}_K^{\phi}, \prgv{E}_G^{\phi}, \prgv{E}_V] \,, 
\end{align}
\end{subequations}
with the fluid kernels in terms of generic lattice operators being
\begin{align}
    {\cal G}^T \equiv&\, -
    \frac{1}{\partial_{\prgv{T}}(\prgv{\rho} -
    a^{2(1-\alpha)}u^2_\LL \,  \prgv{p})}
    \biggl[ 
    \prgv{w} \, \prgv{\nabla}^\LL_i u_i + u_i \bigl[ \prgv{\nabla}^\LL_i \prgv{T}\,
    \partial_{\prgv{T}} + \prgv{\nabla}^\LL_i \prgv{\phi} \,  \partial_{\prgv{\phi}} \bigr]  (\prgv{\rho} - 
    \prgv{p})
    + 2 a^{2(1-\alpha)} u_i
    \mathcal{\prgv F}^i_{\rm tot} \nonumber \\
     &- \bigl( \mathcal{\prgv F}^0_{\rm tot} + {\prgv{F}^0_{\HH}}) \bigl(1 + 
    a^{2(1-\alpha)}u^2_\LL \bigr)  + \prgv{\pi}_\phi \partial_{\prgv{\phi}} (\prgv{\rho} - a^{2(1-\alpha)}u^2_\LL \,  \prgv{p}) \biggr]\,, 
    \\
    \mathcal{G}^{u}_i \equiv&\,
    \frac{u_i}{\prgv{w} \gamma_\LL^2}
    \Biggl\{
    \frac{\partial_{\prgv{T}} \prgv{p}}{\partial_{\prgv{T}} \bigl(\prgv{\rho} -
    a^{2(1-\alpha)}u^2_\LL \,  \prgv{p} \bigr)}
    \Bigl( \,
    \prgv{w}_f \, \prgv{\nabla}^\LL_j u_j + u_j \bigl[ \prgv{\nabla}^\LL_j \prgv{T} \,
    \partial_{\prgv{T}} + \prgv{\nabla}^\LL_j \prgv{\phi} \, \partial_{\prgv{\phi}} \bigr]
    \bigl(\prgv{\rho} - 
    \prgv{p} \bigr)
    + 2 \, a^{2(1 - \alpha)} u_j \mathcal{\prgv{F}}_{\rm tot}^j
    \nonumber 
    \\ 
    &+   \prgv{\pi}_\phi \, \, 
    \partial_{\prgv{\phi}} \bigl(\prgv{\rho} - a^{2(1-\alpha)}u^2_\LL \,  \prgv{p} \bigr)  \Bigr) -
    \frac{\partial_{\prgv{T}} \prgv{w} }{\partial_{\conf{T}} \bigl(\prgv{\rho} - 
    a^{2(1-\alpha)}u^2_\LL \,  \prgv{p} \bigr)} \bigl(\mathcal{\prgv{F}}_{\rm tot}^0 + \mathcal{\prgv{F}}_{\HH}^0 \bigr)
    - \prgv{\pi}_\phi 
    \partial_{\prgv \phi} \, \prgv{p} \Biggr\}
     \nonumber 
     \\
     &- \, a^{2(\alpha-1)} \frac{1}{\prgv{w}  \gamma_\LL^2} \bigl[ \prgv{\nabla}^\LL_i \prgv{T} \,\partial_{\prgv{T}} + \, \prgv{\nabla}^\LL_i \prgv{\phi} \,   
     \partial_{\prgv \phi} \bigr] \prgv{p} 
     + \frac{1}{\prgv{w}  \gamma_\LL^2}
     \bigl(\mathcal{\prgv{F}}^i_{\rm tot} +   \mathcal{\prgv{F}}_{\HH}^i\big) - u_j \prgv{\nabla}^\LL_j u_i
     \, ,
\end{align}
where the label '$\rm L$' denotes some lattice representation of the corresponding object, and $u^2_\LL = \sum_j [u_ju_j]_\LL$. In the above expressions $\prgv{\rho}, \prgv{p}$, and $\prgv{w}$ have to be written in terms of $\prgv{\phi}$ and $\prgv{T}$ using \Eqq{pressure_density}. Moreover, as $\mathcal{\prgv F}^\mu_{\rm tot}$ was already introduced in previous sections, here we only discuss the discretization of the additional terms ${\mathcal{\prgv F}}^\mu_\phi$. The scale factor kernel is the same as in \Eq{scale_factor_kernel_fluidscalar}, with the contributions in \Eqq{fluid_contr} expressed as
\begin{align}
    {\prgv E}_\fl^\rho \equiv \frac{1}{a^4}\left\langle \left(1 + (1+\omega)\gamma_\LL^2a^{2(1-\alpha)}u_\LL^2\right){\prgv \rho}_\LL \right\rangle  \ , & \qquad
{\prgv E}_\fl^p \equiv \frac{1}{3a^{2+2\alpha}}\left\langle \left(3\omega + (1+\omega)\gamma_\LL^2a^{2(1-\alpha)}u_\LL^2\right){\prgv \rho}_\LL  \right\rangle \,,
\label{scalar_fluid_scfac_contrs}
\end{align}
while the scalar field kernel is discretized in the same way as in the non-conservation form [cf.~\Eq{kernel_k_phi_fluid}].

\subsubsubsection*{Collocated implementation}

In the collocated implementation all fields live at lattice sites, hence the discretized versions of the kernels are
\begin{align}
    {\cal G}^T =&\, -
    \frac{1}{\partial_{\prgv{T}}(\prgv{\rho} -
    a^{2(1-\alpha)}u^2 \,  \prgv{p})}
    \biggl[ 
    \prgv{w} \, \prgv{\nabla}^{(0, p)}_i u_i + u_i \bigl( \prgv{\nabla}^{(0, p)}_i \prgv{T}\,
    \partial_{\prgv{T}} + \prgv{\nabla}^{(0,p)}_i \prgv{\phi} \,  \partial_{\prgv{\phi}} \bigr)  (\prgv{\rho} - 
    \prgv{p})
    + 2 a^{2(1-\alpha)} u_i
    \mathcal{\prgv F}^i_{\rm tot} \nonumber \\
     &- \bigl( \mathcal{\prgv F}^0_{\rm tot} + {\prgv{F}^0_{\HH}}) \bigl(1 + 
    a^{2(1-\alpha)}u^2 \bigr)  + \prgv{\pi}_\phi \partial_{\prgv{\phi}} (\prgv{\rho} - a^{2(1-\alpha)}u^2 \,  \prgv{p}) \biggr]\,, \\
 \mathcal{G}^{u}_i =&\,
    \frac{u_i}{\prgv{w} \gamma^2}
    \Biggl\{
    \frac{\partial_{\prgv{T}} \prgv{p}}{\partial_{\prgv{T}} \bigl(\prgv{\rho} -
    a^{2(1-\alpha)}u^2 \,  \prgv{p} \bigr)}
    \Bigl[ \,
    \prgv{w} \, \prgv{\nabla}^{(0, p)}_j u_j + u_j \bigl( \prgv{\nabla}^{(0, p)}_j \prgv{T} \,
    \partial_{\prgv{T}} + \prgv{\nabla}^{(0, p)}_j \prgv{\phi} \, \partial_{\prgv{\phi}} \bigr)
    \bigl(\prgv{\rho} - 
    \prgv{p}\bigr)
    + 2 \, a^{2(1 - \alpha)} u_j \mathcal{\prgv{F}}_{\rm tot}^j
    \nonumber 
    \\ 
    &+   \prgv{\pi}_\phi \, \, 
    \partial_{\prgv{\phi}} \bigl(\prgv{\rho} - a^{2(1-\alpha)}u^2 \,  \prgv{p}\bigr)  \Bigr] -
    \frac{\partial_{\prgv{T}} \prgv{w}}{\partial_{\conf{T}} \bigl(\prgv{\rho} - 
    a^{2(1-\alpha)}u^2 \,  \prgv{p}\bigr)} \bigl(\mathcal{\prgv{F}}_{\rm tot}^0 + \mathcal{\prgv{F}}_{\HH}^0 \bigr)
    - \prgv{\pi}_\phi 
    \partial_{\prgv \phi} \, \prgv{p}\Biggr\}
     \nonumber 
     \\
     &- \, a^{2(\alpha-1)} \frac{1}{\prgv{w} \,  \gamma^2} \bigl[ \prgv{\nabla}^{(0, p)}_i \prgv{T} \,\partial_{\prgv{T}} + \, \prgv{\nabla}^{(0,p)}_i \prgv{\phi} \,   
     \partial_{\prgv \phi} \bigr] \prgv{p}
     + \frac{1}{\prgv{w} \gamma^2}
     \bigl(\mathcal{\prgv{F}}^i_{\rm tot} +   \mathcal{\prgv{F}}_{\HH}^i\big) - u_j \prgv{\nabla}^{(0,p)}_j u_i
     \,, 
\end{align}
with $u^2=\sum_j u_ju_j$. As in the collocated formulation of the conservation form, the discretization of ${\mathcal{\prgv F}}^{\mu}_\phi$ leads to
\be
     \mathcal{\prgv F}^0_\phi = \,  \prgv{\pi_\phi} \biggl[\frac{\partial {\prgv V_T}}{\partial {\prgv \phi}} ({\prgv \phi}, \prgv{T}) + \prgv{\delta_{\phi}} \biggr]\,, \qquad {\rm and} \qquad
   \mathcal{\prgv F}^i_\phi
    = \,  - a^{2(\alpha - 1)}
    \left[\prgv{\nabla}_i^{(0,p)} {\prgv \phi}\right] \, \biggl[\frac{\partial \prgv{V}_T}{\partial {\prgv \phi}} ({\prgv \phi}, \prgv{T}) +
    \prgv{\delta_{\phi}} \biggr]\, ,
\ee
and the discretized version of the $\prgv{\delta}_\phi$-term, as well as $\gamma^2$ are given by,
\be
\prgv{\delta_{\phi}} = a^{1 - \alpha} \gamma \,  \prgv{\eta}_\phi \left[
    \prgv{\pi_\phi} + \sum_{j=1}^3u_j
    \prgv{\nabla}_j^{(0,p)}  \prgv{\phi} \right] \ , \qquad {\rm and} \qquad \gamma^2\equiv\frac{1}{1- a^{2(1 - \alpha )}\sum_j u_j u_j} \, .
\ee 
The fluid contributions to the scale factor kernel are given in \Eq{scalar_fluid_scfac_contrs} with $u_\LL^2 \to u^2, \gamma_\LL^2 \to \gamma^2$ and $\prgv{\rho}_\LL \to \prgv{\rho}$.

\subsubsubsection*{Staggered implementation}
In the staggered implementation 
$\prgv{T}, \prgv{\phi}$, and $\prgv{\pi_{\phi}}$ live at lattice sites, while $u_i({\bf n}+\hat{\imath}/2)$ lives in between lattice sites. Hence, we have
\begin{align}
    {\cal G}^T =&\, -
    \frac{1}{\partial_{\prgv{T}}(\prgv{\rho} -
    a^{2(1-\alpha)}{\cal S}^{(p)}[u^2] \,  \prgv{p})}
    \biggl[ 
    \prgv{w} \, \prgv{\nabla}^{(-, p)}_i u_i + {\cal S}^{(p)}_{-i}[u_i] \bigl( \prgv{\nabla}^{(0, p)}_i \prgv{T}\,
    \partial_{\prgv{T}} + \prgv{\nabla}^{(0,p)}_i \prgv{\phi} \,  \partial_{\prgv{\phi}} \bigr)  (\prgv{\rho} - 
    \prgv{p})\nonumber \\
    & + 2 a^{2(1-\alpha)} \sum_i {\cal S}_{-i}^{(p)} \left[u_i
    \mathcal{\prgv F}^i_{\rm tot} \right] 
     - \bigl( \mathcal{\prgv F}^0_{\rm tot} + {\prgv{F}^0_{\HH}}) \bigl(1 + 
    a^{2(1-\alpha)}{\cal S}^{(p)}[u^2] \bigr)  + \prgv{\pi}_\phi \partial_{\prgv{\phi}} (\prgv{\rho} - a^{2(1-\alpha)}{\cal S}^{(p)}[u^2] \,  \prgv{p}) \biggr]\,, \\
 \mathcal{G}^{u}_i =&\,
    \frac{u_i} {{\cal S}_i^{(p)}[\prgv{w} \gamma^2]} {\cal S}_i^{(p)}
    \Biggl\{
    \frac{\partial_{\prgv{T}} \prgv{p}}{\partial_{\prgv{T}} \bigl(\prgv{\rho} -
    a^{2(1-\alpha)}{\cal S}^{(p)}[u^2] \,  \prgv{p}\bigr)}
    \Biggl[ \,
    \prgv{w} \, \prgv{\nabla}^{(-, p)}_j u_j + {\cal S}_{-j}^{(p)}[u_j] \bigl( \prgv{\nabla}^{(0, p)}_j \prgv{T} \,
    \partial_{\prgv{T}} + \prgv{\nabla}^{(0, p)}_j \prgv{\phi} \, \partial_{\prgv{\phi}} \bigr)
    \bigl(\prgv{\rho} - 
    \prgv{p}\bigr)
    \nonumber \\ &+ 2 \, a^{2(1 - \alpha)} \sum_j {\cal S}_{-j}^{(p)} \left[u_j \mathcal{\prgv{F}}_{\rm tot}^j \right] +   \prgv{\pi}_\phi \, \, 
    \partial_{\prgv{\phi}} \bigl(\prgv{\rho} - a^{2(1-\alpha)}{\cal S}^{(p)}[u^2] \,  \prgv{p}\bigr)  \Biggr] 
    - \prgv{\pi}_\phi 
    \partial_{\prgv \phi} \, \prgv{p}
     \nonumber \\
     &-
    \frac{\partial_{\prgv{T}} \prgv{w}}{\partial_{\conf{T}} \bigl(\prgv{\rho} - 
    a^{2(1-\alpha)}{\cal S}^{(p)}[u^2] \,  \prgv{p} \bigr)} \bigl(\mathcal{\prgv{F}}_{\rm tot}^0 + \mathcal{\prgv{F}}_{\HH}^0 \bigr) \Biggr\}
    - \, a^{2(\alpha-1)} {\cal S}_i^{(p)} \Biggl[ \frac{1}{\prgv{w} \gamma^2} \bigl( \prgv{\nabla}^{(0, p)}_i \prgv{T} \,\partial_{\prgv{T}} + \, \prgv{\nabla}^{(0,p)}_i \prgv{\phi} \,   
     \partial_{\prgv \phi} \bigr) \prgv{p} \Biggr] \nonumber
     \\
     &+ \frac{1}{{\cal S}_i^{(p)}[\prgv{w} \gamma^2]}
     \bigl(\mathcal{\prgv{F}}^i_{\rm tot} +   \mathcal{\prgv{F}}_{\HH}^i\big) - {\cal S}_{i, -j}^{(p)}[u_j] \prgv{\nabla}^{(0,p)}_j u_i
     \, , 
\end{align}
where ${\cal S}^{(p)}[u^2]\equiv \sum_j {\cal S}^{(p)}_{-j}[u_ju_j]$. As in the staggered formulation of the conservation form, the discretized version of ${\mathcal{\prgv F}}^{\mu}_\phi$ can be written as,
\be
     \mathcal{\prgv F}^0_\phi = \,  \prgv{\pi_\phi} \biggl[\frac{\partial {\prgv V_T}}{\partial {\prgv \phi}} ({\prgv \phi}, \prgv{T}) + \prgv{\delta_{\phi}} \biggr]\,, \qquad {\rm and} \qquad
   \mathcal{\prgv F}^i_\phi
    = \,  - a^{2(\alpha - 1)}
    [\prgv{\nabla}_i^{(+,p)} {\prgv \phi}] \, {\cal S}_{i}^{(p)} \biggl[\frac{\partial \prgv{V}_T}{\partial {\prgv \phi}} ({\prgv \phi}, \prgv{T}) +
     \prgv{\delta_{\phi}} \biggr]\, ,
\ee
and the discretized versions of the $\prgv{\delta}_\phi$-term, as well as of $\gamma^2$, are given by,
\be
\prgv{\delta_{\phi}} = a^{1-\alpha} \gamma \,  \prgv{\eta}_\phi \left[
    \prgv{\pi_\phi} + \sum_{j=1}^3{\mathcal S}_{-j}^{(p)}\left[u_j
    \prgv{\nabla}_j^{(+,p)}  \prgv{\phi}\right] \right] \ , \qquad {\rm and} \qquad \gamma^2\equiv\frac{1}{1- a^{2(1 - \alpha )}\sum_j\mathcal{S}_{-j}^{(p)}[u_j u_j]} \, .
\ee 
The fluid contributions to the scale factor kernel are given in \Eq{scalar_fluid_scfac_contrs} with the replacements $u_\LL^2 \to {\cal S}^{(p)}[u^2], \gamma_\LL^2 \to \gamma^2$ and $\prgv{\rho}_\LL \to \prgv{\rho}$.


\newpage
\section{Initial conditions}
\label{sec:init_cond}

In the following we discuss how to set the initial conditions for a scalar-gauge-fluid system in the continuum and on the lattice. On the one hand, we will consider the possibility to initialize the different fields with fluctuations following a given power spectrum on top of a homogeneous mode. This allows to describe for example turbulent velocity or magnetic fields, or to mimic quantum vacuum fluctuations of scalar and gauge fields. On the other hand, one may also initialize the fields with a given spatial profile. This can describe, for example, scalar bubble configurations from, e.g., a first-order phase transition, or laminar flows of velocity and magnetic fields in MHD with low Reynolds numbers.

\subsection{Scalar fields}

We first discuss the initialization of scalar field components from a given power spectrum describing their fluctuations on top of a homogeneous mode. In the case of first-order differential equations, such as the fluid energy conservation equation [cf.~\Eq{eq:EnergyConservation}], this just requires an initial spectrum of field fluctuations, while for second-order differential equations, like the Klein-Gordon equation for a scalar singlet (\ref{eq:klein_gordon_singlet_FLRW}), one needs to provide a spectrum for the field fluctuations as well as for those of its conjugate momentum. In the following we will discuss the latter case, in which we need to provide two spectra (see also section~7 of the {\tt Art\,-\,I}~\cite{Figueroa:2020rrl} and section~6 of the {\tt Art\,-\,II}~\cite{Baeza-Ballesteros:2025tme} for more detailed discussions).

Setting the field fluctuations from an initial power spectrum means that the field amplitude in Fourier space is sampled from a Gaussian distribution with vanishing mean and variance given by the power spectrum.
To do this on the lattice requires a lattice version of the power spectrum in the continuum, which we have previously derived in 
\Eq{eq:LatticePowerSpectrum}. We describe the continuum variances of the field fluctuations ${\tt f}$ and their derivative ${\tt f}'$ respectively as
\be
\langle {\tt f}^2 \rangle = \int d\log k \frac{(k/a)^3}{2\pi^2} {\cal F}_{\tt f}(k/a)\, , \hspace{1cm} {\rm and} \hspace{1cm} \langle {\tt f}'^2 \rangle = a^{2\alpha}\int d\log k \frac{(k/a)^3}{2\pi^2} {\cal G}_{\tt f'}(k/a)\, , 
\ee
where we have adopted the prescription of {\tt The Art\,-\,II} (section~6) \cite{Baeza-Ballesteros:2025tme}, with the spectra being defined as ${\cal{P}_{\tt f}}(k)\equiv a^{-3}{\cal{F}}_{\tt f}(k/a)$ and ${\cal{P}_{\tt f'}}(k)\equiv a^{2\alpha-3}{\cal{G}}_{\tt f'}(k/a)$, such that they only depend on the modulus of the physical momentum $p\equiv k/a$. Using \Eqs{fourier_square_average}{eq:LatticePowerSpectrum}, the lattice equivalent expressions for the fluctuations are then,
\be
\langle | f({\tilde{ \bf n}})|^2 \rangle =  \frac{1}{\Upsilon_{|{\tilde{ \bf n}}|}}\left(\frac{N}{\delta x} \right)^3 \frac{{\cal F}_f(k/a)}{a^3}\,  , \hspace{1cm} {\rm and} \hspace{1cm}
\langle | f'({\tilde{ \bf n}})|^2 \rangle = \frac{1}{\Upsilon_{|{\tilde{ \bf n}}|}}\left(\frac{N}{\delta x} \right)^3 \frac{{\cal G}_{f'}(k/a)}{a^{3-2\alpha}}\, \label{eq:VarInitLat} ,
\ee
where the multiplicity $ \Upsilon_{| \tilde {\bf n}|}\equiv \#_{R(\tilde{{\bf n}})}/4\pi |\tilde{{\bf n}}|$ was introduced below \Eq{eq:LatticePowerSpectrum}.
In terms of program variables we may then write
\be
\langle | \prgv{f}({\tilde{ \bf n}})|^2 \rangle =  \frac{{\cal C}_*}{\Upsilon_{|{\tilde{ \bf n}}|}}\left(\frac{N}{\delta \prgv{x}} \right)^3 \frac{\prgv{{\cal F}}_{\prgv f}(\kappa/a)}{a^3}\, , \hspace{1cm} {\rm and} \hspace{1cm} \langle | \prgv{f}'({\tilde{ \bf n}})|^2 \rangle = \frac{{\cal D}_*}{\Upsilon_{|{\tilde{ \bf n}}|}}\left(\frac{N}{\delta \prgv{x}} \right)^3 \frac{\prgv{{\cal G}_{\prgv f'}}(\kappa/a)}{a^{3-2\alpha}}\, . \label{eq:VarInitLatPrgv}
\ee
where $\kappa\equiv k/\omega_*$, and $\prgv{{\cal F}}$ as well as $\prgv{{\cal G}}$ are dimensionless program power spectra. Depending on the field content, these require a different rescaling. In the case of a singlet scalar field $\prgv{\phi}=\phi/f_*$ they are given by $\prgv{{\cal F}}_\phi\equiv E_* {\cal F}_{\phi}$ and $\prgv{{\cal G}}_{\phi'}\equiv  {\cal G}_{\phi'}/M_*$, where $E_*$ and $M_*$ are two mass scales with $[E_*]=[M_*]=+1$. A standard choice is $E_*=M_*=\omega_*$, by which the factors in front of the spectra become ${\cal C}_*={\cal D}_* =(\omega_*/f_*)^2$ (see section 6.1 of \ArtII~\cite{Baeza-Ballesteros:2025tme}). For the fluid energy density $\prgv{\rho}$ or $\prgv{T}^{00}_\pf$, however, the dimensionless spectrum is given by $\prgv{{\cal F}}_f\equiv K_*^{-5} {\cal F}_f$ for $f=\{\rho, T^{00}_{\rm pf} \}$,  with $K_*$ being a mass scale with $[K_*]=+1$, and the prefactor becomes ${\cal C}_*=K_*^5\omega_*^3/T_*^8$.

With the expressions in \Eq{eq:VarInitLatPrgv} we are now able to initialize the fluctuations on the lattice. For given power spectra of the field fluctuations $\delta f$ and their derivative $\delta f'$, we can initialize the Fourier amplitude of the fluctuations at each reciprocal lattice site $\tilde{\bf n}$ as a sum of {\it left}- and {\it right-movers},
\begin{align}
\delta\prgv{f}({\tilde{\bf n}}) &= \frac{1}{\sqrt{2}} \left[\delta\prgv{f}_{1} ({\tilde{\bf n}}) + \delta\prgv{f}_{2}({\tilde{\bf n}})\right] \ , \\
\delta\prgv{f}'({\tilde{\bf n}}) &= \frac{1}{\sqrt{2}} \left[\delta\prgv{f}'_{1}({\tilde{\bf n}})+ \delta\prgv{f}'_{2}({\tilde{\bf n}})\right] \ ,
\end{align}
where each {\it mover} field is decomposed into real and imaginary parts,  $\delta\prgv{f}_{1,2}(\tilde{\bf n}) = \prgv{R}_{1,2}(\tilde{\bf n}) + \prgv{I}_{1,2}(\tilde{\bf n})$, and $\prgv{f}'_{1,2}(\tilde{\bf n}) = \prgv{ R}'_{1,2}(\tilde{\bf n}) + \prgv{I}'_{1,2}(\tilde{\bf n})$. At each site $\tilde {\bf n}$ we draw independent random realizations of the real and imaginary parts of each {\it mover}, using Gaussian distributions with vanishing mean and variances given by ${1\over2}$ times the expressions in \Eq{eq:VarInitLatPrgv}, respectively.

\subsection{Vortical and compressional vector fields}

We now discuss the initialization of vector fields $f_i$ in Fourier space from a given initial power spectrum. This can be applied, e.g., to the vector components of the fluid velocity field $u_i$ or to the fluid stress-energy components $T^{0i}_{\rm pf}$, but also to other vector fields, such as the electric ${\cal E}_i$ and magnetic field ${\cal B}_i$. We first discuss the continuum properties of such initial conditions and then turn to how to set them on the lattice.

Let us first motivate why such initialization is relevant, e.g., in the case of fluid dynamics.
In fluid dynamics a key quantity characterizing the properties of the flow is represented by the Reynolds number, which measures the ratio of nonlinear to viscous effects. Depending on its value, we can distinguish two completely different fluid dynamics regimes~\cite{MY75}:
\begin{itemize}
    \item the \textit{low Reynolds number}, or \textit{laminar}, regime, in which the motion is ordered; 
    \item the \textit{high Reynolds number}, or \textit{turbulent}, regime, in which the motion is chaotic.
\end{itemize}
In order to account for both, we can think of two different strategies for initializing the fluid variables. For the \textit{laminar} case, one possibility is to initialize the latter with specific spatial profiles. For the \textit{turbulent} case, instead, the velocity field behaves as a random variable, but with well defined statistical properties. 

In the continuum formulation, for a statistically homogeneous and isotropic helical random vector variable ${\tt f}_i$ we have the following decomposition of its two-point function in Fourier space, in terms of physical momenta (which here we make coincide with the comoving momenta $k$, as we fix $a=1$),
\begin{align}
    \langle {\tt f}_i({\bf k}){\tt f}_j^*({\bf k'}) \rangle = (2\pi)^3\delta({\bf k - k'}) \Bigl[ A(k) \, P_{ij}(\hat{{\bf k}})+ B(k) \, \hat{k}_i \hat{k}_j  + i \, C(k) \, \epsilon_{ijl} \, \hat{k}_l \Bigr] \,,
\end{align}
where the projection operator $P_{ij}$ is given by
\be 
P_{ij}(\hat{{\bf k}}) = \delta_{ij}-\hat{k}_i \hat{k}_j \ ,
\ee
with $\hat{k}_i\equiv k_i/k$ and $k\equiv |{\bf k}|$. If we assume that the spectral functions $A(k),B(k),C(k)$ are all proportional to each other, we can write,
\begin{align}
\label{cont_vector_twopointfunction}
    \langle {\tt f}_i({\bf k}){\tt f}_j^*({\bf k'}) \rangle = (2\pi)^3\delta({\bf k - k'}) \, g^2(k) \Bigl[ (1-q)(1+\vartheta^2) \, P_{ij}(\hat{{\bf k}}) + 2q \, \hat{k}_i \hat{k}_j  + i \, 2 \vartheta (1-q) \, \epsilon_{ijl} \, \hat{k}_l \Bigr] \,,
\end{align}
where $q\equiv q(k)$ measures the degree of \textit{compressibility} and $\vartheta\equiv\vartheta(k)$ 
the degree of \textit{helicity} of the fluid, such that
\be  q= \begin{cases}
    \, 1 \hspace{1cm} {\rm fully\, compressional}\  \vspace{0.1cm}\\
    \, 0 \hspace{1cm}{\rm fully\, vortical}\ 
\end{cases}, \hspace{2cm} \vartheta=
\begin{cases}
     \, 0 \hspace{1cm} {\rm non-helical}\  \vspace{0.1cm}\\
     \, 1 \hspace{1cm}{\rm fully\, helical}\ \end{cases}  \ .
\ee
Both $q(k)$ and $\vartheta(k)$ are, in general, scale dependent. In this case, we can initialize a velocity field $f_i$ describing statistically homogeneous and isotropic turbulence with the following Fourier space profile,
\begin{align}
    {\tt f}_i({\bf k}) = g_j ({\bf k}) \, g(k) \Bigl[ \sqrt{1-q} \, (\delta_{ij}-\hat{k}_i \hat{k}_j )  + \sqrt{2q} \, \hat{k}_i \hat{k}_j + i \, \vartheta \sqrt{1-q} \, \epsilon_{ijl} \, \hat{k}_l \Bigr]\,,
\end{align}
where the spectral function $g(k)$ can be chosen as desired, and the factor $g_j$ is a complex gaussian random number, such that $\langle g_i({\bf k}) \, g_j ({\bf k'}) \rangle = \delta^3({\bf k - k'})\, \delta_{ij}$ .

The above description can be used to initialize vortical or compressional vector fields in the continuum. The corresponding lattice implementation requires to replace the $\hat{k}_i$ and the projection operator by the corresponding lattice expressions. In particular, the momentum ${\bf k}$ needs to be replaced by the Fourier transform of the discrete derivative operator that is used in the corresponding discretization scheme, i.e., by the lattice momentum ${\bf k}_\LL$, which we have discussed in \Sec{Sect:LatticeDerivatives}.

In order to introduce the appropriate program variables, let us now consider a vector field $f_i$ with respective dimensionless program variable $\prgv{f}_i=f_i/F_*^p$, where $F_*$ is a mass scale with $[F_*]=+1$, while the power $p$ reflects the mass dimension of the field $f_i$, such that $\prgv{f}_i$ becomes dimensionless. For example, when considering the  velocity field $u_i$ we set $p=0$, while for the vector component of the stress-energy tensor $T^{0i}_\pf$ we set $F_*=T_*$ with $p=4$, and when we consider either the electric ${\cal E}_i$ or the magnetic field ${\cal B}_i$ we set $F_*=\omega_*$ and $p=1$. It then follows that the spectral function $g(k)$ has to be rescaled as $\prgv{g}(\kappa) = \omega_*^{-3/2} F_*^p g(k)$. 

One possible lattice initialization of a vector field in terms of program variables, characterized by a continuum two-point function as in \Eq{cont_vector_twopointfunction}, is then given by
\begin{align}
\label{fourier_vector_field_spectrum_init}
    \prgv{f}_i(\tilde{{\bf n}}) = \prgv{g}_j (\tilde{{\bf n}}) \,  \prgv{g}(\kappa) \Bigl[ \sqrt{1-q} \, (\delta_{ij}-\hat{\kappa}_{\LL,i} \hat{\kappa}_{\LL,j} )  + \sqrt{2q} \, \hat{\kappa}_{\LL,i} \hat{\kappa}_{\LL,j} + i \, \vartheta \sqrt{1-q} \, \epsilon_{ijl} \, \hat{\kappa}_{\LL,l} \Bigr]\, , 
\end{align}
where the program variables associated to the lattice momentum are given by $\kappa_{\LL,i}\equiv k_{\LL,i}/\omega_*$ and $\hat{\kappa}_{\LL,i}\equiv \kappa_{\LL,i}/\kappa$ with $\kappa\equiv |{\vec{ \kappa}_\LL}|$.
The initialization of a vector field as in \Eq{fourier_vector_field_spectrum_init} allows to have as special cases a fully compressional (curl-free) field or a fully vortical (divergence-free) field on the lattice respectively by choosing $q=1$ or $q=0$.

\subsubsection*{Example: von Kármán Spectrum}

For the cases of hydrodynamic and magnetohydrodynamic turbulence, a physically motivated choice for the spectral function $g(k)$ is given by the \textit{von Kármán} spectrum. In the \textit{inertial range} (i.e., between the energy injection scale $k_*$, which is also the peak of the spectrum, and the dissipation scale $k_D$) it follows the \textit{Kolmogorov} spectrum
\begin{align}
    g(k) = C_k \, \bar{\epsilon}^{2/3} \, k^{-5/3} \,\,\, (k_* < k < k_D)\,, 
\end{align}
while for larger scales it follows a different power law
\begin{align}
    g(k) = C_k \, \bar{\epsilon}^{2/3} \, k_*^{-5/3} \Bigl( \frac{k}{k_*} \Bigr)^{\xi} \,\,\, (k < k_*)\,,
\end{align}
where we have $\xi = 4$ (also known as \textit{Batchelor} spectrum) for a fully compressional ($q=1$) or fully vortical ($q=0$) velocity or magnetic field, while $\xi = 2$ for a mixed case ($0<q<1$). In the above spectra $C_k \sim \mathcal{O}(1) $ is a coefficient and $\bar{\epsilon}$ is the constant rate of energy transfer (from large to small scales) in the inertial range.


\newpage
\section{Gravitational waves}
\label{sec:gravitational_waves}

This section is dedicated to gravitational wave production in the context of scalar-gauge-fluid dynamics and its lattice implementation. We first discuss GWs in the continuum, including their definition, equations of motion and the description of the source terms, and then turn to the lattice formulation. Some of the results presented here will not be demonstrated in detail. For a more complete treatment, with more in depth discussions, we refer to section 8 of {\tt The\,Art\,II} \cite{Baeza-Ballesteros:2025tme}.

\subsection{Gravitational waves in the continuum}

Gravitational waves (GWs) are transverse-traceless (TT) tensor perturbations $h_{ij}$ on top of the background metric, which is, in the present context, the FLRW metric introduced in \Eq{eq:FLRW}. Including these linear tensor perturbations, the line element becomes,
\be
\dd s^2=-a^{2\alpha}(\eta)\dd\eta^2+a^2(\eta)(\delta_{ij}+h_{ij})\dd x^i \dd x^j \ .
\ee
Tensor field perturbations that are transverse and traceless fulfill the conditions $\partial_i h_{ij}=0$ and $h_{ii}=0$. The dynamics of the GWs is governed by the following EOM,
\be\label{eq:EOMGW}
h_{ij}''-a^{-2(1-\alpha)}\nabla^2h_{ij}+(3-\alpha)\frac{a'}{a}h_{ij}=\frac{2}{\mpl^2a^{2(1-\alpha)}} \Pi_{ij}^{\rm TT} \,,
\ee
where primes denote derivatives with respect to $\alpha$-time. The GWs are sourced through the transverse-traceless projection of the anisotropic tensor $\Pi_{ij}^{\rm TT}$, i.e., the part of the anisotropic tensor $\Pi_{ij}$ that fulfills the conditions $\partial_i \Pi_{ij}^{\rm TT}(\eta,{\bf x})=0$ and $\Pi_{ii}^{\rm TT}(\eta,{\bf x})=0$. The total anisotropic tensor is defined as,
\be
\Pi_{ij}\equiv T_{ij}-\langle p\rangle g_{ij} \ ,
\ee
where the background pressure $\langle p\rangle$ has been defined in \Eq{eq:Friedmann} and the metric is understood to be the perturbed FLRW metric, $g_{ij}=a^2(\delta_{ij}+h_{ij})$. 
For convenience we introduce a class of effective anisotropic tensors $\Pi_{ij}^{\rm eff}$, such that $(\Pi_{ij}^{\rm eff})^{\rm TT}={\Pi}_{ij}^{\rm TT}$. 
For a system that contains a fluid, as well as scalar and gauge fields, as defined in \Eq{full_lagrangian}, the effective anisotropic tensor is given by,
\be
\Pi_{ij}^{\rm eff}= \partial_i \phi \,  \partial_j \phi 
 - a^{-2\alpha} E_i E_j -a^{-2} B_iB_j 
 + \Pi^{{\rm eff}, \fl}_{ij} \,,
\ee
where the electric and magnetic fields are here defined as $E_i\equiv F_{0i}$ and $B_{i}=\varepsilon_{ijk}F^{jk}/2$ respectively. $\Pi_{ij}^{\rm eff, \fl}$ represents the relevant contributions from the anisotropic tensor stemming from the fluid. 
To $\Pi^{{\rm eff}, \fl}_{ij}$ can contribute both the perfect fluid stress-energy tensor $T_{ij}^{\rm pf}$ [cf.~\Eq{perf_fluid}] and the deviatoric stresses $\delta T_{ij}$ [cf.~\Eq{Pi_visc}] which constitute $T_{ij}^{\fl} = T_{ij}^{\pf} - \delta T_{ij}$ (cf.~\Sec{viscous_term}).
The imperfect fluid contributions $\Pi_{ij}^{\rm eff,\rm ipf}$ can be typically neglected in the early Universe, given the smallness of the viscous transport coefficients \cite{Arnold:2000dr}, hence we focus on the perfect fluid contributions in the following.

For example, when we consider a perfect fluid characterized by an equation of state $\omega$ (cf.~\Sec{perfect_fluids}), the effective anisotropic tensor can be written, in the conservation (C) and non-conservation (NC) form, respectively as
\begin{align}
{\rm C}:\quad \Pi_{ij}^{\rm eff, \pf}&= a^{-2\alpha}\frac{\omega}{z+\omega}\frac{\conf{T}_\pf^{0i}\conf{T}_\pf^{0j}}{\conf{T}_\pf^{00}} \, , \\
{\rm NC}:\quad \Pi_{ij}^{\rm eff, \pf} &= a^{-2\alpha} (1+\omega) \conf{\rho}_\f \gamma^2 u_i u_j \ .
\end{align}
The transverse projection of the anisotropic tensor $\Pi_{ij}$ can be obtained in both Fourier and coordinate space. In the latter, this is a non-local operation. On the other hand, in Fourier space, the transverse-traceless part of $\Pi_{ij}(\eta,{\bf k})$ can be obtained quite easily by projecting the relevant components as follows,
\be
\Pi_{ij}^{\rm TT}(\eta,{\bf k})= \Lambda_{ij,lm}(\hat{{\bf k}})\Pi_{lm}(\eta,{\bf k}) \ , \label{eq:GWProjectionOperator}
\ee
with the projection operator $\Lambda_{ij,lm}(\hat{{\bf k}})$ being defined as,
\be
\Lambda_{ij,lm}(\hat{{\bf k}})=P_{il}(\hat{{\bf k}})P_{jm}(\hat{{\bf k}})-\frac{1}{2}P_{ij}(\hat{{\bf k}})P_{lm}(\hat{{\bf k}})\ ,\hspace{1cm} {\rm with}\hspace{1cm} P_{ij}(\hat{{\bf k}})=\delta_{ij}-\hat{k}_i\hat{k}_j \ ,
\ee
where $\hat{k}_i\equiv k_i/k$ and $k\equiv |\vec{k}|$. By construction $\Pi_{ij}^{\rm TT}(\eta,{\bf k})$ fulfills the TT conditions, i.e., $k_i\Pi_{ij}^{\rm TT}(\eta,{\bf k})=0, \Pi_{ii}^{\rm TT}(\eta,{\bf k})=0$. To obtain the TT projection of the anisotropic stress tensor one has to switch to Fourier space at every time step, which is a computationally expensive task. However, since $\Pi_{ij}^{\rm TT}(\eta,{\bf k})$ is obtained by applying a linear operator on $\Pi_{ij}(\eta,{\bf k})$, and the tensor perturbations follow a linear equation (see \cite{Baeza-Ballesteros:2025tme} for a more detailed discussion), we can express the tensor perturbations $h_{ij}$ as \cite{Garcia-Bellido:2007fiu}, 
\be\label{eq:uijprojection}
h_{ij}({\bf k, \eta})= \Lambda_{ij,lm}(\hat{\bf k})v_{lm}({\bf k, \eta}) \ ,
\ee
where the components $v_{lm}({\bf k, \eta})$ are the Fourier transforms of the symmetric and traceless tensor $v_{ij}({\bf x, \eta})$, i.e., they fulfill $v_{33}=-(v_{11}+v_{22})$. Then we have that these components obey the following EOM,
\be\label{eq:EOMuij}
v_{ij}''-a^{-2(1-\alpha)}\nabla^2v_{ij}+(3-\alpha)\frac{a'}{a}v_{ij}=\frac{2}{\mpl^2a^{2(1-\alpha)}}\left(\Pi_{ij}^{\rm eff}-\frac{1}{3} \delta_{ij}\Pi_{kk}^{\rm eff}\right)\ , 
\ee
which is sourced by the traceless part of the effective anisotropic stress tensor. Instead of solving \Eq{eq:EOMGW} on the lattice, one may then evolve the five auxiliary fields  
$\{v_{11},v_{12},v_{13},v_{22},v_{23}\}$ and recover at any time step the real tensor perturbations $h_{ij}$ by applying the Fourier transform to $v_{ij}({\bf x, \eta})$ and then the projection operation in \Eq{eq:uijprojection}.

\subsubsection*{Gravitational wave observables}

Typical observables of interest are the volume averaged energy density $\rho_{\rm GW}$ and the energy density power spectrum $\Omega_{\rm GW}$ of the GW background (see \cite{Baeza-Ballesteros:2025tme} for a detailed derivation of these quantities). The average energy density of the GW background is determined by the time derivative of the tensor perturbations $h_{ij}$,
\bea
\rho_{\rm GW}(\eta)=\frac{\mpl^2}{4a^{2\alpha}}\frac{1}{V}\int_V h_{ij}'({\bf x, \eta})h_{ij}'({\bf x, \eta})\simeq \frac{\mpl^2}{4a^{2\alpha}}\frac{1}{V}\int_V \frac{d^3{\bf k}}{(2\pi)^3}h_{ij}'({\bf k, \eta})h_{ij}'^{*}({\bf k, \eta}) \equiv\int \frac{d\rho_{\rm GW}}{d\log k}d\log k \ ,
\eea
where we assume that the volume $V$ over which we take the spatial average contains all wavelengths that yield relevant contributions to the energy density. For stochastic processes one can replace the volume average by an ensemble average over random realizations and we may write \cite{Baeza-Ballesteros:2025tme},
\bea
\rho_{\rm GW}(\eta)&=&\frac{\mpl^2}{4a^{2\alpha}}\langle h_{ij}'({\bf x, \eta})h_{ij}'^{*}({\bf x, \eta})\rangle= \frac{\mpl^2}{4a^{2\alpha}}\int \frac{d^3{\bf k}}{(2\pi)^3}\frac{d^3{\bf k}'}{(2\pi)^3}e^{-i {\bf x}({\bf k}-{\bf k}')}\langle h_{ij}'({\bf k, \eta})h_{ij}'^{*}({\bf k', \eta})\rangle \nonumber \\
&=&\frac{\mpl^2}{8\pi^2 a^{2\alpha}}\int \frac{dk}{k}k^3 P_{h'}(k,\eta) \ ,
\eea
where $\langle ... \rangle$ represents an ensemble average. By assuming homogeneity and isotropy, the power spectrum of the time derivative of the tensor perturbations, i.e., $h_{ij}'$, has been defined as,
\be 
\langle h_{ij}'({\bf k, \eta})h_{ij}'^{*}({\bf k', \eta})\rangle \equiv (2\pi)^3\delta^{(3)}({\bf k}-{\bf k}')P_{h'}(k,\eta) \ .
\ee
The energy density power spectrum of the GW background is then given by,
\be
\frac{d\rho_{\rm GW}}{d\log k} = \frac{\mpl^2k^3}{8\pi^2}P_{h'}(k,\eta) \ ,
\ee
which, normalized by the critical energy density $\rho_c\equiv3\mpl^2a^{-2\alpha}{\cal H}^{2}$, gives the fractional GW energy density power spectrum
\be
\Omega_{\rm GW}\equiv \frac{1}{\rho_c}\frac{d\rho_{\rm GW}}{d{\rm log}k} = \frac{k^3}{24\pi^2{\mathcal H}^2}P_{h'}(k,\eta) \ .
\ee

\subsection{Gravitational waves on the lattice}

To investigate gravitational waves sourced by scalar-gauge-fluid dynamics with lattice simulations, we need to discretize the EOMs of the GWs and choose an appropriate evolution algorithm. As described in the previous section, we evolve the five auxiliary fields that are part of the symmetric tensor $v_{ij}({\bf x, \eta})$ and only take the projection leading to the real tensor perturbations $h_{ij}$ when we want to measure some observables. Depending on whether we are working in the collocated or semi-collocated/staggered formulation, the GWs can be interpreted as living at different locations on the lattice. While in the former the ${v}_{ij}$ live at lattice sites $\bf n$, it is more convenient to place them in the middle of the plaquette, at ${\bf n}+\hat{\imath}/2+ \hat{\jmath}/2$, in the two latter cases. This will play a relevant role in the discretization of the EOMs, and in particular of the source term, as we will see in the following.

As we have done for other fields in the lattice, it is convenient to study the fields $v_{ij}$ in terms of rescaled variables $\prgv{v}_{ij}$. Furthermore, we rescale the effective anisotropic stress tensor $\prgv{\Pi}_{ij}^{\rm eff}$ in the same way as the stress-energy tensor in \Eq{eq:FluidProgramVariables}. Both of them, $\prgv{v}_{ij}$ and $\prgv{\Pi}^{\rm eff}_{ij}$, are then defined as,
\be
\prgv{v}_{ij} = \left(\frac{\mpl}{T_*}\right)^2u_{ij} \ , \hspace{1cm} {\rm and} \hspace{1cm}  \prgv{\Pi}_{ij}^{\rm eff}=\frac{\Pi_{ij}^{\rm eff}}{T_*^4}\ .
\ee
To solve the EOMs of the $\prgv{v}_{ij}$-fields on the lattice it turns out to be convenient to define the conjugate momenta as $(\prgv{\pi}_v)_{ij}=a^{3-\alpha} \prgv{v}_{ij}'$. This allows us to bring \Eq{eq:EOMuij} into the form of a Hamiltonian scheme and write it in terms of the following two first-order differential equations, 
\begin{eqnarray}\label{eq:GWscheme}
\left\lbrace
\begin{array}{rcl}
\prgv{v}_{ij}'       &= &a^{\alpha-3} (\prgv{\pi}_v)_{ij} \, , \vspace{0.3cm}\\
(\prgv{\pi}_v)_{ij}' &= &{\mathcal{K}}_{v}[\prgv{v}_{ij},a,\prgv{\Pi}_{ij}^{\rm eff}] \, .
\end{array}
\right.
\end{eqnarray}
The kernels of the $\prgv{v}_{ij}$-fields are given by 
\be
{\mathcal{K}}_{v}^{\rm L}[\prgv{v}_{ij},a,\prgv{\Pi}_{ij}^{\rm eff}] = a^{1+\alpha}\sum_k\prgv{\nabla}_k^{(-,p)}\prgv{\nabla}_k^{(+,p)}\prgv{v}_{ij}+2a^{1+\alpha}\left(\frac{T_*}{\omega_*}\right)^2\left([\prgv{\Pi}_{ij}^{\rm eff}]_{\rm L} -\frac{1}{3} \delta_{ij}\sum_k[\prgv{\Pi}_{kk}^{\rm eff}]_{\rm L}\right)\ ,
\ee
where the label `$\rm L$' represents some lattice version of the corresponding quantity, that has to be separately specified for each discretization scheme. Before we provide the discretized versions of the kernel ${\mathcal{K}}_{v}$ in the different formulations, we quickly discuss how these equations can be numerically advanced in time. This set of equations describes the evolution of the five auxiliary fields $\{v_{11},v_{12},v_{13},v_{22},v_{23}\}$. They have to be solved together with the remaining scalar, gauge and fluid equations, for which we can use non-symplectic integrators such as the Runge-Kutta schemes, as discussed in \Sec{sec:RK}. However, Runge-Kutta algorithms require one or more auxiliary fields for every degree of freedom, which makes the evolution of the five auxiliary fields $v_{ij}$ a memory consuming task. On the other hand, the GW degrees of freedom are passive fields, i.e., they are sourced by the fluid and bosonic fields, but do not backreact on them. This circumstance allows us to evolve the five fields contained in $v_{ij}$ with a different algorithm, while the remaining fields, together with the scale factor, are still evolved by a Runge-Kutta solver. Furthermore, since the kernels ${\mathcal{K}}_{v}$ only depend on the field amplitudes (but not on the corresponding conjugate momenta), we can solve \Eqq{eq:GWscheme} with a second order accurate symplectic algorithm. These algorithms keep the deviations from exact energy conservation bounded. A prominent example of a symplectic algorithm is the Leapfrog method described in \Sec{sec:RK}. Then, we can evolve the scalar-gauge-fluid system with Runge-Kutta integrators, while solving for the gravitational waves using the Leapfrog method, as long as the two algorithms are synchronized correctly, such that the information on the anisotropic stress tensor and the scale factor are available at the relevant time steps for evolving the $v_{ij}$ fields. This procedure of combining two evolution algorithms has the advantage that the five auxiliary fields $\{v_{11},v_{12},v_{13},v_{22},v_{23}\}$ can be evolved in a more memory efficient way, since no additional auxiliary fields are required.

To solve the EOMs of the $v_{ij}$-fields on the lattice we have to discretize the kernel ${\mathcal{K}}_{v}^{\rm L}$, which has different realizations in the three different formulations, which are the collocated, semi-collocated and staggered formulations.

\subsubsection*{Collocated formulation}

In the collocated formulation the kernels of the $\prgv{v}_{ij}$-fields become
\be
{\mathcal{K}}_{v}^{\rm L}[\prgv{v}_{ij},a,\prgv{\Pi}_{ij}^{\rm eff}] = a^{1+\alpha}\prgv{\Delta}^{(p)}\prgv{v}_{ij}+2a^{1+\alpha}\left(\frac{T_*}{\omega_*}\right)^2\left([\prgv{\Pi}_{ij}^{\rm eff}]_{\rm L} -\frac{1}{3} \delta_{ij}\sum_k[\prgv{\Pi}_{kk}^{\rm eff}]_{\rm L}\right)\ ,
\ee
where $\prgv{\Delta}^{(p)}$ denotes the Laplace operator to order $p$ in accuracy introduced in \Eq{eq:laplm}. The lattice version of the rescaled effective anisotropic stress tensor for a system that contains scalars, gauge fields and perfect fluid contributions is given by,
\begin{align}
[\prgv{\Pi}_{ij}^{\rm eff}]_{\rm L}= \left( \frac{f_*^2 \omega_*^2 }{T_*^4} \right) \prgv{\nabla}^{(0,p)}_i \prgv{\phi}  \prgv{\nabla}^{(0,p)}_j \prgv{\phi}  - \left(\frac{\omega_*}{T_*} \right)^4 \left[a^{-2\alpha}\prgv{E}_i\prgv{E}_j +a^{-2}\prgv{B}_i\prgv{B}_j \right]+ [\prgv{\Pi}_{ij}^{\rm eff, pf}]_{\rm L}  \ .
\end{align}
In the conservation form, the perfect fluid part of the effective anisotropic stress tensor is given by,
\be
[\prgv{\Pi}_{ij}^{\rm eff, pf}]_{\rm L} = a^{-2\alpha}\frac{\omega}{z+\omega}\frac{\prgv{T}_\pf^{0i}\prgv{T}_\pf^{0j}}{\prgv{T}_\pf^{00}} \, ,
\ee
where $z$, being a function of $r^2 \equiv a^{2(1-\alpha)}\sum_i(\prgv{T}_{\rm pf}^{0i}/\prgv{T}_{\rm pf}^{00})^2$, is obtained from \Eq{eq:zL} by replacing $\prgv{T}_{\rm pf}^{00}$ and $\prgv{T}_{\rm pf}^{0i}$ by their lattice counterparts at lattice site $\bf n$.

In the non-conservation form the discretized version of the effective anisotropic stress tensor of the perfect fluid has the form,
\be
[\prgv{\Pi}_{ij}^{\rm pf,eff}]_{\rm L} = a^{-2\alpha}(1+\omega)\gamma^2 \prgv{\rho} u_i u_j  \ , \qquad {\rm with} \qquad \gamma^2=\frac{1}{1-a^{2(1-\alpha)}\sum_j u_ju_j} \ .
\ee

\subsubsection*{Semi-collocated/Staggered formulation}

In the semi-collocated (SC) and staggered (ST) formulations the kernels of the $v_{ij}$-fields become
\be
{\mathcal{K}}_{v}^{\rm L}[\prgv{v}_{ij},a,\prgv{\Pi}_{ij}^{\rm eff}] = a^{1+\alpha}\prgv{\Delta}^{(p)}\prgv{v}_{ij}+2 \, a^{1+\alpha}\left(\frac{T_*}{\omega_*}\right)^2\biggl([\prgv{\Pi}_{ij}^{\rm eff}]_{\rm L} -\frac{1}{3} \delta_{ij}{\cal S}_{i,j}^{(p)}\biggl[\sum_k{\cal S}_{-k}^{(p)}[[\prgv{\Pi}_{kk}^{\rm eff}]_{\rm L}]\biggr]\biggr)  \ ,
\ee
with the lattice version of the rescaled effective anisotropic stress tensor for a system that contains scalars, gauge fields and perfect fluid contributions being
\be
[\prgv{\Pi}_{ij}^{\rm eff}]_{\rm L} =  \left( \frac{f_*^2 \omega_*^2 }{T_*^4} \right) \prgv{\nabla}^{(+,p)}_i \prgv{\phi}  \prgv{\nabla}^{(+,p)}_j \prgv{\phi}  - \left(\frac{\omega_*}{T_*} \right)^4 \left[a^{-2\alpha}\prgv{E}_i\prgv{E}_j +a^{-2}\prgv{B}_i\prgv{B}_j \right] + [\prgv{\Pi}_{ij}^{\rm eff, pf}]_{\rm L}  \ .
\ee
In the semi-collocated and staggered formulation of the conservation form, the discretized versions of the perfect fluid part of the effective anisotropic stress tensor are given by,
\begin{align}
{\rm SC}: & \,\, [\prgv{\Pi}_{ij}^{\rm eff, pf}]_{\rm L} = a^{-2\alpha}\frac{\omega}{z_{{\rm sc}, (ij)}+\omega}{\mathcal S}_{i,j}^{(p)}\left[\frac{\prgv{T}_\pf^{0i}\prgv{T}_\pf^{0j}}{\prgv{T}_\pf^{00}} \right] \  , \\
{\rm ST}: & \,\, [\prgv{\Pi}_{ij}^{\rm eff, pf}]_{\rm L} =a^{-2\alpha}\frac{\omega}{z_{{\rm st},(ij)}+\omega}\frac{{\cal S}_{j}^{(p)} [\prgv{T}_\pf^{0i}] {\cal S}_{i}^{(p)}[\prgv{T}_\pf^{0j}]}{{\mathcal S}_{i,j}^{(p)}[\prgv{T}_\pf^{00}]}  \  \, ,
\end{align}
where $z_{{\rm sc}, (ij)} \equiv z(\prgv{r}^2_{{\rm sc}, (ij)})$ and $z_{{\rm st}, (ij)} \equiv z(\prgv{r}^2_{{\rm st}, (ij)})$ can be obtained from \Eq{eq:zL} with the corresponding $r^2$-ratios being given by 
\be
\prgv{r}^2_{{\rm sc}, (ij)}= {\mathcal S}_{i,j}^{(p)}\left[\frac{\sum_{k=1}^{3}\prgv{T}_\pf^{0k}\prgv{T}^{0k}}{[\prgv{T}_\pf^{00}]^2}\right] \, , \hspace{1cm} {\rm and} \hspace{1cm} \prgv{r}^2_{{\rm st}, (ij)} = {\mathcal S}_{i,j}^{(p)} \biggl[ \frac{\sum_{k=1}^{3} {\cal S}_{-k}^{(p)} \prgv{T}_\pf^{0k}\prgv{T}_\pf^{0k}}{[\prgv{T}^{00}]^2} \biggr] \ . 
\ee
In the semi-collocated and staggered formulation of the non-conservation form, the discretized versions of the effective anisotropic stress tensor associated to the perfect fluid can be written as,
\begin{align}
{\rm SC}: & \,\, [\prgv{\Pi}_{ij}^{\rm eff, pf }]_{\rm L} = a^{-2\alpha}(1+\omega)\gamma_{{\rm sc}, (ij)}^2 {\mathcal S}_{i,j}^{(p)}[\prgv{\rho} \,  u_i u_j] \, , \qquad\qquad {\rm with } \quad \gamma_{{\rm sc}, (ij)}^2\equiv\frac{1}{1-a^{2(1-\alpha)}{\mathcal{S}}_{i,j}^{(p)}[\sum_k u_k u_k]} \, , \\
{\rm ST}: & \,\,  [\prgv{\Pi}_{ij}^{\rm eff, pf}]_{\rm L} =a^{-2\alpha}(1+\omega)\gamma_{{\rm st}, (ij)}^2 {\mathcal S}_{i,j}^{(p)}[\prgv{\rho}] {\cal S}_j [u_i] {\cal S}_i [u_j] \, , \quad {\rm with } \quad \gamma_{{\rm st}, (ij)}^2\equiv\frac{1}{1-a^{2(1-\alpha)}{\mathcal{S}}_{i,j}^{(p)}[\sum_k{\mathcal{S}}_{-k}^{(p)}[u_k u_k]] } \ .
\end{align}

\subsubsection*{Tensor perturbations and gravitational wave observables on the lattice}

To extract observables, such as the average energy density of the GW background $\rho_{\rm GW}$ or the energy density power spectrum $\Omega_{\rm GW}$, from the lattice, one needs to recover the tensor perturbations $h_{ij}$. To obtain the true tensor perturbations $h_{ij}$, we need to Fourier transform the components of the $v_{ij}$-tensor and apply the transverse-traceless projection as described in \Eq{eq:uijprojection}. Similarly, the time derivative of the tensor perturbations $h_{ij}'$ can be obtained by projecting the conjugate momenta. On the lattice this can be realized as follows, 
\bea
h_{ij}&=& \Lambda_{ij,lm}^{\rm L}v_{lm}=\left(\frac{T_*}{\mpl}\right)^2\Lambda_{ij,lm}^{\rm L}\prgv{v}_{lm}\ , \\
h_{ij}'&=&\frac{\omega_*}{a^{3-\alpha}}\Lambda_{ij,lm}^{\rm L}(\pi_v)_{lm}=\frac{\omega_*}{a^{3-\alpha}}\left(\frac{T_*}{\mpl}\right)^2\Lambda_{ij,lm}^{\rm L}(\prgv{\pi}_v)_{lm} \, .
\eea
The lattice version of the projection operator $\Lambda_{ij,lm}$ that has been introduced (\ref{eq:GWProjectionOperator}), has different realizations depending on whether we use neutral (`0') or forward/backward (`$\pm$') derivatives, and depending on the accuracy of the derivative operator. Using the lattice momenta ${\bf k}_{\rm L}^0$ and ${\bf k}_{\rm L}^\pm$ that have been introduced in \Sec{Sect:LatticeDerivatives}, the lattice transverse-traceless projection operators can be written as,
\bea
\Lambda_{ij,lm}^{{\rm L},0}=&P_{il}^{{\rm L},0}P_{jm}^{{\rm L},0}-\frac{1}{2}P_{ij}^{{\rm L},0}P_{lm}^{{\rm L},0}\ ,\hspace{1cm} &{\rm with}\hspace{1cm} P_{ij}^{{\rm L},0}=\delta_{ij}-\frac{k_{{\rm L},i}^0 k_{{\rm L},j}^0}{|{\bf k}_{\rm L}^0|^2} \ , \\
\Lambda_{ij,lm}^{\rm L,\pm}=&P_{il}^{\rm L,\pm}P_{jm}^{\rm L,\pm}-\frac{1}{2}P_{ij}^{\rm L,\pm}P_{lm}^{\rm L,\pm}\ ,\hspace{1cm} &{\rm with}\hspace{1cm} P_{ij}^{\rm L,\pm}=\delta_{ij}-\frac{(k_{{\rm L},i})^{\pm} k_{{\rm L},j}^\pm}{|{\bf k}_{\rm L}^\pm|^2} \ .
\eea
After applying the TT-projection to the conjugate momentum of the $v_{ij}$-tensor, the average energy density of the GWs on the lattice is given by,
\be
\rho_{\rm GW}= \frac{\mpl^2}{4a^{2\alpha}N^6}\sum_l 4\pi l^2 \langle h'_{ij} h_{ij}'^*\rangle_{R(l)} = \sum_l \left\{ \frac{\mpl^2\delta x^6}{8\pi^2a^{2\alpha}L^3}k^3(l) \langle h'_{ij} h_{ij}'^*\rangle_{R(l)} \right\} \Delta\log k \ ,
\ee
where the summation goes over spherical bins centered at $k(l)=k_{\rm IR} l$ with regular width $\Delta k \equiv k_{\rm IR}$, such that the approximate amount of points in bin $l$ is given by $\#_l =4\pi l^2$ for $l=1,2,3,...$ (for a discussion on different multiplicities $\#_l$ see \cite{Baeza-Ballesteros:2025tme}). 
Furthermore $\langle (...) \rangle_{R(l)}\equiv (1/\#_l)\sum_{\tilde{{\bf n}}\in R(l)} (...)$ represents an angular average over the spherical shell $R(l)$, and $\Delta \log k\equiv k_{\rm IR}/k(l)$. The GW energy density power spectrum on the lattice is defined as,
\be
\left(\frac{d\rho_{\rm GW}}{d \log k}\right)(l)=\frac{\mpl^2k(l)^3}{8\pi^2a^{2\alpha}L^3}\left\langle \left[\delta x^3 h'_{ij}\right]\left[\delta x^3 h'_{ij}\right]^*\right\rangle_{R(l)} \ ,
\ee
which corresponds to the {\tt Type II} lattice spectrum (for other spectra types see section 8.2 of \cite{Baeza-Ballesteros:2025tme}). Normalizing by the critical energy density $\rho_c\equiv3\mpl^2a^{-2\alpha}{\cal H}^2$ we obtain the fractional GW energy density power spectrum 
\be
\Omega_{\rm GW}(\tilde{\bf n}, \eta)=\frac{1}{\rho_c}\frac{\mpl^2k^3(l)}{8\pi^2a^{2\alpha}}\frac{\delta x^3 }{N^3}\left\langle [h'_{ij}(\tilde{\bf n}', \eta)] [h_{ij}'(\tilde{\bf n}', \eta)]^*\right\rangle_{R(\tilde{\bf n})} \ .
\ee

\newpage
\section{Summary and outlook}

In the current era of precision observational cosmology we are witnessing an increasing necessity for accurate predictions from non-linear field dynamics in early Universe scenarios. The numerical techniques to simulate such scenarios can be referred to as {\bf Lattice Cosmology Techniques} (LCT). While the basic aspects of LCT applied to scalar-singlet and scalar-gauge theories have been presented in {\tt The Art-I} \cite{Figueroa:2020rrl}, and extended to beyond canonical cases in {\tt The Art-II} \cite{Baeza-Ballesteros:2025tme}, the current monograph, which we refer to as {\tt The Art-III}, introduces fluid dynamics, with couplings to both scalar and gauge sectors. This extension of LCT to fully coupled scalar-gauge-fluid dynamics allows for the exploration of physical scenarios where the primordial plasma, modelled as a fluid, plays a fundamental role in the overall dynamics. Primary examples of such scenarios include cosmological phase transitions in the early Universe, and the magnetohydrodynamic (MHD) evolution of primordial magnetic fields across cosmic history.

In this monograph, we discuss lattice implementations of field dynamics in an expanding FLRW background, including: ({\bf i}) fully relativistic perfect fluid dynamics, describing a plasma of particles in local thermal equilibrium (LTE); ({\bf ii}) first-order imperfect fluids, including the leading order effects when considering deviations from LTE; ({\bf iii}) fully relativistic fluids coupled to a U(1) gauge field, including as a special case that of MHD, both in the finite and infinite conductivity limits; ({\bf iv}) fully relativistic fluids coupled to a scalar field, including both equilibrium and first-order out-of-equilibrium interactions between the two sectors.

We now briefly summarize the content of the present work. In \Sec{sec:cont} we introduce scalar-gauge-fluid dynamics in the continuum, first in flat spacetime (cf.~\Sec{subsec:FlatSpace_Dynamics_Cont}), and then generalized to an expanding FLRW background through minimal gravitational coupling (cf.~\Sec{subsec:FLRW_Dynamics_Cont}). We describe in detail first the case of fully relativistic fluid dynamics, including both perfect (cf.~\Sec{perfect_fluids}) and first-order imperfect fluids (cf.~\Sec{viscous_term}), and we then extend the treatment to include interactions of the fluid with IR bosonic degrees of freedom, covering the two fundamental cases of the interactions with a U(1) gauge field (cf.~\Sec{subsec:gauge_fluid_dyns}) and with a scalar field (cf.~\Sec{subsec:scalar_fluid}). We then review, in \Sec{sec:generalLattice}, the basic lattice techniques required for the discretization of the scalar-gauge-fluid system, starting with a basic introduction to the lattice (cf.~\Sec{basic_lattice}), and including high-order numerical schemes for both spatial (cf.~\Secs{Sect:LatticeDerivatives}{sec:LatticeOperators}) and timestepping (cf.~\Sec{sec:RK}) discretization. \SSecs{sec:fluidLattice}{sec:fluid_bosonic} represent the core of the monograph, respectively describing the lattice formulations of pure perfect and imperfect fluid dynamics, and of fluid dynamics coupled to either a U(1) gauge field (cf.~\Sec{sec:fluid_gauge}) or to a scalar field (cf.~\Sec{lattice_fluid_scalar}). The discrete formulation of the scalar-gauge-fluid system is presented considering both the conservation and non-conservation forms, in which the fundamental fluid variables are respectively the $T^{0\mu}_{\rm pf}$ components of the perfect fluid stress-energy tensor or the fluid velocity $u_i$ and, either the fluid energy density $\rho_\f$ or the temperature $T$. Moreover, considering scalar field components as always living at lattice sites, while vector field components as living either at lattice sites or in between lattice sites, we introduce both the collocated and staggered formulations, featuring different properties, and whose implementations require different discretizations. In \Sec{sec:init_cond} we briefly deal with how to set up initial conditions in Fourier space for scalar and vector field components in the scalar-gauge-fluid system, considering a given power spectrum and a given tensorial structure. Finally, in \Sec{sec:gravitational_waves}, we deal with the main properties of Gravitational Waves on the lattice, focusing on the source terms coming from the field content described in the present work.

The techniques presented in this monograph constitute the theoretical basis for the scalar-gauge-fluid module that will be publicly released, after publication of the present monograph, as part of 
{${\mathcal C}$osmo${\mathcal L}$attice}\,\,{\tt v3.0}. For updates on this, please check the ${\mathcal C}$osmo${\mathcal L}$attice website {\href{http://www.cosmolattice.com}{\color{blue}http://www.cosmolattice.com}}. Thanks to the recent developments incorporated in {${\mathcal C}$osmo${\mathcal L}$attice}\,\,{\tt v2.0} \cite{Baeza-Ballesteros:2026uao, Florio:2026vde},  which substantially extended the physics scope and computational capabilities of the code (including, e.g., non-symplectic integrators and GPU acceleration), 
{${\mathcal C}$osmo${\mathcal L}$attice}\,\,{\tt v3.0} will incorporate these upgrades, providing a user-friendly, powerful and modular C++ MPI-based code, fully prepared to perform numerical simulations of perfect or imperfect fluids, in isolation or coupled to scalar or gauge fields, with either CPU or GPU parallelization.

\newpage
\section*{Acknowledgments}

The authors are grateful for the support provided by Nordita (Stockholm, Sweden) during the program
``\href{https://indico.fysik.su.se/event/8805/}{\blue{\em Numerical Simulations of Early Universe Sources of Gravitational Waves}}'', by the Institute of Basic Sciences (Daejeon, South Korea) and the local organizers of the \blue{\href{https://indico.ific.uv.es/event/8110/}{CosmoLattice School 2025}} (Mohammad Ali Gorji, Dong-Won Jung, and Masahide Yamaguchi), where preliminary results of this work were presented, and by CERN (Geneva, Switzerland) during the ``\href{https://indico.cern.ch/event/1548935/}{\blue{\em 1st Pencil Code school on early Universe physics and gravitational waves,}}'' organized by ARP and ASM.

The work of DGF (0000-0002-4005-8915) is supported by the grants PROMETEO/2021/083, EUR2022-134028, PRTR-C17.I01, and ASFAE/2022/020. DGF and KM acknowledge common support from PID2023-148162NB-C22. KM acknowledges support from the Swiss National Science Foundation
(project number P500PT-214466).
ARP and ASM are supported through the Swiss National Science Foundation SNSF Ambizione grant \href{https://data.snf.ch/grants/grant/208807}{208807}.

Computations for this work were performed on the {\tt Graviton} cluster of the SOM group at the Instituto de Física Corpuscular (IFIC), the FinisTerrae III cluster at Centro de Supercomputación de Galicia (CESGA), the Lluis Vives and Tirant II clusters at the University of Valencia, and the {\tt Baobab} HPC cluster at the University
of Geneva.


\newpage 
\appendix

\section{Derivation of the fluid equations of motion in flat spacetime}
\label{app:EOM}

Here we provide a detailed derivation of the fluid equations of motion (EOM)
in flat spacetime, which were presented in
\Sec{subsec:FlatSpace_Dynamics_Cont} [see \Eq{Tmunu_cons_aux}]. 
The total stress-energy tensor of the scalar-gauge-fluid system [see \Eq{eq:StressEnergyTensor}]
can be written as a sum of contributions from the fluid, the gauge,
and the scalar sectors
\begin{align}
    T^{\mu \nu} = T^{\mu \nu}_\fl + T^{\mu \nu}_{U(1)} + T^{\mu \nu}_{SU(N)} + T^{\mu \nu}_{\rm scalar} \,.
\end{align}
The contributions from the Abelian and non-Abelian gauge fields are respectively given by
\begin{subequations}
\begin{align}
    T^{\mu \nu}_{U(1)} &= - \frac{1}{4} \eta^{\mu \nu} F_{\lambda \sigma} F^{\lambda \sigma} + F^{\mu \lambda} F^{\nu}_{\ \, \lambda} \,, \\
    T^{\mu \nu}_{SU(N)} &= - \frac{1}{4} \eta^{\mu \nu} G_{\lambda \sigma}^a G^{\lambda \sigma}_a + G^{\mu \lambda}_a G^{\nu}_{a \, \lambda}\,.
\end{align}
\end{subequations}
For the scalar fields, we have a separate kinetic term for each of them and a common potential $V_0(\phi, |\varphi|, |\Phi|)$, which can in principle include interaction terms between them. Hence, we can write the stress-energy tensor for the scalar sector as  
\begin{align}
\label{t_mu_nu_scalars}
    T^{\mu \nu}_{\rm scalar} = T^{\mu \nu}_{\phi, {\rm kin}}+T^{\mu \nu}_{\varphi, {\rm kin}}+T^{\mu \nu}_{\Phi, {\rm kin}}-\eta^{\mu \nu}\, V_0(\phi, |\varphi|, |\Phi|)\,,
\end{align}
with the kinetic terms respectively given by
\begin{subequations}
\begin{align}
    T^{\mu \nu}_{\phi, {\rm kin}} &= -\frac{1}{2} \eta^{\mu \nu} \partial_{\lambda} \phi \, \partial^{\lambda} \phi +
    \partial^{\mu} \phi \, \partial^{\nu} \phi\,, \\
    T^{\mu \nu}_{\varphi, {\rm kin}} &= - \eta^{\mu \nu} (D_{\lambda}^A \varphi)^{*} (D_A^{\lambda} \varphi) + \bigl[ (D^{\mu}_A \varphi)^* (D^{\nu}_A \varphi) + c.c. \bigr] \,, \\
    T^{\mu \nu}_{\Phi, {\rm kin}} &= - \eta^{\mu \nu} (D_{\lambda} \Phi)^{\dagger} (D^{\lambda} \Phi) + \bigl[ (D^\mu_A \Phi)^{\dagger} (D^{\nu}_A \Phi) + c.c. \bigr] \,.
\end{align}
\end{subequations}
The fluid EOM are computed from the conservation of the total stress-energy tensor as
\begin{align} \label{aux_eom}
    \partial_\mu T^{\mu \nu} = 0 \ \ \ \Leftrightarrow \ \ \ 
    \partial_\mu T^{\mu \nu}_\fl = -\partial_\mu T^{\mu \nu}_{U(1)} - \partial_\mu T^{\mu \nu}_{SU(N)} - \partial_{\mu}T^{\mu \nu}_{\rm scalar}\,.
\end{align}
For clarity, let us separately compute the divergence of the stress-energy tensor from each bosonic sector.
In first place, for the $U(1)$ Abelian gauge field,
\begin{align} \label{term_Abelian}
    \partial_\mu T^\mn_{U(1)} & = - \frac{1}{4} \partial^\nu
    (F_{\lambda \sigma} F^{\lambda \sigma}) + \partial_\mu (F^{\mu\lambda}
    F^{\nu}_{\ \, \lambda}) = (\partial_\mu 
    F^{\mu\lambda}) F^{\nu}_{\ \, \lambda} = - J^\lambda F^{\nu}_{\ \, \lambda} \,,
\end{align}
where we have used the Bianchi
identities, $\partial^\mu F^{\nu \lambda} + \partial^\nu F^{\lambda \mu} + \partial^\lambda F^{\mu \nu} = 0$, together with the antisymmetry of the
field strength 
to obtain
\begin{align}
    - \frac{1}{4} \partial^{\nu}  (F_{\lambda \sigma} F^{\lambda \sigma}) + F^{\mu \lambda} \partial_\mu F^{\nu}_{\, \, \, \lambda}
     &= - \frac{1}{2} F_{\mu \lambda} \partial^\nu F^{\mu \lambda} + F_{\mu \lambda} \partial^{\mu} F^{\nu \lambda}
     = \frac{1}{2} F_{\mu \lambda} \bigl( \partial^{\mu} F^{\lambda \nu} +  \partial^{\lambda} F^{\nu \mu} + 2\, \partial^{\mu} F^{\nu \lambda} \bigr) \nonumber \\
     & = \frac{1}{2} F_{\mu \lambda} \bigl[ \partial^{\mu} F^{\nu \lambda} + (\mu \leftrightarrow \lambda) \bigr] = 0\,,
\end{align}
from which \Eq{term_Abelian} follows, once we include the Abelian gauge field
EOM [\Eq{scalar_fluid_gauge_minkowski1}],
$\partial_\mu F^{\mu \lambda} = -J^{\lambda}$.
Let us follow a similar procedure for the non-Abelian gauge fields,
\begin{align} \label{term_nonAbelian}
    \partial_\mu T^\mn_{SU(N)} & 
    =  - \frac{1}{4} \partial^\nu
    (G_{\lambda \sigma}^a G^{\lambda \sigma}_a) + \partial_\mu (G^{\mu\lambda}_a
    G^{\nu}_{a \, \lambda}) 
    =  - \frac{1}{4} \partial^\nu
    (G_{\lambda \sigma}^a G^{\lambda \sigma}_a) + G^{\mu\lambda}_a
    (\partial_\mu G^{\nu}_{a \, \lambda}) + 
    (\partial_\mu G^{\mu\lambda}_a) G^{\nu}_{a \, \lambda}
    \nonumber \\
    &
    = - \frac{1}{4} \partial^\nu
    (G_{\lambda \sigma}^a G^{\lambda \sigma}_a) + G^{\mu\lambda}_a
    ({\cal D}_\mu)_{ab} G^{\nu}_{b \, \lambda} +  [({\cal D}_\mu)_{ab} G^{\mu\lambda}_b] G^{\nu}_{a \, \lambda}  + f_{abc} \, C_{\mu}^c  (G^{\mu \lambda}_a G^{\nu}_{b \, \lambda} + G^{\mu \lambda}_{b} G^{\nu}_{a \,\lambda})
    \nonumber \\
    & = - \frac{1}{4} \partial^\nu
    (G_{\lambda \sigma}^a G^{\lambda \sigma}_a) + G^{a}_{\mu \lambda}
    ({\cal D}^\mu)_{ab} G^{\nu \lambda}_b +  [({\cal D}_\mu)_{ab} G^{\mu\lambda}_b] G^{\nu}_{a \, \lambda} = -
    {\cal J}_a^{\lambda} \, G^{\nu}_{a \, \lambda} \,.
\end{align}
In the intermediate step
we have used the antisymmetry of the $SU(N)$ structure constants $f_{abc}$
and in the last one, we have used the non-Abelian gauge field EOM [\Eq{scalar_fluid_gauge_minkowski2}],
$({\cal D}_\mu)_{ab} G^{\mu\lambda}_b = -{\cal J}^\lambda_{a}$, and the corresponding Bianchi identities
for non-Abelian gauge fields,
$(\mathcal{D}^{\mu})_{ab} G_b^{\nu \lambda} + (\mathcal{D}^{\nu})_{ab} G_b^{\lambda \mu} + (\mathcal{D}^{\lambda})_{ab} G_b^{\mu \nu}=0$, together with the antisymmetry of the field strengths, leading to the following identity
\begin{align}
    - \frac{1}{4} \partial^{\nu}  (G_{\lambda \sigma}^a G^{\lambda \sigma}_a)
     & + G_{\mu \lambda}^a (\mathcal{D}^\mu)_{ab} G^{\nu \lambda}_{b} 
   = - \frac{1}{2} G_{\mu \lambda}^a (\mathcal{D}^{\nu})_{ab}  G^{\mu \lambda}_b + G_{\mu \lambda}^a (\mathcal{D}^\mu)_{ab} G^{\nu \lambda}_{b} \nonumber \\
    &= \frac{1}{2}  G_{\mu \lambda}^a  \bigl[ (\mathcal{D}^\mu)_{ab} G^{\lambda \nu}_{b} +  (\mathcal{D}^{\lambda})_{ab} G^{\nu \mu}_b +   2\, (\mathcal{D}^\mu)_{ab} G^{\nu \lambda}_b \bigr] = \frac{1}{2} G^a_{\mu \lambda} \bigl[ (\mathcal{D}^{\mu})_{ab} G_b^{\nu \lambda} + (\mu \leftrightarrow \lambda) \bigr]=0 \,.
\end{align}
We now proceed with the scalar sector.
For the singlet
field $\phi$, one gets
\begin{align}
    \partial_\mu T^\mn_{\phi, {\rm kin}}
    &= - \frac{1}{2} \partial^\nu (\partial_\lambda \phi
    \, \partial^\lambda \phi) + \partial_\mu (\partial^\mu \phi \,
    \partial^\nu \phi)= (\partial^\mu \partial_\mu \phi)\, \partial^\nu \phi 
    = (\partial_\phi V_0 - \Omega_\phi)\, \partial^\nu \phi\,,
\end{align}
where we have introduced the scalar field equation of motion [\Eq{scalar_fluid_gauge_minkowski3}]. For the $U(1)-$charged scalar field we have
\begin{align}
    \partial_\mu T^{\mu \nu}_{\varphi, {\rm kin}} & = \bigl[ (D_{\mu}^A \varphi)^*   \bigl( - \partial^\nu D^{\mu}_A \varphi  +  \partial^\mu D^{\nu}_A \varphi \bigr)  + (\partial_{\mu}D^{\mu}_A \varphi)^* (D^{\nu}_A \varphi) + c.c. \bigr] \nonumber \\
    & = \bigl\{ (D_{\mu}^A \varphi)^*   \bigl( - \partial^\nu D^{\mu}_A \varphi  +  \partial^\mu D^{\nu}_A \varphi \bigr) +  \bigl[ (D_\mu^A D^{\mu}_A \varphi)^* -i (D^\mu_A \varphi)^* (g Q_{\varphi} A_\mu) \bigr] D^{\nu}_A \varphi + c.c. \bigr\} \nonumber \\
    & = \bigl\{ (D_{\mu}^A \varphi)^*   \bigl( - \partial^\nu D^{\mu}_A \varphi  +  D^\mu_A D^{\nu}_A \varphi \bigr) +  (D_\mu^A D^{\mu}_A \varphi)^* D^{\nu}_A \varphi + c.c. \bigr\} \nonumber \\
    &= \bigl\{ (D_{\mu}^A \varphi)^*   \bigl( - D^\nu_A D^{\mu}_A \varphi  +  D^\mu_A D^{\nu}_A \varphi \bigr) +  (D_\mu^A D^{\mu}_A \varphi)^* D^{\nu}_A \varphi - i g Q_\varphi A_{\nu} (D_{\mu}^A \varphi)^* (D^{\mu}_A \varphi) + c.c. \bigr\} \nonumber \\
    &= \bigl\{ (D_{\mu}^A \varphi)^*   \bigl( - D^\nu_A D^{\mu}_A \varphi  +  D^\mu_A D^{\nu}_A \varphi \bigr) +  (D_\mu^A D^{\mu}_A \varphi)^* D^{\nu}_A \varphi  + c.c. \bigr\} \nonumber \\
    &= \bigl\{ (D_{\mu}^A \varphi)^*   \bigl[ D^{\mu}_A, D^{\nu}_A \bigr] \varphi +  (D_\mu^A D^{\mu}_A \varphi)^* D^{\nu}_A \varphi  + c.c. \bigr\} \nonumber \\
    &= \bigl\{ -ig Q_{\varphi} F^ {\mu \nu} (D_{\mu}^A \varphi)^* \varphi +  (D_\mu^A D^{\mu}_A \varphi)^* D^{\nu}_A \varphi  + c.c. \bigr\} \nonumber \\
    &=  -2 \,g \, Q_{\varphi} F^{\mu \nu} {\rm Im} [\varphi^* D_{\mu}^A \varphi] + \biggl[ \biggl( \frac{1}{2} \frac{\partial V_0}{\partial |\varphi|} \frac{\varphi^*}{|\varphi|} - \Omega_{\varphi}^{*} \biggr) D^{\nu}_A \varphi  + c.c. \biggr] \nonumber \\
    &= J^{\mu}_{\varphi} F^{\nu}_{\ \, \mu} +\biggl[ \biggl( \frac{1}{2} \frac{\partial V_0}{\partial |\varphi|} \frac{\varphi^*}{|\varphi|} - \Omega_{\varphi}^{*} \biggr) D^{\nu}_A \varphi  + c.c. \biggr] \,,
\end{align}
where we have used the fact that $-i g Q_{\varphi} A_{\nu} (D_{\mu}^A \varphi)^* (D^{\mu}_A \varphi)$ is real, that the commutator of the Abelian gauge covariant derivatives is $[D^{\mu}_A, D^{\nu}_A] = -i \, g \,  Q F^{\mu \nu}$, the scalar field EOM [\Eq{scalar_fluid_gauge_minkowski4}], and we recognize the current $J^{\mu}_{\varphi} = 2 \, g \, Q_{\varphi} {\rm Im} [\varphi^* D^{\mu}_A \varphi]$ introduced in \Sec{subsec:FlatSpace_Dynamics_Cont} [see \Eq{currents}].
Similarly, for the $[SU(N) \times U(1)]-$charged scalar field, we have
\begin{align}
    \partial_\mu T^{\mu \nu}_{\Phi, {\rm kin}}  & =
    \bigl[  (D_{\mu} \Phi)^{\dagger}   \bigl( -\partial^\nu D^{\mu} \Phi  +  \partial^\mu D^{\nu} \Phi \bigr)  + (\partial_{\mu}D^{\mu} \Phi)^{\dagger} D^{\nu} \Phi + c.c. \bigr] \nonumber \\
    & = \bigl\{ (D_{\mu} \Phi)^{\dagger}   \bigl( -\partial^\nu D^{\mu} \Phi  +  \partial^\mu D^{\nu} \Phi \bigr)+ \bigl[ (D_\mu D^{\mu}\Phi)^{\dagger} -i (D^\mu \Phi)^{\dagger} (g Q_{\Phi} A_\mu \mathcal{I} + g' Q'_{\Phi} C^a_{\mu} T_a) \bigr] D^{\nu} \Phi + c. c. \bigr\} \nonumber \\
    & = \bigl\{ (D_{\mu} \Phi)^{\dagger}   \bigl( -D^\nu D^{\mu} \Phi  +  D^\mu D^{\nu} \Phi \bigr)+ (D_\mu D^{\mu}\Phi)^{\dagger} D^{\nu} \Phi -i (D_\mu \Phi)^{\dagger} (g Q_{\Phi} A^\nu \mathcal{I} + g' Q'_{\Phi} C_a^{\nu} T_a) D^{\mu} \Phi + c. c. \bigr\} \nonumber \\
    & = \bigl\{ (D_{\mu} \Phi)^{\dagger}   \bigl[  D^{\mu}, D^{\nu} \bigr] \Phi + \bigl[ (D_\mu D^{\mu}\Phi)^{\dagger} D^{\nu} \Phi + c. c. \bigr\} \nonumber \\
    & = -2 \, g \, Q_{\Phi} F^{\mu \nu} {\rm Im} [\Phi^{\dagger} D_{\mu} \Phi] - 2 \, g' \, Q'_{\Phi} G^{\mu \nu}_a {\rm Im} [\Phi^{\dagger} T_a (D_{\mu} \Phi)] + \bigl[  (D_\mu D^{\mu}\Phi)^{\dagger} D^{\nu} \Phi + c. c. \bigr]
    \nonumber \\
    & = J^{\mu}_{\Phi} F^{\nu}_{\ \, \mu} + {\cal J}^{\mu}_{\Phi, a} G^{\nu}_{a \, \mu} + \biggl[ \biggl( \frac{1}{2} \frac{\partial V_0}{\partial |\Phi|} \frac{\Phi^{\dagger}}{|\Phi|} - \Omega_{\Phi}^{\dagger} \biggr) D^{\nu} \Phi + c.c. \biggr] \,,
\end{align}
where we have used the fact that the generators of $SU(N)$ are hermitian (i.e., $T_a^{\dagger} = T_a$), that the commutator of the $U(1)\times SU(N)$  gauge covariant derivatives is $[D^{\mu}, D^{\nu}] = - i \, g \, Q F^{\mu \nu} - i \, g' \, Q' \, G^{\mu \nu}_a T_a$, the scalar field EOM [cf.~\Eq{scalar_fluid_gauge_minkowski5}], and we recognize the currents $J^{\mu}_{\Phi} = 2 \, g \, Q_{\Phi} {\rm Im} [\Phi^{\dagger} D^{\mu} \Phi]$ and ${\cal J}^{\mu}_{\Phi, a} = 2 \, g' \, Q'_{\Phi} {\rm Im} [\Phi^{\dagger} T_a (D^{\mu} \Phi)]$ introduced in \Sec{subsec:FlatSpace_Dynamics_Cont} [see \Eq{currents}].
Finally, we also have the following term
\begin{align}
    \partial_\mu (V_0 \, \eta^{\mu \nu}) &=  (\partial_{\phi} V_0) \, \partial^\nu \phi + \biggl[\biggl(  \frac{\partial V_0}{\partial \varphi} \biggr) \partial^\nu \varphi + \biggl(  \frac{\partial V_0}{\partial \Phi} \biggr) \partial^\nu \Phi + c.c. \biggr] \nonumber \\
    & =  (\partial_{\phi} V_0) \, \partial^\nu \phi + \biggl[ \biggl( \frac{1}{2} \frac{\partial V_0}{\partial |\varphi|} \frac{\varphi^{*}}{|\varphi|}  \biggr) D^\nu_A \varphi + \biggl( \frac{1}{2}  \frac{\partial V_0}{\partial |\Phi|} \frac{\Phi^{\dagger}}{|\Phi|} \biggr) D^\nu \Phi  + c.c. \biggr] \,,
\end{align}
where, due to the addition of the charge conjugate for the terms inside square brackets, the derivatives acting on the charged scalar fields can be promoted to their gauge covariant counterparts.

The conservation of the total stress-energy tensor of \Eq{aux_eom}
then leads to
\begin{align}
    \partial_\mu T^{\mu \nu}_\fl
    = & \,  J^{\mu} F^{\nu}_{\ \, \mu} +  {\cal J}^{\mu}_a G^{\nu}_{a \, \mu}  - (\partial_\phi V_0 - \Omega_\phi) \partial^\nu \phi  - J^{\mu}_{\varphi} F^{\nu}_{\ \, \mu} - \biggl[  \biggl(\frac{1}{2} \frac{\partial V_0}{\partial |\varphi|} \frac{\varphi^*}{|\varphi|} - \Omega_{\varphi}^* \biggr) D^{\nu}_A \varphi + c.c. \biggr] \nonumber \\
    &-J^{\mu}_{\Phi} F^{\nu}_{\ \, \mu}  - {\cal J}^{\mu}_{\Phi, a} G^{\nu}_{a \, \mu}  - \biggl[ \biggl( \frac{1}{2} \frac{\partial V_0}{\partial |\Phi|} \frac{\Phi^{\dagger}}{|\Phi|} - \Omega_{\Phi}^{\dagger} \biggr) D^{\nu} \Phi + c.c. \biggr] \nonumber \\
    &+ (\partial_{\phi} V_0)  \partial^\nu \phi + \biggl[ \biggl( \frac{1}{2} \frac{\partial V_0}{\partial |\varphi|} \frac{\varphi^{*}}{|\varphi|} \biggr) D^\nu_A \varphi + \biggl( \frac{1}{2} \frac{\partial V_0}{\partial |\Phi|} \frac{\Phi^{\dagger}}{|\Phi|} \biggr) D^\nu \Phi  + c.c. \biggr] \nonumber \\ 
    = &\,  
    J^\mu_\fl F^{\nu}_{\ \, \mu}   +
    {\cal J}^{\mu}_{\fl, a} G^{\nu}_{a \, \mu} 
    +
    \Omega_\phi \, \partial^{\nu}\phi
    + \bigl(\Omega_\varphi^\ast \, D^\nu_A \varphi 
    + \Omega_\Phi^\dagger \, D^\nu \Phi  + c. c. \bigr)\,,
\end{align}
where we recognize the fluid currents $J^{\mu}_\fl = J^{\mu} - J^{\mu}_{\varphi} - J^{\mu}_{\Phi}$ and ${\cal J}^{\mu}_{\fl, a} = {\cal J}^{\mu}_a - {\cal J}^{\mu}_{\Phi,a}$ introduced in \Sec{subsec:FlatSpace_Dynamics_Cont} [see \Eq{gauge_fluid_couplings} and \Eqs{scalar_fluid_gauge_minkowski1}{scalar_fluid_gauge_minkowski2}]. Hence we recover \Eq{Tmunu_cons_aux}.

\section{Rescaling of the fluid equations of motion}
\label{rescaling_fluid_equations}

In this section, we discuss convenient rescaling choices for the fluid stress-energy tensor that
simplify the fluid EOM.
As presented in \Sec{subsec:FLRW_Dynamics_Cont}, precisely in \Eqs{eq:EnergyConservation}{eq:MomentumConservation}, the fluid equations in an expanding FLRW background with metric $g_{\mu \nu} = {\rm diag}(-a^{2\alpha}, a^2, a^2, a^2)$ are
\begin{equation}
\partial_\mu T^{\mu \nu}_\fl + \bigl[(4+2\alpha) {\cal H} T^{00}_\fl + a^{-2\alpha} {\cal H} T_\fl \bigr] \delta^{\nu}_0 + \bigl[(5+\alpha) {\cal H} T^{0i}_\fl \bigr] \delta^{\nu}_i  = {\cal F}_{\rm int}^{\nu}\,,
\label{gravitational_fluid_equations}
\end{equation}
where we have isolated the trace $T_\fl \equiv g_{\mu \nu} T^{\mu \nu}_\fl = -a^{2\alpha} T^{00}_\fl+T^{(3)}_\fl$ of the fluid stress-energy tensor, $\alpha$ indicates the choice of $\alpha$-time
(see discussion at the beginning of \Sec{subsec:FLRW_Dynamics_Cont}),
$\cal{H}$ is the ($\alpha$-time) Hubble rate, and ${\cal F}_{\rm int}^{\nu}$ represents the forces acting on the fluid, introduced at the end of \Sec{subsec:FlatSpace_Dynamics_Cont}.

Let us now consider a general rescaling of the fluid stress-energy tensor as
\begin{align}
    \conf{T}^{\mu \nu}_\fl = a^\beta \, T^{\mu \nu}_\fl\,, \quad \text{with constant } \beta \in {\mathbb R}\,.
\end{align}
The fluid EOM for the rescaled stress-energy tensor then read
\begin{align}
    \partial_\mu {\conf T}^{\mu \nu}_\fl + \bigl[(4+2\alpha - \beta) {\cal H}{\conf T}^{00}_\fl + a^{-2\alpha} {\cal H} {\conf T}_\fl \bigr] \delta^{\nu}_0 + \bigl[(5+\alpha - \beta) {\cal H} {\conf T}^{0i}_\fl \bigr] \delta^{\nu}_i = \conf{\cal F}_{\rm int}^{\nu}\,,
    \label{eq_rescaled_fluid_emt_cons}
\end{align}
where ${\conf T}_\fl \equiv g_{\mu \nu} {\conf T}_\fl^{\mu \nu}$ and we have rescaled the forces as $\conf{\cal F}_{\rm int}^{\nu} = a^\beta {\cal F}_{\rm int}^{\nu}$.
From the equations above,
we see that there are two rescaling choices that
allow us to get rid of one of the Hubble friction terms,
\begin{subequations}
\label{rescaling_aux}
\begin{align}
    \beta &= 4+2\alpha  \quad \Rightarrow \quad \partial_\mu {\conf T}_\fl^{\mu \nu} + a^{-2\alpha} {\cal H} {\conf T}_\fl \, \delta^{\nu}_{~\,0} + (1-\alpha) {\cal H} \conf{T}_\fl^{0i} \, \delta^{\nu}_{~\,i}= \conf{\cal F}_{\rm int}^\nu\,, \\
    \beta &= 5+\alpha \ \, \quad  \Rightarrow \quad \partial_\mu {\conf T}_\fl^{\mu \nu} + \bigl[ (\alpha-1) {\cal H} {\conf T}_\fl^{00} + a^{-2\alpha} {\cal H} \conf{T}_\fl \bigr] \delta^{\nu}_0 = \conf{\cal F}_{\rm int}^{\nu}\,.
\end{align}
\end{subequations}
We see that both rescalings are equivalent for conformal time, i.e.,
$\alpha=1$, in which case they lead to the following equations,
\begin{align}
    \partial_\mu {\conf T}_\fl^{\mu \nu} + a^{-2} {\cal H} \conf{T}_\fl \, \delta^{\nu}_0 = {\conf {\cal F}}_{\rm int}^{\nu}\,, \qquad \text{for} \ \alpha = 1\,,
\end{align}
where ${\conf T}_\fl^{\mu \nu} = a^{6} T^{\mu \nu}_\fl$ and
$\conf{\cal{F}}_{\rm int}^{\nu} = a^6 {\cal F}_{\rm int}^{\nu}$.
In \Eqq{rescaling_aux},
we see that the only Hubble friction term remaining appears in the
temporal component, i.e.,
for $\nu=0$, and it
is proportional to the trace of the stress-energy tensor. If we write $\conf{T}^{\mu \nu}_\fl = \conf{T}^{\mu \nu}_{\rm pf} -
\delta {\conf T}^{\mu \nu}$ with $\conf{T}^{\mu \nu}_{\rm pf}$ the stress-energy tensor of a perfect fluid and $\delta {\conf T}^{\mu \nu}$ the deviatoric
stress-energy tensor, the Hubble friction term becomes
\begin{align}
    {\cal H} a^{-2} {\conf T}_{\fl} =   {\cal H} \, a^4 (3p_\f - \rho_\f)
    - {\cal H} \, a^{-2} \delta {\conf T}\,, \qquad \text{for} \ \alpha = 1\,,
\end{align}
where we have used \Eq{perf_fluid}
for the perfect fluid component.
Since, for $\alpha=1$, the Hubble friction is proportional to the trace of the stress-energy
tensor, which vanishes for perfect fluids (i.e., $\delta T = 0$)
with a constant equation of state
corresponding to radiation particles,
i.e., $p_\f = \frac{1}{3} \rho_\f$,
the fluid EOM become \textit{conformally flat} in this case,
\begin{align}
    \partial_\mu {\conf T}^{\mu \nu}_{\rm pf}  = {\conf {\cal F}}_{\rm int}^{\nu}\,, \qquad \text{for} \ \alpha = 1\,.
\end{align}
This shows that both choices of rescaling $\beta=4+2\alpha$ and $\beta = 5+\alpha$ lead to conformally flat equations for a purely radiation perfect fluid when described with conformal time.
Moreover, as it can be seen from \Eq{eq_rescaled_fluid_emt_cons}, any rescaling choice $\beta(\alpha)$ such that $\beta(1) = 6$ leads to conformally flat equations for $\alpha=1$.

However, an advantage of the choice $\beta = 4+2 \alpha$, which is used in the main text, is that for a perfect fluid it yields
\begin{equation}
\label{rescalings_components_tmunu}
    \conf{T}^{00}_\pf = \conf{w}_\f \gamma^2 - \conf{p}_\f \,, \qquad 
    \conf{T}^{0i}_\pf = \conf{w}_\f \gamma^2 u_i \,, \qquad
    \conf{T}^{ij}_\pf = \conf{w}_\f \gamma^2 u_i u_j + a^{2(\alpha - 1)}
    \conf{p}_\f \delta_{ij}\,,
\end{equation}
with
$\conf{\rho}_\f = a^4 p_\f$,
$\conf{p}_\f = a^4 p_\f$, and $\conf{w}_\f = a^4 w_\f$
respectively the rescaled fluid energy density, pressure, and enthalpy. Hence we have a rescaling of the fluid variables with a factor $a^4$ compensating for the decay of the physical energy density of a purely radiation fluid due to the Universe expansion. This means that, within this choice, the rescaled fluid variables always coincide, for a purely radiation fluid, with the comoving variables. 

\newpage

\begin{multicols}{2}
\footnotesize
  \bibliography{auto,manual}
\bibliographystyle{h-physrev4}
\end{multicols}

\end{document}